\documentstyle[aps,prb,epsf,rotate,bezier,floats,twocolumn]{revtex}
\newif\ifPostScript
\PostScripttrue
\ifx\epsffile\undefined\PostScriptfalse\fi
\ifPostScript
   \def\insertplot#1{\put(0,0){\epsffile{#1.ps}}}
\else
   \def\insertplot#1{\put(0,80){\special{em:graph #1.pcx}}}
\fi

\catcode`@=11  

\def\chaptername{Chapter}
\def\@chapapp{\chaptername}

\ifx\c@chapter\undefined \newcounter{chapter} \fi
\@addtoreset{section}{chapter}
\@addtoreset{equation}{chapter}
\@addtoreset{figure}{chapter}
\@addtoreset{table}{chapter}

\def\thechapter       {\arabic{chapter}}
\def\thesection       {\thechapter.\arabic{section}}

\def\thefigure{\thechapter.\@arabic\c@figure}
\def\thetable{\thechapter.\@arabic\c@table}
\def\theequation{\thechapter.\arabic{equation}}
\def\mathletters{%
\inc@eqnnum  \setcounter{eqletter}{0}%
\edef\@currentlabel{\theequation}%
\def\theequation{\thechapter.\arabic{equation}\alph{eqletter}}%
\def\inc@eqnnum{\addtocounter{eqletter}{1}}%
\def\dec@eqnnum{\addtocounter{eqletter}{-1}}%
}

\def\@makechapterhead#1{%
  \vspace*{50\p@}%
  {\parindent \z@\raggedright
    \ifnum \c@secnumdepth >\m@ne
     \huge\bf \@chapapp{} \thechapter
    \par
    \vskip 20\p@ \fi
    \Huge \bf
    #1\par
    \nobreak
    \vskip 40\p@
  }}

\def\@makeschapterhead#1{%
  \vspace*{50\p@}%
  {\parindent \z@ \raggedright
    \Huge \bf
    #1\par
    \nobreak
    \vskip 40\p@
  }}

\def\chapter{\cleardoublepage
   \thispagestyle{plain}%
   \global\@topnum\z@

   \@afterindentfalse
   \secdef\@chapter\@schapter}

\def\@chapter[#1]#2{\ifnum \c@secnumdepth >\m@ne
        \refstepcounter{chapter}%
        \typeout{\@chapapp\space\thechapter.}%
        \addcontentsline{toc}{chapter}{\protect
        \numberline{\thechapter}#1}\else
      \addcontentsline{toc}{chapter}{#1}\fi
   \chaptermark{#1}%
   \addtocontents{lof}%
      {\protect\addvspace{10\p@}}
   \addtocontents{lot}%
      {\protect\addvspace{10\p@}}
   \if@twocolumn
           \@topnewpage[\@makechapterhead{#2}]%
     \else \@makechapterhead{#2}%
           \@afterheading
     \fi}

\def\@schapter#1{\if@twocolumn \@topnewpage[\@makeschapterhead{#1}]%
        \else \@makeschapterhead{#1}%
              \@afterheading\fi}

\def\ps@headings{\let\@mkboth\markboth
 \def\@oddfoot{}\def\@evenfoot{}
 \def\@evenhead{\rm \thepage\hfil \sl \leftmark}
 \def\@oddhead{{\sl \rightmark}\hfil \rm\thepage}
 \def\chaptermark##1{\markboth {\uppercase{\ifnum \c@secnumdepth >\m@ne
      \@chapapp\ \thechapter. \ \fi ##1}}{}}%
 \def\sectionmark##1{\markright {\uppercase{\ifnum \c@secnumdepth >\z@
   \thesection. \ \fi ##1}}}}

\def\tableofcontents{\@restonecolfalse
  \if@twocolumn\@restonecoltrue\onecolumn\fi
  \chapter*{\contentsname
        \@mkboth{\uppercase{\contentsname}}{\uppercase{\contentsname}}}%
  \@starttoc{toc}\if@restonecol\twocolumn\fi}
\def\l@chapter#1#2{\addpenalty{-\@highpenalty}%
   \vskip 1.0em plus\p@
   \@tempdima 1.5em
   \begingroup
     \parindent \z@ \rightskip \@pnumwidth
     \parfillskip -\@pnumwidth
     \bf
     \leavevmode
      \advance\leftskip\@tempdima
      \hskip -\leftskip
     #1\nobreak\hfil \nobreak\hbox to\@pnumwidth{\hss #2}\par
     \penalty\@highpenalty
   \endgroup}
\def\l@section{\@dottedtocline{1}{1.5em}{2.3em}}
\def\l@subsection{\@dottedtocline{2}{3.8em}{3.2em}}
\def\l@subsubsection{\@dottedtocline{3}{7.0em}{4.1em}}
\def\l@paragraph{\@dottedtocline{4}{10em}{5em}}
\def\l@subparagraph{\@dottedtocline{5}{12em}{6em}}

\catcode`@=12 %

\begin{document}
\renewcommand{\textfraction}{0.1}

\onecolumn\widetext\pagestyle{empty}
\setcounter{page}{0}
\mbox{}
\vfil
\begin{center}%
  {\LARGE \bf
      VERTICAL TRANSPORT
          IN
      SUPERLATTICES
  \par}%
  \vskip 3em
  {\large
    \lineskip .75em
    \begin{tabular}[t]{c}
       \em Thesis submitted for a degree of \\
       Doctor of Philosophy\\
       \em   by \\
       \rm Daniel L. Miller
    \end{tabular}\par}%
   \vskip 1.5em
  {\large  \par}%
\end{center}\par
\vfil
Submitted to the Senate of The Hebrew University of Jerusalem in 1996.

\newpage
\pagestyle{empty}
\setcounter{page}{0}
This work was done under the supervision of Professor Boris Laikhtman.

\newpage
\pagestyle{empty}

To my family.
\twocolumn

\pagestyle{headings}\narrowtext

\chapter*{Preface}

\noindent
\unitlength=1mm
\begin{picture}(00.00,00.00)
\put(90.00,50){\parbox[t]{\columnwidth}{
\begin{flushleft}
\em
Sicut aquae tremulum labris ubi lumen ahenis  \\
Sole repercussum, aut radiantis imagine Lunae \\
Omnia pervolitat late loca jamque sub auras   \\
Erigitur, summique ferit laquearia tecti.
\end{flushleft}
[As when ruffled water in a bronze pot reflects
the light of the sun and the shinning face of the moon,
sending shimmers flying high into the air and striking
against the paneled ceilings.
\mbox{Virgil, {\em{}Aeneid}, VIII, 22. } ]\cite{montaigne-essays}
}}\end{picture}

We are not busy with some particular physical problem. We are
``ranging to and fro over the wastelands of'' the modern physics similar to
Virgil's sunlight reflections.  This thesis consists from a number of parts. 
They are completed separate works motivated by recent experiments. The words
``superlattice'' and ``current'' are common in these works and widely used.
Two chapters, 1 and 3, were written {\em especially} for this thesis. They
are actually unpublished results and discussions. The kind atmosphere of
the Racah Institute of Physics helps us a lot.  I'm grateful to this
place. I would like to acknowledge my and Boris Laikhtman's discussions with
Holger~Grahn, S.~Luryi, Y.~Lyanda-Geller, Yehuda Naveh, M.~Raikh,
Leonid~Shvartzman, and V.~Zevin. I thank F.~Bass for permission to use
Part II, ``Quantum superlattices'', of his book, prior to publication.

All the principal results of this work concern the vertical transport
in generic three dimensional superlattices. In the 1st chapter we make
an historical introduction, then we discuss the geometry of the problem, and
the physical parameters associated with structure of electron minibands and
strength of external fields.  We found the effect of collisionless transverse
magnetoresistance, and we discuss it in the 2nd chapter. This effect is
similar to collisionless Landau damping in a plasma  and we utilize the same
name. In the 3rd chapter we provide quantum mechanical reasons for the above
effect; we show how a magnetic field  bends narrow superlattice minibands,
and we classify the states into Landau-type and Stark-type. In the 4th chapter
we compute longitudinal magnetoresistance of superlattices due to the
imperfections of the interfaces. Correlation length of the interface
roughness can be measured independently by this method.  In the 5th chapter
we discuss the current-voltage characteristic of superlattice when an  
electric field destroys the one-miniband transport. We found that the 
structure of the high-field domains in the superlattices is complicated, but 
can be described analytically with great accuracy. This structure reveals 
itself in the details of the current-voltage characteristics. All our results 
are consistent with existing experiments, and we make careful comparison of 
theoretical predictions and experimental results in all chapters.

\tableofcontents

\chapter{ Introduction }
\label{chap:intr}

\section{ Generalities }
\label{sec:subj}

We consider theoretically semiconductor superlattices. A superlattice is a
one-dimensional periodic potential in monocrystalline semiconductor. The
superlattice potential would be obtained by a periodic variation of alloy
composition introduced during epitaxial growth. It would consist of
alternating ultrathin layers of two semiconductors that closely match in
lattice constant.

We put forth a more specific definition: A superlattice is a stack of
coupled quantum wells. Each quantum well limits electron motion in one
direction and leads to quantization of energy; that is the elementary
example in quantum mechanics textbooks. Coupling of quantum wells means
that the electron can tunnel through the barrier separating adjacent quantum
wells.

Our definition is more accurate because variation of alloy composition leads
to variation of all band structure parameters. They cannot be described by
a simple periodic potential. They modify electron wave functions in potential
wells and tunneling amplitude through potential barriers. Our definition is
particularly useful when variation of superlattice periodic potential is much
larger than mean kinetic energy of electrons. In this case electrons are
really sitting in the wells of the superlattice potential.

The evolution of molecular beam epitaxy has allowed access to superlattices.
High-quality devices started to be available to experimentalists in the early
seventies. For example, effects of energy quantization and tunneling were
observed in 1974, see review of Weisbuch.\cite{weisbuch-87}

Original motivations of superlattice research were to obtain devices with
negative differential conductance\cite{esaki-tsu}, and amplification of
radiation\cite{kazarinov71,kazarinov72}. These effects were predicted for
a superlattice placed in strong electric field perpendicular to the layers
(vertical transport). Unfortunately, instability of electric potential
profile in vertical (growth) direction and formation of high- and low-field
domains made realization of these ideas impossible.\cite{esaki-chang}

Structure and dynamics of the high-field domains in superlattices is a
separate branch of research. Attempts to build a microwave
laser\cite{capasso1} and a microwave radiation detector\cite{levine-sep87}
motivated the investigation of electric field domains. The first project was
not realized for many years due to the relatively small life time of excited
carriers\cite{helm}.  Progress in making a quantum cascade laser was reported
only recently.\cite{capasso-aug96} Structure of the high-field domains, i.e.
potential profile in a vertical direction, remains to be a subject of
intensive research. This potential profile, for example, carries a lot of
information concerning the quality of the device.\cite{grahn-apr96}

There are some similarities between instabilities of the potential profile in
superlattices and in Gunn diodes.\cite{leperson3} Therefore, one hopes to
see high-field domains running through the superlattice. This was one of
the possible explanations of the time dependent oscillations of the potential
profile which had been reported.\cite{merlin-apr95,grahn-aug95,grahn-sep95}
However, numeric simulations show that the strengths of the high- and low-
field domains change periodically in time with the domain boundary being
pinned within a few quantum wells.\cite{Bonilla-jan97,Grahn-jan97}

Modern technology uses superlattices for construction of optical devices.
The physics of such devices is tightly related to properties of excitons in
quantum wells. We will not touch the world of optics in the present
work. Besides the device application, growth  of superlattices is still a
challenge for labrotaries employing molecular beam epitaxy.

\begin{figure}
\unitlength=1.00mm
\linethickness{0.4pt}
\begin{picture}(90.00,138.00)(3.00,0.00)
\put(10.00,130.00){\vector(1,0){8.00}}
\put(10.00,130.00){\vector(0,1){8.00}}
\put(10.00,130.00){\vector(-1,-1){5.00}}
\put(11.00,137.00){\makebox(0,0)[lt]{$y$}}
\put(17.00,131.00){\makebox(0,0)[rb]{$z$}}
\put(7.00,126.00){\makebox(0,0)[lt]{$x$}}
\put(0.00,70.00){\line(1,0){50.00}}
\put(50.00,70.00){\line(3,2){30.00}}
\put(80.00,90.00){\line(0,1){10.00}}
\put(80.00,100.00){\line(-3,-2){30.00}}
\put(50.00,80.00){\line(-1,0){50.00}}
\put(0.00,80.00){\line(0,-1){10.00}}
\put(50.00,80.00){\rule{0.00\unitlength}{-10.00\unitlength}}
\put(50.00,80.00){\line(0,-1){10.00}}
\put(46.00,84.00){\line(-1,0){30.00}}
\put(16.00,84.00){\line(0,1){6.00}}
\put(16.00,90.00){\line(1,0){30.00}}
\put(46.00,90.00){\line(0,-1){6.00}}
\put(46.00,84.00){\line(3,2){18.00}}
\put(64.00,96.00){\line(0,1){6.00}}
\put(64.00,102.00){\line(-3,-2){18.00}}
\put(16.00,92.00){\line(1,0){30.00}}
\put(46.00,92.00){\line(3,2){18.00}}
\put(64.00,106.00){\line(-3,-2){18.00}}
\put(46.00,94.00){\line(-1,0){30.00}}
\put(64.00,100.00){\line(1,0){16.00}}
\put(0.00,80.00){\line(5,3){16.00}}
\put(16.00,96.00){\line(1,0){30.00}}
\put(46.00,96.00){\line(3,2){18.00}}
\put(16.00,98.00){\line(1,0){30.00}}
\put(46.00,98.00){\line(3,2){18.00}}
\put(64.00,112.00){\line(-3,-2){18.00}}
\put(46.00,100.00){\line(-1,0){30.00}}
\put(16.00,102.00){\line(1,0){30.00}}
\put(46.00,102.00){\line(3,2){18.00}}
\put(16.00,101.00){\line(1,0){30.00}}
\put(46.00,101.00){\line(3,2){18.00}}
\put(16.00,99.00){\line(1,0){30.00}}
\put(46.00,99.00){\line(3,2){18.00}}
\put(64.00,109.00){\line(-3,-2){18.00}}
\put(46.00,97.00){\line(-1,0){30.00}}
\put(16.00,95.00){\line(1,0){30.00}}
\put(46.00,95.00){\line(3,2){18.00}}
\put(64.00,105.00){\line(-3,-2){18.00}}
\put(46.00,93.00){\line(-1,0){30.00}}
\put(16.00,91.00){\line(1,0){30.00}}
\put(46.00,91.00){\line(3,2){18.00}}
\put(16.00,103.00){\line(0,1){6.00}}
\put(16.00,109.00){\line(1,0){30.00}}
\put(46.00,109.00){\line(0,-1){6.00}}
\put(46.00,103.00){\line(-1,0){30.00}}
\put(46.00,103.00){\line(3,2){18.00}}
\put(64.00,115.00){\line(0,1){6.00}}
\put(64.00,121.00){\line(-3,-2){18.00}}
\put(64.00,121.00){\line(-1,0){30.00}}
\put(34.00,121.00){\line(-3,-2){18.00}}
\put(40.00,115.00){\line(0,1){15.00}}
\put(72.00,98.00){\line(0,1){12.00}}
\put(72.00,120.00){\line(0,1){18.00}}
\put(72.00,138.00){\line(-4,-1){32.00}}
\put(72.00,110.00){\circle{2.00}}
\put(72.00,120.00){\circle{2.00}}
\put(73.00,115.00){\makebox(0,0)[lc]{$V$}}
\put(0.00,115.00){\vector(1,0){15.00}}
\put(7.00,117.00){\makebox(0,0)[cb]{$B$}}
\put(88.00,138.00){\makebox(0,0)[rt]{(a)}}
\put(88.00,60.00){\makebox(0,0)[rc]{(b)}}
\put(15.00,10.00){\vector(0,1){48.00}}
\put(30.00,20.00){\vector(-3,-2){30.00}}
\put(25.00,35.00){\line(1,0){5.00}}
\put(30.00,3.00){\line(1,0){5.00}}
\put(35.00,35.00){\line(1,0){5.00}}
\put(40.00,3.00){\line(1,0){5.00}}
\put(45.00,35.00){\line(1,0){5.00}}
\put(50.00,3.00){\line(1,0){5.00}}
\put(55.00,35.00){\line(1,0){5.00}}
\put(60.00,3.00){\line(1,0){5.00}}
\put(65.00,35.00){\line(1,0){5.00}}
\put(30.00,8.00){\rule{0.00\unitlength}{1.00\unitlength}}
\put(30.00,9.00){\rule{5.00\unitlength}{1.00\unitlength}}
\put(30.00,30.00){\rule{5.00\unitlength}{1.00\unitlength}}
\put(40.00,9.00){\rule{5.00\unitlength}{1.00\unitlength}}
\put(40.00,30.00){\rule{5.00\unitlength}{1.00\unitlength}}
\put(50.00,9.00){\rule{5.00\unitlength}{1.00\unitlength}}
\put(50.00,30.00){\rule{5.00\unitlength}{1.00\unitlength}}
\put(60.00,9.00){\rule{5.00\unitlength}{1.00\unitlength}}
\put(60.00,30.00){\rule{5.00\unitlength}{1.00\unitlength}}
\put(15.00,14.00){\dashbox{3.00}(65.00,1.00)[cc]{}}
\put(25.00,35.00){\line(3,2){6.00}}
\put(30.00,35.00){\line(3,2){6.00}}
\put(35.00,35.00){\line(3,2){6.00}}
\put(40.00,35.00){\line(3,2){6.00}}
\put(45.00,35.00){\line(3,2){6.00}}
\put(50.00,35.00){\line(3,2){6.00}}
\put(55.00,35.00){\line(3,2){6.00}}
\put(60.00,35.00){\line(3,2){6.00}}
\put(65.00,35.00){\line(3,2){6.00}}
\put(70.00,35.00){\line(3,2){6.00}}
\put(30.00,50.00){\vector(-3,-2){12.00}}
\put(22.00,47.00){\makebox(0,0)[rb]{$B$}}
\put(75.00,35.00){\vector(1,0){15.00}}
\put(83.00,37.00){\makebox(0,0)[cb]{$F$}}
\put(75.00,20.00){\vector(1,0){15.00}}
\put(82.00,21.00){\makebox(0,0)[cb]{$j$}}
\put(85.00,8.00){\makebox(0,0)[ct]{$y$}}
\put(1.00,3.00){\makebox(0,0)[rb]{$z$}}
\put(15.00,10.00){\vector(1,0){75.00}}
\put(25.00,3.00){\line(0,1){32.00}}
\put(30.00,35.00){\line(0,-1){32.00}}
\put(35.00,3.00){\line(0,1){32.00}}
\put(40.00,35.00){\line(0,-1){32.00}}
\put(45.00,3.00){\line(0,1){32.00}}
\put(50.00,35.00){\line(0,-1){32.00}}
\put(55.00,3.00){\line(0,1){32.00}}
\put(60.00,35.00){\line(0,-1){32.00}}
\put(65.00,3.00){\line(0,1){32.00}}
\put(70.00,35.00){\line(0,-1){32.00}}
\bezier{60}(15.00,10.00)(22.00,10.00)(25.00,3.00)
\bezier{60}(70.00,3.00)(73.00,10.00)(80.00,10.00)
\put(15.00,31.00){\line(1,0){3.00}}
\put(13.00,31.00){\makebox(0,0)[rc]{$E_g$}}
\put(13.00,15.00){\makebox(0,0)[rc]{$E_F$}}
\put(13.00,55.00){\makebox(0,0)[rc]{$E$}}
\put(58.00,37.00){\line(0,1){8.00}}
\put(48.00,37.00){\line(0,1){8.00}}
\put(43.00,43.00){\line(1,0){20.00}}
\put(43.00,43.00){\vector(1,0){5.00}}
\put(63.00,43.00){\vector(-1,0){5.00}}
\put(53.00,44.00){\makebox(0,0)[cb]{$l$}}
\end{picture}
   \caption{(a) -- Measurement of the vertical current in a superlattice.
   Current goes in the $y$-direction, transverse magnetic field is applied in
   the $z$-direction. (b) -- One miniband transport through a superlattice.
   The Fermi energy (dashed lines), $E_F$, lies slightly above the lowest
   level in the quantum well. Energy of the lowest level is zero. Energy of
   the second level is $E_g\gg E_F$.
} \label{fig:geometry} \end{figure}
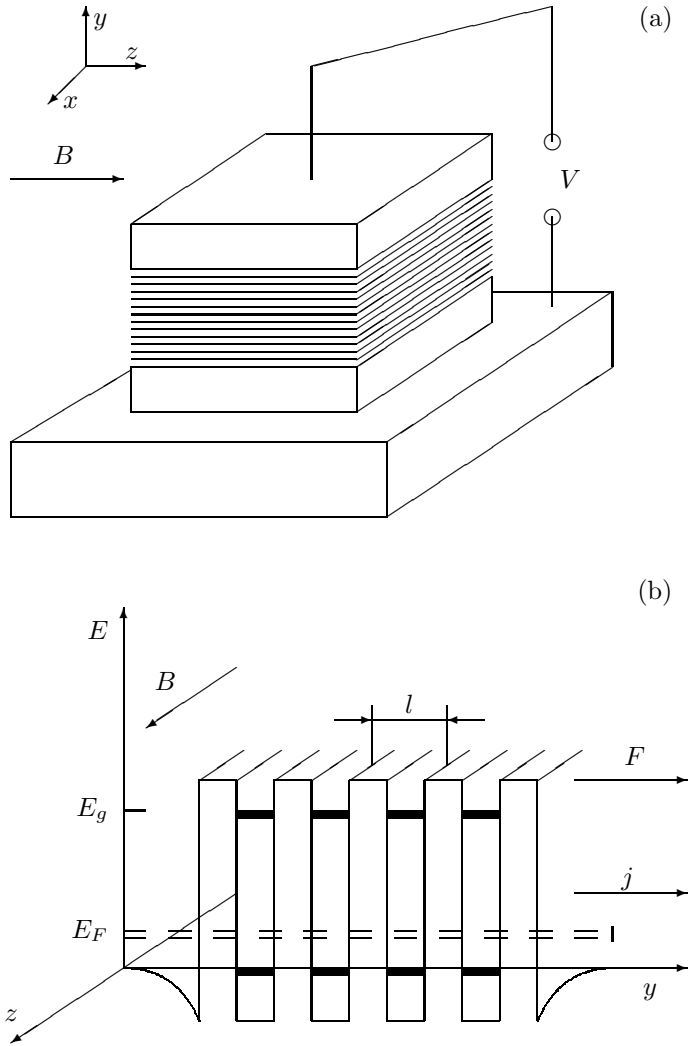

Typical superlattice geometry is shown in Fig.~\ref{fig:geometry}(a). The
horizontal sheets schematically show a superlattice sandwiched between
materials providing electrical contacts. The superlattice itself is formed by
alternating layers of wide and narrow band gap materials. The real
potential profile of such a device is complicated, and depends on the
materials and doping level.  The part of the potential profile related to the
superlattice is shown schematically  in  Fig.~\ref{fig:geometry}(b).
The potential energy of electrons changes periodically in the $y$ direction
and remains constant in the $x$ and $z$ directions.

This potential defines the sequence of the quantum wells separated by the
barriers. Potential wells limit electron motion in the $y$ direction and lead
to energy quantization. We have chosen the energy of lowest electron state to
be the zero of energy. We  show the case of two electron states in a well,
they are marked by black rectangles in Fig.~\ref{fig:geometry}(b). The energy
gap between these two states is $E_g$, and it is one of three main
parameters characterizing the superlattice potential.

The superlattice structure brings to semiconductor physics a new length scale,
which is the period of the superlattice, $l$, and a new energy scale, which is
tunneling energy, $\Lambda$. These are two other parameters mentioned above,
and their typical values in 1974 were 500\AA{} and 1meV correspondingly.
Modern technology allows growth of superlattices having very fine structure.
For example, the barriers of only  three monatomic layers width were reported
in Ref.\onlinecite{sibille7}. The typical superlattice potential period in
1996 is about 50\AA{} and tunneling energy may reach 40meV.

The resonance between energy levels in adjacent quantum wells leads to
formation of minibands, and one can associate the tunneling energy with
a quarter miniband width. This energy can be measured by photocurrent
spectroscopy\cite{fujiwara94} and results are in agreement with prediction of
Kronig-Penney model.\cite{sibille-inbook95} However, calculation of minibands
of holes is a much more complicated problem, because wave functions of holes
are four component spinors.\cite{bastard-91,smith-mailhiot} We will not
consider holes in this work.

The energy gap $E_g$ is assumed to be very large in the theory of the
one-miniband transport, much larger than the Fermi energy, temperature,
potential drop of electric field per period, energy uncertainty due to
scattering and miniband width. In Fig.~\ref{fig:geometry}(b) we showed the
Fermi energy between two minibands, i.e., the second miniband is empty. A
high electric field destroys these minibands and leads to formation of low-
and high-field domains.  Inside the high-field domain $E_g$
is of the order of the electric potential drop per period. For this reason,
the physics of high-field domains stands separately from the physics of
one-miniband transport.

The tunneling energy is usually so small, that the coherence of tunneling
through subsequent barriers can be easily destroyed by external fields,
sample imperfections or impurities. If it is known that sequential tunneling
is incoherent, one can introduce the conductivity of the barrier or  the
transition time of an electron through the barrier.\cite{laikhtman1}  We will
use this approach in Chapters~\ref{chap:long} and \ref{chap:domains}. If it
is known that the tunneling is coherent, one can write down the usual kinetic
equation for an electron gas with anisotropic dispersion. This dispersion
contains all information about quantum mechanical effects(tunneling), and the
kinetic equation is already a classical object. We will use it in
Chapter~\ref{chap:trans}. Very often, it is not known whether tunneling is
coherent or not, and it is possible to scan all intermediate situations by
changing external fields. For this reason we will consider the quantum
kinetic equation derived by Keldysh's technique. It is the only tool which
can describe the destruction of a miniband by a high electric field. We
will derive it in Chapter~\ref{chap:deriv}.

\begin{figure}
\unitlength=1mm
\begin{picture}(88.00,88.00)(2.0,0.0)
   \put(7.0,7.0){\insertplot{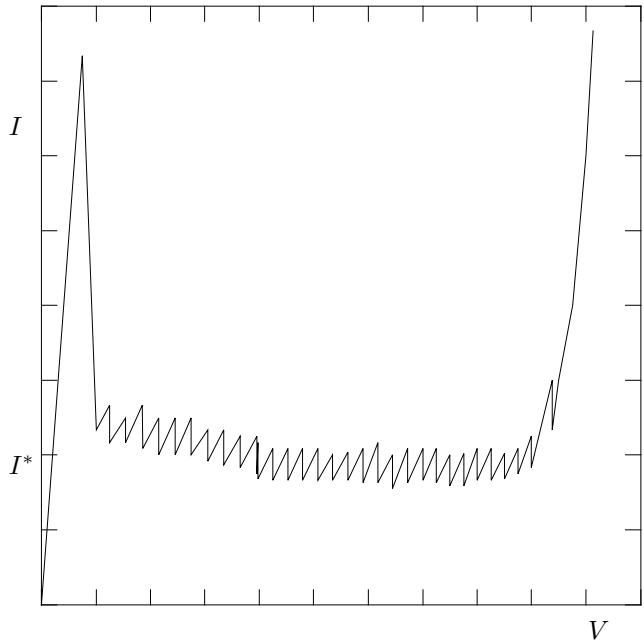}}
   \put(80.00,3.00){$V$}
   \put(3.00,70.00){$I$}
   \put(3.00,25.00){$I^\ast$}
\end{picture}
   \caption{
    Measured\protect{\cite{kawamura1}} $I$-$V$ characteristics of the
    superlattice.  The first discontinuity is the instability point of the
    one-miniband transport and corresponds to the formation of the high-field
    domain.  Further increasing of the bias leads to the expansion of domains
    and small current oscillations near the value $I^\ast$. This is the first
    plato, see details in Chap.~\protect{\ref{chap:domains}}.
   }
\label{fig:iv}
\end{figure}

\section{ Motivations and the program }
\label{sec:prog}

This project is actually a collection of a few works concerning
vertical transport in superlattices which were motivated by recent
experiments.  Motivation for each of these individual works and related
experiments are considered in their introductions, but we need to put all of
them in the proper context here.

A lot of experiments and theoretical problems concerning one-miniband
transport were discussed in the literature in the years 1970-1985,
however interpretation of the recent experiments with high-quality
devices is not always possible in the framework of the old theories. Most of
the recent experimental results are collected in the book of
Grahn\cite{grahn-book95}.  The modern theories of superlattice physics are
considered in this book only briefly.

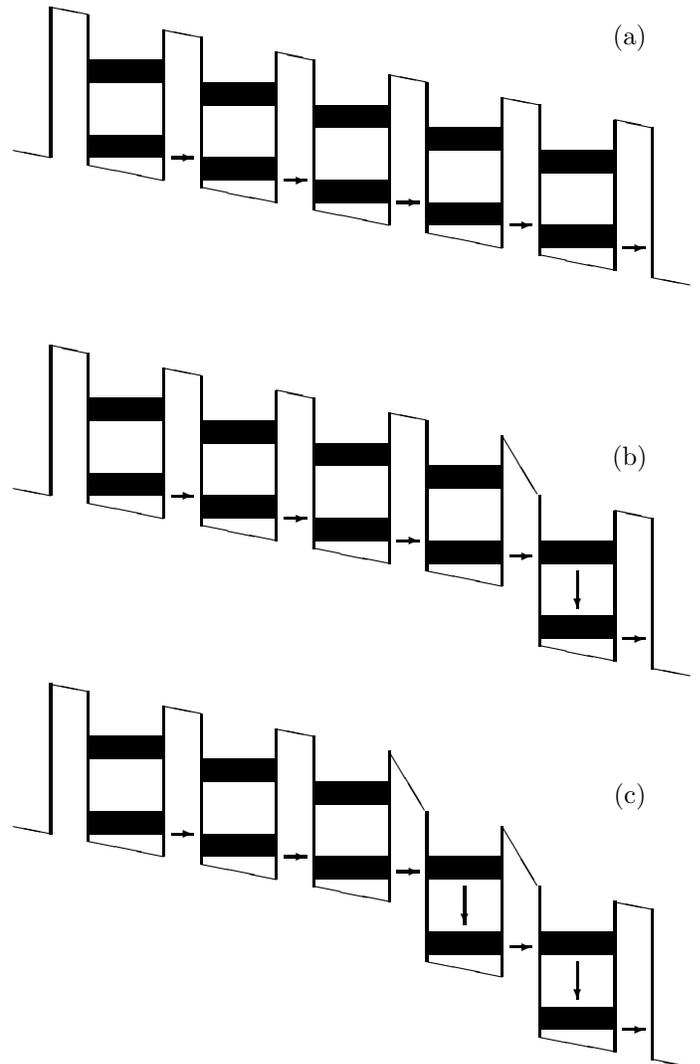
\begin{figure}
\unitlength=1.00mm
\linethickness{0.4pt}
\begin{picture}(90.00,142.00)(2.0,0.0)
\linethickness{0.8pt}
\put(5.00,142.00){\line(0,-1){20.00}}
\put(5.00,142.00){\line(5,-1){5.00}}
\put(10.00,121.00){\line(0,1){20.00}}
\put(10.00,121.00){\line(5,-1){10.00}}
\put(20.00,139.00){\line(0,-1){20.00}}
\put(20.00,139.00){\line(5,-1){5.00}}
\put(25.00,118.00){\line(0,1){20.00}}
\put(25.00,118.00){\line(5,-1){10.00}}
\put(35.00,136.00){\line(0,-1){20.00}}
\put(35.00,136.00){\line(5,-1){5.00}}
\put(40.00,115.00){\line(0,1){20.00}}
\put(40.00,115.00){\line(5,-1){10.00}}
\put(50.00,133.00){\line(0,-1){20.00}}
\put(50.00,133.00){\line(5,-1){5.00}}
\put(55.00,112.00){\line(0,1){20.00}}
\put(55.00,112.00){\line(5,-1){10.00}}
\put(65.00,130.00){\line(0,-1){20.00}}
\put(65.00,130.00){\line(5,-1){5.00}}
\put(70.00,109.00){\line(0,1){20.00}}
\put(70.00,109.00){\line(5,-1){10.00}}
\put(80.00,127.00){\line(0,-1){20.00}}
\put(80.00,127.00){\line(5,-1){5.00}}
\put(85.00,106.00){\line(0,1){20.00}}
\put(10.00,122.00){\rule{10.00\unitlength}{3.00\unitlength}}
\put(25.00,119.00){\rule{10.00\unitlength}{3.00\unitlength}}
\put(10.00,132.00){\rule{10.00\unitlength}{3.00\unitlength}}
\put(40.00,116.00){\rule{10.00\unitlength}{3.00\unitlength}}
\put(55.00,113.00){\rule{10.00\unitlength}{3.00\unitlength}}
\put(70.00,110.00){\rule{10.00\unitlength}{3.00\unitlength}}
\put(25.00,129.00){\rule{10.00\unitlength}{3.00\unitlength}}
\put(40.00,126.00){\rule{10.00\unitlength}{3.00\unitlength}}
\put(55.00,123.00){\rule{10.00\unitlength}{3.00\unitlength}}
\put(70.00,120.00){\rule{10.00\unitlength}{3.00\unitlength}}
\put(5.00,97.00){\line(0,-1){20.00}}
\put(5.00,97.00){\line(5,-1){5.00}}
\put(10.00,76.00){\line(0,1){20.00}}
\put(10.00,76.00){\line(5,-1){10.00}}
\put(20.00,94.00){\line(0,-1){20.00}}
\put(20.00,94.00){\line(5,-1){5.00}}
\put(25.00,73.00){\line(0,1){20.00}}
\put(25.00,73.00){\line(5,-1){10.00}}
\put(35.00,91.00){\line(0,-1){20.00}}
\put(35.00,91.00){\line(5,-1){5.00}}
\put(40.00,70.00){\line(0,1){20.00}}
\put(40.00,70.00){\line(5,-1){10.00}}
\put(50.00,88.00){\line(0,-1){20.00}}
\put(50.00,88.00){\line(5,-1){5.00}}
\put(55.00,67.00){\line(0,1){20.00}}
\put(55.00,67.00){\line(5,-1){10.00}}
\put(65.00,85.00){\line(0,-1){20.00}}
\put(70.00,57.00){\line(0,1){20.00}}
\put(70.00,57.00){\line(5,-1){10.00}}
\put(80.00,75.00){\line(0,-1){20.00}}
\put(80.00,75.00){\line(5,-1){5.00}}
\put(85.00,54.00){\line(0,1){20.00}}
\put(10.00,77.00){\rule{10.00\unitlength}{3.00\unitlength}}
\put(25.00,74.00){\rule{10.00\unitlength}{3.00\unitlength}}
\put(10.00,87.00){\rule{10.00\unitlength}{3.00\unitlength}}
\put(40.00,71.00){\rule{10.00\unitlength}{3.00\unitlength}}
\put(55.00,68.00){\rule{10.00\unitlength}{3.00\unitlength}}
\put(70.00,58.00){\rule{10.00\unitlength}{3.00\unitlength}}
\put(25.00,84.00){\rule{10.00\unitlength}{3.00\unitlength}}
\put(40.00,81.00){\rule{10.00\unitlength}{3.00\unitlength}}
\put(55.00,78.00){\rule{10.00\unitlength}{3.00\unitlength}}
\put(70.00,68.00){\rule{10.00\unitlength}{3.00\unitlength}}
\put(65.00,85.00){\line(3,-5){4.67}}
\put(75.00,67.00){\vector(0,-1){5.00}}
\put(66.00,69.00){\vector(1,0){3.00}}
\put(5.00,52.00){\line(0,-1){20.00}}
\put(5.00,52.00){\line(5,-1){5.00}}
\put(10.00,31.00){\line(0,1){20.00}}
\put(10.00,31.00){\line(5,-1){10.00}}
\put(20.00,49.00){\line(0,-1){20.00}}
\put(20.00,49.00){\line(5,-1){5.00}}
\put(25.00,28.00){\line(0,1){20.00}}
\put(25.00,28.00){\line(5,-1){10.00}}
\put(35.00,46.00){\line(0,-1){20.00}}
\put(35.00,46.00){\line(5,-1){5.00}}
\put(40.00,25.00){\line(0,1){20.00}}
\put(40.00,25.00){\line(5,-1){10.00}}
\put(50.00,43.00){\line(0,-1){20.00}}
\put(55.00,15.00){\line(0,1){20.00}}
\put(55.00,15.00){\line(5,-1){10.00}}
\put(65.00,33.00){\line(0,-1){20.00}}
\put(70.00,5.00){\line(0,1){20.00}}
\put(70.00,5.00){\line(5,-1){10.00}}
\put(80.00,23.00){\line(0,-1){20.00}}
\put(80.00,23.00){\line(5,-1){5.00}}
\put(85.00,2.00){\line(0,1){20.00}}
\put(10.00,32.00){\rule{10.00\unitlength}{3.00\unitlength}}
\put(25.00,29.00){\rule{10.00\unitlength}{3.00\unitlength}}
\put(10.00,42.00){\rule{10.00\unitlength}{3.00\unitlength}}
\put(40.00,26.00){\rule{10.00\unitlength}{3.00\unitlength}}
\put(55.00,16.00){\rule{10.00\unitlength}{3.00\unitlength}}
\put(70.00,6.00){\rule{10.00\unitlength}{3.00\unitlength}}
\put(25.00,39.00){\rule{10.00\unitlength}{3.00\unitlength}}
\put(40.00,36.00){\rule{10.00\unitlength}{3.00\unitlength}}
\put(55.00,26.00){\rule{10.00\unitlength}{3.00\unitlength}}
\put(70.00,16.00){\rule{10.00\unitlength}{3.00\unitlength}}
\put(50.00,43.00){\line(3,-5){4.67}}
\put(65.00,33.00){\line(3,-5){4.67}}
\put(60.00,25.00){\vector(0,-1){5.00}}
\put(75.00,15.00){\vector(0,-1){5.00}}
\put(51.00,27.00){\vector(1,0){3.00}}
\put(66.00,17.00){\vector(1,0){3.00}}
\put(51.00,71.00){\vector(1,0){3.00}}
\put(81.00,58.00){\vector(1,0){3.00}}
\put(81.00,6.00){\vector(1,0){3.00}}
\put(36.00,29.00){\vector(1,0){3.00}}
\put(21.00,32.00){\vector(1,0){3.00}}
\put(21.00,77.00){\vector(1,0){3.00}}
\put(36.00,74.00){\vector(1,0){3.00}}
\put(21.00,122.00){\vector(1,0){3.00}}
\put(36.00,119.00){\vector(1,0){3.00}}
\put(51.00,116.00){\vector(1,0){3.00}}
\put(66.00,113.00){\vector(1,0){3.00}}
\put(81.00,110.00){\vector(1,0){3.00}}
\put(82.00,138.00){\makebox(0,0)[cc]{(a)}}
\put(82.00,82.00){\makebox(0,0)[cc]{(b)}}
\put(82.00,37.00){\makebox(0,0)[cc]{(c)}}
\put(0.00,123.00){\line(5,-1){5.00}}
\put(0.00,78.00){\line(5,-1){5.00}}
\put(0.00,33.00){\line(5,-1){5.00}}
\put(85.00,106.00){\line(5,-1){5.00}}
\put(85.00,54.00){\line(5,-1){5.00}}
\put(85.00,2.00){\line(5,-1){5.00}}
\end{picture}
   \caption{
      Superlattice potential is shown schematically for several values of
      applied bias: before the first instability point (a), at the formation
      of the high-field domain (b), corresponding to expansion of the domain
      (c).
   }
\label{fig:steps}\end{figure}

The simplest transport measurement is detection of the $I$-$V$ curve. The
typically observed $I$-$V$ curve is very nonlinear,  and it exhibits a
sequence of discontinuities, see Fig.~\ref{fig:iv}.  The first discontinuity
corresponds to the destruction of single miniband. The nature of this
instability point was considered for the first time by
Laikhtman\cite{laikhtman1} and Laikhtman and Miller\cite{laikhtman2}. The
latter work contains the derivation of the quantum kinetic equation which is
valid near the first instability point. The instability development leads to
high-field domain formation. At least two states in each quantum well become
involved in transport, see Fig.~\ref{fig:steps}.

The structure of the high-field domain was calculated for the first time by
Miller and Laikhtman\cite{miller-dec94}. We reproduce this work in
Chap.~\ref{chap:domains}. Since this publication, the structure of the
domains has been intensively considered in the literature. For example,
additional effects, like imperfections and fluctuation of barrier size, were
considered recently by Schwarz et al\cite{grahn-apr96}.
Chapter~\ref{chap:domains} considers the parameters of the typical $I$-$V$
curve, Fig.~\ref{fig:iv}, in great detail.  Recent
measurements\cite{merlin-jul94,kastrup94,grahn-book95} are also in
agreement with this theory. The physical problems considered in
Chapter~\ref{chap:domains} are
\begin{itemize}
\item limiting of the current by diffusion,
\item minimal domain size,
\item nonresonant tunneling of electrons inside the domain,
\item time resolved process of the domain formation near the instability
      threshold,
\item mechanism of the domain expansion,
\item possibility of domain formation near the cathode and therefore
      injection of electrons into the second miniband and two-miniband
      transport.
\end{itemize}

It is very well known that transverse magnetic field shifts
discontinuities of the superlattice $I$-$V$ curve to higher bias.
Concerning the first instability point we can mention the recent experiment
of Aristone et al\cite{aristone-nov95} and much more data can be found in the
book of Grahn\cite{grahn-book95}. It is believed that this discontinuity
corresponds to the maximum of the $I$-$V$ curve of the uniform superlattice.
The position of the current peak is usually calculated by numeric
solution of the kinetic equation in one miniband.\cite{grahn-book95}
Analytical results were obtained first by
Epshtein\cite{Epshtein-79,bass-book86}. They show a quadratic shift with
magnetic field of the $I$-$V$ curve peak. However, when magnetic field
becomes stronger, the peak shifts linearly with magnetic field. The
explanation was given by Miller and Laikhtman\cite{miller-oct95} in terms of
resonant group of electrons. We reproduce this work in
Chapter~\ref{chap:trans}.  The new results of Chapter~\ref{chap:trans} are
\begin{itemize}
\item resonant group of electrons which give the main contribution to the
      current,
\item ``collisionless'' current -- analog of collisionless Landau damping
      of plasma waves.
\item difference between transverse magnetoresistance and magnetoconductance.
\end{itemize}

The use of the Boltzmann kinetic equation in Chapter~\ref{chap:trans} has to
be justified. In certain cases one has to use the
more general quantum kinetic equation as it was done by Levinson and
Yasevichute\cite{levinson-may72}, Suris and Shchamkhalova\cite{suris-jul84}
and Laikhtman and Miller\cite{laikhtman2}. We will follow the last work in
order to show that
\begin{itemize}
\item electron states in the narrow miniband in the presence of the parabolic
      potential created by a magnetic field are of two types Landau-like and
      Stark-like,
\item the quantum kinetic equation supports qualitatively results of the
      semiclassical approach of Chapter~\ref{chap:trans},
\item heating of the electron gas near the current peak can be important
      for wide interval of magnetic fields.
\end{itemize}

The solution to the kinetic equation in the miniband is possible only if we
are in the regime of the miniband transport. Four conditions of the miniband
transport are given in the next section. It may happen, that some of these
conditions are violated and we have to solve the quantum kinetic equation. In
this case we have to justify also the effective Hamiltonian method and that
is done in Sec.~\ref{sec:ham}.

\begin{figure}
\unitlength=1.00mm
\linethickness{0.4pt}
\begin{picture}(88.00,115.00)
\put(0.00,65.00){\vector(1,0){60.00}}
\put(0.00,65.00){\vector(0,1){50.00}}
\put(2.00,115.00){\makebox(0,0)[lc]{$E$}}
\put(60.00,67.00){\makebox(0,0)[cb]{$y$}}
\put(0.00,70.00){\rule{56.00\unitlength}{1.00\unitlength}}
\put(0.00,85.00){\rule{56.00\unitlength}{3.00\unitlength}}
\put(35.00,100.00){\vector(1,0){15.00}}
\put(10.00,65.00){\line(0,1){30.00}}
\put(10.00,95.00){\line(1,0){5.00}}
\put(15.00,95.00){\line(0,-1){30.00}}
\put(25.00,65.00){\line(0,1){30.00}}
\put(25.00,95.00){\line(1,0){5.00}}
\put(30.00,95.00){\line(0,-1){30.00}}
\put(40.00,65.00){\line(0,1){30.00}}
\put(40.00,95.00){\line(1,0){5.00}}
\put(45.00,95.00){\line(0,-1){30.00}}
\put(55.00,65.00){\line(0,1){30.00}}
\put(55.00,95.00){\line(1,0){3.00}}
\put(45.00,105.00){\vector(1,0){15.00}}
\put(48.00,107.00){\makebox(0,0)[cb]{$\vec F$}}
\put(38.00,102.00){\makebox(0,0)[cb]{$\vec H$}}
\put(0.00,0.00){\line(0,1){50.00}}
\put(10.00,0.00){\line(0,1){10.00}}
\put(10.00,10.00){\line(-1,0){1.00}}
\put(9.00,10.00){\line(0,1){20.00}}
\put(9.00,30.00){\line(1,0){1.00}}
\put(10.00,30.00){\line(0,1){20.00}}
\put(15.00,50.00){\line(0,-1){12.00}}
\put(15.00,38.00){\line(-1,0){1.00}}
\put(14.00,38.00){\line(0,-1){18.00}}
\put(14.00,20.00){\line(1,0){1.00}}
\put(15.00,20.00){\line(0,-1){20.00}}
\put(26.00,50.00){\line(0,-1){36.00}}
\put(26.00,14.00){\line(-1,0){1.00}}
\put(25.00,14.00){\line(0,-1){14.00}}
\put(30.00,50.00){\line(0,-1){10.00}}
\put(30.00,40.00){\line(1,0){1.00}}
\put(31.00,40.00){\line(0,-1){15.00}}
\put(31.00,25.00){\line(-1,0){1.00}}
\put(30.00,25.00){\line(0,-1){25.00}}
\put(40.00,50.00){\line(0,-1){50.00}}
\put(45.00,50.00){\line(0,-1){34.00}}
\put(45.00,16.00){\line(-1,0){1.00}}
\put(44.00,16.00){\line(0,-1){16.00}}
\put(55.00,50.00){\line(0,-1){15.00}}
\put(55.00,35.00){\line(-1,0){1.00}}
\put(54.00,35.00){\line(0,-1){22.00}}
\put(54.00,13.00){\line(1,0){1.00}}
\put(55.00,13.00){\line(0,-1){13.00}}
\put(60.00,50.00){\line(0,-1){30.00}}
\put(60.00,20.00){\line(1,0){1.00}}
\put(61.00,20.00){\line(0,-1){20.00}}
\put(10.00,35.00){\line(1,1){5.00}}
\put(15.00,45.00){\line(-1,-1){5.00}}
\put(10.00,45.00){\line(1,1){5.00}}
\put(9.00,29.00){\line(1,1){5.00}}
\put(14.00,29.00){\line(-1,-1){5.00}}
\put(9.00,19.00){\line(1,1){5.00}}
\put(9.00,14.00){\line(1,1){6.00}}
\put(15.00,15.00){\line(-1,-1){5.00}}
\put(15.00,10.00){\line(-1,-1){5.00}}
\put(15.00,5.00){\line(-1,-1){5.00}}
\put(26.00,46.00){\line(1,1){4.00}}
\put(30.00,45.00){\line(-1,-1){4.00}}
\put(30.00,40.00){\line(-1,-1){4.00}}
\put(26.00,31.00){\line(1,1){5.00}}
\put(26.00,26.00){\line(1,1){5.00}}
\put(26.00,21.00){\line(1,1){5.00}}
\put(26.00,16.00){\line(1,1){4.00}}
\put(30.00,15.00){\line(-1,-1){5.00}}
\put(25.00,5.00){\line(1,1){5.00}}
\put(30.00,5.00){\line(-1,-1){5.00}}
\put(44.00,50.00){\line(-1,-1){4.00}}
\put(40.00,41.00){\line(1,1){5.00}}
\put(40.00,36.00){\line(1,1){5.00}}
\put(40.00,31.00){\line(1,1){5.00}}
\put(40.00,26.00){\line(1,1){5.00}}
\put(40.00,21.00){\line(1,1){5.00}}
\put(40.00,16.00){\line(1,1){5.00}}
\put(40.00,11.00){\line(1,1){4.00}}
\put(40.00,6.00){\line(1,1){4.00}}
\put(40.00,1.00){\line(1,1){4.00}}
\put(60.00,50.00){\line(-1,-1){5.00}}
\put(55.00,40.00){\line(1,1){5.00}}
\put(60.00,40.00){\line(-1,-1){6.00}}
\put(54.00,29.00){\line(1,1){6.00}}
\put(60.00,30.00){\line(-1,-1){6.00}}
\put(54.00,19.00){\line(1,1){6.00}}
\put(60.00,20.00){\line(-1,-1){6.00}}
\put(55.00,10.00){\line(1,1){6.00}}
\put(61.00,11.00){\line(-1,-1){6.00}}
\put(55.00,0.00){\line(1,1){6.00}}
\put(7.00,35.00){\vector(1,0){11.00}}
\put(22.00,46.00){\vector(1,0){11.00}}
\put(22.00,20.00){\vector(1,0){11.00}}
\put(37.00,7.00){\vector(1,0){11.00}}
\put(52.00,44.00){\vector(1,0){11.00}}
\put(50.00,20.00){\vector(0,1){15.00}}
\put(35.00,25.00){\vector(0,-1){10.00}}
\put(21.00,33.00){\line(0,-1){10.00}}
\put(21.00,33.00){\vector(0,-1){10.00}}
\put(21.00,35.00){\vector(0,1){10.00}}
\put(88.00,80.00){\makebox(0,0)[rc]{(a)}}
\put(88.00,25.00){\makebox(0,0)[rc]{(b)}}
\end{picture}
   \caption{ Vertical current through superlattice in presence of
   longitudinal magnetic field.(a) Current goes around thick areas of the
   barriers and becomes sensitive to the magnetic field.(b)
}
\label{fig:fenomenol}
\end{figure}

The quality of the superlattices is determined basically by the quality of
the interfaces. Surface roughness destroys the coherence of electron
tunneling, and measurement of roughness parameters is very important
technologically. It was suggested recently, that surface roughness can be
responsible for longitudinal magnetoresistance of
superlattice.\cite{lee-aug94} If the superlattice layers are isotropic, there
should not be classical longitudinal magnetoresistance, (see the geometry of
the experiment in Fig.~\ref{fig:fenomenol}(a)). Magnetoresistance was
explained by the roughness induced in-plane currents, see
Fig.~\ref{fig:fenomenol}(b), which are sensitive to the magnetic field
perpendicular to the layers. Theory of this effect was developed by Miller
and Laikhtman\cite{miller-oct96}. We reproduce this work in
Chapter~\ref{chap:long}. The main result is that longitudinal
magnetoresistance of superlattices is very sensitive to the correlation
length of the roughness and allows calculation of this length.

\section{ Four conditions of miniband transport }
\label{sec:oneband}

\begin{figure}
\unitlength=1mm
\linethickness{0.4pt}
\begin{picture}(86.00,60.00)
\put(5.00,0.00){\vector(0,1){60.00}}
\put(3.00,56.00){\makebox(0,0)[rc]{$E$}}
\put(5.00,30.00){\line(1,0){2.00}}
\put(3.00,30.00){\makebox(0,0)[rc]{$0$}}
\put(15.00,29.50){\rule{10.00\unitlength}{1.00\unitlength}}
\put(35.00,28.00){\line(1,0){10.00}}
\put(35.00,32.00){\line(1,0){10.00}}
\put(35.00,35.00){\line(1,0){10.00}}
\put(35.00,25.00){\line(1,0){10.00}}
\put(48.00,38.00){\line(1,0){5.00}}
\put(48.00,22.00){\line(1,0){5.00}}
\put(50.00,30.00){\vector(0,1){8.00}}
\put(50.00,30.00){\vector(0,-1){8.00}}
\put(51.00,30.00){\makebox(0,0)[lc]{$4\Lambda$}}
\put(65.00,30.00){\line(1,0){10.00}}
\put(65.00,40.00){\line(1,0){10.00}}
\put(65.00,50.00){\line(1,0){10.00}}
\put(65.00,20.00){\line(1,0){10.00}}
\put(70.00,20.00){\vector(0,1){10.00}}
\put(70.00,30.00){\vector(0,-1){10.00}}
\put(72.00,25.00){\makebox(0,0)[lc]{$eFl$}}
\put(20.00,10.00){\makebox(0,0)[cc]{$\Lambda=0$}}
\put(40.00,5.00){\makebox(0,0)[cc]{$\Lambda\ne0$}}
\put(70.00,5.00){\makebox(0,0)[cc]{$F\ne0$}}
\put(20.00,3.00){\makebox(0,0)[cc]{$F=0$}}
\put(20.00,57.00){\makebox(0,0)[cc]{(a)}}
\put(40.00,57.00){\makebox(0,0)[cc]{(b)}}
\put(70.00,57.00){\makebox(0,0)[cc]{(c)}}
\end{picture}
\caption{ Structure of energy levels in superlattice. (a) - Zero coupling and
zero electric field; (b) - Coupling of quantum wells lifts degeneracy; (c) -
strong electric field breaks miniband to Stark ladder.}
\label{fig:stark.eff}
\end{figure}
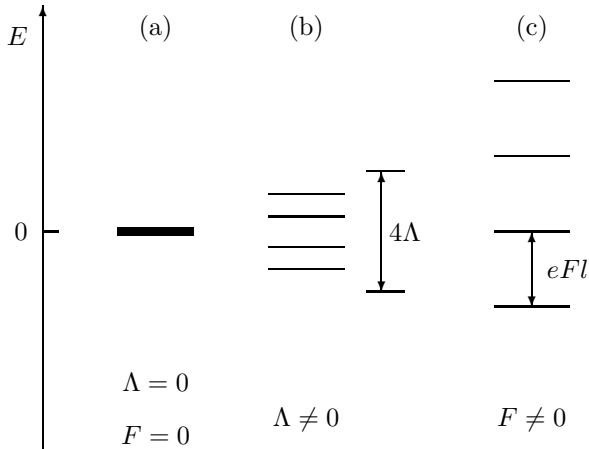

In this section we introduce the main parameters characterizing electron
states and electron transport in a superlattice. An example of  a superlattice
having $N=4$ periods is shown in Fig.~\ref{fig:geometry}. Bold horizontal
lines mark the positions of the electron levels in the quantum wells as if
these wells are isolated. Then, the lowest states in all four wells have the
same energy $E=0$, and we mark it by single bold line in
Fig.~\ref{fig:stark.eff}(a).

Coupling of quantum wells lifts this degeneracy. The tunneling energy,
$\Lambda$, depends strongly on the height of the potential barrier separating
adjacent wells and on the effective mass of an electron under the barrier.
Coupling of quantum wells spreads energies of electron states in the interval
of width $4\Lambda$, see Fig.~\ref{fig:stark.eff}(b). In the limit of the
infinitely long superlattice $N\rightarrow\infty$, this interval is covered
densely by the levels.  The energy interval $[-2\Lambda,2\Lambda]$ is called
a {\em miniband}.  Therefore, by definition, a miniband exists in an
infinitely long idealized superlattice, without impurities, interface
roughness, phonons and external fields.

An electric field, $F$, applied in the vertical direction, see
Fig.~\ref{fig:geometry}, causes a relative potential shift of quantum wells.
The drop of the electron potential energy per period is $eFl = eV/N$, where
$e$ is the electron charge, $l$ is the superlattice period, $N$ is the number
of periods, and $V$ is the applied voltage. If the coupling energy is zero,
then this potential drop destroys the degeneracy of electron states and makes
them equally spaced with the interval $eFl$, see Fig.~\ref{fig:stark.eff}(c).
This structure of the electron levels is called a {\em Stark ladder}.

However, if both energies $eFl$ and $\Lambda$ are not zero, one can
observe the transition from the level structure of Fig.~\ref{fig:stark.eff}(b)
to the level structure of Fig.~\ref{fig:stark.eff}(c).  This transition is
called the Stark effect. It has been intensively studied during the whole
superlattice history.\cite{agullo-rueda1,kawashima2,Cohen-94} Stark states
have a finite size in the vertical direction $\Delta y \sim \Lambda/(eF)$.

A Stark ladder appears when the electric field satisfies the condition
\begin{equation}
   eFl \gtrsim {\Lambda \over N } \;,
\label{eq:oneband.1}
\end{equation}
i.e. when the mean spacing  of the Stark ladder exceeds the mean spacing of
the miniband. We see, therefore, that in the infinitely long superlattice the
infinitesimal field destroys the miniband. Nevertheless, the picture of
miniband transport is justified under the conditions which will be discussed
in the rest of this section.

\begin{figure}
\unitlength=1.00mm
\linethickness{0.4pt}
\begin{picture}(80.00,65.00)
\put(0.00,5.00){\vector(1,0){80.00}}
\put(0.00,5.00){\vector(0,1){60.00}}
\put(3.00,65.00){\makebox(0,0)[lt]{$\displaystyle{(E_F,T)\tau_p\over\hbar}$}}
\put(78.00,9.00){\makebox(0,0)[cb]{$\displaystyle{\Lambda\tau_p\over\hbar}$}}
\put(40.00,65.00){\line(0,-1){30.00}}
\put(40.00,35.00){\line(1,0){40.00}}
\put(20.00,50.00){\makebox(0,0)[cc]{\shortstack{sequential \\ tunneling}}}
\put(60.00,50.00){\makebox(0,0)[cc]{\shortstack{miniband \\ transport}}}
\put(50.00,20.00){\makebox(0,0)[cc]{\shortstack{hopping \\ conductivity}}}
\end{picture}
   \caption{ Mechanisms, which govern vertical transport, are shown
   in the diagram. }
\label{fig:diagrams}
\end{figure}
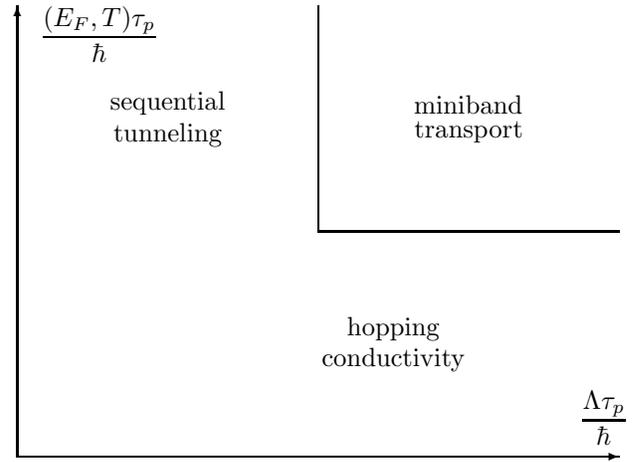

The scattering of the electrons by charged impurities and by phonons can be
characterized by the elastic $\tau_p$ and the inelastic $\tau_\varepsilon$
relaxation times. The scattering destroys plane waves of the free electrons
and also Bloch waves of the electrons in the superlattice. However, the
plane waves approximate the electron wave function well if the mean
kinetic energy of the electron gas is large enough
\begin{equation}
     E_F,T \gg \hbar/\tau_p, \hbar/\tau_\varepsilon\;.
\label{eq:oneband.2}
\end{equation}
This is the first condition of miniband transport. If it is violated one
can talk only about hopping between spatially localized states, which leads
to {\em hopping conductivity}.

The condition Eq.~(\ref{eq:oneband.2}) compares the kinetic energy of the
electrons with the energy uncertainty due to scattering. However, it may
happen that the miniband width is smaller than this energy uncertainty. In
this case the subsequent tunneling events lose their coherence and there are
no more plane waves propagating in the vertical direction. Therefore, we
obtain the second condition of miniband transport
\begin{equation}
     \Lambda \gg \hbar/\tau_p, \hbar/\tau_\varepsilon\;.
\label{eq:oneband.3}
\end{equation}
If this condition is violated then electron propagation in the vertical
direction is called {\em sequential tunneling}.

The two inequalities Eqs.~(\ref{eq:oneband.2}) and (\ref{eq:oneband.3}) define
the phase diagram Fig.~\ref{fig:diagrams}. In this diagram we show the three
main regimes of superlattice transport. In the regime of miniband
transport one can safely make use of the classical kinetic equation. In the
regime of sequential tunneling there are no plane waves propagating
in the vertical direction and therefore one should compute transport
coefficients from the quantum kinetic equation. Various percolation type
models can be used for investigation of transport in the hopping regime.

The strength of the electric field is also limited in the regime of
miniband transport. It is clear that the Stark effect can be observed if
$eFl$ is much larger than $\hbar/\tau_p, \hbar/\tau_\varepsilon$, and that is
very well known from the optical measurements\cite{agullo-rueda1}. In the
opposite case we can consider the electric field as a force driving
electrons and that is precisely what we want.  Therefore, we obtain the third
condition of miniband transport
\begin{equation}
      \lesssim {\hbar\over\tau_p} ,{\hbar\over\tau_\varepsilon}\;.
\label{eq:oneband.4}
\end{equation}
Violation of this condition leads to the so-called {\em field-induced
localization} of the electrons in the Stark states and to the transport of
the { \em sequential hopping} type. This means the hopping between the
subsequent Stark states.\cite{tsu-dohler} This regime cannot be observed
experimentally because of the potential profile instability.

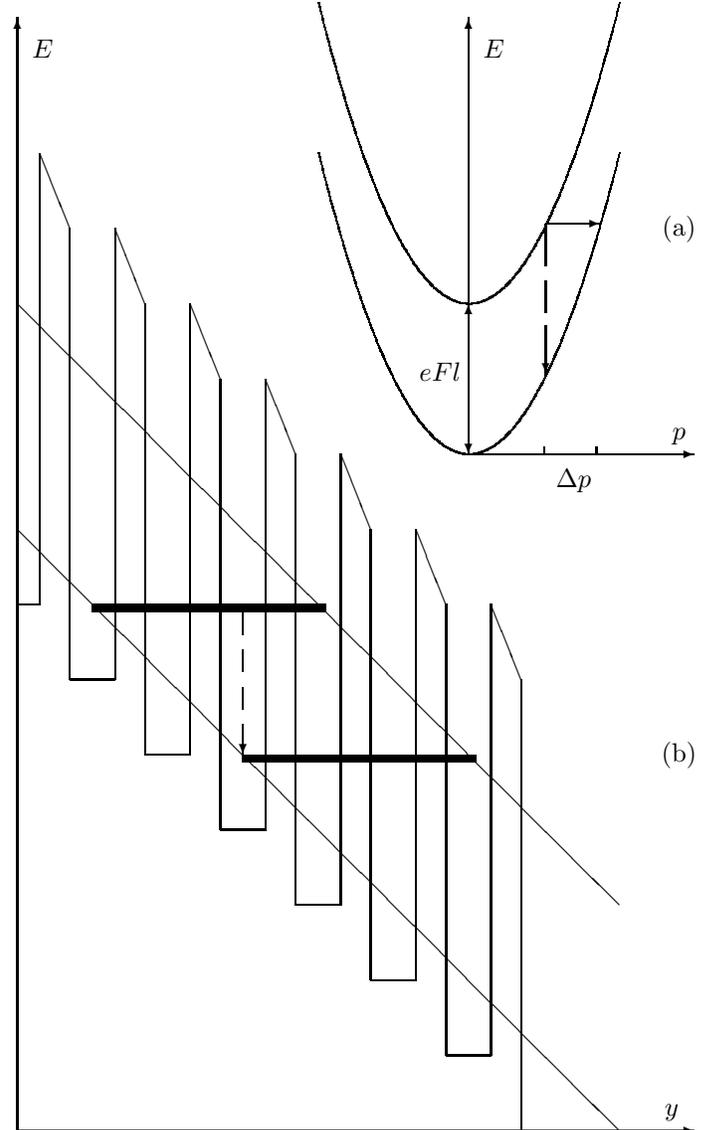
\begin{figure}
\unitlength=1.00mm
\linethickness{0.4pt}
\begin{picture}(88.00,150.00)
\put(0.00,80.00){\line(1,-1){80.00}}
\put(0.00,110.00){\line(1,-1){80.00}}
\put(0.00,70.00){\line(1,0){3.00}}
\put(3.00,70.00){\line(0,1){60.00}}
\put(3.00,130.00){\line(2,-5){4.00}}
\put(7.00,120.00){\line(0,-1){60.00}}
\put(7.00,60.00){\line(1,0){6.00}}
\put(13.00,60.00){\line(0,1){60.00}}
\put(13.00,120.00){\line(2,-5){4.00}}
\put(17.00,110.00){\line(0,-1){60.00}}
\put(17.00,50.00){\line(1,0){6.00}}
\put(23.00,50.00){\line(0,1){60.00}}
\put(23.00,110.00){\line(2,-5){4.00}}
\put(27.00,100.00){\line(0,-1){60.00}}
\put(27.00,40.00){\line(1,0){6.00}}
\put(33.00,40.00){\line(0,1){60.00}}
\put(33.00,100.00){\line(2,-5){4.00}}
\put(37.00,90.00){\line(0,-1){60.00}}
\put(37.00,30.00){\line(1,0){6.00}}
\put(43.00,30.00){\line(0,1){60.00}}
\put(43.00,90.00){\line(2,-5){4.00}}
\put(47.00,80.00){\line(0,-1){60.00}}
\put(47.00,20.00){\line(1,0){6.00}}
\put(53.00,20.00){\line(0,1){60.00}}
\put(53.00,80.00){\line(2,-5){4.00}}
\put(57.00,70.00){\line(0,-1){60.00}}
\put(57.00,10.00){\line(1,0){6.00}}
\put(63.00,10.00){\line(0,1){60.00}}
\put(63.00,70.00){\line(2,-5){4.00}}
\put(67.00,60.00){\line(0,-1){60.00}}
\put(67.00,0.00){\line(1,0){3.00}}
\put(10.00,69.00){\rule{31.00\unitlength}{1.00\unitlength}}
\put(30.00,49.00){\rule{31.00\unitlength}{1.00\unitlength}}
\put(30.00,69.00){\line(0,-1){3.00}}
\put(30.00,64.00){\line(0,-1){3.00}}
\put(30.00,59.00){\line(0,-1){3.00}}
\put(30.00,54.00){\vector(0,-1){4.00}}
\put(0.00,0.00){\vector(1,0){90.00}}
\put(0.00,0.00){\vector(0,1){148.00}}
\put(2.00,144.00){\makebox(0,0)[lc]{$E$}}
\put(87.00,2.00){\makebox(0,0)[cb]{$y$}}
\put(60.00,90.00){\vector(1,0){30.00}}
\bezier{660}(40.00,130.00)(60.00,50.00)(80.00,130.00)
\bezier{660}(40.00,150.00)(60.00,70.00)(80.00,150.00)
\put(70.31,120.61){\vector(1,0){7.18}}
\put(88.00,92.00){\makebox(0,0)[cb]{$p$}}
\put(88.00,120.00){\makebox(0,0)[cc]{(a)}}
\put(88.00,50.00){\makebox(0,0)[cc]{(b)}}
\put(70.28,120.60){\line(0,-1){5.56}}
\put(70.28,112.97){\line(0,-1){4.13}}
\put(70.28,107.05){\vector(0,-1){6.42}}
\put(60.00,90.00){\vector(0,1){58.00}}
\put(62.00,144.00){\makebox(0,0)[lc]{$E$}}
\put(70.00,90.00){\line(0,1){1.00}}
\put(77.00,90.00){\line(0,1){1.00}}
\put(74.00,88.00){\makebox(0,0)[ct]{$\Delta p$}}
\put(60.00,107.00){\vector(0,1){3.00}}
\put(60.00,93.00){\vector(0,-1){3.00}}
\put(59.00,101.00){\makebox(0,0)[rc]{$eFl$}}
\end{picture}
   \caption{Electrons reflects many times from the miniband boundaries before
   being scattered. The electron orbits are shown by thick lines (b).
   The Elastic scattering is accompanied by the large momentum transfer
   $\Delta p$ (a), and therefore the inelastic process may be important.
   The inelastic transition is shown by the dashed lines, see (a) and (b). }
\label{fig:inelast}
\end{figure}

The field induced localization is an interesting physical phenomenon. It has,
of course, something to do with a Stark transition, but we would like to
discuss it also in different terms. The main purpose of the following
discussion is to obtain the condition of miniband transport in the
presence of a transverse magnetic field.

Let us assume that all conditions of the miniband transport
Eqs.~(\ref{eq:oneband.2}), (\ref{eq:oneband.3}), and (\ref{eq:oneband.4})
are fulfilled. In this case one can introduce the semiclassical picture of
the miniband inclined by an electric field. The energy of the electron
having zero in-plane momenta $p_x=p_z=0$ lies in the interval
\begin{equation}
       eFy - 2\Lambda \le E \le eFy + 2\Lambda\;.
\label{eq:oneband.4b}
\end{equation}
This equation is semiclassically correct, because it contains both the
coordinate $y$ of the electron and the kinetic energy of the vertical motion
of the electron.

We show an inclined miniband in Fig.~\ref{fig:inelast} on the background of the
superlattice potential. The relative shift of the subsequent quantum wells is
$eFl$. In semiclassical language, the electron sitting at the bottom of the
miniband is accelerated by the electric field. It crosses the minibnad having
passed the distance $\Delta y\approx 4\Lambda/(eF)$ and hits the upper
boundary of the miniband. After reflection it goes back to the bottom of the
miniband.

The period of such a trip can be estimated by dividing $\Delta y$ by the
mean electron velocity in the miniband $\Lambda l/\hbar$, which gives $\sim
\hbar/(eFl)$. The periodic motion of the electron between the bottom and the
top of the miniband is called a Bloch oscillator. It has been observed in
optical four-wave mixing experiments.\cite{leisching-nov94,meier-may94}
The semiclassical language gives another physical meaning to
Eq.~(\ref{eq:oneband.4}), namely that the electron does not swing a long time
between the miniband boundaries because of the scattering.

The swinging of the electron means that there is field induced localization,
because the electric field is applied but cannot drive the electric current.
This picture is very clear and there is no reason not to use the
semiclassical theory of the miniband transport at least near the transition
point
\begin{equation}
   {eFl\tau_p} \sim \hbar\;.
\label{eq:oneband.4a}
\end{equation}
Far from the transition point, where ${eFl\tau_p} \gg \hbar$ the main
contribution to the current is given by the hopping between subsequent Stark
levels, and the contribution of the next neighbor hopping contains the small
parameter $\hbar / (eFl\tau)$. This is not true at the instability point
given by Eq.~(\ref{eq:oneband.4a}), where hopping between all Stark levels
contributes to the current. A number of works, which ignore summation over all
Stark levels near the instability point, have been published and show a
wrong condition for instability and current peak\cite{tsu-dohler}.

It is known that elastic relaxation processes are usually faster than
inelastic ones. For this reason we put $\tau_p$ in Eq.~(\ref{eq:oneband.4a})
and not $\tau_\varepsilon$. However, inelastic scattering processes can be
important near the field induced localization transition. In
Fig.~\ref{fig:inelast}(a) we showed the in-plane dispersion of the electron
energy for two subsequent Stark states. The electron may need to transfer a
large in-plane momentum to the impurity in order to jump between these states
elastically. At the same time the inelastic process may be very effective.

\begin{figure}
\unitlength=1.00mm
\linethickness{0.4pt}
\begin{picture}(85.00,150.00)
\put(5.00,10.00){\vector(1,0){60.00}}
\put(5.00,10.00){\vector(0,1){60.00}}
\put(38.00,64.00){\line(0,-1){60.00}}
\bezier{816}(18.00,72.00)(38.00,-28.00)(58.00,72.00)
\bezier{816}(18.00,52.00)(38.00,-48.00)(58.00,52.00)
\put(63.00,8.00){\makebox(0,0)[ct]{$y$}}
\put(28.00,8.00){\makebox(0,0)[ct]{$y_0$}}
\put(3.00,68.00){\makebox(0,0)[rc]{$E$}}
\put(38.00,12.00){\vector(0,1){10.00}}
\put(40.00,26.00){\makebox(0,0)[lc]{$4\Lambda$}}
\put(20.00,16.00){\vector(1,0){7.00}}
\put(56.00,16.00){\vector(-1,0){7.00}}
\put(53.00,18.00){\makebox(0,0)[cb]{$\Delta y$}}
\put(5.00,90.00){\vector(1,0){60.00}}
\put(5.00,90.00){\vector(0,1){60.00}}
\put(25.00,150.00){\line(0,-1){60.00}}
\bezier{816}(22.00,150.00)(42.00,50.00)(62.00,150.00)
\bezier{816}(22.00,130.00)(42.00,30.00)(62.00,130.00)
\put(22.00,127.00){\rule{6.00\unitlength}{2.00\unitlength}}
\put(63.00,88.00){\makebox(0,0)[ct]{$y$}}
\put(25.00,88.00){\makebox(0,0)[ct]{$y_0$}}
\put(3.00,148.00){\makebox(0,0)[rc]{$E$}}
\put(25.00,145.00){\vector(0,-1){8.00}}
\put(25.00,106.00){\vector(0,1){10.00}}
\put(27.00,141.00){\makebox(0,0)[lc]{$4\Lambda$}}
\put(15.00,128.00){\vector(1,0){7.00}}
\put(35.00,128.00){\vector(-1,0){7.00}}
\put(32.00,130.00){\makebox(0,0)[cb]{$\Delta y$}}
\put(27.00,15.00){\rule{22.00\unitlength}{2.00\unitlength}}
\put(38.00,10.00){\vector(0,-1){8.00}}
\put(85.00,110.00){\makebox(0,0)[cc]{(a)}}
\put(85.00,40.00){\makebox(0,0)[cc]{(b)}}
\end{picture}
   \caption{ The semiclassical picture of the superlattice miniband bended by
   the transverse magnetic field. (a) -- Bloch oscillator like orbits. (b) --
   Usual cyclotron orbits.
   }
\label{fig:parabolic}
\end{figure}
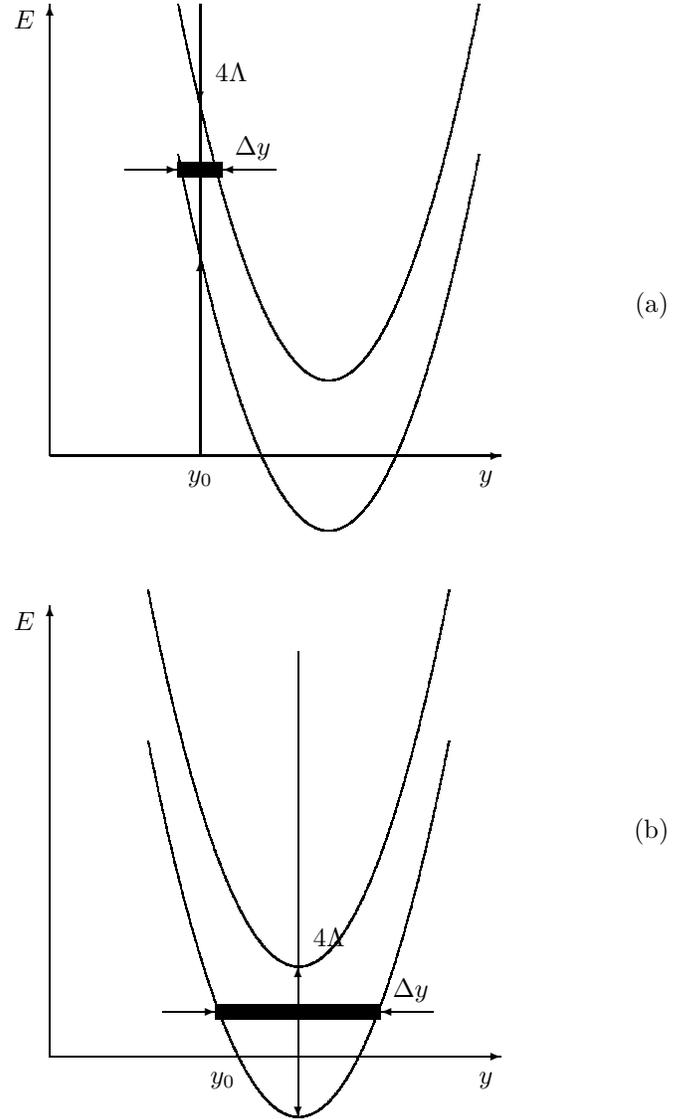

The acceleration of the  electron by the electric field is equivalent to
the presence of a linear potential, and to the motion of the electron in the
tilted miniband, as it is shown in Fig.~\ref{fig:inelast}. The transverse
magnetic field creates a parabolic potential, see Fig.~\ref{fig:parabolic}.
In this case, the size of the electron orbit depends very much on the
in-plane electron characteristic velocity $\bar v_x$ (let the magnetic field
be $z$-directed see Fig.~\ref{fig:geometry}(a) ).  In order to see how
it happens, we introduce the so-called effective Hamiltonian; see the next
section for the justification of this approach.
\begin{eqnarray}
   {\cal H} & = &
   {p_z^2\over 2m} + {p_x^2\over 2m} + 2\Lambda
   \left[ 1 - \cos\left(  {p_y l\over \hbar} \right)\right]
   - eFy
\nonumber\\
   p_x &=& P_x  + eBy\;,
\label{eq:oneband.5}
\end{eqnarray}
where we used the MKS units and the tight-binding approximation for the
dispersion law of the superlattice miniband. The coordinates here correspond
to the Fig.~\ref{fig:geometry}.

Two integrals of motion $P_x$ and guiding center of the orbit $y_0$ allow us
to introduce another conserving quantity, the in-plane characteristic velocity
$\bar v_x \equiv ( P_x  + eBy_0 )/ m$. We can obtain the one-dimensional
Hamiltonian for the particle with defined values of $\bar v_x$ and $y_0$:
\begin{eqnarray}
   {\cal H}' & = &
   2\Lambda
   \left[ 1 - \cos\left(  {p_y l\over \hbar} \right)\right]
\nonumber\\
   &+ &
   (e\bar v_x B - eF)(y-y_0) +
   {\bigl[eB(y-y_0)\bigr]^2\over 2m}\;.
\label{eq:oneband.6}
\end{eqnarray}
This equation shows that the kinetic energy of the vertical motion is limited
by two inequalities
\begin{eqnarray}
   E & \le &
   2\Lambda
   +
   (e\bar v_x B - eF)(y-y_0) +
   {\bigl[eB(y-y_0)\bigr]^2\over 2m}\;,
\label{eq:oneband.6a}\\
   E & \ge &
   - 2\Lambda
   +
   (e\bar v_x B - eF)(y-y_0) +
   {\bigl[eB(y-y_0)\bigr]^2\over 2m}\;,
\label{eq:oneband.6b}
\end{eqnarray}
which are quite analogous to the case of Eq.~(\ref{eq:oneband.4b}). This
energy lies between two parabolas, see Fig.~\ref{fig:parabolic}.

There is a basic difference between two parallel lines and two ``parallel
parabolas''.  Whereas the energy distance between them is constant
$4\Lambda$, the size of the orbit, $\Delta y$, in the real space depends very
much on $\bar v_x$. The expression for the size of the orbit at the origin
of the parabolas Fig.~\ref{fig:parabolic}(b) is different from the expression
for the orbit in the branches Fig.~\ref{fig:parabolic}(a). Therefore, we have
two cases:
\begin{equation}
   \Delta y =
   \left\{ \begin{array}{ll}
      \displaystyle
      {2\Lambda \over |eF - e\bar v_x B|}\;\;
      &
      |F/B - \bar v_x |  >  \sqrt{\Lambda\over2m}
      \\[10pt]
      \displaystyle
      {\sqrt{ 2\Lambda m} \over |e B|}\;\;
      &
      |F/B - \bar v_x |  <  \sqrt{\Lambda\over2m}
   \end{array}\right.\;.
\label{eq:oneband.7}
\end{equation}
The first case corresponds to a slightly modified Bloch oscillator and the
second case corresponds to the usual anisotropic cyclotron orbits.

The periods of these orbits can be obtained by the
division of $\Delta y$ by the mean electron velocity in the miniband
$\sim\Lambda l/\hbar$. The magnetic and electric fields induce localization
of the electrons when these periods for most of electrons are of the order of
the relaxation time. This happens either when
\begin{equation}
   \Omega_B\tau_p \sim 1\;, \;\;\;   \hbar\Omega_B > |eFl - ev_FBl|\;,
\label{eq:oneband.8}
\end{equation}
or when
\begin{equation}
    |eFl - ev_FBl|\tau_p \sim \hbar\;,\;\;\;
    \hbar\Omega_B < |eFl - ev_FBl|\;.
\label{eq:oneband.9}
\end{equation}
Here we replaced $\bar v_x$ by Fermi velocity $v_F$. In the non-degenerate
case one has to substitute here the thermal velocity $\sqrt{mT}$. The
cyclotron frequency introduced here is defined by
\begin{equation}
   \Omega_B \equiv {eB\over \sqrt{mm_\perp} }
   = {eBl \over \hbar }\sqrt{\Lambda\over 2m}\;.
\label{eq:oneband.10}
\end{equation}
Note that the condition of the magnetic field induced localization
Eq.~(\ref{eq:oneband.8}) is valid for bulk semiconductors, while
the condition Eq.~(\ref{eq:oneband.9}) is specific for superlattices.

The analogy between Eqs.~(\ref{eq:oneband.8}) and (\ref{eq:oneband.9}) from
one side and Eq.~(\ref{eq:oneband.4a}) from other side leads to the {\em
fourth} condition of the miniband transport:
\begin{equation}
   \text{max}\; \hbar\Omega_B ,\; |eFl - eBv_Fl|
   \lesssim
   \text{max}\;{\hbar\over\tau_p},\; {\hbar\over\tau_\varepsilon}
   \;.
\label{eq:oneband.11}
\end{equation}
In order to provide the quantum mechanical arguments, as we did for
Eq.~(\ref{eq:oneband.4}), we have to quantize the effective Hamiltonian
and compare the level spacings with relaxation rates. The structure of the
eigenenergies of the effective Hamiltonian is discussed in
Sec.~\ref{sec:quant} and one can check that conditions
Eq.~(\ref{eq:oneband.11}) are correct.

\section{ Effective Hamiltonian }
\label{sec:ham}

In the previous section we derived conditions of the miniband transport
Eqs.~(\ref{eq:oneband.2}), (\ref{eq:oneband.3}), (\ref{eq:oneband.4}), and
(\ref{eq:oneband.11}). The last one was obtained by making use of the
effective Hamiltonian method. In this section we derive conditions
justifying this method.

The dynamics of a Bloch electron in the presence of a magnetic field
can be described approximately by an effective Hamiltonian, introduced
first by Peierls\cite{peierls33}. The effective Hamiltonian is obtained from
the spectrum of the Bloch electron $E({\bf p})$ by the replacement of
the momentum $\bf p$ by ${\bf p} - e{\bf A}$, where $\bf A$ is the
vector potential of the magnetic field. Luttinger\cite{luttinger51}
justified this approach for a weak magnetic field. Blount\cite{blount62}
calculated corrections to the effective Hamiltonian for a general three
dimensional lattice. Berezhkovskii and Suris\cite{suris-jan84} made the
accurate calculation of the electron spectrum in a superlattice and the
external magnetic field applied perpendicular to the growth direction.
They assumed that the effective Hamiltonian can be applied if the magnetic
field does not affect much the electron wave functions in separate wells of
the superlattice. This results in the condition
\begin{equation}
   \hbar\Omega_c E_F\ll E_g^2\;,
\label{eq:ham.0}
\end{equation}
where $\Omega_c=eB/m$, $E_F$ is Fermi energy, and $E_g$ is the energy gap
between minibands, see Fig.~\ref{fig:geometry}(b). The condition of the
applicability of the effective Hamiltonian Eq.~(\ref{eq:ham.0})  can be
considered as generally accepted, for the case of metals, see
Ref.~\onlinecite{abrikosov-mb}.

Lifshitz and Pitaevsky\cite{landau-ms} have assumed that the effective
Hamiltonian method can be used if the space between energy levels in the
presence of magnetic field is much smaller than the minimal energy scale
of the band structure. We find a different condition.

In this section we show that the condition Eq.~(\ref{eq:ham.0}) is not
enough. We carry out the derivation of the effective Hamiltonian and find
another necessary condition which can be more severe. We obtain our
condition by the calculation of corrections to the effective Hamiltonian.

In the absence of magnetic field, the eigenfunctions of the Hamiltonian
${\cal H}_{0} = (\hbar^{2}\nabla^{2}/2m) + U(y)$, where $U(y)$ is a
periodic potential, are
\begin{equation}
   \psi_{s{\bf k}} = {1\over \sqrt{{\cal V}}}e^{i{\bf kr}} u_s(k_y, y)\;,
\label{eq:ham.1}
\end{equation}
where $s$ is the miniband number and the Bloch amplitude $u_s$ is periodic
in $y$.  So, with the help of the unitary matrix
\begin{equation}
   S_s(r', r) = \sum_{\bf k}
	{1\over \sqrt{{\cal V}}}e^{-i{\bf kr}}\psi_{s{\bf k}}(\bf r')
\label{eq:ham.2}
\end{equation}
the Hamiltonian can be transformed to the form where its wave functions are
plane waves and the spectrum is the electron spectrum in the periodic
potential, $E_s({\bf k})$,
\begin{equation}
   S_s^{\dag}
	\bigl[\frac{\hbar^2{\hat{\bf k}}^2}{2m} + U(y)\bigr]S_s
        = E_s(-i\nabla)\;.
\label{eq:ham.3}
\end{equation}
In the presence of a constant vector potential this equation becomes
\begin{equation}
   S_{s,\bf A}^{\dag}\bigl[
      \frac{(\hbar{\hat{\bf k}-e{\bf A}})^2}{2m} + U(y)
	\bigr]S_{s,\bf A}
   = E_s\left(-i{\bf\nabla}-\frac{e}{\hbar}{\bf A}\right)\;.
\label{eq:ham.4}
\end{equation}
The matrix $S_{s,\bf A}$ is obtained from
Eqs.~(\ref{eq:ham.1},\ref{eq:ham.2}) by the replacement of
$u_s(k_y,y)$ by $u_s(k_y-eA_y/\hbar,y)$.\cite{landau-ms}

If the vector potential is not a constant then equation Eq.~(\ref{eq:ham.4})
is not exact. First of all, the matrix $S_{s,\bf A}$ introduced above, in
general, is not unitary and has to be corrected. However, in a one
dimensional superlattice and a uniform magnetic field it is possible to
choose a gauge where $A_y=0$, and the transformation matrix $S_s$ is
independent of the vector potential. Even with such a transformation matrix
there are corrections to the right-hand side of Eq.~(\ref{eq:ham.4}) because $S_s$ does
not commute with the coordinate dependent vector potential.

It's well known from the theory of Bloch wave functions that
\begin{equation}
	\langle s'k_y'|y|s k_y \rangle =
	\left( i\frac{\partial}{\partial k_y} + \langle s'|Y|s\rangle
	\right)\delta(k_y-k_y')\;,
\label{eq:ham.5}
\end{equation}
where $\langle s \vert Y \vert s^{\prime} \rangle = \langle u_{s}
\vert i(\partial /\partial k_y) \vert u_{s^{\prime}} \rangle$ is a
function of $k_y$.  Here states $\langle{u}_{s}\vert$ are chosen in such a
way that $\langle{s}\vert{Y}\vert{s}\rangle=0$. Thus, the diagonal
matrix element of the Hamiltonian is
\begin{equation}
   {\cal H}_{ss} = E_s\left(-i{\bf\nabla}-\frac{e}{\hbar}{\bf A}\right)
   + \frac{e^2B^2}{2m}(Y^2)_{ss}\;,
\label{eq:ham.6}
\end{equation}
where $B$ is the component of the magnetic field perpendicular to the
superlattice axis. The first term on the right-hand side of this equation is
called the effective Hamiltonian. In order to estimate the second term we
make use of an equation for the operator $Y$.  This equation can be obtained
by calculating the commutator $[[{\cal H}_{0}, y], y]$ directly and with the
help of Eq.(\ref{eq:ham.5}),
\begin{eqnarray}
	\sum_{s''}(E_{s}+E_{s'}-2E_{s''})Y_{ss''}Y_{s''s'}
	+i(E_s-E_{s'})\frac{d}{dk}Y_{ss'}
\nonumber\\
	+2iY_{ss'}\frac{d}{dk}(E_s-E_{s'})
	+(\frac{\hbar^2}{m}-\frac{d^2E_s}{dk^2})\delta_{ss'} = 0
\label{eq:ham.6a}
\end{eqnarray}
If there are no band crossings, that is
$E_1<E_2<E_3<\ldots$ for any $k$, the diagonal part of this equation
gives the following inequality
\begin{equation}
   (Y^2)_{11}\le\frac{\hbar^2}{2}\frac{1/m - 1/m_\perp}{E_2-E_1}\;,
\label{eq:ham.6b}
\end{equation}
where $1/m_\perp(k) = d^2E_1/(dk^2)$. In the case of the smooth potential
$U(y)$ the right-hand side of this inequality can be considered as estimation for the
matrix element $(Y^2)_{11}$ in the left-hand side, see Appendix
Sec~\ref{sec:hill}. The variance of this matrix element in the Brillouin zone
is, therefore, of the order of $(\hbar^2/m)(\Lambda/E_g^2)$, i. e. it's
proportional to the miniband width divided by the square of the energy gap
between the first and second minibands, $E_g=\text{min}(E_2-E_1)$.

The last term in Eq.~(\ref{eq:ham.6}) in the case $s=1$ can be neglected if
it is much smaller than the miniband width, which gives the condition for the
magnetic field
\begin{equation}
   \hbar\Omega_c \ll E_g\;,
\label{eq:ham.7}
\end{equation}
Note that this condition is stronger than Eq.~(\ref{eq:ham.0})
unless the Fermi energy is comparable or larger than the $E_g$.

The offdiagonal with respect to the miniband matrix element of the effective
Hamiltonian can be neglected under the condition that is similar to
Eq.~(\ref{eq:ham.0}).

\section{ Appendix: matrix elements of \protect{$Y$} in the tight and weak
          binding approximations. }
\label{sec:hill}

In the tight binding approximation, $\Lambda\ll{E_g}$, both sides of
Eq.~(\ref{eq:ham.6b}) can be represented as a power series of the parameter
$1/\Theta\sim\Lambda/E_g$. For example for a superlattice with rectangular
wells, narrow barriers, and period $\pi$
\begin{eqnarray}
   (Y^2)_{11} &=& {\pi^2\over12} - {1\over2} +
   {1\over\Theta} \Bigl[ - {\pi\over6} - {2\over\pi}
\nonumber\\
   &+&  \left({2\pi\over3} +{2\over\pi}\right)\sin^2(k\pi/2) \Bigr]
    + O\left({1\over\Theta^2}\right)\;
\label{eq:hill.1}
\end{eqnarray}
For the same case the right-hand side of Eq.~(\ref{eq:ham.6b}) is
\begin{eqnarray}
   &&{1\over3}  + {1\over\Theta}
   \Bigl[ - {\pi\over3} - {4\over9\pi}
\nonumber\\
   &+& \left({2\pi\over3} +{20\over9\pi}\right)\sin^2(k\pi/2) \bigr]
   + O\left({1\over\Theta^2}\right)\;.
\label{eq:hill.2}
\end{eqnarray}
So, in this particular case the calculated coefficients are different by
about 3\%.

In the weak binding approximation, when band width is much larger than the
gap, let
\begin{equation}
   U_n = {1\over\pi}\int_0^\pi\cos(2ny)U(y)\;,
\label{eq:hill.3}
\end{equation}
where the superlattice potential $U(y)$ is an even function with period $\pi$,
and $U_0=0$. We assume that this potential is smooth; in other words we
assume that the amplitude of the first harmonic $U_1$ is larger than all other
harmonics. In the gap region $|k|\sim1$ Eq.~(\ref{eq:ham.6b}) becomes
equality in the zero order perturbation theory:
\begin{equation}
   (Y^2)_{11}=\left[
      {1\over2}\;{U_1\over (1-|k|)^2+U_1^2}
   \right]^2\;.
\label{eq:hill.4}
\end{equation}
where we put $\hbar=m=1$. For small and intermediate $k$ the second order
perturbation theory gives
\begin{equation}
   (Y^2)_{11} = \sum_{s=2}^\infty
   U_{\left[s\over2\right]}^2
   {1\over E_1(k)-E_s(k) }
   {d^2\over dk^2}\left(
      {1\over E_1(k)-E_s(k)}
   \right)
\label{eq:hill.5}
\end{equation}
For the same case the right-hand side of Eq.~(\ref{eq:ham.6b}) is
\begin{equation}
   {1\over E_1(k)-E_2(k)}
   \sum_{s=2}^\infty
   U_{\left[s\over2\right]}^2
   {d^2\over dk^2}\left(
      {1\over E_1(k) - E_s(k) }
   \right)\;.
\label{eq:hill.6}
\end{equation}
Under the assumption that $U_1\ge|U_n|$ for any $n$, the difference between
the left hand side and right hand side of Eq.~(\ref{eq:ham.6b}) is less than
15\%.

\chapter{ Transverse magnetoresistance. I }
\label{chap:trans}

   The motion of the Bloch electron in the crossed electric, $F$, and
   magnetic, $B$, fields is studied for the case of the anisotropic band
   structure that usually exists in superlattices. The electron trajectories
   are calculated from the so-called effective Hamiltonian with an electric
   field applied in the growth direction of a superlattice and a magnetic
   field parallel to layers. We solve the kinetic equation and calculate the
   electric current for the case when the miniband width is much smaller than
   the characteristic kinetic energy of electrons. If the magnetic field is
   strong enough, only electrons that have a velocity component perpendicular
   to both electric and magnetic fields equal to the drift velocity,
   $v_A=F/B$, contribute into the current. The current reaches its maximum
   value when $v_A$ is close to the Fermi velocity, $v_F$, or to the thermal
   velocity, $v_T$, whatever is larger.  However, as the magnetic field goes
   to zero the current peak position goes to its limit $F_{\text{th}}$ like
   $B^2$.  The magnetoresistance at low magnetic field changes its sign at
   the field $F_{\text{th}}/\sqrt{3}$. The quantitative agreement with
   experiment is obtained for these results without fitting parameters. We
   also find that electric current is independent on the relaxation time in
   some interval of the applied field.  This can be called the effect of
   collisionless conductivity.


\section{ Introduction}
\label{sec:int}

The nonlinear transport phenomena in superlattices have attracted much
interest in recent years. These phenomena come from the narrow width of
electronic mini-bands in superlattices\cite{esaki-tsu}. Experiments
show that with the increase of the applied bias the electric current first
grows linearly, then reaches a maximum\cite{chang-ploog}. At this point the
uniform potential distribution across the superlattice becomes unstable, and
the superlattice breaks into low- and high-field domains\cite{esaki-chang}.
Further increase of the applied voltage leads to the expansion of the
high-field domain, which appears as a series of peaks on the $I$-$V$
characteristic.

Measurements of the vertical current in the presence of a magnetic field
parallel to the
layers\cite{choi4,vuong,davies,sibilleA,lee93,grahn93,%
aristone-93,herbert-jan94} reveal a new interesting effect. It was found that
all peaks on the $I$-$V$ characteristic of a superlattice are shifted by
magnetic field toward the higher bias. A linear theory of the
magneto-conductivity of superlattices (i.e. a theory for the Ohm's law
region) developed by Shik\cite{shik-aug73}, Ando\cite{ando-sep81} and a
theory of electron transport in the Harper band developed by Suris and
Shchamkhalova\cite{suris90} cannot explain the behavior of current
peaks. A qualitative explanation of this phenomenon was suggested by
Movaghar\cite{movaghar}.  Numeric calculations performed by Sibille et
al.\cite{sibilleA}, Palmier et al.\cite{palmier-92}, and Hutchinson et
al.\cite{herbert-jan94} also lead to the same qualitative behavior and some
of them can be fitted to experiment\cite{palmier-92}. An analytical solution
of the kinetic equation and results for magneto-conductivity of superlattice
have been published\cite{bass-book86}. The final results were expanded in
powers of magnetic field and they are reproduced by Eq.~(\ref{eq:con.6}) here.
In the present paper we calculate analytically the $I$-$V$ characteristic of
a superlattice in strong electric and magnetic fields. The theory
agrees quantitatively with the behavior of the current peaks
measured in different experiments.

The qualitative arguments suggested by Movaghar\cite{movaghar} are based on
the quantum mechanical explanation of negative differential conductivity
(NDC) in superlattices.  NDC in an electric field applied parallel to the
superlattice axis ($y$ axis) comes about because this field localizes
electrons at Stark levels. A magnetic field applied parallel to the layers (in
the $z$ direction) causes the localization of electrons in the $(x,y)$ plane.
In a weak electric field the spatial size of electron states is limited by
the magnetic field and it is smaller than the length of possible Stark
states, $\Lambda/(eF)$. Therefore the electric field does not affect the
electron states, and usual positive magnetoresistance is observed. In other
words the $I$-$V$ characteristic in a magnetic field goes below that
without a magnetic field.  With the increase of the electric field the size
of Stark states eventually becomes smaller than the length of the possible
magnetic states.  At this electric field the effect of the magnetic field can
be neglected and the $I$-$V$ characteristic passes to the falling branch of
the $I$-$V$ curve without magnetic field.  Because of the positive
magnetoresistance, the peak of the $I$-$V$ characteristic is shifted to a
higher electric field.

Sibille et al.\cite{sibilleA} calculated the $I$-$V$ characteristic making
use of the hydrodynamical model similar to that used by Esaki and
Tsu\cite{esaki-tsu}. The classical equation of motion for electrons in a
superlattice in electric and magnetic fields has been solved numerically. The
electron drift velocity was calculated by averaging of the electron velocity
along the trajectory with some relaxation time. Although the hydrodynamic
model gives a qualitatively correct $I$-$V$ characteristic it does not
describe the experimental results quantitatively. In a more recent work of
Palmier et al.\cite{palmier-92} the $I$-$V$ characteristic was obtained from
the numeric solution of the Boltzmann kinetic equation in good agreement with
the experiment.

In this work we make an analytic calculation of the current in a
superlattice in the presence of electric and magnetic fields with the help of
the classical Boltzmann equation.  The equations of electron motion in the
crossed electric and magnetic fields in the superlattice are obtained in
Sec.~\ref{sec:din} from the effective Hamiltonian. These equations are
similar to the equations of motion of a nonlinear pendulum\cite{zaslavsky88},
but in our case two degrees of freedom are mixed and this complicates the
solution. Electron trajectories are expressed in terms of Jacobian elliptic
functions.  In the $v_x$-$v_y$ plane trajectories are
closed orbits confined in a strip $|v_y|\le2\Lambda l/\hbar$ due to
reflection from miniband boundaries.  Here $\Lambda$ is the tunneling matrix
element between electron wave functions in adjacent quantum wells and $l$ is
the superlattice period.  The centers of the trajectories are located at the
$v_x$ axis. The size of each trajectory in the $v_x$ direction is also limited
from above, $\text{Var}(v_x)\le4\sqrt{2\Lambda/m}$, where $m$ is the in-plane
effective mass.  The frequency of electron motion along an orbit has a
minimum, when the center of the orbit is located at $v_x = v_A \equiv F/B$.
The frequency increases with the distance of the center from this point.

In Sec.~\ref{sec:kin} we solve the Boltzmann equation in a superlattice
miniband to calculate the electron distribution function, $f$. We assume
that the characteristic electron energy (temperature or the Fermi energy) is
much larger than the miniband width and the main scattering mechanism is
elastic scattering by impurities or surface roughness. We do not make any
assumption concerning the relative importance of external fields and
collisions for the shape of the distribution function. This means that, in
general, electrons may travel rather a long distance in  physical space as
well as in momentum space before being scattered. To solve the Boltzmann
equation in this case we introduce as coordinates two integrals of motion
characterizing a trajectory in the momentum space and a coordinate
characterizing electron position along the trajectory. The electron
distribution is asymmetric along the trajectory with respect to the direction
of the electric field. The asymmetric part is proportional to the relaxation
time $\tau$ when it is smaller than the period of motion along the trajectory
and is proportional to $1/\tau$ in the opposite case.

The electric current is calculated in Sec.~\ref{sec:con}. The contribution of
an electron to the current is proportional to its velocity. The velocity in
the growth direction, $v_{y}$, is an oscillating function of the electron
momentum in the same direction, $p_{y}$. This momentum is changing in time
under the electric field and the Lorentz force. For that reason $v_{y}$,
oscillates and the average contribution to the current is zero. However, the
Lorentz force in $y$ direction is proportional to the velocity in the
direction perpendicular to the electric and magnetic fields, $v_{x}$, and for
some value of this velocity the force due to the electric field and the
Lorentz force cancel each other. This means that there is a resonant group of
electrons with a time independent $p_{y}$ that contributes to the current.
Scattering complicates this picture because for a nonzero contribution to the
current it is enough that an electron does not change $v_{y}$ between two
scattering events. The width of the resonant region in the momentum space is
determined by competition between the electron motion in the external fields
and scattering. For some values of the fields a resonance region appears at
the exponential tail of the electron distribution function.  In this case a
nonresonant contribution to the current, which is proportional to the
probability for an electron to be scattered during one period of the orbital
motion has to be taken into account.

Application of a magnetic field perpendicular to the electric current
generates an  electric field parallel to the layers (in the $x$-direction).
This field is introduced into the kinetic equation in Sec.~\ref{sec:hall}. We
show that the correction to the current in the superlattice direction due to
this field (Hall effect) can be neglected.

In a certain range of the magnetic field we found that the electric
current is independent of the scattering rate. This result suggests that the
kinetic equation can be solved without considering the collision term, as
is done for plasma, see Ref.~\onlinecite{landau-dmp}. That is the regime
of the collisionless conductivity considered specifically in
Sec.~\ref{sec:dmp}. In Sec.~\ref{sec:comp} we compare the theoretical
predictions with available experimental data. The summary of the results is
given in Sec.~\ref{sec:cncl}.


\section{ Electron dynamics }
\label{sec:din}

\begin{figure*}
\unitlength=1mm
\begin{center}
\begin{picture}(150.00,75.00)
  \put(00.00, 10.00){
     \insertplot{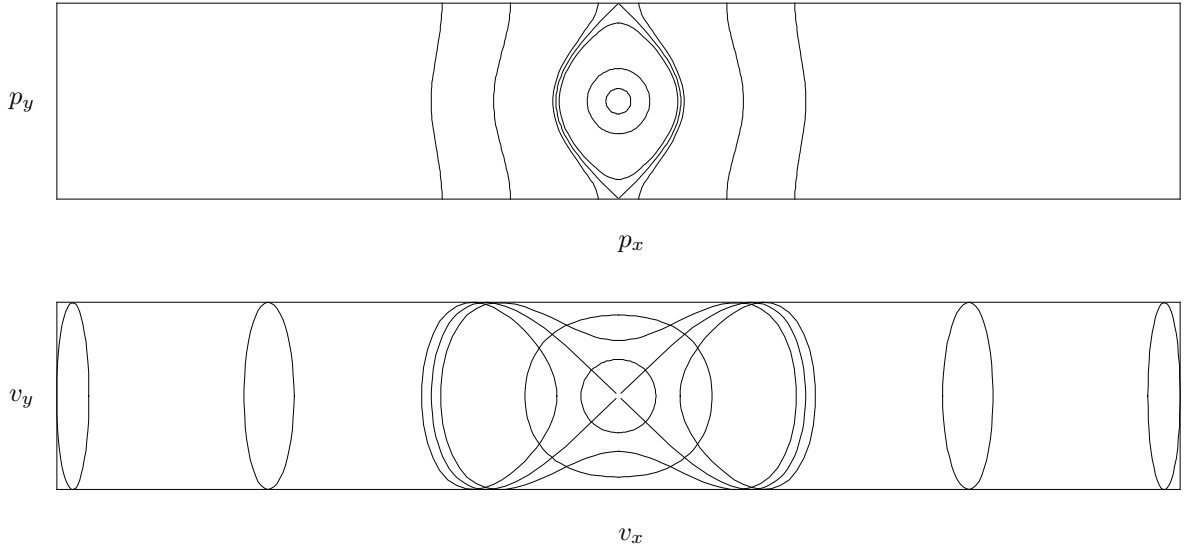}
     \put(-6.00, 52.00){$p_y$}
     \put(-6.00, 12.50){$v_y$}
     \put(75.00, 33.00){$p_x$}
     \put(75.00, -6.00){$v_x$}
  }
\end{picture}
\end{center}
   \caption{
      The trajectories in $p$- and in $v$-space. The quantity $p_y$ ranges
      from $-\pi\hbar/d$ to $\pi\hbar/d$, $p_x$ ranges from $mF/B-18p_\Lambda$
      to $mF/B+18p_\Lambda$, $v_y$ ranges from $-2v_\perp$ to $2v_\perp$, and
      $v_x$ ranges from $F/B-6p_\Lambda/m$ to $F/B+6p_\Lambda/m$.
      Trajectories in the $p$ space are closed for $\Omega < \Omega_B$ and
      open in the opposite case.  The values of the ratio $\Omega/\Omega_B$
      for the trajectories shown at the figure are 0.2, 0.5, 0.95, 1.0, 1.05,
      2, and 3.
   }
\label{fig:traject}
\end{figure*}

We consider a superlattice in the tight binding approximation when, without
external fields, the electron spectrum in the first miniband is
\begin{equation}
   E_{\bf p} = \frac{p_x^2+p_z^2}{2m} + 2\Lambda
   \left[1-\cos\left(
      \frac{p_yl}{\hbar}\right)\right]\;,
\label{eq:din.1}
\end{equation}
where $m$ is the electron effective mass, $\Lambda$ is the overlap integral
between adjacent wells, and  $l$ is the superlattice period. In an electric
field in the $y$ direction and a magnetic field in $z$ direction it is
convenient to choose a gauge where the vector potential depends only on $y$,
${\bf A}=(-By, 0, 0)$.

The external fields can be approximately taken in to account by the effective
Hamiltonian method introduced first by Peierls\cite{peierls33}. The effective
Hamiltonian is obtained from the spectrum of the Bloch electron $E_{\bf p}$
by the replacement of the momentum $\bf p$ by ${\bf p} - e{\bf A}$, (see
justification of this approach in Sec.~\ref{sec:ham}). The Hamiltonian has the
form
\begin{eqnarray}
   {\cal H} &=& \frac{p_z^2}{2m} + \frac{1}{2m}
   \left(P_x+eBy\right)^2 \nonumber\\
	&+& 2\Lambda
   \left[1-\cos\left(
      \frac{p_yl}{\hbar}\right)\right] - eFy\;,
\label{eq:din.1aa}
\end{eqnarray}
and gives the following expressions for the velocity components,
\begin{equation}
	v_x = {p_x\over m}\ ,
	\ v_y = 2v_\perp\sin\left(
      \frac{p_yl}{\hbar}\right)\ ,
	v_z = {p_z\over m}\ ,
\label{eq:din.1a}
\end{equation}
where $p_x=P_x+eBy$, and $v_\perp={\Lambda l/\hbar}$. The equations of
motion resulting from the Hamiltonian Eq.~(\ref{eq:din.1aa}) are
\begin{mathletters}
\begin{eqnarray}
   \dot p_x & = & ev_yB =  \Omega_Bp_\Lambda\sin\left(
      \frac{p_yl}{\hbar}\right)\ ,
\label{eq:din.2a}\\
   \dot p_y & = & eF - ev_xB =  eF - {eB\over m}p_x \ ,
\label{eq:din.2b}
\end{eqnarray}
\label{eq:din.2}
\end{mathletters}
where
\begin{equation}
   \Omega_B = {eB\over \sqrt{mm_\perp}}\;,
\label{eq:din.3}
\end{equation}
$p_\Lambda = \sqrt{2m\Lambda}$, and  $m_\perp = \hbar^2/(2{\Lambda}l^2)$ is
the electron effective mass at the bottom of the miniband. Equations
(\ref{eq:din.2}) have the following integral
\begin{equation}
	\sin^2\left(
      \frac{p_yl}{2\hbar}\right)\
   + \frac{(mF/B - p_x)^2}{4p_\Lambda^2}
   = \left(\frac{\Omega}{\Omega_B}\right)^2 \ ,
\label{eq:din.5}
\end{equation}
where $\Omega>0$ is a constant of the integration. Equations~(\ref{eq:din.2})
are pendulum equations and their solution is expressed in terms of
elliptic functions, see Fig.~\ref{fig:traject}. In the case $\Omega>\Omega_B$
there are two solutions for each value of $\Omega$. We have for first
solution $p_x<mF/B$ and
\begin{mathletters}
\begin{eqnarray}
   p_y(t) &=& {2\hbar\over l}\mbox{am}(\Omega{t})
\label{eq:din.9a}\\
   p_x(t) &=& {mF\over B}
   - 2p_\Lambda{\Omega\over\Omega_B}\mbox{dn}(\Omega{t})
   \;,
\label{eq:din.9b}
\end{eqnarray}
where the modulus of the elliptic functions is $k=\Omega_B/\Omega$.
These equations describe open trajectories in momentum space. Since
the values of the electron quasimomentum $p_y$  are limited by the interval
$-\pi\hbar/l<p_y<\pi\hbar/l$ only  a fractional part
$\mbox{am}(\Omega{t})/\pi$ should be kept in Eq.~(\ref{eq:din.9a}). The
second solution, which corresponds to $p_x>mF/B$, can be obtained from
Eqs.~(\ref{eq:din.9a},\ref{eq:din.9b}) by changing $\Omega$ to $-\Omega$.

In the case $\Omega<\Omega_B$ electron trajectories are closed and there is
only one trajectory for each value of $\Omega$. Electron momenta $p_y(t)$ and
$p_x(t)$ are again expressed in terms of elliptic functions, but the modulus
is different, $k=\Omega/\Omega_B$,
\begin{eqnarray}
   p_y(t) &=& {2\hbar\over l}
   \mbox{arcsin}\left[{\Omega\over\Omega_B}
      \mbox{sn}(\Omega_B{t})\right]
\label{eq:din.9c}\\
   p_x(t) &=& {mF\over B}
   - 2p_\Lambda{\Omega\over\Omega_B}\mbox{cn}(\Omega_B t) \ .
\label{eq:din.9d}
\end{eqnarray}
In the case $\Omega=\Omega_B$ the solution is expressed in the elementary
functions
\begin{eqnarray}
   p_y(t) &=&\pm
        {2\hbar\over l} \mbox{arcsin}\left[ \mbox{tanh}( \Omega t)\right]
\label{eq:din.9e}\\
   p_x(t) &=& {mF\over B} \mp 2p_\Lambda\mbox{sech}( \Omega t ) \ .
\label{eq:din.9f}
\end{eqnarray}
\label{eq:din.9}
\end{mathletters}

For further calculations we also need expressions for the $y$-component of
the electron velocity,
\begin{equation}
	v_y(t) =
   \left\{
      \begin{array}[c]{ll}
         4v_\perp\mbox{sn}(\Omega{t})\mbox{cn}(\Omega{t})\;, &
            \;k=\Omega_B/\Omega\;<\;1 \;, \\
         4v_\perp k\mbox{sn}(\Omega_Bt)\mbox{dn}(\Omega_Bt) \;,&
            \;k=\Omega/\Omega_B\;<\;1\;, \\
         4v_\perp\mbox{tanh}(\Omega{t})\mbox{sech}(\Omega{t})\;, &
            \;\Omega=\Omega_B\;.
      \end{array}
   \right.
\label{eq:din.10}
\end{equation}
The motion of the electron in real space can be obtained by the integration
of electron velocities. The integration of Eq.~(\ref{eq:din.10}) shows that
the $y$-coordinate oscillates around some value $y_0$, which can be
considered as the third integral of the motion after $p_z$ and $\Omega$. The
total electron energy is conserved and can be represented as a function of
the integrals of the motion
\begin{equation}
   E =
   \frac{p_z^2}{2m}
   + \frac{1}{2m}
   \left( {mF\over B}-2p_\Lambda\frac{\Omega}{\Omega_B}\right)^2
        - eFy_0 \ .
\label{eq:din.12a}
\end{equation}


\section{ Solution of the Boltzmann equation}
\label{sec:kin}

For the calculation of the current we make use of the Boltzmann equation for
one electron distribution function $f({\bf p})$
\begin{equation}
	eBv_y\frac{\partial f}{\partial p_x}
	+\left(eF-eBv_x\right)
	\frac{\partial f}{\partial p_y}
	= \left(\frac{\partial f}{\partial t}\right)_{\text{coll}}\;.
\label{eq:kin.1}
\end{equation}
The collision integral on the right hand side of this equation includes
elastic scattering mechanisms (impurity scattering, surface roughness
scattering and, possibly, absorption and immediate emission of an optical
phonon) and inelastic ones (acoustic and optical phonons). Elastic relaxation
time is typically about $10^{-13}$sec. Concerning inelastic relaxation, we
assume that both the Fermi energy and temperature are smaller than the
optical phonon energy $\hbar\Omega_{LO}$ so that the energy relaxation due to
optical phonon emission can be neglected.\cite{laikhtman2,meier-may94} The
inelastic relaxation time due to acoustic phonon scattering typically is
about $10^{-10}$sec or larger.  This means that the main relaxation mechanism
in Eq.~(\ref{eq:kin.1}) is elastic scattering.

We will consider the elastic scattering of the electrons as point
scattering, when the electron position and electron kinetic energy,
$E_{\bf{p}}$, are conserved. This approximation is justified if the electron
wavelength is much larger than the length scale of the scatter. For
electron motion in the growth direction the electron wavelength ranges
from one superlattice period to the superlattice length. At the same time,
scattering in the adjacent wells can be considered as uncorrelated. The
electron in-plane wavelength depends on the electron kinetic energy, that is
typically about or less than $10$meV, and leads to the wavelength
$\gtrsim5000${\AA}. The length scale of the in-plane elastic scattering  is
very short because in superlattices the spacer separating impurities from
electrons in quantum wells is smaller than the superlattice period and the
correlation length of the surface roughness typically is about 500{\AA}
(see e.g. Ref.~\onlinecite{ruf-94} and Section~\ref{sec:summ}).

It is important to note that $E_{\bf p}$ is not an integral of motion and is
not conserved along a trajectory. Elastic collisions lead to relaxation of
the electron distribution to its average over the surface of constant
kinetic energy,
\begin{equation}
   \bar f(E) = {
      \int d{\bf p}f({\bf p})\delta(E-E_{\bf p})
      \over\int d{\bf p}\delta(E-E_{\bf p})
   }\;.
\label{eq:kin.2}
\end{equation}
The locality of elastic scattering means also that the matrix element of the
scattering is a constant, which allows us to describe the elastic collision
operator with the help of the constant relaxation time,
$[\bar f(E_{\bf p}) - f({\bf p})]/\tau$.

The electron gas can be appreciably heated by a strong electric field
$\Omega_E\tau\sim1$. In this case $\bar f(E)$ can be significantly different
from the equilibrium function. In this work we will elaborate two limiting
cases when the difference between $\bar f(E)$ and the Fermi function with
some effective temperature $T$, $f_0(E)$, is not important for the
calculation of the current. In the first case, $T\ll{E_F}$, the electron gas
is degenerate and a detailed structure of the electron distribution near the
Fermi surface does not make any difference. In the second case,
$E_F\ll{T}\approx{T_s}$ where $T_s$ is the lattice temperature, electron
heating can be neglected.\cite{laikhtman2} It is worth noting that at strong
magnetic fields, quantum corrections to the classical Boltzmann equation
(\ref{eq:kin.1}) can be neglected only at high enough electron
temperature,\cite{esaki-chang-77}
\begin{equation}
   \frac{\hbar\Omega_B}{2\pi^2T}\ll1\;.
\label{eq:kin.jus}
\end{equation}

It is convenient to transform Eq.~(\ref{eq:kin.1}) to other variables
connected with electron trajectories\cite{landau-mf}. These variables are the
integrals of motion $\Omega$ and $p_z$, and time $t_{\bf p}$ necessary for an
electron to move from some fixed point in momentum space to the point
$\bf p$. In the new variables the kinetic equation becomes
\begin{equation}
	\frac{\partial f}{\partial t_{\bf p}}
   = \frac{f_0(E_{\bf p})-f({\bf p})}{\tau}\;.
\label{eq:kin.3}
\end{equation}
The formal solution to this equation is
$f({\bf p}) = f_0(E_{\bf p}) + f_{1}({\bf p})$ where
\begin{equation}
   f_1({\bf p})
   =
   - \int_{-\infty}^{t_{\bf p}} eFv_{y}(t)
	f_{0}^{\prime}(E_{\bf p}(t))
   {\rm e}^{- (t_{\bf p} - t)/\tau} dt \;.
\label{eq:kin.4}
\end{equation}
Here $f_0'$ is the derivative of the distribution function. Terms depending
on time in $E_{\bf p}$ are of the order of $\Lambda$. If these terms are
neglected then $f_{0}^{\prime}$ can be considered time independent and
Eq.~(\ref{eq:kin.4}) becomes
\begin{equation}
   f_{1}({\bf p}) = - eF\tau Q_{\bf p} {df_0\over dE_{\bf p}} \;,
\label{eq:kin.6}
\end{equation}
where
\begin{equation}
    = \int_{-\infty}^{t_{\bf p}} v_{y}(t)
	{\rm e}^{- (t_{\bf p} - t)/\tau} dt =
	(1+\tau\partial/\partial{t}_{\bf p})^{-1} v_{y} \ .
\label{eq:kin.6a}
\end{equation}
is a periodic function of time. The Fourier series  representation of
$Q_{\bf p}$ is given in Appendix~\ref{sec:app3}.

Corrections to Eq.~(\ref{eq:kin.6}) contain $\Lambda^{n}f_{0}^{(n+1)}$ which
formally is of the order of $(\Lambda/T)^{n}$. The electric current, however,
contains integrals of this function and after an integration by parts,
$f_{0}^{(n+1)}$ in the integrand is replaced by a quantity of the order of
$f_{0}^{\prime}/\max(E_F,T)^n$. Therefore, for the calculation of current,
Eq.~(\ref{eq:kin.6}) is justified when
\begin{equation}
   \Lambda \ll \max(E_F,T) \;,
\label{eq:kin.5}
\end{equation}
which is assumed to be satisfied in this work.

The factor $Q_{\bf p}$ can be separated into two parts,
$Q_{\bf p} = Q_{\bf p}^x + Q_{\bf p}^y$ where $Q_{\bf p}^x$ is even and
$Q_{\bf p}^y$ is odd with respect to $v_y$. Only the odd part contributes to
the current in the $y$ direction. From Eqs.~(\ref{eq:din.1a}) and
(\ref{eq:din.2}) it is obvious that $v_y$ is odd in time so that the
parity with respect to $v_y$ is the same as the parity with respect to
time. For that reason $Q_{\bf p}^y$ is obtained from the second
expression in Eq.~(\ref{eq:kin.6a}) keeping only even derivatives of $v_y$.
Therefore
\begin{equation}
   Q_{\bf p}^y = {
      1
   \over
      1- \tau^2\partial^2/\partial{t}_{\bf p}^2
   } v_y \;.
\label{eq:kin.7}
\end{equation}

This function is easily calculated in two limiting cases. Far from the region
of closed trajectories given by the condition $\Omega\gg\Omega_B$, we have
$v_y=2v_\perp\sin(2\Omega{t})$.  Then
$\partial^2\sin(2\Omega{t})/\partial{t}^2 = -(2\Omega)^2\sin(2\Omega{t})$
and we obtain
\begin{equation}
    Q_{\bf p}^y =  {v_y\over1+\left(2\Omega\tau\right)^2}\;.
\label{eq:kin.8}
\end{equation}
Under the same condition, $2\Omega \approx \Omega_B|mF/B-p_x|/p_\Lambda$,
which means that $Q_{\bf p}^y$ as a function of $p_x$ has a resonance
at $p_x = mF/B$. The width of the resonance is $p_\Lambda/(\Omega_B\tau)$. If
this width is much larger than the width of the region of closed
trajectories, $p_\Lambda$, that is $\Omega_B\tau\ll1$, the latter can be
neglected and Eq.~(\ref{eq:kin.8}) can be used for all values of $p_x$.

The second limiting case is a strong magnetic field when $\Omega_B\tau\gg1$,
i.e., the width of the resonance in Eq.~(\ref{eq:kin.8}) is smaller than the
width of the region of closed trajectories. Then the unit in
Eq.~(\ref{eq:kin.7}) can be neglected compared to the second derivative, so
that
$-\tau^{2}(\partial^2Q_{\bf p}^y/\partial{t}^2) = v_y$.
The integration is carried out with the help of
the equation of motion, Eq.~(\ref{eq:din.2}), and the result is
\begin{equation}
    Q_{\bf p}^y =  {2v_{\perp}\over\left(\Omega_B\tau\right)^2}
    \left\{
       \begin{array}{ll}
          \displaystyle
          {p_yd\over\hbar} - {\pi t_{\bf p}\over T_{\bf p}},
          & \Omega > \Omega_B \\
          \displaystyle
          {p_yd\over\hbar} , & \Omega < \Omega_B \\
       \end{array}
    \right.
    \;,
\label{eq:kin.9}
\end{equation}
where half of the period $T_{\bf p}$ is given by Eq.~(\ref{eq:app3.6}).
One can verify that Eqs.~(\ref{eq:kin.8}) and (\ref{eq:kin.9}) have the same
asymptotic behavior when $1/\tau\ll\Omega_B\ll\Omega$.

These two limits show that the perturbation of the distribution
function due to an electric field has a resonant behavior. The physical
reasons of the resonance have been discussed in Sec.~\ref{sec:int}. If the
resonance region is wider than the region of closed trajectories the maximum
value of $Q_{\bf p}^y$ is of the order of $v_{\perp}$. If the relaxation time
is very long, $\Omega_B\tau\gg1$, the value of $Q_{\bf p}^y$ is reduced by
the factor of $(\Omega_B\tau)^{2}$ due to oscillations of $v_y$ along the
electron trajectory during the time $\tau$.


\section{ Calculation of the electric current}
\label{sec:con}

The part of the distribution function which depends only on $E_{\bf p}$ does
not contribute to the electric current, and the current in the field
direction is
\begin{equation}
   j = - 2e^{2}F\tau
   \int\frac{d{\bf p}}{(2\pi\hbar)^3}
   {df_0\over dE_{\bf p}}
   Q_{\bf p}^y v_y({\bf p})\;.
\label{eq:con.1}
\end{equation}
Two factors in the integrand, $df_{0}/dE_{\bf p}$ and $Q_{\bf p}^y$, have
maxima and the value of the integral crucially depends
on the relation between the widths of these maxima. The width of the
$Q_{\bf p}^y$ maximum depends on the value of $\Omega_B\tau$. If this product
is small then the width is $p_{\Lambda}/(\Omega_B\tau)$. In the opposite
case it is $p_{\Lambda}$. The width of the distribution function derivative
is of the order of $p_T\equiv\sqrt{2mT}$. Three cases are possible,
$(\Omega_B\tau)^2 \ll (p_\Lambda/p_T)^2 = \Lambda/T$ when the region
of closed trajectories is not important and the width of $Q_{\bf p}^y$ is
much larger than the width of $df_{0}/dE_{\bf p}$,
$\Lambda/T \ll (\Omega_B\tau)^{2} \ll 1$ when the region
of closed trajectories also is not important but the width of $Q_{\bf p}^y$
is much smaller than the width of $df_{0}/dE_{\bf p}$, and
$\Omega_B\tau \gg 1$ when only the vicinity of closed trajectories
contributes to the resonance. We consider these three cases separately.

\subsection{ Weak magnetic field, $(\Omega_B\tau)^2\ll\Lambda/T$}

In this case Eq.~(\ref{eq:kin.8}) can be used for computation of $Q_{\bf
p}^y$ on the right hand side of Eq.~(\ref{eq:con.1}). After the
integration with respect to $p_y$ one obtains
\begin{eqnarray}
   j_{y}
   &=&
   - { 4e^2Fl\tau\Lambda^2 \over \hbar^2}
	\int {dp_{x}dp_{z} \over (2\pi\hbar)^{2}}
\nonumber\\
   &\times&
	{df_0\over dE_{\bf p}}
   \left[1 + \left(\Omega_F\tau - \Omega_B\tau \
		{p_x \over p_{\Lambda}}\right)^{2}
   \right]^{-1} \;,
\label{eq:con.3}
\end{eqnarray}
where we introduced frequency of the Bloch oscillations
\[
   \Omega_F = {eFl \over \hbar}\;.
\]
Under the condition in the title of this subsection the
absolute value of the momentum in the integral is controlled by
$df_{0}/dE_{\bf p}$ and
\begin{eqnarray}
   j_{y}
   &=& {e^2Fl\tau\Lambda^2m \over \pi^2\hbar^4}
   \int_0^{2\pi}
   {
      d\phi
   \over
      1 + \left(
         \Omega_F\tau - \Omega_B\tau p_F\cos\phi/p_\Lambda
      \right)^{2}
   }
\nonumber \\
   &=& {\sqrt{2}e^{2}Fl\tau\Lambda^2m \over \pi\hbar^{4}}
   {\sqrt{r - \eta} \over r} \;,
\label{eq:con.4}
\end{eqnarray}
where
\begin{eqnarray}
   \eta
   &=&
   \Omega_F^2\tau^2 - \Omega_B^2\tau^2{ p_F^2\over p_\Lambda^2} - 1 \;,
\nonumber \\
   r^2
   &=&
   (\eta+2)^2 + 4\Omega_B^2\tau^2{p_F^2\over p_\Lambda^2}
\label{eq:con.5}
\end{eqnarray}
In the nondegenerate case the main contribution into the integral
Eq.~(\ref{eq:con.3}) is given by $p_x\lesssim{p}_T$ which is small compared to
$p_{\Lambda}/\Omega_B\tau$. If $p_x$ is neglected in the square brackets in
Eq.~(\ref{eq:con.3}) the result is independent of the magnetic field. So in
this case the magnetic field introduces just a small correction computed for
the first time by Epshtein\cite{Epshtein-79},
\begin{equation}
   j_{y} = {4e\Lambda^{2}n \over \hbar T}
   {\Omega_F\tau \over 1 + (\Omega_F\tau)^{2}}
		\left\{
   1 + \Omega_B^{2}\tau^{2} \ {T \over 2\Lambda} \
   {3(\Omega_F\tau)^{2} - 1 \over
      [1 + (\Omega_F\tau)^{2}]^{2}}
      \right\}\;,
\label{eq:con.6}
\end{equation}
where $n$ is the sheet concentration. Both of the results,
Eqs.~(\ref{eq:con.4}) and (\ref{eq:con.6}) describe an $I$-$V$ characteristic
that has a peak. At zero magnetic field
the peak position is given by the condition $\Omega_F\tau=1$.  With the
increase of the magnetic field this peak shifts to a higher bias.

Equation~(\ref{eq:con.6}) shows that magnetoresistance changes its sign
at $\Omega_F^2\tau^2 = 1/3$, or $F = \hbar l/(e\tau\sqrt{3})$. The
magnetoresistance is positive for a weaker electric field and negative for a
stronger one. For the Fermi statistics the same effect takes place when the
magnetic field goes to zero, which can be proved by expansion of
Eq.~(\ref{eq:con.4}).

\subsection{ Intermediate field, $\Lambda/T\ll(\Omega_B\tau)^2\ll1$ }

\begin{figure}
\begin{center}
\unitlength=1mm
\linethickness{0.4pt}
\begin{picture}(80.00,60.00)
\put(0.00,10.00){\vector(1,0){80.00}}
\put(40.00,10.00){\vector(0,1){50.00}}
\bezier{192}(30.00,29.00)(40.00,51.00)(50.00,29.00)
\bezier{156}(50.00,29.00)(59.00,10.00)(77.00,10.00)
\bezier{152}(3.00,10.00)(21.00,10.00)(30.00,29.00)
\put(52.00,10.00){\line(0,1){22.00}}
\put(58.00,10.00){\line(0,1){22.00}}
\linethickness{0.2pt}
\put(52.00,10.00){\line(1,1){6.01}}
\put(53.00,10.00){\line(1,1){5.01}}
\put(54.00,10.00){\line(1,1){4.01}}
\put(55.00,10.00){\line(1,1){3.01}}
\put(56.00,10.00){\line(1,1){2.01}}
\put(57.00,10.00){\line(1,1){1.01}}
\put(52.00,11.00){\line(1,1){6.01}}
\put(51.98,12.01){\line(1,1){5.72}}
\put(51.98,13.02){\line(1,1){5.19}}
\put(51.98,13.99){\line(1,1){4.71}}
\put(51.98,15.00){\line(1,1){4.23}}
\put(51.98,16.01){\line(1,1){3.79}}
\put(51.98,16.98){\line(1,1){3.35}}
\put(51.98,17.99){\line(1,1){2.91}}
\put(51.98,19.01){\line(1,1){2.47}}
\put(51.98,20.02){\line(1,1){2.03}}
\put(51.98,20.99){\line(1,1){1.67}}
\put(51.98,22.00){\line(1,1){1.23}}
\put(51.98,23.01){\line(1,1){0.84}}
\put(51.98,23.98){\line(1,1){0.44}}
\put(51.98,24.99){\line(1,1){0.13}}
\linethickness{0.4pt}
\put(52.00,30.00){\vector(1,0){6.00}}
\put(70.00,30.00){\vector(-1,0){18.00}}
\put(42.00,55.00){\makebox(0,0)[lc]{$\displaystyle{f}_0\left({p_x^2\over2m}\right)$}}
\put(64.00,32.00){\makebox(0,0)[cb]{$\Delta{p}$}}
\put(77.00,8.00){\makebox(0,0)[ct]{$p_x$}}
\put(55.00,8.00){\makebox(0,0)[ct]{$mF/B$}}
\put(55.00,10.00){\line(0,1){1.00}}
\end{picture}
\end{center}
   \caption{
       The division of the electron distribution into a nonresonant part and
       a resonant part(hatching area) in an intermediate and strong magnetic
       field. The width of the resonance, $\Delta{p}$, is $p_\Lambda$ or
       $p_\Lambda/(\Omega_B\tau)$, whichever is larger.
   }
\label{fig:resgroup}
\end{figure}
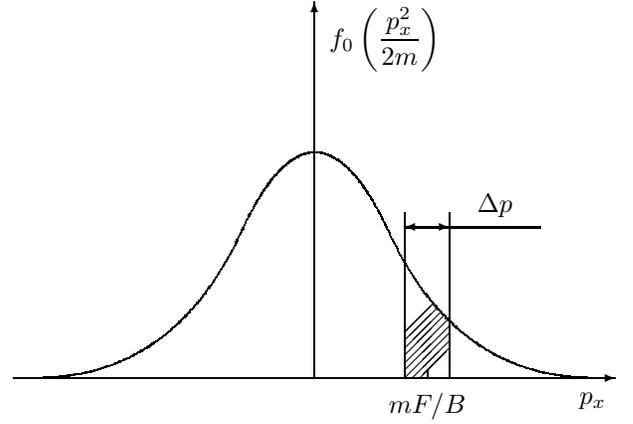

\begin{mathletters}
In this case Eq.~(\ref{eq:con.3}) also can be used. However, the value of
$p_x$ is controlled now by the Lorentz factor. This factor picks up
particles from the phase space, which have values of momentum near $p_x =
p_\Lambda \Omega_F/\Omega_B = mF/B$, see Fig.~\ref{fig:resgroup}.  Then the
integration with respect to $p_x$ gives
\begin{eqnarray}
   j &=& {
      e^{2}F l \Lambda^2 m
   \over
      \pi\hbar^4 \Omega_B
   }  \,\Phi(F,B)\;,
\nonumber\\ \relax
\label{eq:con.7a}
\\
   \Phi(F,B) &=& {p_{\Lambda}\over m}
   \int dp_z
   \left(
      {\partial f_0 \over \partial E}
   \right)_{p_x=mF/B}
\nonumber\\
   &=& \left\{
      \begin{array}{ll}
         \displaystyle
         {\sqrt{2}\hbar^2n\over (mT/\pi)^{3/2}} p_{\Lambda}
         {\rm e}^{-(mF/B)^2/p_T^2}\;,
         & T \gg E_F
         \\[7pt]
         \displaystyle
         { p_{\Lambda} \over \sqrt{
            p_F^2 - (mF/B)^2
         }}\;,
         & T \ll E_F
      \end{array}
   \right.
\nonumber\\ \relax
\label{eq:con.7b}
\end{eqnarray}
In the case of Fermi statistics, Eq.~(\ref{eq:con.7b}) is valid only if
$p_F-mF/B\gg{p}_T$. This result is also the limiting form of
Eq.~(\ref{eq:con.4}) when $p_F-mF/B\gg{p}_\Lambda/(\Omega_B\tau)$. Therefore
the reciprocal square root singularity obtained near the point $mF/B=p_F$ is
smeared either due to the temperature or due to the width of the resonance,
whichever is larger.  When $mF/B>p_F$ the current drops off and the position
of the current maximum is given by $mF/B=p_F$, i.e., $F=(p_{F}/m)B$. In the
case of Boltzmann statistics, the position of the current maximum is $mF/B
= p_T/\sqrt{2}$, i.e., $F = (p_{T}/m\sqrt{2})B$, meaning that in the
intermediate magnetic field limit the current peak shifts linearly with $B$.
\label{eq:con.7}\end{mathletters}

\subsection{ Strong magnetic field, $\Omega_B\tau \gg 1$ }

The calculation in this case is similar to that in the previous section. The
only difference is that it is necessary to use Eq.~(\ref{eq:kin.9}) for
$Q_{\bf{p}}^y$ and the integration with respect to $p_y$ can be carried out
in a different way.  After the separation of the integration with respect to
$p_{x}$
\begin{equation}
   j = {8R \over \pi^2} \
   {e^{2}m\Lambda^2 l \over \hbar^4\Omega_B^{2}\tau} \;
   F \ \Phi(F,B)\;,
\label{eq:con.9}
\end{equation}
where the dimensionless constant of the order of unity,
\begin{equation}
   R = \left(32\pi \ {p_{\Lambda}\hbar \over l} \
   {v_{\perp}^{2} \over (\Omega_B\tau)^{2}} \right)^{-1}
	\int Q_{\bf p}^y v_y dp_{x}dp_{y} \ ,
\label{eq:con.10}
\end{equation}
is calculated in Appendix.~\ref{sec:app2}.

It is instructive to note that Eq.~(\ref{eq:con.9}) can be written down in the
form similar to that in the bulk,
\begin{equation}
   j = {\tilde n e^2 \over m_\perp} \; {F\over\Omega_B^2\tau}\;.
\label{eq:con.11}
\end{equation}
Here $\tilde n$ is the effective electron concentration, i.e., the number of
the electrons inside the resonance region
\begin{equation}
   {\tilde n}
   = {4R\over\pi^2} \ {m\Lambda \over \hbar^{2}l} \ \Phi(F,B)
\label{eq:con.12}
\end{equation}
%


\section{ Estimation of the role of the Hall effect.}
\label{sec:hall}

Typically, in the current-voltage characteristic measurement in
superlattices, Hall contacts are not fabricated. Then, in a magnetic field, a
Hall voltage comes about which can affect the value of the current. As a
result of this effect in a weak electric field the quantity measured in the
experiment is not the conductivity $\sigma_{xx}$ but the resistivity
$\rho_{xx}$. In this section we show that the effect of the Hall voltage on
the current-voltage characteristic in a superlattice can be neglected.

The Hall field $F_{x}$ and the current in the growth direction can be
calculated from the equations
\begin{eqnarray}
   j_y &=& j_{y}^{(0)} + \sigma_{\text{yx}}F_x\;,
\label{eq:hall.1a} \\
   j_x &=& j_H + \sigma_{\text{xx}}F_x = 0 \;.
\label{eq:hall.1b}
\end{eqnarray}
where $j_{y}^{(0)}$ is the value of the current for zero Hall field and $j_H$
is the Hall current under the same condition. All the components of the
conductivity may depend on the field $F_{y}$. Both $j_{y}^{(0)}$ and $j_H$
nontrivially depend on $F_{y}$. The conductivity components
$\sigma_{\text{yx}}$ and $\sigma_{\text{xx}}$ also can depend on $F_{y}$.
This dependence results from a narrow miniband characterizing the electron
motion in this direction.  There is no such small energy scale for the
in-plane electron motion, so that the effect of $F_{x}$ on $j_{y}^{(0)}$,
$j_H$, and the conductivity components can be neglected.

In the calculation of the Hall current
\begin{equation}
   j_H = 2e \int\frac{d{\bf p}}{(2\pi\hbar)^3} \; v_x f_{1}({\bf p}) \;,
\label{eq:hall.2}
\end{equation}
it is convenient to make use of the identity
\begin{equation}
   \int_{-\pi\hbar/d}^{\pi\hbar/d} f_{1}({\bf p}) dp_{y} =
	- eB\tau {\partial \over \partial p_{x}}
   \int_{-\pi\hbar/d}^{\pi\hbar/d} v_{y} f_{1}({\bf p}) dp_{y} \;,
\label{eq:hall.3}
\end{equation}
which is obtained by the integration of the Boltzmann equation
(\ref{eq:kin.1}) with respect to $p_{y}$. The substitution of
Eq.~(\ref{eq:hall.3}) into Eq.~(\ref{eq:hall.2}) gives the result
\begin{equation}
   j_{H} = {eB\tau \over m} \ j_{y}^{(0)}
\label{eq:hall.4}
\end{equation}
which is very well known for the bulk semiconductors with a constant
relaxation time.

In order to calculate $\sigma_{xx}$ and $\sigma_{yx}$ it is necessary to
introduce into the Boltzmann equation a small $x$-component of the
electric field and calculate the perturbation of the distribution function,
$f_{2}$, caused by this field. After the linearization with respect to this
field, the Boltzmann equation becomes
\begin{equation}
   eF_x{\partial f \over \partial p_x}
   + {\partial f_{2} \over \partial t_{\bf p}}
   = - {f_{2} \over \tau}\;,
\label{eq:hall.5}
\end{equation}
where $f = f_{0} + f_{1}$ is the distribution function obtained in
Sec.~\ref{sec:kin} for the case $F_x=0$.

The integration of Eq.~(\ref{eq:hall.5}) with respect to $p_{y}$ gives the
identity
\begin{eqnarray}
\int_{-\pi\hbar/d}^{\pi\hbar/d} f_{2}({\bf p})
   dp_{y} &=&
\nonumber\\
   - eF_x\tau{\partial f \over \partial p_x}
   &-& eB\tau {\partial \over \partial p_{x}}
   \int_{-\pi\hbar/d}^{\pi\hbar/d} v_{y} f_{2}({\bf p})
   dp_{y}
   \;,
\label{eq:hall.6}
\end{eqnarray}
that results in the relation between the conductivities,
\begin{equation}
   \sigma_{\text{xx}}
   =
   {ne^2\tau\over md}
   + {eB\tau\over m}
   \sigma_{\text{yx}}\;.
\label{eq:hall.7}
\end{equation}
Therefore only one of them, say $\sigma_{yx}$, has to be calculated.

A formal solution to Eq.~(\ref{eq:hall.5}) is
\begin{equation}
   f_{2} =
   {-eF_x\tau \over 1+ \tau\partial/\partial t_{\bf p}}
   {\partial \over \partial p_x}
   { 1 \over 1+ \tau\partial/\partial t_{\bf p}}
   f_0(E_{\bf p})
\label{eq:hall.8}
\end{equation}
Then
\begin{equation}
\sigma_{\text{yx}} =
      \left\langle
         v_y
         { 1 \over 1+ \tau\partial/\partial t_{\bf p}} \
         {\partial \over \partial p_x} \
         { 1 \over 1+ \tau\partial/\partial t_{\bf p}}
         f_0(E_{\bf p})
      \right\rangle \ ,
\label{eq:hall.9}
\end{equation}
where
\begin{equation}
\langle \ ... \ \rangle
   = - 2e^2\tau \int\frac{d{\bf p}}{(2\pi\hbar)^3} \ .
\label{eq:hall.10}
\end{equation}

We will show now that $\sigma_{\text{yx}}$ can be expressed in terms of
$\sigma_{\text{yy}}$ which is defined by the relation
$j_y^{(0)}=\sigma_{\text{yy}}F_y$. With the help of the definition
Eq.~(\ref{eq:hall.10})
\begin{equation}
   \sigma_{\text{yy}}
   =
   \left\langle
      v_y { 1 \over 1+ \tau\partial/\partial t_{\bf p}} v_y
      {\partial f_0 \over \partial E}
   \right\rangle \;.
\label{eq:hall.11}
\end{equation}

With the help of the commutation relation
\begin{equation}
	\left[
      {\partial \over \partial t_{\bf p}},\;
      {\partial \over \partial p_{x}}
   \right]
   = {eB \over m} \; {\partial \over \partial p_{y}}
\label{eq:hall.12}
\end{equation}
it is easy to show that
\begin{eqnarray}
   &&
   \left(1 + \tau \ {\partial \over \partial t_{\bf p}} \right)^{-1}
	{\partial \over \partial p_{x}} =
	{\partial \over \partial p_{x}}
   \left(1 + \tau \ {\partial \over \partial t_{\bf p}} \right)^{-1}
\nonumber\\
   &-&
	{eB\tau \over m}
	\left(1 + \tau \ {\partial \over \partial t_{\bf p}} \right)^{-1}
		{\partial \over \partial p_{y}}
	\left(1 + \tau \ {\partial \over \partial t_{\bf p}} \right)^{-1} \ .
\label{eq:hall.13}
\end{eqnarray}
The substitution of Eq.~(\ref{eq:hall.13}) into Eq.~(\ref{eq:hall.9}) gives
\begin{eqnarray}
   \sigma_{\text{yx}}
   &=& - {eB\tau\over m}
\nonumber\\
   &\times&
      \left\langle
         v_y
         { 1 \over 1+ \tau\partial/\partial t_{\bf p}}
         {\partial \over \partial p_y}
         \left(
            { 1 \over 1+ \tau\partial/\partial t_{\bf p}}
         \right)^2
         f_0(E_{\bf p})
      \right\rangle
\nonumber\\
\relax
\label{eq:hall.14}
\end{eqnarray}
Now we make use of the commutation relation
\begin{equation}
	\left[
{\partial \over \partial t_{\bf p}} , \	{\partial \over \partial p_{y}}
	\right] = - {eB \over m} \ {dv_{y} \over dp_{y}} \
	{\partial \over \partial p_{y}}
\label{eq:hall.15}
\end{equation}
to estimate the commutator of the operator factors in Eq.~(\ref{eq:hall.14}).
If we take into account that $\partial/\partial t_{\bf p} \sim \Omega$ then
the ratio of the commutator of $\partial/\partial p_{y}$ and
$(1+ \tau\partial/\partial t_{\bf p})^{-1}$ to the product of these operators
is of the order of
$[{\Omega_B\tau/( 1 + \Omega\tau})] \; (p_\Lambda/p_x)$. The last quantity
is much smaller than unity due to Eq.~(\ref{eq:kin.5}) and $\Omega_B
\lesssim \Omega$, which means the commutator of $\partial/\partial{p}_{y}$ and
$(1+\tau\partial/\partial{t}_{\bf{p}})^{-1}$ can be neglected in
Eq.~(\ref{eq:hall.14}). Then Eq.~(\ref{eq:hall.14}) becomes
\begin{equation}
   \sigma_{\text{yx}}
   = - {eB\tau\over m}
   \left\langle
      v_y
      \left(
         { 1 \over 1+ \tau\partial/\partial t_{\bf p}}
      \right)^3
      v_y
      {\partial f_0 \over \partial E}
   \right\rangle
\label{eq:hall.16}
\end{equation}
The expression above is proportional to the second derivative of the
$\sigma_{\text{yy}}$ with respect to the field, which can be proved as
follows. First of all Eq.~(\ref{eq:hall.11}) gives
\begin{eqnarray}
   && {\partial^2 \over \partial F_y^2 }\;
   \sigma_{\text{yy}} = 2e^2\tau^2\;\Biggl\langle v_y
   { 1 \over 1+ \tau\partial/\partial t_{\bf p}}
\nonumber\\
   &\times&
   {\partial \over \partial p_{y}}\;
   { 1 \over 1+ \tau\partial/\partial t_{\bf p}}\;
   {\partial \over \partial p_{y}}\;
   { 1 \over 1+ \tau\partial/\partial t_{\bf p}}\;
   v_y
   {\partial f_0 \over \partial E}
   \Biggr\rangle
\nonumber\\
   &=& - 2\left({e\tau d\over\hbar }\right)^2
   \left\langle
      v_y
      \left(
         { 1 \over 1+ \tau\partial/\partial t_{\bf p}}
      \right)^3
      v_y
      {\partial f_0 \over \partial E}
   \right\rangle\;.
\label{eq:hall.17}
\end{eqnarray}
Here we neglect the commutator of $\partial/\partial{p}_{y}$ and
$(1+\tau\partial/\partial{t}_{\bf{p}})^{-1}$ once again, and
make use of $\partial^2v_y/\partial{p}_{y}^2  = -(d^2/\hbar^2)v_y$.
The comparison of Eq.~(\ref{eq:hall.16}) with Eq.~(\ref{eq:hall.17}) gives
\begin{equation}
   \sigma_{\text{yx}}
   = {eB\tau\over 2m}
   \left({\hbar\over e\tau l}\right)^2
      {\partial^2\sigma_{\text{yy}}\over \partial F_y^2 } \ .
\label{eq:hall.18}
\end{equation}
In the limit of a small $F_y$ one can easy verify that
$-\sigma_{\text{yx}}=\sigma_{\text{xy}}\equiv{}j_H/F_y$.

Now we can estimate the value of the second term in Eq.~(\ref{eq:hall.1a}).
Making use of Sec.~\ref{sec:con} we have the relation of the second term
in Eq.~(\ref{eq:hall.1a}) to the first one
\begin{eqnarray}
   { \sigma_{\text{yx}} F_x \over j_0 }
   &=&
   {eB\tau\over m} {\sigma_{\text{yx}}\over\sigma_{\text{xx}}}
\nonumber\\
   &\sim& \left\{
   \begin{array}{ll}
      \displaystyle
      {\Omega_B^2\tau^2}{\Lambda\over E_F,T}\;,
   &  \displaystyle
      {\Omega_B^2\tau^2}\ll{\Lambda\over E_F,T}
   \\[7pt]
      \displaystyle
      {1\over\Omega_B\tau}\left({\Lambda\over E_F,T}\right)^{5/2}\;,
   &  \displaystyle
      {\Lambda\over E_F,T} \ll{\Omega_B^2\tau^2} \ll 1
   \\[7pt]
      \displaystyle
      {1\over\Omega_B^2\tau^2}\left({\Lambda\over E_F,T}\right)^{5/2}\;,
   &  \displaystyle
      1\ll{\Omega_B^2\tau^2}
   \end{array}   \right.
\nonumber\\
   \relax
\label{eq:hall.19}
\end{eqnarray}
and one can see that this ratio is small in all considered limits.


\section{ Discussion}
\label{sec:dmp}

One of the most interesting results of Sec.~\ref{sec:con} is that in some
region of magnetic field, $\Lambda/T\ll(\Omega_B\tau)^2\ll1$, the current
does not depend on the relaxation time, see Eq.~(\ref{eq:con.7}). That means
that a finite resistivity of a superlattice exists even without scattering.
To understand the physical reason of this result it is instructive to note
that the considered problem is quite similar to the propagation of a
longitudinal wave in a collisionless plasma. Indeed, if we consider the
miniband width as a small parameter
($v_{y} = [2\Lambda l/\hbar]\sin[p_{y}l/\hbar]$ and
for $\Lambda = 0$ the current in the growth direction is zero) then in
the linear approximation in $\Lambda$ and neglecting collisions,
Eq.~(\ref{eq:kin.1}) takes the form
\begin{equation}
eB\left(v_x - {F \over B}\right)
	\frac{\partial f}{\partial p_y} -
   \frac{2eB\Lambda l}{\hbar} \ \sin {p_{y}l \over\hbar} \
	\frac{\partial f_{0}(E_{\bf p})}{\partial p_x} = 0\;.
\label{eq:dmp.1}
\end{equation}
On the other hand, the Boltzmann equation for a collisionless plasma with
an electric field in the wave $F_{0}\sin k(x - v_{p}t)$ is\cite{landau-dmp}
\begin{equation}
(v_{x} - v_{p}) \frac{\partial f}{\partial x} +
	eF_{0} \sin k(x - v_{p}t) \
	\frac{\partial f_{0}(E_{\bf p})}{\partial p_x} = 0\;.
\label{eq:dmp.2}
\end{equation}
The comparison of these equations shows that there is a one to one
correspondence
\begin{center}
\begin{tabular}{cc|cc}
     plasma            &~~~&~~~& superlattice             \\
   $x - v_{p}t$        &   &   & ${p}_y/eB$               \\
   $k$                 &   &   & $eBl/\hbar$              \\
   $v_{p}$             &   &   & $F/B$                    \\
   $eF_{0}$            &   &   & $- 2eB\Lambda l/\hbar $  \\
\end{tabular}
\end{center}

In a plasma there exists a resonant group of electrons moving with the 
velocity close to the wave velocity, $v_{p}$. This group of electrons 
strongly interacts with the wave, resulting in the collisionless Landau 
damping.  A similar resonant group of electrons exists in a superlattice. 
Those are electrons moving in the $x$ direction with the Hall drift velocity, 
$F/B$.  For those electrons the Lorentz force cancels out with the force of 
the electric field and their velocity in the $y$ direction does not oscillate 
in time. As a result their contribution to the current is nonzero.

The correspondence between a monochromatic plasma wave and a superlattice in
crossed electric and magnetic fields can be extended even beyond the linear
Landau damping in plasma and the linear in $\Lambda$ theory in a superlattice.
Because of some specifics of collisions in a gas plasma, a close
correspondence in this case exists between the superlattice and the solid
state plasma\cite{galperin-72}. We show here only the correspondence
between the applicability conditions for the linear theory.

In plasma a wave with a finite amplitude traps the resonant electrons and
they oscillate in potential wells of the wave.  The linear theory is
applicable if the period of the oscillation is larger than a scattering time
which can be written as $eF_{0}k\tau^{2}/m \ll 1$.  The above correspondence
between the plasma and superlattice quantities immediately gives the
condition for the superlattice, $\Omega_B^{2}\tau^{2} \ll 1$. The physical 
meaning of that condition is quite similar to the plasma one. A finite width 
of the miniband in the superlattice leads to Bloch oscillations of resonant 
electrons. Then the theory which is linear in $\Lambda$ is justified if the 
period of the oscillation is larger than the relaxation time.

\begin{figure}
\begin{center}
\unitlength=1mm
\linethickness{0.4pt}
\begin{picture}(86.50,86.50)
\put(5.00, 5.00){
   \put(00.00,00.00){\insertplot{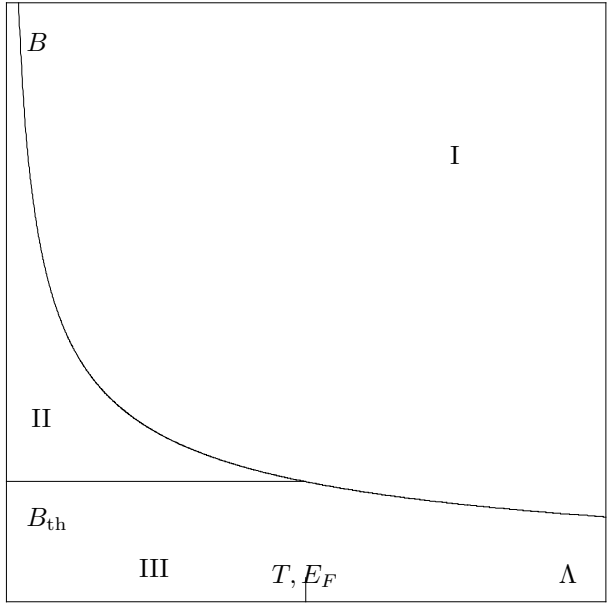}}
   \put(75.00,3.00){\makebox(0,0)[cb]{$\Lambda$}}
   \put(40.00,3.00){\makebox(0,0)[cb]{$T,E_F$}}
   \put(3.00,75.00){\makebox(0,0)[lc]{$B$}}
   \put(3.00,13.00){\makebox(0,0)[lt]{$B_{\rm th}$}}
   \put(05.00,25.00){\makebox(0,0)[cc]{II}}
   \put(60.00,60.00){\makebox(0,0)[cc]{I}}
   \put(20.00,05.00){\makebox(0,0)[cc]{III}}
}
\end{picture}
\end{center}
   \caption{
      I,-region of the strong magnetic field, II,-region of the collisionless
      conductivity, and III,-region of the weak magnetic field. The threshold
      value of the magnetic field,
      $B_{\rm{th}}\sim\hbar{m}/[el\tau(p_F,p_T)]$, is only a few T for
      superlattices.  }
\label{fig:bllimits}
\end{figure}

The situation when the collision term can be neglected in the kinetic
equation is typical for plasma and given by the condition\cite{landau-dmp}
$kp_T\tau/m\gg1$.  Regarding the superlattice, this condition becomes the
condition of the intermediate magnetic field
$\Omega_B\tau(p_T/p_\Lambda)\gg1$.  In the case of the degenerate electron
gas important for superlattices, the collisions can be neglected under a more
weak condition.  Equation.~(\ref{eq:con.3}) takes the form of the
Eq.~(\ref{eq:con.7}) for $E_F\gg{T}$, when $\Omega_B\tau(p_F/p_\Lambda)\gg1$, 
i.e. the range of the magnetic field when the collisionless 
conductivity takes place in superlattices is 
\begin{equation}
   {\Lambda\over E_F, T} \ll \Omega_B^2\tau^2 \ll 1\;.
\label{eq:dmp.3}
\end{equation}
The existence of this effect depends on the relation between miniband width
and characteristic electron energy which is schematically shown in the diagram
Fig.~\ref{fig:bllimits}.  The plane $B$-$\Lambda$ in this diagram is divided
into three parts corresponding to the different regimes. The collisionless
conductivity (intermediate magnetic field regime) can be observed only if
$\Lambda\ll{E_F,T}$, region II in the diagram Fig.~\ref{fig:bllimits}. The
strong magnetic field limit, region I in Fig.~\ref{fig:bllimits}, is the same
for small and large $\Lambda$; electric current exhibits $1/\tau$ dependence.
The positive or negative magnetoresistance can be observed in the regime of
the weak magnetic field, region III in Fig.~\ref{fig:bllimits},  depending on
the electric field. Due to the large period of the superlattice the boundary
between regions II and III lies in the experimentally available range of the
magnetic fields.


\section{ Comparison with experiment}
\label{sec:comp}

\begin{figure*}
\begin{picture}(87,95)
\put(6.,14.){
   \insertplot{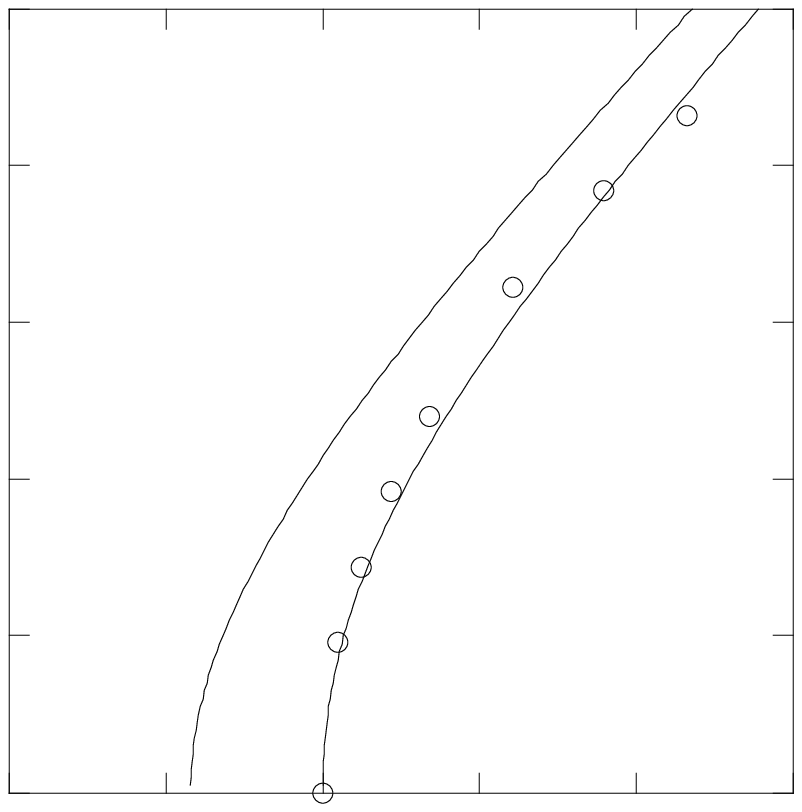}
   {\footnotesize
   \put(-1.50,40.00){\makebox(0,0)[rc]{$\displaystyle{B\over B_{\rm th}}$}}
   \put(40.00,-6.00){\makebox(0,0)[ct]{$F/F_{\rm th}$}}
   \put(29.00, 45.00){\makebox(0,0)[cc]{%
   $\displaystyle {\partial j\over\partial B} (F,B) = 0$}}
   \put(48.00, 25.00){\makebox(0,0)[cc]{%
   $\displaystyle {\partial j\over\partial F} (F,B) = 0$}}
   \put(00.00,-0.20){\makebox(0,0)[ct]{0.0}}
   \put(16.00,-0.20){\makebox(0,0)[ct]{0.5}}
   \put(32.00,-0.20){\makebox(0,0)[ct]{1.0}}
   \put(48.00,-0.20){\makebox(0,0)[ct]{1.5}}
   \put(64.00,-0.20){\makebox(0,0)[ct]{2.0}}
   \put(80.00,-0.20){\makebox(0,0)[ct]{2.5}}
   \put(-0.20,16.00){\makebox(0,0)[rc]{0.5}}
   \put(-0.20,32.00){\makebox(0,0)[rc]{1.0}}
   \put(-0.20,48.00){\makebox(0,0)[rc]{1.5}}
   \put(-0.20,64.00){\makebox(0,0)[rc]{2.0}}
   \put(-0.20,80.00){\makebox(0,0)[rc]{2.5}}
   }
   \put(40.00,77.00){\makebox(0,0)[ct]{(a)}}
}
\end{picture}~\begin{picture}(87,95)
\put(6.00,14.00){
   \insertplot{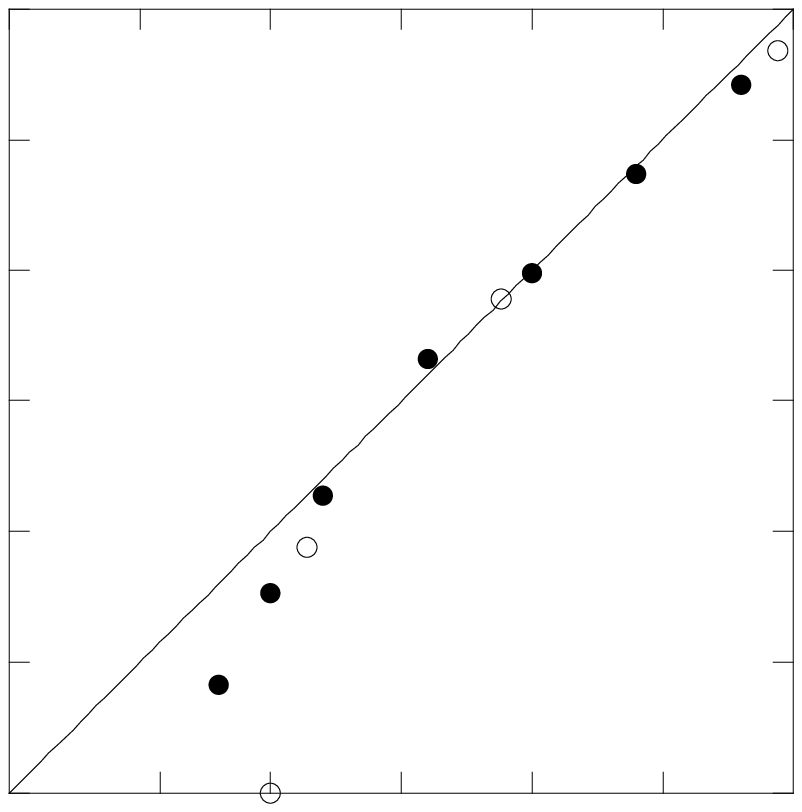}
   {\footnotesize
   \put(-1.5,44){\makebox(0,0)[rb]{
      $\displaystyle{B\over B_{\rm th}}$}}
   \put(40.00,-6.00){\makebox(0,0)[ct]{$F/F_{\rm th}$}}
   \put(00.00,-0.20){\makebox(0,0)[ct]{0.0}}
   \put(15.39,-0.20){\makebox(0,0)[ct]{$1/\sqrt{3}$}}
   \put(26.66,-0.20){\makebox(0,0)[ct]{1.0}}
   \put(40.00,-0.20){\makebox(0,0)[ct]{1.5}}
   \put(53.33,-0.20){\makebox(0,0)[ct]{2.0}}
   \put(66.66,-0.20){\makebox(0,0)[ct]{2.5}}
   \put(80.00,-0.20){\makebox(0,0)[ct]{3.0}}
   \put(-0.20,13.33){\makebox(0,0)[rc]{0.5}}
   \put(-0.20,26.66){\makebox(0,0)[rc]{1.0}}
   \put(-0.20,40.00){\makebox(0,0)[rc]{1.5}}
   \put(-0.20,53.33){\makebox(0,0)[rc]{2.0}}
   \put(-0.20,66.66){\makebox(0,0)[rc]{2.5}}
   \put(-0.20,80.00){\makebox(0,0)[rc]{3.0}}
   }
   \put(40.00,77.00){\makebox(0,0)[ct]{(b)}}
}
\end{picture}
   \caption{
      The position of the maximum in \protect{$I$}-\protect{$V$} 
      characteristics of the superlattice, empty circles, and on 
      \protect{$I$}-\protect{$B$} characteristics, filled circles. The solid 
      lines are theoretical predictions.  
   } 
\label{fig:bfpfermi} \label{fig:bfpbolts}
\end{figure*}

The influence of the electric and magnetic fields on the current in the
superlattice was measured in two types of experiments. In the first kind of
experiments\cite{choi4,aristone-93} the $I$-$V$ curve of the superlattice
was measured in the presence of the constant magnetic field. The position of
the current peak depends on the magnetic field and this dependence can be
plotted in the $B$-$F$ diagram. In the other kind of
experiments\cite{aristone-93,herbert-jan94} the dependence of the electric
current on the magnetic field was measured in the presence of a constant
electric field. A peak in the $I$-$B$ curve has been detected, and its 
position can be plotted as a function of the electric field in the $B$-$F$ 
diagram.

We describe now the predictions of the present theory for such experiments. 
The peak on the $I$-$V$ curve can be obtained from the equation $dj/(dF)=0$. 
In the absence of a magnetic field the position of the peak is given by 
$\Omega_F\tau=1$, or $F=F_{\rm{th}}\equiv\hbar/(el\tau)$. When a weak 
magnetic field is applied, $B{\ll}B_{\rm{th}}$, where
\begin{equation}
   B_{\rm{th}} =
   \left\{ \begin{array}{ll}
      \displaystyle {\hbar m \over e p_F l \tau}\;,
      & E_F \gg T\;,
      \\[7pt]
      \displaystyle \sqrt{2}{\hbar m \over e p_T l \tau}\;,
      & E_F \ll T\;,
   \end{array}\right.
\label{eq:comp.1}
\end{equation}
the peak position is shifted by small amount $F-F_{\rm{th}} \ll F_{\rm{th}}$.
In the lowest order of the magnetic field the shift obtained from
Eqs.~(\ref{eq:con.4}) and (\ref{eq:con.6}) is
\begin{equation}
   F - F_{\rm{th}}
   \approx
   {eB^2d\over2m}
   {\max\left(E_F, T \right)\over\hbar/\tau}\;.
\label{eq:comp.7}
\end{equation}
With the increase of the magnetic field $F-F_{\rm{th}}$ becomes comparable
with $F_{\rm{th}}$. This happens when $B{\gtrsim}B_{\rm{th}}$ and in this
limit current is given by Eq.~(\ref{eq:con.7}). In this interval of the
magnetic field the peak position changes linearly
\begin{equation}
   F = \max\left(v_F, v_T/\sqrt{2}\right)B
     = {F_{\rm{th}} \over B_{\rm{th}}}B\;.
\label{eq:comp.8}
\end{equation}

The peak in the $I$ -- $B$ characteristic is given by the equation
$\partial{j}/\partial{B}=0$. In the limit $B$ goes to zero, 
Eqs.~(\ref{eq:con.4},\ref{eq:con.6}) give the solution
\begin{equation}
   \left({F \over F_{\rm th}}\right)^2
   = \left(\Omega_F\tau\right)^2  = {1\over3}
\label{eq:comp.10}
\end{equation}
When $B{\gtrsim}B_{\rm{th}}$, the solution to equation 
$\partial{j}/\partial{B}=0$ coincides with the solution to the equation 
$\partial{j}/\partial{F}=0$, Eq.~(\ref{eq:comp.8}), because the current is a 
function of $F/B$, see Eq.~(\ref{eq:con.7}). That is true unless 
$\Omega_B\tau\gtrsim1$. In the last case the current acquires the factor 
$1/B$, Eq.~(\ref{eq:con.9}), and the peak in the $I$ -- $B$ curve is shifted 
to weaker $B$. In the case of the nondegenerate electron gas and in the limit 
$\Omega_B\tau\gg1$ we have the peak position given by $F=v_TB$.  Therefore, 
the peak in the $I$ -- $B$ curve lies between $F/(\sqrt{2}v_T)$ and $F/v_T$ 
depending on $\Omega_B\tau$.  In the experiment the peak position in the 
$I$ -- $B$ curve  is measured as a function of the applied bias, $V=LF$, 
where $L$ is the superlattice length, and we have the following useful 
inequality for the bias corresponding to the peak position 
\begin{equation}
\sqrt{T \over m}LB \le V \le \sqrt{2T \over m}LB
   \;,\;\;  B{\gtrsim}B_{\rm{th}}
   \;,\;\;  E_F\ll T\;.
\label{eq:comp.11}
\end{equation}

In the work of Choi et al\cite{choi4} the  $B$ -- $F$ diagram was reported. 
The superlattice investigated in this
experiment has $N=49$ periods, the length of the period is $l=166${\AA}, the
width of the first miniband is $4\Lambda=0.47$meV, and the concentration of
the carriers corresponds to the Fermi energy $E_F=6$meV. At zero magnetic
field the current peak appears near the bias $V_{\rm{th}} = NlF_{\rm{th}} =
590$mV, and therefore the momentum relaxation time is
$\tau = \hbar{N}/(eV_{\rm{th}}) = 0.54\times10^{-13}$sec. The critical value
of the magnetic field is $B_{\rm th}=4.2$T. The experimental data from
Fig.~7 of this work is plotted in Fig.~\ref{fig:bfpfermi}(a), empty circles.
The solid lines on Fig.~\ref{fig:bfpfermi}(a) are obtained from
Eq.~(\ref{eq:con.4}), since in this experiment $\Omega_B\tau\ll1$ for the
whole range of magnetic fields. The upper curve represents the maximum
position on the $I$ -- $B$ characteristic and the lower curve represents the
maximum position on the $I$ -- $F$ characteristic. One can see a good 
agreement between the lower curve and the experimental points without any 
fitting parameters.

The both types of measurements (peak position on the $I$ -- $B$ curve versus
electric field and maximum of the $I$ -- $F$ characteristic versus magnetic
field) have been reported in the work of Aristone et al\cite{aristone-93}.
This experiment has been done on low doped samples with $E_F$ much
smaller than the temperature $T=23$meV.  The narrowest miniband sample
$4\Lambda=4$meV has a period $l=74${\AA} and a length $Nl=1\mu$m. The
critical value of the applied bias is $V_{\rm{th}}=500$mV, and therefore the
momentum relaxation time is
$\tau=\hbar{N}/(eV_{\rm{th}})=1.7\times10^{-13}$sec. Thus, the critical value
of the magnetic field is $B_{\rm{th}}=2.1$T. The experimental data from
this work is plotted on Fig.~\ref{fig:bfpfermi}(b).  The empty circles
represent the peak position on the $I$-$V$ characteristic taken from Fig.~4
of Ref.~\onlinecite{aristone-93}, and the filled circles represent the peak
position on the $I$-$B$ characteristic taken from Fig.~6 of
Ref.~\onlinecite{aristone-93}. On the same graph, Fig.~\ref{fig:bfpfermi}(b),
we draw the line Eq.~(\ref{eq:comp.8}) that is asymptotic to the peak
positions for intermediate magnetic field. The theoretical prediction
for the weak magnetic field regime is not shown on the graph, but one can see 
that the filled circles (the peak positions on the $I$-$B$ characteristic 
measured experimentally) converge to the point $B=0,\;F=F_{\rm th}/\sqrt{3}$ 
in agreement with Eq.~(\ref{eq:comp.10}). The good agreement between theory 
and experiment is obtained without any fitting parameters.

In the experiment of Aristone the samples with the wider minibands were
investigated too.  The position of the peak on the $I$-$B$ characteristics of
these samples exhibits a weak dependence on the miniband width, that is in
agreement with the present theory, since for these samples the limit
$\Omega_B\tau\sim1$ is reached.  One can check that the inequality
Eq.~(\ref{eq:comp.11}) is fulfilled for this data.


\section{ Conclusions}
\label{sec:cncl}

In conclusion, we solved analytically the kinetic equation for the
superlattice in crossed electric and magnetic fields, under the assumption
that the miniband width is small compared to the Fermi energy or the 
temperature.  Our calculations of the position of the maximum on the $I$-$V$ 
and $I$-$B$ characteristics are in quantitative agreement with 
experiment. We also found that in a certain range of the magnetic field the 
current is independent of the electron relaxation time.


\section{ Appendix A}
\label{sec:app3}

The Fourier series expansion of the velocity is
\begin{equation}
   v_y(t) = 4v_\perp
   \left(\frac{\pi}{\Omega_BT_{\bf p}}\right)^2
	\sum_{n=1}^{\infty}l_n\mbox{sech}
	\left(\frac{\pi l_n K'}{K}\right)
   \sin\left(\frac{\pi l_n t}{T_{\bf p}}\right)
\label{eq:app3.1}
\end{equation}
The application of the identity
\begin{equation}
   {1\over1+\tau\partial/\partial{t}} \sin(\omega t) =
   {\sin(\omega t) - \omega\tau \cos(\omega t) \over 1 + (\omega\tau)^2}\;
\label{eq:app3.2}
\end{equation}
to each term in the series Eq.~(\ref{eq:app3.1}) results in the following
series representation for the factor $Q_{\bf{p}}$:
\begin{eqnarray}
   Q_{\bf p}
   &=&
   4v_{\perp}\left({\pi\over\Omega_BT_{\bf p}}\right)^2
      \sum_{n=1}^\infty
      l_n
      \text{sech}\left(\pi l_nK'(k)\over K(k)\right)
\nonumber\\
   &\times&
   \left[
      { \sin\left(\pi l_n t_{\bf p}/T_{\bf p}\right)
           \over
        1 + \left(\pi l_n \tau/T_{\bf p}\right)^2}
   -  {\pi l_n \tau\over T_{\bf p}}\;
      { \cos\left(\pi l_n t_{\bf p}/T_{\bf p}\right)
           \over
        1 + \left(\pi l_n \tau/T_{\bf p}\right)^2}
   \right],
\label{eq:app3.4}
\\
  l_n &=& \left\{
  \begin{array}[c]{ll}
     n & \Omega\; > \;\Omega_B \\
     n - 1/2  &  \Omega\; < \;\Omega_B
  \end{array}\right.\;,
\label{eq:app3.5}
\\
   T_{\bf p}
   &=&
   { K(k) \over \text{max}\left(\Omega,\Omega_B\right)}\;,\;\;\;
   k = \text{min}\left({\Omega\over\Omega_B}, {\Omega_B\over\Omega}\right)\;.
\label{eq:app3.6}
\end{eqnarray}
Here $K(k)$ is a Complete Elliptic Integral of the first kind, and
$K'(k)=K(\sqrt{1-k^2})$. The division of the factor $Q_{\bf{p}}$ into parts
odd and even with respect to $v_y$ corresponds to the two terms in the
square brackets in Eq.~(\ref{eq:app3.4}). If one keeps only the left term
then the sum will result in $Q_{\bf{p}}^y$, and if one keeps only the
right term then the sum will result in $Q_{\bf{p}}^x$.

\section{ Appendix B}
\label{sec:app2}

The substitution of Eq.~(\ref{eq:kin.9}) into Eq.~(\ref{eq:con.10}) gives
\begin{eqnarray}
   R &=& \left(16\pi \ \Lambda p_{\Lambda} \right)^{-1}
\nonumber\\
   &\times&
   \int \left[
   {p_y l\over\hbar} -
      {\pi t_{\bf p}\over T_{\bf p}} \ \theta(\Omega^{2} - \Omega_B^{2})
	 \right] v_y dp_{x}dp_{y} \ ,
\label{eq:app2.1}
\end{eqnarray}
where $\theta(x)$ is the step function. The integral with respect to $p_{y}$
of the first term in the integrand is trivial but the next integration of
this term with respect to $p_{x}$ diverges. In the second term it is
convenient to carry out the integration in variables $\Omega$ and $t$. The
integral with respect to $\Omega$ also diverges. The sum of the integrals has
to converge.  So it is convenient to put the limits of the integration in the
first term $p_{x} = - p_{M}$ and $p_{x} = p_{M}$. In the second term the
limits are
$\Omega = \Omega_F/2 + \Omega_{M}$ and
$\Omega = \Omega_F/2 - \Omega_{M}$ respectively, where
$\Omega_{M} = \Omega_Bp_{M}/2p_{\Lambda}$. The result is obtained in the
limit $p_{M} \rightarrow \infty$.  Then keeping in mind that the Jacobian of
the transformation from variables $p_{x},\;p_{y}$ to $\Omega,\;t$ is
$2(\Omega/\Omega_B) (p_{\Lambda}\hbar/l)$ and calculating the integrals of
the elliptic functions with the help of Eqs.~(5.136.1) and (5.133.3)  of
Ref.~\onlinecite{gradsteyn} we have
\begin{eqnarray}
   R
   & = & {p_{M} \over 2p_{\Lambda}} -
   \int_{\Omega_B}^{\Omega_{M} - \Omega_F/2 }
   {\Omega^{2} d\Omega \over \Omega_B^{3}K(\Omega_B/\Omega)}
\nonumber\\
   &\times&
   \left[ \pi - 2K(\Omega_B/\Omega)
      \sqrt{1 - {\Omega_B^{2} \over \Omega^{2}}} \right]
\nonumber \\ && \hspace{1cm} -
   \int_{\Omega_B}^{\Omega_{M} + \Omega_F/2}
   {\Omega^{2} d \Omega \over \Omega_B^{3}K(\Omega_B/\Omega)}
\nonumber\\
   &\times&
   \left[ \pi - 2K(\Omega_B/\Omega)
      \sqrt{1 - {\Omega_B^{2} \over \Omega^{2}}} \right] \ .
\label{eq:app2.2}
\end{eqnarray}
When $\Omega/\Omega_B$ goes to $\infty$ the integrand here goes to $1/2$.
This limit value can be integrated separately and then
\begin{equation}
   R = 1 + \int_{1}^{\infty}
		\left[
	1 + 4k \sqrt{k^{2} - 1} - {2\pi k^{2} \over K(1/k)}
      \right] dk \approx 0.747
\label{eq:app2.3}
\end{equation}

\chapter{ Transverse magnetoresistance. II }
\label{chap:deriv}

In this chapter we derive the quantum kinetic equation, which is a valid
approach for any relation between four energies $\Lambda$, $\hbar/\tau_p$,
$eFl$, $\hbar\Omega_B$. Such a general case can be described quantitatively
with the help of the density matrix if an effective electron temperature is
large enough. The qualitative results are similar to the results of
Chap.~\ref{chap:trans}; there exists a resonant group of electrons with $v_x$
near $F/B$ which gives the main contribution to the current.

We showed in Chap.~\ref{chap:trans} that the current has a maximum when $F/B$
is close to the characteristic electron in-plane velocity. Therefore, the
ratio $F/B$ near the current peak allows one to measure such properties of the
in-plane electron distribution function as electron temperature or Fermi
energy. If the width of the miniband is smaller than the energy of optical
phonons, the cooling of electron gas becomes
slow\cite{suris-jul84,laikhtman2}.  The effective electron temperature in the
region of applied fields near the current peak can be significantly larger
than the crystal temperature\cite{laikhtman2}. Therefore, we need to have
theoretical predictions for this temperature; it can be compared with
experimentally measured $m(F/B)^2$ near the current peak.

Therefore, the purpose of this chapter is two-fold. We want to repeat part of
the calculations of Chap.~\ref{chap:trans} starting from the quantum kinetic
equation, and we want to see the effect of heating of the electron gas.

Levinson and Yasevichute\cite{levinson-may72} considered for the first time
the heating of an electron gas in an anisotropic semiconductor or
superlattice.  They wrote down the quantum kinetic equation and considered
simultaneously two effects:  Stark localization and the heating of electron
gas.  Their results were obtained in the limit $eFl\gg\hbar/\tau$, far away
from the current peak; this region of fields {\bf cannot} be observed
experimentally in superlattices due to the {\bf instability} of the potential
distribution, see Fig.~\ref{fig:iv} and Chap.~\ref{chap:domains}. The same
shortage is in the work of Suris and Shchamkhalova\cite{suris-jul84}, who
concentrated on the interplay between the energy of optical phonons and
miniband width.

The calculations in this Chapter follow the work of Laikhtman and
Miller\cite{laikhtman2}. The presence of the magnetic field is taken into
account in the appropriate places. Estimations for electron temperature and
heating conditions near the current peak remained the same.

We use the Keldysh technique in order to derive the kinetic equation for the
density matrix. We need the Fermi energy or the temperature to be much larger
than the energy uncertainty $\Gamma$, that covers the upper part of the
diagram in Fig.~\ref{fig:diagrams}. In order to obtain analytical results we
need to expand the kinetic equation with respect to $\Lambda/T_e$, and
therefore the region of the miniband transport in Fig.~\ref{fig:diagrams} is
covered only partially. Scattering is considered in the Born approximation.
We do not use a translationary and gauge invariant Green
function\cite{laikhtman-altshuler}, however we use a translationally and gauge
invariant counterpart of the density matrix.


\section{ Quantum kinetic equation. }
\label{sec:keld}

\subsection{ Quantization of the effective Hamiltonian. }
\label{sec:quant}

In this subsection we discuss the wave function of an electron moving
in a one dimensional periodic superlattice potential $U(y)$ in presence of the
crossed electric and magnetic fields. For the purpose of this work we
want to know how the external fields modify the wave functions of the lowest
miniband produced by the superlattice potential. This problem can be solved by
the method of effective Hamiltonian already discussed in Sec.~\ref{sec:ham}.

The method of effective Hamiltonian does not contain limitations on the
shape of the potential $U(y)$, and in this chapter we consider tight-binding
case, sinse it is most relevant to experiment. In order to quantize  the
effective Hamiltonian in this case, it is convenient to start calculations in
the Wannier representation, which is diagonal with respect to electron states
in different potential wells of $U(y)$. The explicit form of the effective
Hamiltonian obtained in this way will be equivalent to that which was
derived in Sec.~\ref{sec:ham}. This can be proved by transformation to Bloch
waves.

Electron wave function in the Wannier representation is
$S^{-1/2}e^{ip_zz/\hbar+iP_xx/\hbar}w(y-\alpha l)$; this is the wave function
of the first level in the $\alpha$-th well of $U(y)$, $l$ is the period of the
superlattice potential $U(y)$, $(x,z)$ are in-plane coordinates, $S$ is
normalization area.  The electron Hamiltonian of a perfect superlattice in a
uniform electric field $F$ and $z$-directed magnetic field $B$, see
Fig.~\ref{fig:geometry}, is
\begin{eqnarray}
   {\cal H}_{0}
   &=&
   \left(
        {p_z^2\over 2m} + {(P_x + eB\alpha l)^2 \over 2m }
        - eFl\alpha
   \right)
   \delta_{\alpha \alpha^\prime}
\nonumber\\
   &-&
   \Lambda
   (\delta_{\alpha \alpha^\prime + 1}
   + \delta_{\alpha \alpha^\prime - 1})\;.
\label{eq:keld.1}
\end{eqnarray}
Here $e$ and $m$ are the electron charge and mass respectively, and
$\Lambda$ is the tunneling matrix element between first levels in nearest
wells; we neglect tunneling on the distance more than one period. We
neglect inter-miniband matrix elements of coordinate $y$, and presence of
other bands; we also neglect field induced corrections to the tunneling. That
is possible if the fields $F$ and $B$ are so small that $eFl$ and $eBv_Fl$
($v_F$--Fermi velocity) are much smaller than the distance between energy
levels in a well.

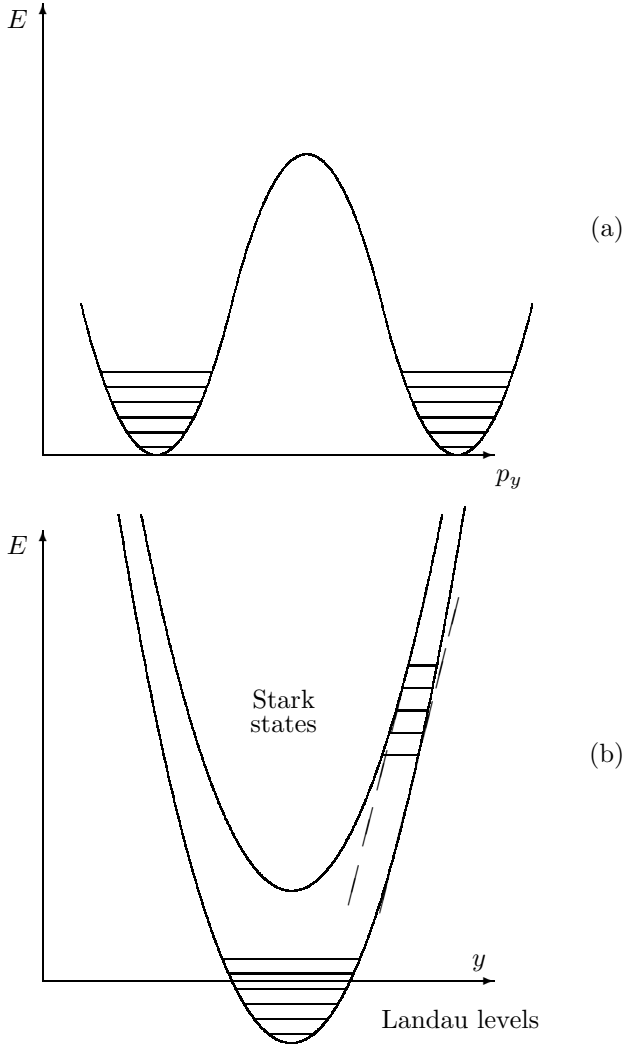
\begin{figure}
\unitlength=1.00mm
\linethickness{0.4pt}
\begin{picture}(85.00,140.00)
\put(5.00,10.00){\vector(1,0){60.00}}
\put(5.00,10.00){\vector(0,1){60.00}}
\bezier{816}(18.00,72.00)(38.00,-28.00)(58.00,72.00)
\put(63.00,12.00){\makebox(0,0)[cb]{$y$}}
\put(3.00,68.00){\makebox(0,0)[rc]{$E$}}
\put(80.00,40.00){\makebox(0,0)[cc]{(b)}}
\bezier{1148}(15.00,72.00)(38.00,-69.00)(61.00,73.00)
\put(34.87,2.97){\line(1,0){6.08}}
\put(32.94,4.97){\line(1,0){9.87}}
\put(31.63,6.97){\line(1,0){12.63}}
\put(30.53,8.97){\line(1,0){14.84}}
\put(29.63,10.98){\line(1,0){16.64}}
\put(28.73,12.98){\line(1,0){18.36}}
\put(50.00,5.00){\makebox(0,0)[lc]{Landau~levels}}
\put(60.28,61.06){\line(-1,-4){1.30}}
\put(56.13,62.21){\line(-1,-4){1.30}}
\put(58.58,54.23){\line(-1,-4){1.30}}
\put(54.44,55.39){\line(-1,-4){1.30}}
\put(56.82,47.22){\line(-1,-4){1.30}}
\put(52.68,48.38){\line(-1,-4){1.30}}
\put(54.91,39.47){\line(-1,-4){1.30}}
\put(50.76,40.63){\line(-1,-4){1.30}}
\put(53.00,32.01){\line(-1,-4){1.30}}
\put(48.85,33.17){\line(-1,-4){1.30}}
\put(51.08,24.26){\line(-1,-4){1.30}}
\put(46.94,25.41){\line(-1,-4){1.30}}
\put(50.06,40.03){\line(1,0){4.80}}
\put(50.91,42.98){\line(1,0){4.60}}
\put(51.90,46.00){\line(1,0){4.20}}
\put(52.69,49.03){\line(1,0){4.14}}
\put(53.54,51.98){\line(1,0){3.74}}
\put(37.00,46.00){\makebox(0,0)[cc]{\shortstack{Stark \\ states}}}
\bezier{328}(10.00,100.00)(20.00,60.00)(30.00,100.00)
\bezier{328}(30.00,100.00)(40.00,140.00)(50.00,100.00)
\put(5.00,80.00){\vector(1,0){60.00}}
\put(5.00,80.00){\vector(0,1){60.00}}
\put(3.00,138.00){\makebox(0,0)[rc]{$E$}}
\put(80.00,110.00){\makebox(0,0)[cc]{(a)}}
\put(67.00,78.00){\makebox(0,0)[ct]{$p_y$}}
\put(17.70,81.01){\line(1,0){4.63}}
\put(16.11,83.01){\line(1,0){7.80}}
\put(15.01,85.01){\line(1,0){10.08}}
\put(14.11,87.01){\line(1,0){11.87}}
\put(13.28,89.01){\line(1,0){13.46}}
\put(12.52,91.02){\line(1,0){14.84}}
\bezier{328}(50.00,100.00)(60.00,60.00)(70.00,100.00)
\put(57.70,81.01){\line(1,0){4.63}}
\put(56.11,83.01){\line(1,0){7.80}}
\put(55.01,85.01){\line(1,0){10.08}}
\put(54.11,87.01){\line(1,0){11.87}}
\put(53.28,89.01){\line(1,0){13.46}}
\put(52.52,91.02){\line(1,0){14.84}}
\end{picture}
   \caption{Two ways to quantize the effective Hamiltonian. See explanation in
   text.}
\label{fig:twoways} \end{figure}

The traditional way to quantize this Hamiltonian is\cite{suris-jan84}
replacing the operator $y$ or $\alpha l$ by $-i\hbar\partial/\partial p_y$.
The field part of Hamiltonian Eq.~(\ref{eq:keld.1}) becomes a parabolic
``kinetic'' term proportional to $-\partial^2/\partial p_y^2$ and the
tunneling term of Eq.~(\ref{eq:keld.1}) becomes a periodic ``potential'',
$\Lambda\sin(p_yl/\hbar)$.  It is shown in Fig.~\ref{fig:twoways}(a), that
quantization of such Hamiltonian leads to formation of Landau levels at the
parabolic bottom of each Brillouin zone. However, the tunneling between
neighbor Brillouin zones results in the broadening of Landau levels into
so-called Harper bands. These Landau levels type states are important for
transport if the Fermi energy or temperature of the electron gas are much
smaller than the miniband width $\Lambda$. Conductivity of the lowest Harper
band was computed by Suris and Shchamkhalova\cite{suris90} by making use of
the Kubo formula.

In the opposite case of the large mean kinetic energy of electrons, the
greatest contribution to transport is given by the states far above the
periodic ``potential'' shown in Fig.~\ref{fig:twoways}(a). These states are
not similar to Landau levels at all. Berezhkovskii and Suris compute them
from a weak binding model for ``potential'' $\Lambda\sin(p_yl/\hbar)$. We
will use another approach in order to give new physical interpretation to
these states.

In the framework of the semi-classical approach, one can draw in the same
picture miniband boundaries and the potential of an external field bending
this miniband, see Fig.~\ref{fig:twoways}(b). We immediately resolve in this
picture two types of states: Landau-like states and Stark-like
states.  Near the bottom of the low parabola, the Landau-like states are
formed (states of the harmonic oscillator). They are slightly broadened by
the presence of the top of the miniband, nevertheless they are shown by
parallel horizontal lines with an energy spacing $\hbar\Omega_B$.  Far away
from the center of the parabolas, the two parabolas are almost parallel near
some particular energy, see dashed lines in Fig.~\ref{fig:twoways}(b).
Electron wavefunctions are Stark-like near such an energy. Energy separation
between Stark-like states is the slope of these parabolas multiplied by
the superlattice period.

A Stark-like state is localized around a certain well. The position or number
of such a well is a good quantum number, and let's call it the guiding center
of the electron orbit $\nu$. After adding and extracting a few terms
containing $\nu$, the Hamiltonian Eq.~(\ref{eq:keld.1}) can be rewritten in
the form
\begin{eqnarray}
  {\cal H}_{0} &=&
  \left(
     {p_z^2\over 2m}
     + m{ v_x^2\over 2 } - eFl\nu
  \right)
  \delta_{\alpha \alpha^\prime} + {\cal H}_0'
\\
  {\cal H}_0'
  & = &
  \left(
      {[eBl(\alpha-\nu)]^2 \over 2m }
      +
      \left[ eBv_xl - eFl\right]
      ( \alpha - \nu )
  \right)
  \delta_{\alpha \alpha^\prime}
\nonumber\\
  &-&\Lambda
  (\delta_{\alpha \alpha^\prime + 1} + \delta_{\alpha \alpha^\prime - 1}).
\label{eq:keld.2}
\end{eqnarray}
Here $v_x\equiv p_x/m \equiv P_x/m + eB\nu l/m$. The Hamiltonian ${\cal H}_0'$
contains three terms; a quadratic term, a linear term and a hopping term. If
the quadratic term can be neglected, we can use $\nu$ as a quantum number,
the eigenvalues of ${\cal H}_0'$ are zero and the eigenfunctions are
expressed in terms of Bessel functions
\begin{eqnarray}
   \psi^{S}_{\nu,{\bf p}} &=&  S^{-1/2}
   e^{ip_zz/\hbar + i(p_x-eB\nu l)x/\hbar}
\nonumber\\
   &\times&
   \sum_{\alpha}w(y-\alpha l)
      J_{\nu-\alpha} \left(
      {2\Lambda\over eFl - eBv_xl }
   \right)\;.
\label{eq:keld.3a}
\end{eqnarray}
The superscript $S$ here means that this is a Stark-like solution, and
the in-plane momentum ${\bf p}=(p_z, p_x)$. The size of this state is
$|\nu-\alpha|\sim \Lambda/|eBv_xl-eFl|$.

We can neglect the quadratic term in Eq.~(\ref{eq:keld.2}) if the
obtained state is not too extended. Comparison of the quadratic term
with the linear term leads to the condition $|\nu-\alpha| \ll |eFl -
eBv_xl|/[(eBl)^2/m]$. Therefore, the Stark-like wave function,
Eq.~(\ref{eq:keld.3a}) is justified when
\begin{equation}
   | eBv_xl - eFl | \gg \hbar\Omega_B\;,
\label{eq:keld.3b}
\end{equation}
where the cyclotron frequency $\Omega_B$ in our anisotropic case is given by
Eq.~(\ref{eq:oneband.10}). In the Stark-like representation, the Hamiltonian
${\cal H}_0$ becomes diagonal and looks like a Stark ladder
\begin{equation}
   {\cal H}^{S} =  ( E_{\bf p} - eFl\nu )\delta_{\nu\nu'}\;,
\label{eq:keld.3c}
\end{equation}
where $E_{\bf p}=(p_x^2+p_z^2)/(2m)$. This new interpretation of the states
far above the ``periodic potential'' is shown in Fig.~\ref{fig:twoways}(a).

The wave functions of the system are of Landau levels type if the linear term
in Eq.~(\ref{eq:keld.2}) is much smaller than the quadratic term. This
happens in the narrow strip of the phase space $p_x,p_z$, where condition
Eq.~(\ref{eq:keld.3b}) is violated. Inside this strip $p_x$ is not a good
quantum number, and it is convenient to define the continuous quantum number
$Y \equiv mF/(eB^2) - P_x/(eB)$, which is a position of electron orbit
(guiding center). Therefore we don't need the index $\nu$, and it will be
used for labeling Landau type states in the rest of this section.
Diagonalization of ${\cal H}_0$ leads to
\begin{equation}
   {\cal H}^{L} = \Bigl\{ E_{(p_z,mF/B)} - eFY
                          + \hbar\Omega_B(\nu+{1\over 2})
                          + \tilde\Lambda_{\nu}(Y)
                  \Bigr\} \delta_{\nu\nu'}\;,
\label{eq:keld.4a}
\end{equation}
where $\tilde\Lambda_{\nu}(Y)$ is a periodic function of $Y$,
$\tilde\Lambda_{\nu}(Y+l)=\tilde\Lambda_{\nu}(Y)$; this function describes
the spreading of Landau levels into Harper bands.\cite{harper55,suris-jan84}
The cyclotron frequency in Eq.~(\ref{eq:keld.4a}) is defined as in
Eq.~(\ref{eq:oneband.10}). The widths of Harper bands can be estimated from
the quasiclassical expression for the tunneling amplitude between neighboring
Brillouin zones:
\begin{eqnarray}
    \text{Var}\;\tilde\Lambda_\nu
    &\sim& \hbar\Omega_B\exp
    \Biggl\{
       -{4\Lambda\over\hbar\Omega_B}
\nonumber\\
   &\times&
       \int
       \sqrt{
          {1+\cos(t)\over2}-{\hbar\Omega_B(\nu+1/2)\over4\Lambda}
       } dt
    \Biggr\}\;,
\label{eq:keld.4b}
\end{eqnarray}
where integration has to be performed over positive values of the expression
inside the square root. The number of such states is limited by the condition
\begin{equation}
   \hbar\Omega_B(\nu+1/2) \lesssim 4\Lambda\;,
\label{eq:keld.4c}
\end{equation}
and for the strong enough magnetic field there are no Landau type states at
all.

The width of Landau type states can be estimated from the usual Gaussian
shape of the linear oscillator wave function as $|\alpha-Y/l|\sim
\Lambda\nu/(\hbar\Omega_B)$. These states are more extended in the $y$
direction than the Stark-type states and can give important contribution to
the vertical current under very special conditions.

\subsection{ Orthogonality conditions for Stark type wave functions. }
\label{sec:ortog}

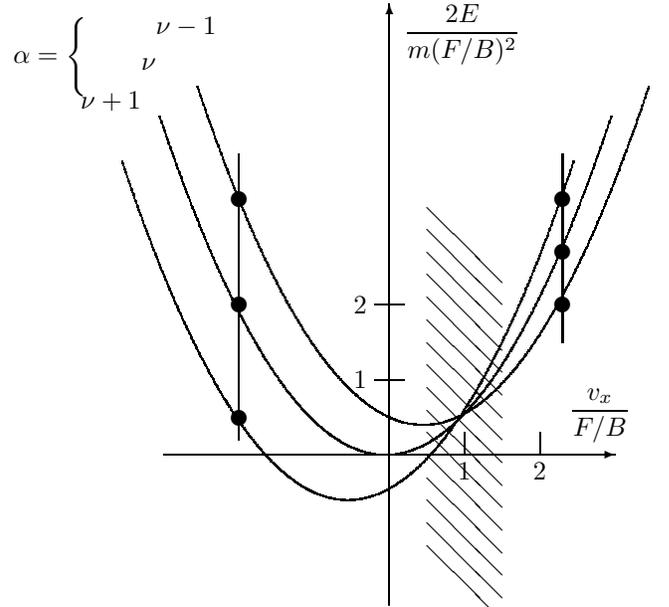
\begin{figure}
\unitlength=1.00mm
\linethickness{0.4pt}
\newsavebox{\parabola}
\begin{picture}(85.00,80.00)
\put(20.00,20.00){\vector(1,0){60.00}}
\put(50.00,0.00){\vector(0,1){80.00}}
\savebox{\parabola}(00.00,00.00){
   \begin{picture}(00.00,00.00)
   \bezier{760}(30.00,45.00)(00.00,-45.00)(-30.00,45.00)
   \end{picture}
}
\put(50.00,20.00){\usebox{\parabola}}
\put(45.00,14.00){\usebox{\parabola}}
\put(55.00,24.00){\usebox{\parabola}}
\put(78.00,25.00){\makebox(0,0)[cb]{$\displaystyle {v_x\over F/B}$}}
\put(52.00,76.00){\makebox(0,0)[lc]{$\displaystyle {2E\over m(F/B)^2}$}}
\put(23.00,77.00){\makebox(0,0)[cc]{$\nu-1$}}
\put(18.00,72.00){\makebox(0,0)[cc]{$\nu$}}
\put(13.00,67.00){\makebox(0,0)[cc]{$\nu+1$}}
\put(0.00,73.00){\makebox(0,0)[lc]{$\alpha = \Biggl\{ $}}
\put(30.00,60.00){\line(0,-1){38.00}}
\put(30.00,25.00){\circle*{2.00}}
\put(30.00,40.00){\circle*{2.00}}
\put(30.00,54.00){\circle*{2.00}}
\put(73.00,60.00){\line(0,-1){25.00}}
\put(73.00,54.00){\circle*{2.00}}
\put(73.00,47.00){\circle*{2.00}}
\put(73.00,40.00){\circle*{2.00}}
\put(55.00,8.00){\line(1,-1){10.00}}
\put(55.00,17.00){\line(1,-1){10.00}}
\put(55.00,11.00){\line(1,-1){10.00}}
\put(55.00,14.00){\line(1,-1){10.00}}
\put(55.00,20.00){\line(1,-1){10.00}}
\put(55.00,29.00){\line(1,-1){10.00}}
\put(55.00,23.00){\line(1,-1){10.00}}
\put(55.00,26.00){\line(1,-1){10.00}}
\put(55.00,32.00){\line(1,-1){10.00}}
\put(55.00,41.00){\line(1,-1){10.00}}
\put(55.00,35.00){\line(1,-1){10.00}}
\put(55.00,38.00){\line(1,-1){10.00}}
\put(55.00,44.00){\line(1,-1){10.00}}
\put(55.00,53.00){\line(1,-1){10.00}}
\put(55.00,47.00){\line(1,-1){10.00}}
\put(55.00,50.00){\line(1,-1){10.00}}
\put(60.00,20.00){\line(0,1){3.00}}
\put(70.00,20.00){\line(0,1){3.00}}
\put(48.00,30.00){\line(1,0){4.00}}
\put(48.00,40.00){\line(1,0){4.00}}
\put(60.00,19.00){\makebox(0,0)[ct]{$1$}}
\put(70.00,19.00){\makebox(0,0)[ct]{$2$}}
\put(47.00,30.00){\makebox(0,0)[rc]{$1$}}
\put(47.00,40.00){\makebox(0,0)[rc]{$2$}}
\end{picture}
   \caption{
      Approximation by Stark type states is not valid in hatched area
      only. See explanation in the text. }
\label{fig:diagram3}
\end{figure}

Stark like states constructed above are very important for further
calculations and we explain now more details about the approximation which
has been made. For fixed $\nu$, the Stark like state is a mixture of Wannier
states from different wells $\alpha$. The energies of these Wannier states
are in the diagonal part of the Hamiltonian ${\cal H}_0$ minus $p_z^2/(2m)$
\begin{eqnarray}
   E_{\alpha v_x}
   &=&
   { m(F/B)^2 \over 2}
   \biggl\{
      {1\over 2} \left( {v_x\over F/B} \right)^2
\nonumber\\
   &+&
   \left( {v_x\over F/B}  - 1 \right)
   {\alpha-\nu \over C}
   +
   {1\over 2}
   \left( {\alpha-\nu \over C}
   \right)^2
   \biggr\}\;,
\label{eq:ortog.1}
\end{eqnarray}
where $C=mF/(eB^2l)$. We wrote this expression in dimensionless form, and
plotted it in Fig.~\ref{fig:diagram3} for $C=2$ and
$\alpha=\nu,\nu\pm1$. So, in this figure one
can see the electron energy in three subsequent wells. The vertical shift of
the parabolas is due to the electric field and the horizontal shift is due to
the magnetic field. In dimensionless units both shifts are $1/C$.

Tunneling between subsequent wells mixes Wannier states. Two examples of such
mixing are shown in Fig.~\ref{fig:diagram3}. Three states marked by filled
circles in the vertical line $v_x={}$const form a new Stark-like state with
energy close to the energy of central Wannier state. In order to get wave
functions Eq.~(\ref{eq:keld.3a}) we have to neglect the last term on the right
hand side of Eq.~(\ref{eq:ortog.1}). One may think that this approximation is
valid under the condition $C\gtrsim1$, but that is not true. This
approximation is not valid for states in the hatched area in
Fig.~\ref{fig:diagram3}, which is given by the inequality opposite to
Eq.~(\ref{eq:keld.3b}).

Let us show that Eq.~(\ref{eq:keld.3b}) provides a condition
for orthogonality of the wave functions Eq.~(\ref{eq:keld.3a}). We have
\begin{eqnarray}
\lefteqn{
   \int d{\bf r} \sum_{\alpha}
   \psi^{S\ast}_{\nu,{\bf p}}(\alpha,{\bf r})
   \psi^{S}_{\nu',{\bf p}'}(\alpha,{\bf r})
}
\nonumber\\
   &=&
   {(2\pi\hbar)^2\over S}
   \delta(p_z'-p_z)
   \delta(p_x-eB\nu l -p_x' + eB\nu' l)
\nonumber\\
   &\times&
   J_{\nu-\nu'}
   \left(
      {2\Lambda eBl(v_x'-v_x)\over [eFl - eBv_xl][eFl - eBv_x'l] }
   \right)
\nonumber\\
   &\approx&
   {(2\pi\hbar)^2\over S}
   \delta(p_z'-p_z)
   \delta(p_x' - p_x)
   \delta_{\nu'\nu}
\label{eq:ortog.5}
\end{eqnarray}
We would like also to check the orthogonality condition by integration over
momenta and summation over index $\nu$. That is impossible to do, however,
because of the pole near $v_x=F/B$.

\subsection{ Matrix elements in the Stark representation.}
\label{sec:matrel}

The operator of the electron velocity in the $y$ direction in the Wannier
representation has the form
\[
   \hat v_y =
   (i \Lambda l / \hbar)
   (\delta_{\alpha,\alpha' +1} - \delta_{\alpha,\alpha' -1})
   \delta( x - x')
   \delta( z - z')\;.
\]
It becomes in our Stark-like representation
\begin{eqnarray}
   \hat v &=&
   {(2\pi\hbar)^2\over S}
   \delta(p_z-p_z')\delta(p_x-eB\nu l-p_x' + eB\nu'l)
\nonumber\\
   &\times&
   { i \Lambda d \over \hbar }\sum_\alpha
   \left(
      J_{\nu-\alpha}^{p_x} J_{\nu'-\alpha+1}^{p_x'}
      -
      J_{\nu-\alpha}^{p_x} J_{\nu'-\alpha-1}^{p_x'}
   \right)
\label{eq:matrel.1}
\\
   J_{\nu-\alpha}^{p_x}
   & \equiv &
   J_{\nu-\alpha}
      \left(
         {2\Lambda\over eFl - eBv_xl}
      \right)
\label{eq:matrel.2}
\end{eqnarray}

The total Hamiltonian contains ${\cal H}_0$, and a scattering potential,
$\hat U$. We consider that the main scattering mechanism is impurity
scattering or scattering on the other type of static potential. The overlap
of electron wave functions in different wells is small and only diagonal
elements with respect to the wells, $U^{\alpha\alpha}_{{\bf p}{\bf p}'}$,
will be taken into account.  Matrix elements of the operators $\hat{U}$ in
the Stark representation have the form
\begin{equation}
   U^{\nu \nu^\prime}_{{\bf p} {\bf p}' }
   =
   \sum_\alpha
   J_{\nu-\alpha}^{p_x}
   J_{\nu^\prime-\alpha}^{p_x'}
   U^{\alpha \alpha}_{p_z,p_x-eB\nu l; p_z',p_x'-eB\nu' l } \;,
\label{eq:keld.5}
\end{equation}
and $J_{\nu-\alpha}^{p_x}$ are defined above. The product of two matrix
elements averaged over the impurity configuration is
\begin{eqnarray}
\lefteqn{
   \overline{
      U^{\nu \mu}_{{\bf p} {\bf q} }
      U^{\nu' \mu'}_{{\bf p}' {\bf q}' }
   }
   =
   N_I (2\pi\hbar)^2 \delta(p_z-q_z+p_z'-q_z')
}
\nonumber\\
   &\times&
   \delta(p_x-eB\nu l-q_x+eB\mu l+p_x'-eB\nu' l-q_x'+eB\mu' l)
\nonumber\\
   &\times&
       \tilde U_{p_z,p_x-eB\nu l; q_z,q_x-eB\mu l }
       \tilde U_{p_z',p_x'-eB\nu' l; q_z',q_x'-eB\mu' l }
\nonumber\\
   &\times&
   \sum_{\alpha}
   J_{\nu-\alpha}^{p_x} J_{\mu-\alpha}^{q_x}
   J_{\nu'-\alpha}^{p_x'}J_{\mu'-\alpha}^{q_x'}\;,
\label{eq:matrel.3}
\end{eqnarray}
where $N_I$ is the sheet density of the impurities and $\tilde U$ is the
matrix element of the single impurity potential.

\subsection{ Equation for density matrix }
\label{sec:denmatr}

\begin{mathletters}
We will assume that the Fermi energy and temperature are much larger than
the level spacing of ${\cal H}_0$. This assumption allows us to
derive a kinetic equation for the electron density matrix. We will make use
of the Keldysh technique and derive this equation in the same way as the
Boltzmann equation is usually derived.\cite{keldysh-diagrams} In the Keldysh
technique the kinetic equation results from the Dyson equation, which can be
written in two equivalent forms
\begin{eqnarray}
    {\hat G}_{01}^{-1} G_{12} = \sigma_z
    + \sigma_z \Sigma_{13} G_{32}                          \;,
\label{eq:keld.6} \\
   {\hat G}_{02}^{-1 \ast } G_{12} = \sigma_z
   + \Sigma_{13} G_{32} \sigma_z                          \;.
\label{eq:keld.7}
\end{eqnarray}
The matrix Keldysh Green function, $G_{12}$, depends on two sets of
variables, $\{\nu, {\bf p}, t \}$ and a sum and integral with respect to the
variables with the subscript 3 is implied in Eqs.~(\ref{eq:keld.6}) and
(\ref{eq:keld.7}).  The operators on the left hand sides, ${\hat G}_{01}^{-1}$
and ${\hat G}_{02}^{-1}$ are obtained from
\begin{equation}
   {\hat G}_0^{-1} =
   \Bigl(i \hbar {{\partial} \over {\partial t}} - {\cal H}^S
   \Bigr)\sigma_z
\label{eq:keld.8}
\end{equation}
by substitution sets $\{\nu_1, {\bf p}_1, t_1 \}$ or  $\{\nu_2, {\bf p}_2, t_2
\}$ correspondingly. Definitions of the Green functions
Eqs.~(\ref{eq:keld.8th}) contain also the Pauli matrix $\sigma_z$.
\label{eq:keld.8th}
\end{mathletters}

Green functions depend on two times, $t_1$ and $t_2$. Instead of them
the difference, $t_1 - t_2$, and the sum, $t = (t_1+t_2)/2$, times can be
introduced.  The characteristic values of the difference time is of the order
of $\hbar$ divided by the characteristic energy, i.e. the maximum of the
Fermi energy and temperature. The time $t$ characterizes much more slow
variation of occupation numbers. So that in the right-hand sides  of
Eqs.~(\ref{eq:keld.6}) and (\ref{eq:keld.7}) all functions can be considered
dependent on the same time $t$ and the integration with respect to $t_3$ is
reduced to the integration with respect to difference times. Then the
Fourier transform with respect to the difference time leads to the Green
functions depending on $t$ and a frequency $\omega$, which is a spectral
variable.

The explicit form of the matrix Green function as it is defined in
Ref.\onlinecite{landau-diagrams} is
\begin{equation}
   G \equiv
   \left(\begin{array}{cc}
      G^{--} & G^{-+} \\
      G^{+-} & G^{++}
   \end{array}\right)
   =
   \left(\begin{array}{cc}
      G^{-+}+ G^r  & G^{-+} \\
      G^{-+}+ G^r- G^a  & G^{-+} - G^a
   \end{array}\right)
\label{eq:keld.7a}
\end{equation}
The same pattern holds for self-energies. The difference and half of the sum
of $(-+)$ matrix elements of Eqs.~(\ref{eq:keld.6}) and
(\ref{eq:keld.7}) has the form
\begin{eqnarray}
    i \hbar {{\partial} \over {\partial t}} G_{12}^{-+}
    - [ {\cal H}^S , G^{-+} ]_{12}
    &=&
    \Sigma_{13}^{r} G_{32}^{-+} - \Sigma_{13}^{-+} G_{32}^{a}
\nonumber\\
  &+& G_{13}^{r} \Sigma_{32}^{-+} - G_{13}^{-+} \Sigma_{32}^{a}   \;,
\label{eq:keld.10}  \\
   \hbar \omega G_{12}^{-+}
   - {1 \over 2} \{ {\cal H}^S , G^{-+} \}_{12}
    &=&
  {1 \over 2}
  ( \Sigma_{13}^{r} G_{32}^{-+} - \Sigma_{13}^{-+} G_{32}^{a}
\nonumber\\
  &-& G_{13}^{r} \Sigma_{32}^{-+} + G_{13}^{-+} \Sigma_{32}^{a} )   \;.
\label{eq:keld.11}
\end{eqnarray}
For the calculation of self-energies we need also advanced and retarded
Green functions. Equations for them are obtained from Eqs.~(\ref{eq:keld.6})
and (\ref{eq:keld.7})
\begin{eqnarray}
   \hbar \omega G_{12}^{r} - {1 \over 2} \{ {\cal H}^S , G^{r} \}_{12}
   &=& 1 + {1 \over 2} \{ \Sigma^{r} , G^{r} \}_{12}              \;,
\label{eq:keld.12} \\
   G_{12}^{a} &=& G_{21}^{r \ast}                                 \;.
\label{eq:keld.13}
\end{eqnarray}
Here $[a,b] = ab - ba$ and $\{ a,b \} = ab + ba$.

The retarded Green function can be found from Eqs.~(\ref{eq:keld.12}) and
(\ref{eq:keld.14}) (see below).  Equations~(\ref{eq:keld.11}) and
(\ref{eq:keld.16}) (see below) are linear and uniform with respect to $G_{\nu
\nu^\prime}^{-+}({\bf p})$ and this function cannot be found from them.
Actually, Eqs.~(\ref{eq:keld.11}) and (\ref{eq:keld.16}) can be used to find
only the dependence of $G^{-+}$ on $\omega$. The integral of $G^{-+}$ with
respect to $\omega$, the density matrix
\begin{equation}
   \rho^{\nu \nu^\prime}_{{\bf p}{\bf p}'}(t)
   = \hbar \int
   {d\omega \over {2\pi i}}\,
   G^{-+}_{\nu{\bf p};\nu'{\bf p}'}(\omega,t) \;,
\label{eq:keld.18}
\end{equation}
has to be determined with the help of Eq.~(\ref{eq:keld.10}). The equation for the
density matrix can be obtained by the integration of Eq.~(\ref{eq:keld.10}) with
respect to $\omega$,
\begin{eqnarray}
   {{\partial \rho} \over {\partial t}}
    &-& {1 \over i\hbar} [ {\cal H}^S , \rho ]
    = - \int {d\omega \over {2\pi}}\,
    \Bigl(\Sigma^{r} G^{-+}
\nonumber\\
    &-&
    \Sigma^{-+} G^{a}
    + G^{r} \Sigma^{-+} - G^{-+} \Sigma^{a}
    \Bigr)  \;.
\label{eq:keld.19}
\end{eqnarray}
The superlattice is uniform in the $x,z$ plane and one may expect Green
functions as well as self-energies to be diagonal with respect to $\bf p$.
However, this is not true because we are looking for a solution, which
describes electric current, and therefore the density matrix has to contain
terms proportional to the matrix element of the velocity $y$-component. This
matrix element Eq.~(\ref{eq:matrel.1}) is not diagonal with respect to ${\bf
p}$, and therefore all elements of the Green function are non-diagonal too.

We consider the case of weak scattering when the energy uncertainty due to
the scattering is much smaller then the width of the in-plane electron energy
distribution. Then self-energies can be calculated in the first
approximation.  For simplicity we assume that electrons in different wells
are scattered by different impurities. This implies that the screening radius
is smaller than the period of the superlattice. Then
\begin{eqnarray}
   \Sigma^{r}_{\nu{\bf p};\nu'{\bf p}'}(\omega)
   & = &
   \sum_{\mu\mu'}
   \int { d {\bf q} d{\bf q}' \over (2 \pi\hbar)^4 }
   \overline{
      U^{\nu \mu}_{{\bf p} {\bf q} }
      U^{\mu' \nu'}_{{\bf q}' {\bf p}' }
   }
   G^{r}_{\mu{\bf q};\mu'{\bf q}'}(\omega)     \;,
\nonumber\\
\label{eq:keld.14} \\
   \Sigma^{a}_{\nu{\bf p};\nu'{\bf p}'}(\omega)
   &=&
   \Sigma^{r\ast}_{\nu{\bf p};\nu'{\bf p}'}(\omega)
\label{eq:keld.15} \\
   \Sigma^{-+}_{\nu{\bf p};\nu'{\bf p}'}(\omega)
   & = &
   -\sum_{\mu\mu'}
   \int { d {\bf q} d{\bf q}' \over (2 \pi\hbar)^4 }
   \overline{
      U^{\nu \mu}_{{\bf p} {\bf q} }
      U^{\mu' \nu'}_{{\bf q}' {\bf p}' }
   }
   G^{-+}_{\mu{\bf q};\mu'{\bf q}'}(\omega)     \;.
\nonumber\\
\label{eq:keld.16}
\end{eqnarray}
These self-energies have to be substituted into the right hand side of
Eq.~(\ref{eq:keld.19}) which can be solved together with
Eqs.~(\ref{eq:keld.10}) and (\ref{eq:keld.11}).

The next important approximation is the weak scattering limit, which allows
one to neglect terms containing self-energies in Eqs.~(\ref{eq:keld.11}) and
(\ref{eq:keld.12}).  Therefore we are going to take
into account scattering in the so-called Born approximation. Computation of
the right hand side of Eq.~(\ref{eq:keld.19}) is cumbersome and we put it
into Appendix~\ref{sec:App.coll}.


\section{ Calculation of the current }
\label{sec:generic}

In order to compute the current we have to find a stationary,
translational, and gauge invariant solution to the kinetic equation
Eq.~(\ref{eq:keld.19}) with the right hand side given by
Eqs.~(\ref{eq:generic.2}).  Equation~(\ref{eq:keld.19}) is not gauge
invariant, and the density matrix $\rho^{\nu\nu'}_{{\bf p}{\bf p}'}$ also is
not gauge invariant, because wave functions were chosen in the specific
gauge. However, in Wannier representation we can introduce translational
and gauge invariant\cite{laikhtman-altshuler} counterpart of the density
matrix $\wp_{\alpha-\alpha'}({\bf q})$
\begin{equation}
   \rho^{\alpha\alpha'}_{{\bf r}{\bf r}'}
   =
   e^{-ieBl{\alpha+\alpha' \over2\hbar}(x-x')}
   \int {d{\bf q}\over(2\pi\hbar)^2}
   e^{i{\bf q}({\bf r}-{\bf r}')}
   \wp_{\alpha-\alpha'}({\bf q})\;.
\label{eq:generic.3a}
\end{equation}
Here $\rho^{\alpha\alpha'}_{{\bf r}{\bf r}'}$ is not an invariant density
matrix defined in the Wannier representation. The first task is to convert
both sides of Eq.~(\ref{eq:generic.3a}) from Wannier to our Stark-like
representation. We have to make another approximation in order to perform
this transformation.

We will solve Eq.~(\ref{eq:keld.19}) only in the case when
\begin{equation}
   \Lambda , \ |eBv_x-eFl| \ll T_e\;,
\label{eq:generic.5}
\end{equation}
where $T_e$ is the effective electron temperature, which characterizes the
width of the electron energy distribution. (We do not assume that a real
distribution is the equilibrium one with the temperature $T_e$. This quantity
is used only for estimates.) In the case of an effective energy relaxation
$T_e\sim T$. In the case of low temperature and an appreciable heating of the
electron gas $T_e > T$. The estimate of $T_e$ under different conditions can
be found from the balance of the heating $\sim jF$ and cooling
$n(T_e-T)\over\tau_{\varepsilon}$, see Eqs.~(\ref{eq:generic.17}) and
(\ref{eq:generic.18}) below. In the case of zero magnetic field the
answer can be found in Ref.\onlinecite{laikhtman2}.

Under the condition Eq.~(\ref{eq:generic.5}) we can convert
$\rho^{\alpha\alpha'}_{{\bf r}{\bf r}'}$ and $\wp_{\alpha-\alpha'}({\bf q})$
to $\rho^{\nu\nu'}_{{\bf p}{\bf p}'}$ and $\wp_{\nu-\nu'}({\bf q})$.
Equation~(\ref{eq:generic.3a}) is converted into Eq.~(\ref{eq:generic.3b}),
see Appendix~\ref{sec:App.inv} for more details. Substitution of
Eq.~(\ref{eq:generic.3b}) into the quantum kinetic equation
Eq.~(\ref{eq:keld.19}) and averaging of the scattering probability over energy
surface leads to the following resulting equations
\begin{eqnarray}
   & & \hbar {\hat I}_E \wp_0 (E)
   + \Lambda  {d \over {dE}}
     \Gamma (E) [ R(E)
         + 2 \Lambda {{d\wp_0} \over {dE}}] = 0   \;,
\label{eq:generic.10} \\
    & &\pm i(eBv_xl- eFl) \wp_{\pm 1} ({\bf p})
    \pm i\Lambda eBv_x l
    {{d\wp_0} \over {dE_{\bf p}}}
\nonumber\\
    &=&
  - \Gamma (E_{\bf p}) \wp_{\pm 1} ({\bf p})
  - \Lambda\Gamma (E_{\bf p}) {{d\wp_0} \over {dE_{\bf p}}}   \;,
\label{eq:generic.11}\\
   && R(E) =
     \int { d{\bf p}\over 2\pi m }
    \delta (E - E_p)
     \big[
       \wp_{1}({\bf p}) + \wp_{-1}({\bf p})
     \big]
\label{eq:generic.11a}
\end{eqnarray}
where ${\hat I}_E$ is the operator of energy relaxation. Calculations are
actually performed in Appendix~\ref{sec:App.inv}. Condition
Eq.~(\ref{eq:generic.5}) allows us to keep the Stark-like levels difference
$\nu-\nu'=0,\pm1$. Quantum uncertainty of the energy levels due to scattering
$\Gamma(E_p)$ is defined in Eq.~(\ref{eq:generic.9}).
Equations~(\ref{eq:generic.10}) and (\ref{eq:generic.11}) are very
remeniscent of the Boltzmann equations for the parts of the electron
distribution function even and odd with respect to the electron momentum in
the theory of hot electrons in wide band semiconductors
\cite{gantsevich65,gurevich65}.

The main energy relaxation mechanism in the narrow band superlattices is
acoustic phonon scattering.\cite{laikhtman2,meier-may94} We have
\begin{equation}
   \hbar {\hat I}_E \wp_0 (E)
   = {d \over {dE}} Q (E)
     \left[ \wp_0 (1 -\wp_0 ) + T {{d\wp_0} \over {dE}} \right]  \;,
\label{eq:generic.12}
\end{equation}
where
\begin{equation}
   Q(E) = {{\pi^2 m \Xi^2} \over {2 \rho_0 d_{w}^3}}
   \left( 1 + {{3Emd_{w}^2} \over {\pi^2 \hbar^2}} \right)          \;,
\label{eq:generic.13}
\end{equation}
$\rho_0$ is the crystal density, $m$ is the in-plane effective mass, $d_w$ is
the width of a well, and $\Xi$ is the deformation potential.

Equation~(\ref{eq:generic.11}) can be written as
\begin{equation}
  \wp_{\pm 1}({\bf p}) =
  - \Lambda{ \Gamma (E_{\bf p}) \pm i eBv_x l
\over
  {\Gamma (E_{\bf p}) \pm i(eBv_xl-eFl)}}
   {{d\wp_0} \over {dE_{\bf p}}}                                \;.
\label{eq:generic.14}
\end{equation}
This solution has a pole at $v_x=F/B$, which is smeared by scattering. The
width of this smearing is $\Gamma/(eBl)$ and it has to cover the hatched area
in the diagram Fig.~\ref{fig:diagram3}. The condition justifying our whole
theory is, therefore,
\begin{equation}
   \Omega_B\tau_p \ll 1\;,
\label{generic.14a}
\end{equation}
where $\tau_p\sim \hbar/\Gamma(E_p)$. Equation~(\ref{eq:generic.10}) with
the help of Eq.~(\ref{eq:generic.12}) is reduced to
\begin{equation}
    Q(E) \left[ \wp_0 (1 -\wp_0 ) + T {{d\wp_0} \over {dE}} \right]
   + 2 \Lambda^2
    {\Gamma (E)} \tilde R(E)
    {{d\wp_0} \over {dE}}
    = 0                              \;,
\label{eq:generic.15}
\end{equation}
where
\begin{equation}
  \tilde R(E_{\bf p})
  = \int {d\phi \over 2\pi}
  { (eFl)^2
  \over
  \Gamma^2(E_{\bf p}) + (eBlp\cos(\phi)/m - eFl)^2  }\;.
\label{eq:generic.15a}
\end{equation}
Integration can be performed, the same integral was calculated in
Chap.~\ref{chap:trans}, see Eq.~(\ref{eq:con.4}).

The solution to Eq.~(\ref{eq:generic.15}) is
\begin{equation}
   \wp_0 = \left\{
         \exp
	  \left[
	   \int_0^E
	       {  dE \over
             T + { 2 \Lambda^2\Gamma (E)\tilde R(E)  / Q(E) }
	      } - \zeta
	  \right] + 1
	 \right\}^{-1}     \;,
\label{eq:generic.16}
\end{equation}
where $\zeta$ is a normalization constant. For zero magnetic field this
solution was obtained by Laikhtman and Miller.\cite{laikhtman2}
Even without an
exact calculation Eq.~(\ref{eq:generic.16}) allows us to estimate the importance of
the heating of the electron gas near the current maximum. We can
estimate the momentum and energy relaxation times as $\tau_p \sim \hbar /
\Gamma$ and $\tau_\varepsilon \sim \hbar T_e/Q$ respectively. Near the
current maximum, when $eFl \sim \Gamma, eBlv_F, eBlv_T$, we have $\tilde
R\sim 1$, and the width of the electron energy distribution $T_e$ in the
case of weak heating and the condition for weak heating are
\begin{equation}
   T_e \sim T \ , \ \ \
  {{\Lambda^2} \over {T^2}} {{\tau_\varepsilon} \over {\tau_p}} \ll 1 \;.
\label{eq:generic.17}
\end{equation}
Because $\tau_\varepsilon \gg \tau_p$, the condition
Eq.~(\ref{eq:generic.17}) is satisfied only for $\Lambda \ll T$. For the band
width $4 \Lambda \sim 0.5$ meV the last condition is satisfied for
temperatures about 30 K and higher. This estimate shows that for lower
temperature or for a wider band one can expect an appreciable heating of the
electron gas. In such a case
\begin{equation}
    T_e \sim {\Lambda^2 \tau_\varepsilon \over T \tau_p} \ , \ \ \
    {{\Lambda^2} \over {T^2}} {{\tau_\varepsilon} \over {\tau_p}} \gg 1 \;.
\label{eq:generic.18}
\end{equation}

Now it is possible to justify Eq.~(\ref{eq:generic.6}). In the estimate we
assume that $eFl,|eFl-eBv_F| \sim \Gamma$ because this region of the electric
field is close to the current maximum, and $\Gamma^2 /\Lambda^2 \ll \tau_\varepsilon
/\tau_p$ because in the opposite case the resonance tunneling is smeared so
much that the width of the resonance can be of the order of the separation
between the levels in a well. Then for weak heating,
Eq.~(\ref{eq:generic.17}), $T_e \sim T \geq \Lambda (\tau_\varepsilon /\tau_p
)^{1/2} \gg \Lambda , \Gamma $.  In the case of strong heating,
Eq.~(\ref{eq:generic.18}), $\Gamma /T_e \ll (T/\Lambda ) (\tau_p
/\tau_\varepsilon )^{1/2} \ll 1$ and $\Lambda / T_e \sim (T /\Lambda )
(\tau_p /\tau_\varepsilon ) \ll (T/\Lambda ) (\tau_p /\tau_\varepsilon
)^{1/2} \ll 1$; that is, in both cases Eq.~(\ref{eq:generic.5}) is satisfied.

\onecolumn\widetext

For the uniform electron distribution, the expression for current density
is proportional to the trace of the density matrix times the velocity
operator, Eq.~(\ref{eq:matrel.1})
\begin{eqnarray}
    j & = & {{2ie \Lambda } \over {\hbar}}
    \int {{d{\bf p}} \over {(2 \pi\hbar)^2}}
	 \big( \wp_{-1} - \wp_{1} \big)
     =
     - {{4 e \Lambda^2 } \over {\hbar}}
    \int {{d{\bf p}} \over {(2 \pi\hbar)^2}}
    {
     \Gamma (E_{\bf p}) eFl
     \over
     \Gamma^2 (E_{\bf p}) + (eBv_xl-eFl)^2
     }
     {d\wp_0 \over dE_{\bf p} }      \;.
\nonumber \\
\label{eq:generic.19}
\end{eqnarray}
This expression shows that Ohm's law is satisfied for $eFl, eBv_Fl \ll
\Gamma $. The collisionless current is recovered in the opposite case, and
\begin{eqnarray}
    j
     =  - {{4\pi eF \Lambda^2 } \over {\hbar}}
    \int {{d{\bf p}} \over {(2 \pi\hbar)^2}}
     \delta(Bv_x-F)
     {d\wp_0 \over dE_{\bf p} }      \;.
\label{eq:generic.20}
\end{eqnarray}
This result is similar to Eq.~(\ref{eq:con.7}). The important difference is
the form of $\wp_0$; in Eq.~(\ref{eq:generic.20}) it has to be substituted
from Eq.~(\ref{eq:generic.16}), whereas in Eq.~(\ref{eq:con.7}) we use the
equilibrium distribution function $f_0$.



\section{Appendix: Collision integral of the quantum kinetic equation.}
\label{sec:App.coll}

For weak scattering, terms containing self-energies in
Eqs.~(\ref{eq:keld.11}) and (\ref{eq:keld.12}) can be neglected compared to
$\hbar \omega$ and solutions to them are
\begin{eqnarray}
   G^r_{\nu{\bf p}; \nu'{\bf p}'}(\omega)
   & = &
   {(2\pi\hbar)^2}
   {
    \delta_{\nu \nu^\prime}
    \delta({\bf p}- {\bf p}')
    \over
    \hbar\omega - E_{\bf p} + eFl\nu + i 0
   }              \;,
\label{eq:generic.1a} \\
   \Sigma^{r}_{\nu{\bf p};\nu'{\bf p}'}(\omega)
   & = &
   {(2\pi\hbar)^2}
   \delta(p_z-p_z')
   \delta(p_x-eB\nu l-p_x'+eB\nu' l)
\nonumber\\
   &\times&
   N_I
   \sum_{\mu}
   \int { d {\bf q} \over (2 \pi\hbar)^2 }
       |\tilde U_{p_z,p_x-eB\nu l; q_z,q_x-eB\mu l }|^2
   {
    \sum_{\alpha}
    J_{\nu-\alpha}^{p_x} J_{\mu-\alpha}^{q_x}
    J_{\mu-\alpha}^{q_x}J_{\nu'-\alpha}^{p_x'}
    \over
    \hbar\omega - E_{\bf q} + eFl\mu + i 0
   }
\label{eq:generic.1b}\\
   G^{-+}_{\nu{\bf p}; \nu'{\bf p}'}(\omega)
   & = &
   2 \pi i \rho^{\nu \nu'}_{{\bf p}{\bf p}'}
   \delta \left(
        \hbar\omega - { E_{p} + E_{p'} \over 2  }
        + eFl{{\nu + \nu'} \over 2 }
   \right) \;.
\label{eq:generic.1c}\\
   \Sigma^{-+}_{\nu{\bf p};\nu'{\bf p}'}(\omega)
   & = &
   -2 \pi i\sum_{\mu\mu'}
   \int { d {\bf q} d{\bf q}' \over (2 \pi\hbar)^4 }
   \overline{
      U^{\nu \mu}_{{\bf p} {\bf q} }
      U^{\mu' \nu'}_{{\bf q}' {\bf p}' }
   }
   \rho^{\mu \mu'}_{{\bf q}{\bf q}'}
   \delta \left(
        \hbar\omega - { E_{q} + E_{q'} \over 2  }
        + eFl{{\mu + \mu'} \over 2 }
   \right) \;.
\label{eq:generic.1d}
\end{eqnarray}

\begin{mathletters}
All four terms on the right hand side of Eq.~(\ref{eq:keld.19}) can be
expressed
explicitly in terms of the density matrix:
\begin{eqnarray}
   -\int{d\omega\over 2\pi}
   (\Sigma^{r}G^{-+})^{\nu\nu'}_{{\bf p}{\bf p}'}
   &=&
   -{iN_I\over\hbar}
   \sum_{\mu\mu'}
   \int { d {\bf q} \over (2 \pi\hbar)^2 }
       |\tilde U_{p_z,p_x-eB\nu l; q_z,q_x-eB\mu l }|^2
       \rho^{\mu' \nu'}_{(p_z, p_x-eB(\nu-\mu') l);{\bf p}'}
\nonumber\\
   &\times&
   {
    \sum_{\alpha}
    J_{\nu-\alpha}^{p_x} J_{\mu-\alpha}^{q_x}
    J_{\mu-\alpha}^{q_x} J_{\mu'-\alpha}^{p_x-eB(\nu-\mu') l}
    \over
    \Bigl(
       E_{(p_z, p_x-eB(\nu-\mu') l)} + E_{p'}
    \Bigr)/ 2
        - eFl{ \mu' + \nu' \over 2}  - E_{\bf q} + eFl\mu + i 0
   }
\label{eq:generic.2a}\\[20pt]
   \int{d\omega\over 2\pi}
   (G^{-+}\Sigma^{a})^{\nu\nu'}_{{\bf p}{\bf p}'}
   &=&
   {iN_I\over\hbar}
   \sum_{\mu\mu'}
   \int { d {\bf q} \over (2 \pi\hbar)^2 }
       |\tilde U_{p_z',p_x'-eB\nu' l; q_z,q_x-eB\mu l }|^2
       \rho^{\nu \mu'}_{{\bf p};(p_z',p_x'-eB(\nu'-\mu')l)}
\nonumber\\
   &\times&
   {
    \sum_{\alpha}
    J_{\mu'-\alpha}^{p_x'-eB(\nu'-\mu')l} J_{\mu-\alpha}^{q_x}
    J_{\mu-\alpha}^{q_x}J_{\nu'-\alpha}^{p_x'}
    \over
    \Bigl( E_{p} + E_{ (p_z',p_x'-eB(\nu'-\mu')l) }\Bigr) / 2
        - eFl{{\nu + \mu'} \over 2 } - E_{\bf q} + eFl\mu - i 0
   }
\label{eq:generic.2b}\\[20pt]
   \int{d\omega\over 2\pi}
   (\Sigma^{-+}G^{a})^{\nu\nu'}_{{\bf p}{\bf p}'}
   &=&
   - {iN_I\over\hbar}
   \sum_{\mu\mu'}
   \int { d {\bf q} \over (2 \pi\hbar)^2}
       |\tilde U_{q}|^2
   \rho^{\mu \mu'}_{(q_z+p_z,q_x+p_x-eB(\nu-\mu) l);
   (q_z+p_z', q_x+p_x'-eB(\nu'-\mu') l)}
\nonumber\\
   &\times&
   \biggl\{
      { E_{(q_z+p_z,q_x+p_x-eB(\nu-\mu) l )}
        + E_{(q_z+p_z', q_x+p_x'-eB(\nu'-\mu') l)}
       \over 2 }
   -
      E_{{\bf p}'}
      - eFl{{\mu + \mu'} \over 2 }  + eFl\nu' - i 0
   \biggr\}^{-1}
\nonumber\\
   &\times&
      \sum_{\alpha}
      J_{\nu-\alpha}^{p_x} J_{\mu-\alpha}^{q_x+p_x-eB(\nu-\mu) l }
      J_{\mu'-\alpha}^{q_x+p_x'-eB(\nu'-\mu') l}J_{\nu'-\alpha}^{p_x'}
      \;,
\label{eq:generic.2c}\\[20pt]
   -\int{d\omega\over 2\pi}
   (G^r\Sigma^{-+})^{\nu\nu'}_{{\bf p}{\bf p}'}
   &=&
   {iN_I\over\hbar }
   \sum_{\mu\mu'}
   \int { d {\bf q} \over (2 \pi\hbar)^2 }
       |\tilde U_q|^2
   \rho^{\mu \mu'}_{(q_z+p_z,q_x+p_x-eB(\nu-\mu) l);
                    (q_z+p_z',q_x+p_x'-eB(\nu'-\mu') l)}
\nonumber\\
   &\times&
   \biggl\{
    { E_{(q_z+p_z,q_x+p_x-eB(\nu-\mu) l)}
    + E_{(q_z+p_z',q_x+p_x'-eB(\nu'-\mu') l)} \over 2  }
    - E_{\bf p}
    - eFl{{\mu + \mu'} \over 2 }  + eFl\nu + i 0
   \biggr\}^{-1}
\nonumber\\
   &\times&
   \sum_{\alpha}
   J_{\nu-\alpha}^{p_x} J_{\mu-\alpha}^{q_x+p_x-eB(\nu-\mu) l}
   J_{\mu'-\alpha}^{q_x+p_x'-eB(\nu'-\mu') l}
   J_{\nu'-\alpha}^{p_x'}\;,
\label{eq:generic.2d}
\end{eqnarray}
and the  expression for the commutator on the left hand side of
Eq.~(\ref{eq:keld.19}) does not change.
\label{eq:generic.2}
\end{mathletters}


\section{Appendix: Gauge and translationally invariant equations.}
\label{sec:App.inv}

The derivative of the density matrix with respect to $p_x$ is proportional
to $1/\sqrt{mT_e}$ in the high temperature limit. The expansion with respect
to this parameter allow us to obtain the expression for the density matrix in
the Stark representation. We have from Eqs.~(\ref{eq:generic.5})
and (\ref{eq:keld.3a})
\begin{eqnarray*}
   &&\rho^{\nu\nu'}_{{\bf p}{\bf p}'}
   = \sum_{\alpha\alpha'}
   J_{\nu-\alpha}^{p_x}
   J_{\nu'-\alpha'}^{p_x'}
   (2\pi\hbar)^2\int d{\bf q}
   \wp_{\alpha-\alpha'}({\bf q})
   \delta(p_z- q_z)
   \delta(p_z'- q_z)
\nonumber\\
   &\times&
   \delta\left(p_x-eB\left[\nu - {\alpha+\alpha'\over 2}\right] l -q_x \right)
   \delta
   \left(p_x'-eB\left[\nu' - {\alpha+\alpha'\over 2}\right] l -q_x \right)
\end{eqnarray*}
and after shift of $p_x$ and $p_x'$, the above expression becomes
\begin{eqnarray}
   &&\rho^{\nu\nu'}_{
      (p_z,p_x+eB(\nu-\nu')l/2);
      (p_z',p_x'+eB(\nu'-\nu)l/2)
   }
   =
   (2\pi\hbar)^2 \delta({\bf p} - {\bf p}')
\nonumber\\
   &&\times
   \sum_{\alpha\alpha'}
   J_{\nu-\alpha}^{p_x+eB(\nu-\nu')l/2}
   J_{\nu'-\alpha'}^{p_x+eB(\nu'-\nu)l/2}
   \wp_{\alpha-\alpha'}
   \left(p_z,p_x + eBl{\alpha+\alpha'\over 2} - eBl{\nu+\nu'\over 2}\right)
\nonumber\\
   &&=
   (2\pi\hbar)^2 \delta({\bf p} - {\bf p}')
   \sum_{\alpha\alpha'}
   J_{\nu-\alpha}^{p_x+eB(\nu-\nu')l/2}
   J_{\nu'-\alpha'}^{p_x+eB(\nu'-\nu)l/2}
   \biggl\{
      1 -  eBl{\nu-\alpha+\nu'-\alpha'\over 2}
      {\partial \over \partial p_x}
   \biggr\}
   \wp_{\alpha-\alpha'}({\bf p})
\nonumber\\
   &&\approx
   (2\pi\hbar)^2 \delta({\bf p} - {\bf p}')
   \Biggl[
   \wp_{\nu-\nu'}({\bf p})
   +
      { \Lambda eBl \over eBp_x l/m - eFl }
      {\partial \over \partial p_x}
   \biggl\{
      \wp_{\nu-\nu'+1}({\bf p})
      +
      \wp_{\nu-\nu'-1}({\bf p})
   \biggr\}
   \Biggr]
\label{eq:generic.3b}
\end{eqnarray}
where high order derivatives with respect to $p_x$ are neglected.
The commutator on the left hand side of Eq.~(\ref{eq:keld.19}) becomes
\begin{eqnarray}
   &&
   \left( [H^S,\rho] \right)^{\nu\nu'}
   _{(p_z,p_x+eB(\nu-\nu')l/2);(p_z',p_x'+eB(\nu'-\nu) l/2)}
\nonumber\\
   & =&
   (2\pi\hbar)^2
   \delta({\bf p}-{\bf p}')
   \Biggl[
      (eBv_xl-eFl)(\nu-\nu')
      \wp_{\nu-\nu'}({\bf p})
 +
      \Lambda eBl (\nu-\nu')
      {\partial \over \partial p_x}
   \biggl\{
      \wp_{\nu-\nu'+1}({\bf p})
      +
      \wp_{\nu-\nu'-1}({\bf p})
   \biggr\}
   \Biggr]
\label{eq:generic.3}
\end{eqnarray}

The energy differences $E_p - E_q$ in the collision operator
Eq.~(\ref{eq:generic.2}) are of the order of $T_e$. Therefore, we can
neglect the terms containing $\partial \over \partial p_x$ in
Eq.~(\ref{eq:generic.3b}) for the density matrix when we substitute it in
Eq.~(\ref{eq:generic.2}). We have to expand first collision integrals
in terms of $(eBv_xl-eFl)/T_e$ too. After the
expansion, sums with respect to numbers of the Stark levels can be calculated
explicitly. Keeping only terms of the first and the second
order in Eq.~(\ref{eq:keld.19}) we get
\begin{eqnarray}
      i (eBv_xl - eFl)  \nu \wp_{\nu}({\bf p})
 &+&  i\Lambda eBl \nu
      {\partial \over \partial p_x}
   \biggl\{
      \wp_{\nu+1}({\bf p})
      +
      \wp_{\nu-1}({\bf p})
   \biggr\}
  =
   {2 \pi N_I}
   \int {{d{\bf q}} \over {(2 \pi\hbar)^2}}
   |V_{\bf p q}|^2
   \biggl\{
      \Bigl[\delta_{\nu,0} \wp_0 ({\bf q}) -\wp_{\nu} ({\bf p})\Bigr]
   \delta (E_p - E_q)
\nonumber\\
   & + & \Lambda \Bigl[
      \big( \wp_1({\bf q}) + \wp_{-1}({\bf q}) \big) \delta_{\nu,0}
      - \wp_0 ({\bf q}) (\delta_{\nu,1} + \delta_{\nu,-1})
      + \wp_{\nu + 1}({\bf p}) + \wp_{\nu - 1}({\bf p})
   \Bigr]\delta^{\prime} (E_p - E_q)
\nonumber\\
   &+& 2 \Lambda^2 \delta_{\nu , 0}
   \Bigl[\wp_0 ({\bf q}) -\wp_0 ({\bf p})\Bigr]
   \delta'' (E_p - E_q)
   \biggr\}
\label{eq:generic.6}
\end{eqnarray}

We neglected the term on the collision integral, which is different from the
first one on the left hand side of Eq.~(\ref{eq:generic.6}) only by a factor.
It can be considered as a renormalization of electron charge and can be
neglected.  Then one can see that if the terms of the order of $\Lambda /
T_e$ are neglected at all, the equation for $\wp_0$ is separated from
equations for $\wp_{\nu}$ with $\nu \neq 0$ and does not contain fields. This
is natural because $\wp_0$ is the distribution function in a layer and
without tunneling it does not 'know' about the electric field. Along with the
tunneling in this equation it is necessary to take into account inelastic
scattering, which was not considered so far. The operator of the inelastic
collisions, $I_{\rm in} \wp_0$ , can be added into Eq.(\ref{eq:generic.6})
without making use of the Keldysh technique.

Equations~({\ref{eq:generic.6}) with $\nu \neq 0$ show that $\wp_{\nu} \sim
(\Lambda /T_e)^{|\nu|} \wp_0$, i.e., $\wp_{ |\nu |}$ with $ \vert \nu \vert
\geq 2$ can be neglected and we come up with the following equations:
\begin{eqnarray}
& &{{2 \pi N_I} \over \hbar}
   \int { d{\bf q} \over (2 \pi \hbar)^2 }
    |V_{{\bf p}{\bf q}} |^2
   [\wp_0 ({\bf q}) -\wp_0 ({\bf p})]
   [\delta (E_p - E_q)
     + 2 \Lambda^2  \delta'' (E_p - E_q) ]
\nonumber \\
& + & {{2 \pi N_I \Lambda} \over \hbar}
   \int {{d{\bf q}} \over {(2 \pi\hbar)^2}}
    |V_{{\bf p q}} |^2
     \big[
       \wp_{1}({\bf q}) + \wp_{-1}({\bf q})
 +  \wp_{1}({\bf p}) + \wp_{- 1}({\bf p})
     \big]
      \delta' (E_p - E_q)
   + {\hat I}_{\rm in} \wp_0 ({\bf p})
      = 0
\label{eq:generic.7} \\[7pt]
 &\pm&  i { eBv_xl-eFl \over \hbar} \wp_{\pm 1} ({\bf p})
      \pm\Lambda eBl
      {\partial \over \partial p_x}
      \wp_{0}({\bf p})
 =
  - {1 \over {\hbar}} \Gamma ( E_p ) \wp_{\pm 1} ({\bf p})
\nonumber \\
& + &
   {{2 \pi N_I \Lambda} \over \hbar}
   \int { d{\bf q} \over (2 \pi\hbar)^2}
    |V_{\bf p q} |^2
     \big(
       \wp_{0}({\bf p}) - \wp_{0}({\bf q})
     \big)
      \delta' (E_p - E_q)    \;,
\label{eq:generic.8}
\end{eqnarray}
where for an isotropic energy spectrum and scattering (i.e. for
$ |V_{\bf p q}|^2$ depending on
$ |{\bf p} - {\bf q} | $ )
\begin{equation}
   \Gamma ( E_p ) = 2 \pi N_I
   \int {{d{\bf q}} \over (2 \pi\hbar)^2}
    | V_{\bf p q} |^2 \delta (E_p - E_q)
\label{eq:generic.9}
\end{equation}
depends only on the energy. For zero electric field Eqs.~(\ref{eq:generic.7})
and (\ref{eq:generic.8}) have isotropic solution.

The first term in Eq.~(\ref{eq:generic.7}) describes an elastic relaxation in
separate wells and is the largest one. If all other terms are neglected
it leads to a distribution function depending only on the energy, i.e.
$\wp_0 ({\bf p}) = \wp_0 ( E_p )$. An equation for this function can
be obtained by the averaging of Eq.~(\ref{eq:generic.7}) with respect to the
energy.\cite{gantsevich65,gurevich65} For example
\begin{eqnarray*}
\lefteqn{
   \int { d{\bf p}\over 2\pi m }\delta(\varepsilon-E_p)
   {2 \pi N_I \over \hbar}
   \int { d{\bf q} \over (2 \pi\hbar)^2}
    |V_{{\bf p q}} |^2
     \big[
       \wp_{1}({\bf q}) + \wp_{-1}({\bf q})
    +  \wp_{1}({\bf p}) + \wp_{- 1}({\bf p})
     \big]
     \delta' (E_p - E_q)
}
\nonumber \\
     &=&
   \int { d{\bf p}\over 2\pi m }\delta(\varepsilon-E_p)
     \big[
       \wp_{1}({\bf p}) + \wp_{- 1}({\bf p})
     \big]
     {d\over d\varepsilon}
   {2 \pi N_I \over \hbar}
   \int { d{\bf q} \over (2 \pi\hbar)^2}
    |V_{{\bf p q}} |^2  \delta(\varepsilon-E_q)
\nonumber \\
    &+&
     \int { d{\bf q}\over 2\pi m }
    \delta' (\varepsilon - E_q)
     \big[
       \wp_{1}({\bf q}) + \wp_{-1}({\bf q})
     \big]
   {2 \pi N_I \over \hbar}
   \int { d{\bf p} \over (2 \pi\hbar)^2}
    |V_{{\bf p q}} |^2
   \delta(\varepsilon-E_p)
\nonumber \\
    &=&
    {d\over d\varepsilon}
     \int { d{\bf p}\over 2\pi m }
    \delta (\varepsilon - E_p)
     \big[
       \wp_{1}({\bf p}) + \wp_{-1}({\bf p})
     \big]
   {2 \pi N_I \over \hbar}
   \int { d{\bf q} \over (2 \pi\hbar)^2}
    |V_{{\bf p q}} |^2
   \delta(\varepsilon-E_q)
\nonumber \\
    &=&
    {d\over d\varepsilon}
     \int { d{\bf p}\over 2\pi m }
    \delta (\varepsilon - E_p)
     \big[
       \wp_{1}({\bf p}) + \wp_{-1}({\bf p})
     \big]
   { \Gamma(E_p) \over \hbar}
    =
    {d\over d\varepsilon}
    {\Gamma(\varepsilon)R(\varepsilon)}\;,
\end{eqnarray*}
where $R(E)$ was defined in Eq.~(\ref{eq:generic.11a}).
The elastic relaxation is averaged out. On the right hand side of
Eq.~(\ref{eq:generic.8}) we have
\begin{eqnarray*}
   &&
   {{2 \pi N_I } \over \hbar}
   \int { d{\bf q} \over (2 \pi\hbar)^2}
    |V_{\bf p q} |^2
     \big(
       \wp_{0}({\bf p}) - \wp_{0}({\bf q})
     \big)
      \delta' (E_p - E_q)
\\
   &=&
   {{2 \pi N_I } \over \hbar}
   \int d\varepsilon\int { d{\bf q} \over (2 \pi\hbar)^2}
    |V_{\bf p q} |^2
     \big(
       \wp_{0}(E_p) - \wp_{0}(\varepsilon)
     \big)
      \delta' (E_p - \varepsilon)
      \delta(E_q-\varepsilon)
\\
   &=&
   {{2 \pi N_I } \over \hbar}
   \int d\varepsilon
   \delta(E_p - \varepsilon)
   { d \over d \varepsilon }
   \int { d{\bf q} \over (2 \pi\hbar)^2}
     |V_{\bf p q} |^2
     \big(
       \wp_{0}(E_p) - \wp_{0}(\varepsilon)
     \big)
     \delta(E_q-\varepsilon)
   =
   -{\Gamma(E_p)\over\hbar} {d\wp_0\over dE_p}
   \;.
\end{eqnarray*}

\twocolumn

\chapter{ Longitudinal magnetoresistance }
\label{chap:long}

   Classical longitudinal magnetoresistance of superlattices is calculated
   in the framework of a model which includes fluctuations of barrier
   conductivity. We found that the result depends very significantly on the
   fluctuations correlation length. We also found that fluctuations of
   the electron potential are not uniform along the superlattice, and depend
   on the superlattice length. A good agreement between theory and
   experiment is obtained.


\section{ Introduction}
\label{sec:int-long}

In this work we consider the vertical longitudinal magnetoresistance (LMR) of
a superlattice; that is the magnetoresistance in the geometry when both
electric and magnetic fields are along the growth direction. Purely classical
(i.e., without any quantum effects) LMR was observed many times in
experiments, but only recently has a qualitative explanation been
suggested.  It is obvious that in an ideal superlattice, classical LMR has to
be zero, because the magnetic field does not affect electron motion parallel
to it. For this reason experimentally observed LMR
(Refs.~\onlinecite{choi4,vuong,davies,skromme90,pavesi93,herbert-jan94,aristone-93})
has not been explained for a rather long time.  The qualitative
explanation suggested by Lee {\it et al}..\cite{lee-aug94} attributes this to
nonuniform fluctuations of the superlattice barriers width.

We present resistance calculations for a superlattice with
nonuniform barriers.  We consider the case of a narrow-miniband superlattice
when the vertical transport can be considered as sequential tunneling.
Each barrier in this case can be characterized by a conductivity
fluctuating around some average value.  The opposite case of wide-band
superlattices where an electron tunnels across a few barriers between two
successive scattering events seems to be less interesting. The effect
of the barrier width fluctuations is averaged out as a result of tunneling
across a few barriers.

The qualitative picture of the longitudinal magnetoresistance of
superlattices with nonuniform barriers suggested by Lee {\it et
al.}~\cite{lee-aug94} is as follows. A current across each barrier is
larger in places where the conductivity is larger. If high-conductivity
regions of adjacent barriers are not positioned against each other, then
nonuniform currents across barriers induce in-plane currents between
barriers. The magnetic field perpendicular to the layers brings about a
transverse magnetoresistance, reducing these in-plane currents. As a result
the current across barriers cannot pass through places with maximal
conductivity. In this way the magnetic field in the growth direction
increases the superlattice resistance in this direction.

The effective conductivity of a spatially inhomogeneous medium has been
considered many times in the literature; see, e.g., the review paper by
Landauer.\cite{landauer-78} A superlattice is just another example of such
a medium with a specific geometrical structure of the inhomogeneities. We
consider this problem for weak fluctuations of the barrier conductivity (the
exact parameter will be shown below).  We also assume that the conductivity
fluctuations of different barriers are not correlated.  These assumptions
allow us to obtain an analytic expression for the superlattice resistance.


\section{ Perturbation theory for potential fluctuations }
\label{sec:fluc}
\sectionmark{Perturbation theory}

The superlattice consists of $N+1$ wells separated by $N$ barriers, and
the electric potential in the $\nu$th well is $\phi_\nu(\bf r)$,
where ${\bf r}=(x,z)$ is the in-plane coordinate, see
Fig.~\ref{fig:long.geom}. The electric current $j_{\nu,\nu+1}(\bf r)$ from
well $\nu$ to the well $\nu+1$ is given by Ohm's law,
\begin{equation}
   j_{\nu,\nu+1} = \sigma_{\nu,\nu+1}^\perp
   (\phi_{\nu} - \phi_{\nu+1}) \;,
\label{eq:fluc.1}
\end{equation}
where $\sigma_{\nu,\nu+1}^\perp({\bf r})$ is the conductance per unit area of
the barrier following the $\nu$th well. The formulation of the problem will
be completed with the charge conservation law
\begin{equation}
   j_{\nu,\nu+1}  - j_{\nu-1,\nu} =
   {\bf \nabla}{\hat \sigma} {\bf \nabla} \phi_{\nu}\;,
\label{eq:fluc.2}
\end{equation}
where ${\hat \sigma} {\bf \nabla} \phi_{\nu}$ is an in-plane
electric current in the $\nu$th well, and ${\hat \sigma}$ is a
two-dimensional conductivity tensor of the well. We will assume that this
tensor depends on the magnetic field but does not depend on coordinates.  We
will also assume that the conductivity in the wells is isotropic, so that
$\sigma_{xx}=\sigma_{zz}=\sigma^\parallel$, and
$\sigma_{xz}=-\sigma_{zx}$.  This assumption gives
\begin{equation}
   {\bf \nabla}{\hat \sigma} {\bf \nabla}
   = \sigma^\parallel{\bf \nabla}^2\;,
\label{eq:fluc.3}
\end{equation}
that is, the Hall conductivity does not enter into the problem.
Equations~(\ref{eq:fluc.1})-(\ref{eq:fluc.3}) have to be solved with some
boundary conditions. We will assume that potentials in the first and last
wells are independent of $\bf r$ due to the presence of the highly doped
uniform plane contacts, $\phi_0 = NU = $const and $\phi_N = 0$.

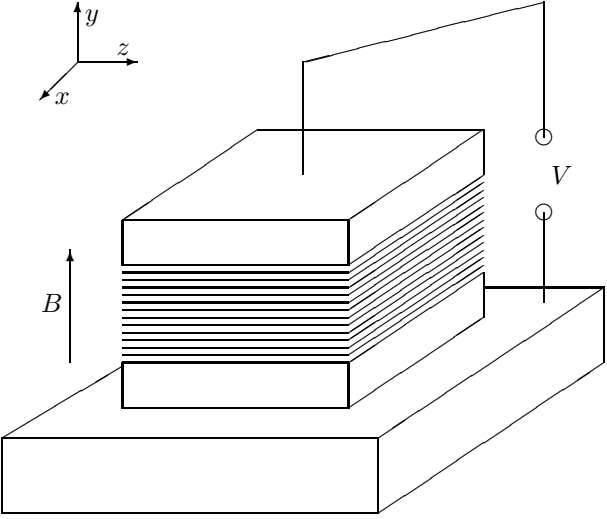
\begin{figure}
\unitlength=1.00mm
\linethickness{0.4pt}
\begin{picture}(80.00,73.00)
\put(10.00,65.00){\vector(1,0){8.00}}
\put(10.00,65.00){\vector(0,1){8.00}}
\put(10.00,65.00){\vector(-1,-1){5.00}}
\put(11.00,72.00){\makebox(0,0)[lt]{$y$}}
\put(17.00,66.00){\makebox(0,0)[rb]{$z$}}
\put(7.00,61.00){\makebox(0,0)[lt]{$x$}}
\put(0.00,5.00){\line(1,0){50.00}}
\put(50.00,5.00){\line(3,2){30.00}}
\put(80.00,25.00){\line(0,1){10.00}}
\put(80.00,35.00){\line(-3,-2){30.00}}
\put(50.00,15.00){\line(-1,0){50.00}}
\put(0.00,15.00){\line(0,-1){10.00}}
\put(50.00,15.00){\rule{0.00\unitlength}{-10.00\unitlength}}
\put(50.00,15.00){\line(0,-1){10.00}}
\put(46.00,19.00){\line(-1,0){30.00}}
\put(16.00,19.00){\line(0,1){6.00}}
\put(16.00,25.00){\line(1,0){30.00}}
\put(46.00,25.00){\line(0,-1){6.00}}
\put(46.00,19.00){\line(3,2){18.00}}
\put(64.00,31.00){\line(0,1){6.00}}
\put(64.00,37.00){\line(-3,-2){18.00}}
\put(16.00,27.00){\line(1,0){30.00}}
\put(46.00,27.00){\line(3,2){18.00}}
\put(64.00,41.00){\line(-3,-2){18.00}}
\put(46.00,29.00){\line(-1,0){30.00}}
\put(64.00,35.00){\line(1,0){16.00}}
\put(0.00,15.00){\line(5,3){16.00}}
\put(16.00,31.00){\line(1,0){30.00}}
\put(46.00,31.00){\line(3,2){18.00}}
\put(16.00,33.00){\line(1,0){30.00}}
\put(46.00,33.00){\line(3,2){18.00}}
\put(64.00,47.00){\line(-3,-2){18.00}}
\put(46.00,35.00){\line(-1,0){30.00}}
\put(16.00,37.00){\line(1,0){30.00}}
\put(46.00,37.00){\line(3,2){18.00}}
\put(16.00,36.00){\line(1,0){30.00}}
\put(46.00,36.00){\line(3,2){18.00}}
\put(16.00,34.00){\line(1,0){30.00}}
\put(46.00,34.00){\line(3,2){18.00}}
\put(64.00,44.00){\line(-3,-2){18.00}}
\put(46.00,32.00){\line(-1,0){30.00}}
\put(16.00,30.00){\line(1,0){30.00}}
\put(46.00,30.00){\line(3,2){18.00}}
\put(64.00,40.00){\line(-3,-2){18.00}}
\put(46.00,28.00){\line(-1,0){30.00}}
\put(16.00,26.00){\line(1,0){30.00}}
\put(46.00,26.00){\line(3,2){18.00}}
\put(16.00,38.00){\line(0,1){6.00}}
\put(16.00,44.00){\line(1,0){30.00}}
\put(46.00,44.00){\line(0,-1){6.00}}
\put(46.00,38.00){\line(-1,0){30.00}}
\put(46.00,38.00){\line(3,2){18.00}}
\put(64.00,50.00){\line(0,1){6.00}}
\put(64.00,56.00){\line(-3,-2){18.00}}
\put(64.00,56.00){\line(-1,0){30.00}}
\put(34.00,56.00){\line(-3,-2){18.00}}
\put(40.00,50.00){\line(0,1){15.00}}
\put(72.00,33.00){\line(0,1){12.00}}
\put(72.00,55.00){\line(0,1){18.00}}
\put(72.00,73.00){\line(-4,-1){32.00}}
\put(72.00,45.00){\circle{2.00}}
\put(72.00,55.00){\circle{2.00}}
\put(73.00,50.00){\makebox(0,0)[lc]{$V$}}
\put(9.00,25.00){\vector(0,1){15.00}}
\put(8.00,33.00){\makebox(0,0)[rc]{$B$}}
\end{picture}
   \caption{Measurements of LMR in superlattice.}
\label{fig:long.geom}
\end{figure}

Equations.~(\ref{eq:fluc.1})~and~(\ref{eq:fluc.2}) can be solved by means of
perturbation theory with respect to the fluctuations of
$\sigma_{\nu,\nu+1}^\perp$. In the Fourier representation,
\begin{equation}
   \sigma_{\nu,\nu+1}^\perp({\bf r})
      = \sigma^\perp
      + \sum_{\bf q}
      \delta\sigma_{\nu, {\bf q} }
      e^{-i{\bf q}\cdot{ r}}  \;,
\label{eq:fluc.4}
\end{equation}
where $\delta\sigma_{\nu, {\bf q} }$ is considered a small quantity and
$\delta\sigma_{\nu, {\bf 0} }\equiv0$. The conductivity fluctuations are
assumed to be uniform, with a correlation length much shorter than the
superlattice plane size, so that
\begin{equation}
   {
      \langle \delta\sigma_{\nu, {\bf q} }
      \delta\sigma_{\nu', {\bf q'} }^\ast \rangle
   \over (\sigma^\perp)^2
   }
   = \xi_q\delta_{\nu,\nu'}\delta_{\bf q, q'}\;,
\label{eq:fluc.4a}
\end{equation}
where $\langle\rangle$ means an ensemble average over all possible 
fluctuation configurations. The function $\xi_q$ is inversely proportional to 
the barrier area.  These conductivity fluctuations induce fluctuations of the
potential
\begin{equation}
   \phi_{\nu}({\bf r})
   = (N-\nu)U
   + \sum_{\bf q}
     \delta\phi_{\nu, {\bf q} }
     e^{-i{\bf q}\cdot{\bf r}}  \;.
\label{eq:fluc.5}
\end{equation}

\begin{mathletters}
The Fourier transform of Eqs.~(\ref{eq:fluc.1})-(\ref{eq:fluc.3})
can be linearized with respect to fluctuations if $\bf q \ne 0$,
\begin{eqnarray}
   && \Bigl[(\sigma^\parallel/\sigma^\perp) q^2 + 2\Bigr]
      \delta\phi_{\nu, {\bf q}}
      - \delta\phi_{\nu-1, {\bf q}} - \delta\phi_{{\nu+1}, {\bf q}}
\nonumber\\
   &=&  (U/\sigma^\perp)
   (\delta\sigma_{\nu-1, {\bf q} }-\delta\sigma_{\nu, {\bf q} })\;.
\label{eq:fluc.6a}
\end{eqnarray}
The second-order terms have to be kept in the same equations for $\bf q = 0$
\begin{eqnarray}
   && \Bigl[
      2 \delta\phi_{\nu, {\bf 0}}
      - \delta\phi_{\nu-1, {\bf 0}} - \delta\phi_{{\nu+1}, {\bf 0}}
   \Bigr]  \sigma^\perp
\nonumber\\
   &=&  \sum_{\bf q}
   \Bigl[
      ( \delta\phi_{\nu+1, {-\bf q}} - \delta\phi_{\nu, {-\bf q}})
      \delta\sigma_{\nu, {\bf q} }
\nonumber\\
   &-&
      ( \delta\phi_{\nu, {\bf -q}} - \delta\phi_{\nu-1, {\bf -q}} )
      \delta\sigma_{\nu-1, {\bf q} }
   \Bigr]\;.
\label{eq:fluc.6b}
\end{eqnarray}
\label{eq:fluc.6}
\end{mathletters}

The solution to Eq.~(\ref{eq:fluc.6a}) with the boundary conditions
$\delta\phi_{0, {\bf q}}=\delta\phi_{N, {\bf q}}=0$ can be expressed in terms
of the Green function,
\begin{eqnarray}
   G_{\nu,\nu'}(q) &=& {1\over N}
   \sum_{j=1}^{N-1}
      { \sin(\pi j \nu/ N)\sin(\pi j \nu'/N)
           \over
        \cosh(a_q) - \cos(\pi j /N)
       }
\label{eq:fluc.7a}
\\
   &=& { 1 \over \sinh(a_q)\sinh(a_qN) }
\nonumber\\
   & \times &
      \left\{\begin{array}{ll}
         \sinh(a_q\nu)\sinh[a_q(N-\nu')]
         \;,& \nu\le\nu'\;,\\
         \sinh(a_q\nu')\sinh[a_q(N-\nu)]
         \;,& \nu\ge\nu'\;,
      \end{array}
   \right.
\label{eq:fluc.7b}
\end{eqnarray}
where
\begin{equation}
   \cosh(a_q) = \left(1+{\sigma^\parallel q^2\over2\sigma^\perp}\right)\;.
\label{eq:fluc.7c}
\end{equation}
One has
\begin{equation}
   \delta\phi_{\nu, {\bf q} } =
    U\sum_{\nu'=1}^{N-1}  G_{\nu,\nu'}(q)
   {\delta\sigma_{\nu'-1, {\bf q} }-\delta\sigma_{\nu', {\bf q} }
   \over \sigma^\perp}\;,
\label{eq:fluc.8a}
\end{equation}
with $\bf q \ne 0$. The derivation of the above expression for the
Green function is given in Appendix~\ref{sec:appA}.

Physical properties of the result Eq.~(\ref{eq:fluc.8a}) can be seen from the
average value of the potential fluctuations squared,
\begin{eqnarray}
   \langle |\delta\phi_{\nu, {\bf q} }|^2 \rangle &=&
   U^2\xi_q
   \sum_{\nu'=1}^{N-1}
   G_{\nu,\nu'}(q)
\nonumber \\
   &\times&
   \Bigl[
      2G_{\nu,\nu'}(q) - G_{\nu,\nu'-1}(q) - G_{\nu,\nu'+1}(q)
   \Bigr]
\nonumber\\
   &=&  U^2\xi_q
   \Bigl[
      G_{\nu,\nu}(q) + {q\over2}\;{\partial\over\partial q} G_{\nu,\nu}(q)
   \Bigr]\;,
\label{eq:fluc.8b}
\end{eqnarray}
where $G_{\nu,\nu}(q)$ is given by Eq.~(\ref{eq:fluc.7b}).

Equation~(\ref{eq:fluc.8b}) describes the increase of the potential
fluctuations from the contacts toward the middle of the superlattice.
For a limited region of $q$, three situations are conceivable. The first is
the case of strong in-plane conductivity, when the potential fluctuations are
limited by in-plane currents, and $\langle |\delta\phi_{\nu, {\bf q} }|^2
\rangle\;, \nu\ne0,N$
nearly does not depend on $\nu$.  The second is the opposite case, when
in-plane currents are not important, and the fluctuations of the potential
are similar to the fluctuations in a series of random resistors.
In the third intermediate case the fluctuations
increase with the distance from the contacts, but in the internal part of the
superlattice they are limited by in-plane currents.

\begin{mathletters}
The first case is realized under the condition $\sinh(a_q)\gg1$; then
Eqs.~(\ref{eq:fluc.7b}) and (\ref{eq:fluc.8b}) give
\begin{equation}
   \langle |\delta\phi_{\nu, {\bf q} }|^2 \rangle
   =  2\xi_q
   \left({U\sigma^\perp \over\sigma^\parallel q^2}\right)^2\;,\;\;
   \nu\ne0,N
\label{eq:fluc.9b}
\end{equation}
This result is independent of $\nu$ and inverse proportional to the in-plane
conductivity. In the second case, $a_q\ll1/N$, and we have
\begin{equation}
   \langle |\delta\phi_{\nu, {\bf q} }|^2 \rangle =
   U^2\xi_q
   { \nu (N-\nu)\over N} \;.
\label{eq:fluc.9a}
\end{equation}
In the third intermediate  case, $1/N \lesssim a_q\ll1$, and we have to
consider separately the internal region of superlattice and regions near the 
contacts,
\begin{equation}
   \langle |\delta\phi_{\nu, {\bf q} }|^2 \rangle
   = U^2\xi_q
   \times\left\{\begin{array}{ll}
      \nu\;,\;\; &  a_q\nu\ll1\;,\\
      N-\nu\;,\;\;& a_q(N-\nu)\ll1\;,\\
      1/a_q\;,\;\;& \text{otherwise}\;.
   \end{array}
   \right.
\label{eq:fluc.9c}
\end{equation}
\label{eq:fluc.9}
\end{mathletters}

In order to calculate the correction to the average potential, we have to
substitute Eq.~(\ref{eq:fluc.8a}) into Eq.~(\ref{eq:fluc.6b}). This leads to
an equation for $\delta\phi_{\nu,\bf 0}$, which should be averaged
with the help of Eq.~(\ref{eq:fluc.4a}). The solution to the obtained equation
is
\begin{eqnarray}
   \langle\delta\phi_{\nu,{\bf{0}}}\rangle
   &=& - U\sum_{\bf q} { \xi_q \over \sigma^\parallel q^2/\sigma^\perp + 4}
\nonumber\\
   &\times&
   \left[
      {\sinh[a_q(N-2\nu)] \over \sinh(a_qN) }
      +
      {2\nu-N\over N}
   \right]\;.
\label{eq:fluc.10}
\end{eqnarray}
The $\nu$ dependence of the averaged potential,
$(N-\nu)U + \langle\delta\phi_{\nu,{\bf{0}}}\rangle$,
becomes more smooth near
the contacts; that is near the contacts the electric field is weaker.

One can prove that the perturbation theory developed here is justified if
the fluctuations of the potential are small,
\begin{equation}
   \sum_{\bf q}\langle|\delta\phi_{\nu,{\bf q}}|^2\rangle \ll U^2
\label{eq:fluc.11a}
\end{equation}
The substitution of the results obtained above,
Eqs.~(\ref{eq:fluc.9}) and (\ref{eq:fluc.10}), to this condition leads to
\begin{equation}
   \Xi\equiv
   \sum_q\xi_q
   \ll
   \mbox{max}
   \biggl\{
      { 1\over N },\;
      q_0\left(\sigma^\parallel\over\sigma^\perp\right)^{1/2},\;
      {\sigma q_0^2\over\sigma^\perp}
   \biggr\}\;.
\label{eq:fluc.11b}
\end{equation}
Here $q_0$ is the characteristic wave number of the function
$\xi_q$, and the quantity $1 / q_0$ can be considered as a
conductivity fluctuations characteristic correlation length.


\section{ Averaged current and longitudinal magnetoresistance }
\label{sec:aver}
\sectionmark{Averaged current}

The total current across a barrier,
\begin{equation}
   j =  \sigma^\perp U  - \delta j \;,
\label{eq:aver.0}
\end{equation}
is the same for all barriers, and can be calculated for the first barrier.
Substitution of Eqs.~(\ref{eq:fluc.4}), (\ref{eq:fluc.6}), and
(\ref{eq:fluc.8a}) in Eq.~(\ref{eq:fluc.1}) with $\nu=0$, averaging over all
possible fluctuations of the barrier conductivity and summation over all $q$,
give
\begin{eqnarray}
   \delta j  &=& \sigma^\perp \left\{ U \sum_{\bf q}
   \xi_q G_{1,1}(q)
   + \langle\delta\phi_{1,{\bf 0}}\rangle \right\}
\\
   &=&
   \sigma^\perp U \sum_{\bf q}
   { 2\xi_q\over \sigma^\parallel q^2/\sigma^\perp + 4 }
\nonumber
\\ &\times&
   \left[
      {N-1\over N} + {\sinh[a_q(N-1)]\over\sinh(a_qN)}
   \right] \;.
\label{eq:aver.1}
\end{eqnarray}
This equation is the main result of our paper. It describes the change of
the current due to barrier resistance fluctuations. The sign of $\delta j$ is
equal to the sign of the current without fluctuations; that is, fluctuations
lead to an increase of the superlattice resistance.

We are particularly interested in the application of this result to a
calculation of the magnetoresistance of the superlattices. Here we consider
only a weak magnetic field when $\Omega_c\tau\ll1$, where
\[
   \Omega_c = {eB\over m}
\] is the
cyclotron frequency and $\tau$ is the relaxation time. In this case the
magnetic-field-induced change of the in-plane conductivity is
$\sigma^\parallel(0)-\sigma^\parallel(B) \approx \Omega_c^2\tau^2
\sigma^\parallel(0)$, and the superlattice magnetoresistance becomes
\begin{equation}
   {R(B)-R(0) \over R(0)}
   = - { \Omega_c^2\tau^2 \sigma^\parallel\over U \sigma^\perp}
   {\partial\over\partial\sigma^\parallel} \delta j\;.
\label{eq:aver.2}
\end{equation}

In transport theory, surface roughness is often approximated by a Gaussian
function. Such an approximation immediately gives the Gaussian form for
the barrier conductivity fluctuations correlation function, i.e.,
\begin{equation}
   \xi_q = {4\pi \Xi\over q_0^2 S}e^{-q^2/q_0^2}\;,
\label{eq:aver.3}
\end{equation}
where $S$ is the area of the barrier, and $\Xi$ is the standard deviation of 
the normalized barrier conductivity, which is defined generally in 
Eq.~(\ref{eq:fluc.11b}). Substitution of Eq.~(\ref{eq:aver.3}) into
Eq.~(\ref{eq:aver.2}) allows us to evaluate the superlattice
magnetoresistance
\begin{eqnarray}
   {R(B)-R(0) \over R(0)}
   &=& \Omega_c^2\tau^2
   \sum_{\bf q}\xi_q
\nonumber\\
   &\times&\left\{ \begin{array}{ll}
      \displaystyle
      \sigma^\parallel q^2N / (6\sigma_\perp)
      \;,\;\;&\gamma N^2\ll1\;, \\ 
      \displaystyle
      { 2\left[\sigma^\parallel q^2/\sigma^\perp \right]^{1/2}
         \over
        \left[ \sigma^\parallel q^2/\sigma^\perp + 4 \right]^{3/2}
      } \;,\;\;&\gamma N^2\gg1\;,
   \end{array}\right.
\label{eq:aver.4}
\\
   &\sim&
   \Xi \Omega_c^2\tau^2
   \left\{ \begin{array}{ll}
      \gamma N,\;&\gamma N^2\ll1\;,    \\ 
      \sqrt{\gamma},\;&\gamma\ll1\;, \\ 
      \ln{\gamma}/\gamma,\;&\gamma\gg1\;,
   \end{array}\right.
\nonumber\\
   \relax
\label{eq:aver.5}
\end{eqnarray}
where $\gamma=\sigma^\parallel q_0^2/\sigma^\perp$. One can see from
Eq.~(\ref{eq:aver.5}) that the magnetoresistance disappears for both very
small and very large $q_0$. The reasons for this, however, are different.
The former is a case of effectively ``metallic'' superlattice wells. They
are almost equipotential planes, in-plane currents are small, and
magnetoresistance is also small. In addition, in this case barrier
conductivity fluctuations are averaged out and $\delta j$ itself goes to
zero. The latter case is a case of effectively ``dielectric'' planes. The
high-conductivity regions of adjacent barriers are located far from each
other. In this case the conductances of the in-plane path are small and the
in-plane currents are also small.


\section{ Discussion and summary }
\label{sec:summ}

Beside the correction to the current, the surface roughness of superlattice
barriers leads to a quite unexpected result: distribution of an electric field
along the superlattice appears to be nonuniform. Indeed, for the correction
to the average potential drop across one barrier, Eq.~(\ref{eq:fluc.10})
gives
\begin{eqnarray}
\delta\phi_{\nu-1,0}-\delta\phi_{\nu,0} & = &
   U\sum_{\bf q} { \xi_q \over \sigma^\parallel q^2/\sigma^\perp + 4}
\nonumber\\ && \hspace{-2cm}
   \times
   \left[
      {2\over N}
      -
      {\sinh(a_q)\cosh[a_q(2\nu-1-N)] \over \sinh(a_qN) }
   \right]\;.
\label{eq:aver.10}
\end{eqnarray}
In the middle of the superlattice the sign of this quantity is the same as
that of the potential drop without surface roughness, $U$, and near the
contacts it is the opposite, see Fig~\ref{fig:inhf}. Because of
surface roughness the field becomes stronger at the middle and weaker near
the contacts. The size of the contact regions is about $1/a_{q}$ periods. The
redistribution of the field along the superlattice is not a large effect, but
it can be stronger for more pronounced surface roughness. The physical reason
for the field redistribution is that the current is trying to go across the
least resistive regions of the barriers.  Because of the lack of
correlation of the surface roughness in different barriers, this produces an
in-plane current which makes the overall resistance larger.  Near the
contacts where the in-plane potential redistribution is not fully developed,
this effect is suppressed.

\begin{figure}
\begin{center}
\unitlength=1mm
\linethickness{0.4pt}
\begin{picture}(86.00,100.00)
\ifPostScript
   \put(6.00,10.00){\epsffile{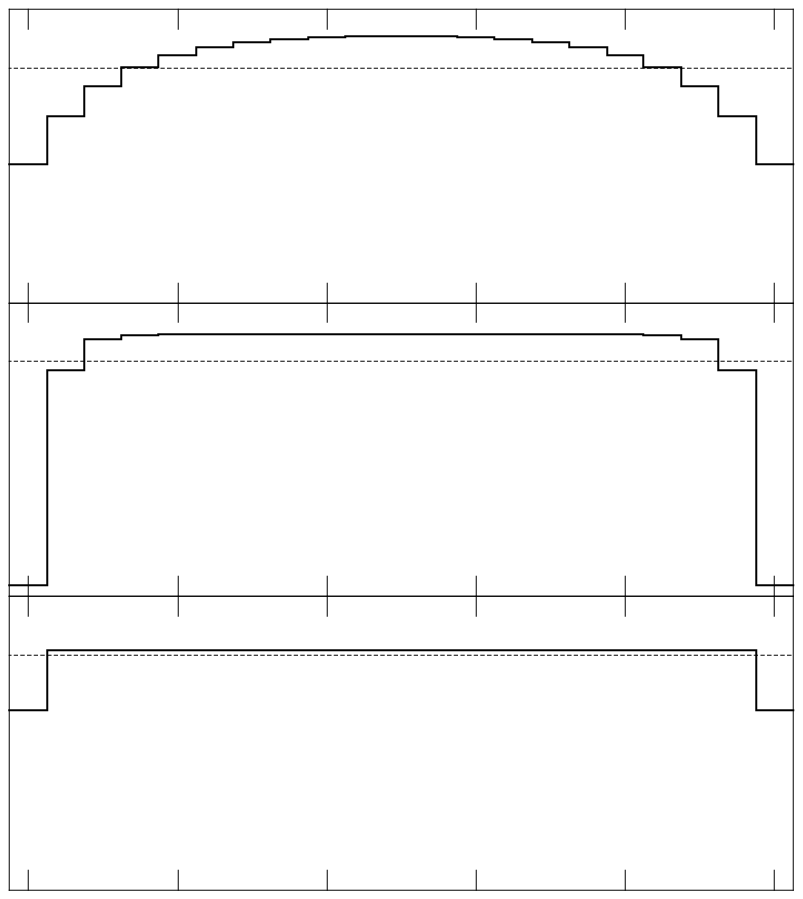}}
\else
   \put(0.0,105){\special{em:graph inhfield.pcx}}
\fi
\put(84.00,8.00){\makebox(0,0)[ct]{20}}
\put(69.00,8.00){\makebox(0,0)[ct]{16}}
\put(54.00,8.00){\makebox(0,0)[ct]{12}}
\put(39.00,8.00){\makebox(0,0)[ct]{8}}
\put(24.00,8.00){\makebox(0,0)[ct]{4}}
\put(9.00,8.00){\makebox(0,0)[ct]{0}}
\put(46.00,3.00){\makebox(0,0)[cb]{$\nu$}}
\put(46.00,12.00){\makebox(0,0)[cb]{$\sigma^\parallel q^2/\sigma^\perp = 20$}}
\put(46.00,42.00){\makebox(0,0)[cb]{$\sigma^\parallel q^2/\sigma^\perp = 1$}}
\put(46.00,72.00){\makebox(0,0)[cb]{$\sigma^\parallel q^2/\sigma^\perp = 1/20$}}
\put(3.00,55.00){\makebox(0,0)[cc]{
\ifPostScript
   \rotate[l]{Change of the field (arb. units)}
\else
   $\Delta F$
\fi
}}
\end{picture}

\end{center}
   \caption{ Contribution of the terms with different wave vectors to
   the change of the averaged field. One can see that the field becomes
   higher at the middle of superlattice and lower near the contacts.  The
   scales on all the graphs are the same, and the dashed lines show zeros of
   the field change.
}
\label{fig:inhf}
\end{figure}

\begin{figure}
\begin{center}
\unitlength=1mm
\begin{picture}(86.00,90.00)
\put(6.00,10.00){
   \ifPostScript
      \put(0.00,0.00){\epsffile{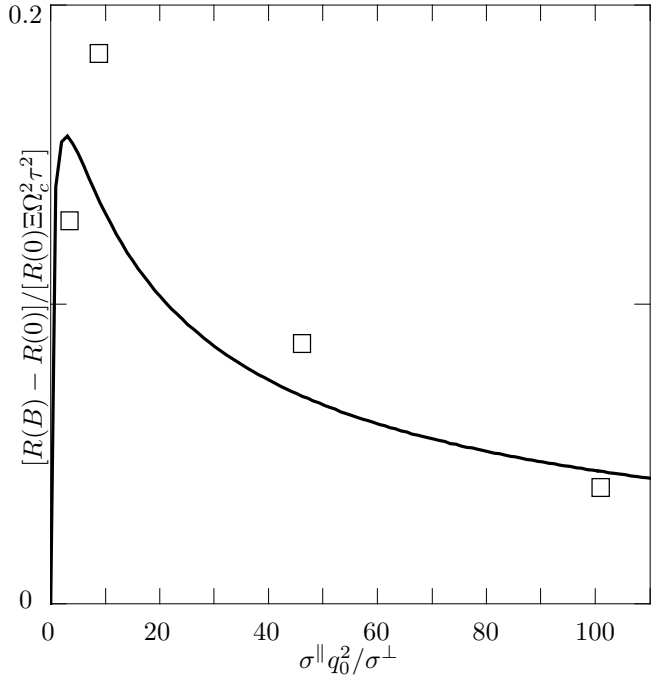}}
   \else
      \put(-6.00,86.00){\special{em:graph inhfitto.pcx}}
   \fi
   \put(72.7,-2.00){\makebox(0,0)[ct]{100}}
   \put(58.2,-2.00){\makebox(0,0)[ct]{80}}
   \put(43.6,-2.00){\makebox(0,0)[ct]{60}}
   \put(29.0,-2.00){\makebox(0,0)[ct]{40}}
   \put(14.5,-2.00){\makebox(0,0)[ct]{20}}
   \put(0.00,-2.00){\makebox(0,0)[ct]{0}}
   \put(40.00,-8.00){\makebox(0,0)[cb]{$\sigma^\parallel q_0^2/\sigma^\perp$}}
   \put(-3.00,40.00){
   \ifPostScript
      \makebox(0,0)[cc]{
         \rotate[l]{$[R(B)-R(0)]/[R(0)\Xi\Omega_c^2\tau^2]$}
      }
   \else
      \makebox(0,0)[rc]{
         $\displaystyle { R(B)-R(0) \over R(0)\Xi\Omega_c^2\tau^2}$
      }
   \fi}
   \put(-3.00,00.00){\makebox(0,0)[cb]{0}}
   \put(-3.00,80.00){\makebox(0,0)[ct]{0.2}}
}
\end{picture}

\end{center}
   \caption{
      Comparison of the experimental data of
   Ref.~\protect{\onlinecite{lee-aug94}},
   (open squares), with the theoretical prediction of
   Eq.~(\protect{\ref{eq:aver.8}}), solid line. Combinations
   $[R(B)-R(0)]/[\Xi R(0)\Omega_c^2\tau^2]$ and $\sigma^\parallel q_0^2
   /\sigma^\perp$ are calculated from the experimental data summarized in
   Table~\protect{\ref{table:expdata}}, with two fitting parameters
   $\tau=5.3\times10^{-13}$ s and $q_0=3.3\;\mu$m$^{-1}$.
}
\label{fig:inhfit}
\end{figure}

The correction to the current due to surface roughness strongly depends on
$\sigma^\parallel q_0^2/\sigma^\perp$. This parameter may significantly
vary in experiments. Its value can be estimated in terms of microscopic
parameters of the superlattice. We can estimate
$\sigma^\perp\sim(me^{2}\Lambda^{2}\tau\hbar^{-4})[1 - \exp(E_{F}/T)]$
(Ref.\onlinecite{laikhtman2}), where $e$ and $m$ are the electron charge and
mass, respectively, $\Lambda$ is the transition amplitude between adjacent
wells, $\tau$ is the relaxation time, $E_F$ is the Fermi energy, $T$ is the
temperature, and $\Lambda$, $\hbar/\tau$ are assumed to be much less than the
maximum of $E_F$, $T$. In this case we have
\begin{equation}
   \sigma^\parallel q_0^2 /\sigma_\perp
   \approx {\hbar^2q^2\over m\Lambda^2}\text{max}(E_F,T)
   \approx { (lq)^2\hbar^2 \over (\Lambda\tau)^2 }\;,
\label{eq:aver.6}
\end{equation}
where $l$ is the in-plane mean free path. For the conductivity in this
expression we used the classical expression $\sigma^\parallel \sim
ne^{2}\tau/m$, where $n$ is the two-dimensional electron concentration. This
expression, as well as the phenomenological Eq.~(\ref{eq:fluc.2}), is correct
under the condition $ql\ll1$. The other basic equation,
Eq.~(\ref{eq:fluc.1}), is justified only under the condition of sequential
tunneling, i.e., $\hbar/(\Lambda\tau)\gg1$.  That is, the right-hand side  in
Eq.~(\ref{eq:aver.6}) is the product of a large factor and a small factor, so
that all cases in Eq.~(\ref{eq:aver.5}) are possible. In these three cases
the temperature dependences of the magnetoresistance are  $T$, $T^{1/2}$,
and $\ln(T)/T$, respectively.

The condition $ql\ll1$ means that the characteristic scale of the surface
roughness is much larger than the mean free path. The theory can be easily
generalized for the case when this condition is not satisfied. The in-plane
conductivity in Eq.~(\ref{eq:fluc.2}) is a response to a uniform electric
field. If the electric field is nonuniform at the scale of the mean free
path, then the current conservation law, Eq.~(\ref{eq:fluc.2}), holds but the
in-plane conductivity cannot be taken from the phenomenological theory and
should be calculated with the help of the Boltzmann equation. The calculation
is carried out in Appendix~\ref{sec:appB}, and the resulting conductivity
depends on $q$. The only modification in the previous theory is that the
expression $\sigma^\parallel q^{2}$ is determined now by
Eq.~(\ref{eq:appB.5}).  The estimate of the magnetoresistance,
Eq.~(\ref{eq:aver.5}), which was done before for the phenomenological case,
is replaced in the limit $q_0l\gtrsim1$ by
\begin{equation}
   {R(B) - R(0) \over R(0)}
   = \Xi \Omega_c^2\tau^2
   {\Lambda^2\tau^2\over\hbar^2q_0^2l^2}
   \ln\left( {\hbar^2\over\Lambda^2\tau^2}\right)\;.
\label{eq:aver.7}
\end{equation}

In the calculation of the conductivity we neglect quantum corrections. That
is justified when the magnetic quantization is smeared by
scattering, $\Omega_c\tau\ll 1$, or at high enough temperature, when
$\hbar\Omega_c\lesssim T$.

\begin{table}
\label{table:expdata}
\caption{ Summary of experimental data, which was used in
Fig.~\protect{\ref{fig:inhfit}}.}
\begin{center}
\begin{tabular}{cccc}
   $(d_w/d_B)$ &
   $\Xi$ &
   $\displaystyle { \Delta R/( RB^2)}$ &
   $\sigma^\perp={N/(RS)}$
   \\
   (\AA/\AA) & & T$^{-2}$  & S/cm$^2$ \\
\hline \\
   (50/50) & 0.36 & 0.027 & 1200\\
   (20/80) & 0.09 & 0.032 & 5500\\
   (80/20) & 0.25 & 0.042 & 4200\\
   (20/40) & 0.19 & 0.047 & 14000\\
\end{tabular}
\end{center}
\end{table}

The comparison of the our result with available experimental
data is difficult because not all parameters necessary for theoretical
calculations are known. Here we compare our results with the measurements of
Ref.~\onlinecite{lee-aug94}, taking the relaxation time and the
characteristic length of interface roughness as adjustable parameters. The
widths of wells ($d_{w}$) and barriers ($d_{B}$) and
measured magnetoresistance and barrier conductances in four measured samples
are summarized in Table~\ref{table:expdata}.  We assume that fluctuations of
the transition amplitude between adjacent wells, $\Delta\Lambda$, result
from the fluctuations of the width of the barrier by $\pm1$ monolayer. The
known geometry of the structures allowed us to calculate $\Lambda$, and the
fluctuations of the barrier conductance were evaluated according to
$\Xi\approx(2\Delta\Lambda/\Lambda)^2$. The conductivities in wells were
calculated according to $\sigma^\parallel=ne^{2}\tau/m$, where the
two-dimensional electron concentration $n=d_w\times10^{17}$ cm$^{-3}$. The
dependence of the magnetoresistance on parameters of the samples can be
written in the form
\begin{equation}
   {R(B)-R(0) \over R(0)B^2}
   {m^2 \over e^2\tau^2 \Xi}
   =
   f\left( {\sigma^\parallel q_0^2 \over \sigma^\perp }\right)\;.
\label{eq:aver.8}
\end{equation}
For the function in the right-hand side Eq.~(\ref{eq:aver.4}) gives
\begin{equation}
   f(\gamma) = \int_0^\infty dx e^{-x}
      \left( 4\gamma x\over(\gamma x + 4)^3\right)^{1/2}\;.
\label{eq:aver.9}
\end{equation}
In Fig.~\ref{fig:inhfit} we show the theoretical curve for this function and
the experimental results for four samples. The best fit
is obtained for $\tau=5.3\times10^{-13}$ s and $q_0=3.3\;\mu$m$^{-1}$. We have
to note an obvious qualitative and good quantitative agreement between the
theory and experiment.  The discrepancy (no more than 10\%) can be attributed
to slightly different relaxation times $\tau$ and surface roughness
characteristics in different samples.

The results of the fitting give $q_0l\approx0.2$, which justifies the
phenomenological expression for $\sigma^\parallel$ used in the calculations.
The estimate of the surface roughness relaxation time for a quantum well with
$d_{w}=50$ \AA{} and $q_0=3.3\;\mu$m$^{-1}$ gives a value of
$2\times10^{-10}$ s.  That is, the dominant scattering mechanism is
probably impurity scattering, which explains approximately equal relaxation
times in samples with different values of $d_{w}$. The surface roughness
correlation length of 3000 \AA{} seems large, but even values larger by an 
order of magnitude have been reported.

In summary, we calculated the correction to the superlattice resistance due
to nonuniform fluctuations of the conductivity of each of the barriers. Our
results explain the classical longitudinal magnetoresistance of
superlattices.  We found that the magnetoresistance has a nontrivial
dependence on the characteristic length scale of fluctuations; it goes to zero
for both very large-scale and very-small scale fluctuations.  The
fluctuations of the barrier conductivity lead also to a nonuniform
distribution of the electric field along the superlattice. The results of the 
theory give a good quantitative agreement with experimental data.

\section{ Derivation of the Green function}
\label{sec:appA}

Trigonometric sums can be calculated with the help of the equation
\begin{equation}
   2\cosh(\lambda)u(\nu) - u(\nu+1) - u(\nu-1) = 0 \;,\;\;
   0 < \nu < N \;.
\label{eq:appA.1}
\end{equation}
The general solution to Eq.~(\ref{eq:appA.1}) has the form
\begin{equation}
   u(n) = C_{1} e^{\lambda\nu} + C_{2} e^{-\lambda\nu} \;,
\label{eq:appA.3}
\end{equation}
and the eigenvalues and the eigenfunctions of Eq.~(\ref{eq:appA.1})
with the boundary conditions $u_{0} = u_{N} = 0$ are
\begin{eqnarray}
   \cosh(\lambda_j) &=& \cos(\pi j /N)\;,\;\; 1 \le j \le N - 1 \;,
\nonumber\\
   u_{j}(\nu) &=& \left(2 \over N\right)^{1/2} \sin(\pi j\nu / N) \;.
\label{eq:appA.4}
\end{eqnarray}

Let us consider the function $G_{\nu,\nu'}$ satisfying the equation
\begin{equation}
   2\cosh(\lambda)G(\nu,\nu') - G_{\nu+1,\nu'} - G_{\nu-1,\nu'}  =
   \delta_{\nu,\nu'}
\label{eq:appA.5}
\end{equation}
with the boundary conditions $G_{0,\nu'} = G_{N,\nu'} = 0$. This can be
expressed in terms of the eigenfunctions Eq.~(\ref{eq:appA.4}),
\begin{equation}
   G_{\nu,\nu'} = {1\over N}
   \sum_{j=1}^{N-1}
      { \sin(\pi j \nu/ N)\sin(\pi j \nu'/N)
           \over
        \cosh(\lambda) - \cosh(\lambda_j)
       }
\label{eq:appA.6}
\end{equation}
and this is the first line in Eq.~(\ref{eq:fluc.7a}),
with $\lambda \equiv a_q$.

The same function can be calculated in another way with the help of the
general solution Eq.~(\ref{eq:appA.3}). Apparently, Eq.~(\ref{eq:appA.5}) with the
boundary conditions is satisfied by the function
\begin{eqnarray}
   G_{\nu,\nu'} &=& A_1\; \sinh( \lambda\nu) \;,\;\;
      \nu < \nu'  \;,
\nonumber\\
   G_{\nu,\nu'} &= & A_2\; \sinh[\lambda (N - \nu)] \;,\;\;
      \nu > \nu'  \;.
\label{eq:appA.7}
\end{eqnarray}
The only problem is with the values $\nu = \nu'$. The values of $A,\;B$, and
$G_{\nu',\nu'}$ can be obtained by substitution of the solutions
Eq.~(\ref{eq:appA.7}) into Eq.~(\ref{eq:appA.5}) for $\nu = \nu', \nu'\pm 1$,
that leads to the system of equations
\begin{eqnarray}
   A_1\; \sinh(\lambda\nu') - G_{\nu',\nu'} &=& 0 \;,
\nonumber\\
   A_1\; \sinh[ \lambda (\nu' - 1)] - 2\cosh(\lambda) G_{\nu',\nu'} &&
\nonumber\\
   + A_2\; \sinh[ \lambda (N - \nu' - 1)] &=& -1 \;,
\nonumber\\
   G_{\nu',\nu'} - A_2\; \sinh[ \lambda (N - \nu')] &=& 0 \;,
\nonumber\\
   \relax
\label{eq:appA.10}
\end{eqnarray}
which has a solution
\begin{eqnarray}
    A_1 &=&  {  \sinh[ \lambda (N - \nu')]
            \over \sinh(\lambda) \sinh(\lambda N)
         } \;,
\nonumber\\
    G_{\nu',\nu'} &=&
    {\sinh( \lambda \nu') \ \sinh[ \lambda (N - \nu')]
       \over \sinh(\lambda) \ \sinh( \lambda N)
    } \;,
\nonumber\\
    A_2 &=& {  \sinh( \lambda \nu')
             \over \sinh( \lambda) \ \sinh (\lambda N)
          } \;,
\label{eq:appA.16}
\end{eqnarray}
and the final expression for the Green function is
\begin{equation}
   G_{\nu,\nu'} = \left\{
\begin{array}{ll}
	{\displaystyle
   \sinh \lambda \nu \ \sinh \lambda (N - \nu')
	\over\displaystyle
   \sinh \lambda \ \sinh \lambda N} \ ,
\hspace{1cm} & \nu \le \nu' \ , \\[10pt]
	{\displaystyle
   \sinh \lambda \nu'2 \ \sinh \lambda (N - \nu)
	\over\displaystyle
   \sinh \lambda \ \sinh \lambda N} \ ,
\hspace{1cm} & \nu \ge \nu' \ ,
\end{array}	\right.
\label{eq:appA.17}
\end{equation}
This result is also an explicit expression for the sum in
Eq.~(\ref{eq:appA.6}) and this is the second expression in
Eq.~(\ref{eq:fluc.7b}).

\section{ Wave-vector-dependent in-plane magnetoconductivity }
\label{sec:appB}
\sectionmark{In-plane magnetoconductivity }

The Fourier transformation of the Boltzmann kinetic equation for the
two-dimensional electron gas in the conducting layer is
\begin{equation}
  i{\bf q}\cdot{\bf v}f_{\bf q} +
  e [{\bf v} \times {\bf B}]
  {\partial f_{\bf q}\over \partial {\bf p}}
  -ie{\bf q}\cdot{\bf v}\phi_{\bf q} {\partial f\over \partial E} =
  - {f_{\bf q}\over \tau}\;,
\label{eq:appB.1}
\end{equation}
where $f(E)$ is the equilibrium distribution function, $f_{\bf q}$ is the
Fourier transform of the distribution function perturbation,
$\phi_{\bf q}$ is the fluctuation of the electron potential, and $\bf B$ is
the magnetic field, which is considered to be perpendicular to the layers,
see Fig.~\ref{fig:long.geom}
i.e., ${\bf q}\perp{\bf B}$. By introducing ${\bf q} = (q,0,0)$, ${\bf v} =
(v\cos(\theta), 0, v\sin(\theta))$, and ${\bf B} = (0, m\Omega_c/e, 0)$
we reduce the kinetic equation to
\begin{equation}
   \Omega_c{\partial f_{\bf q}\over \partial \theta}
   - \left[ {1\over \tau} + iqv\cos(\theta) \right]f_{\bf q}
   = -iqve\phi_{\bf q}\cos(\theta){\partial f\over\partial E}\;.
\label{eq:appB.2}
\end{equation}
It is easy to solve this first-order differential equation, and the answer is
a function of $v$ and $\theta$:
\begin{eqnarray}
   f_{\bf q} &=& e\phi_{\bf q}{\partial f\over\partial E}
   \biggl\{ 1- \int_0^\infty{d\theta'\over\Omega_c\tau}
\nonumber\\
   &\times&
   \exp\Bigl({-\theta'\over\Omega_c\tau}
   +i{qv\over\Omega_c}
   \bigl[\sin(\theta) - \sin(\theta'+\theta)\bigr]\Bigr)
   \biggr\}
\label{eq:appB.3}
\end{eqnarray}

For Eq.~(\ref{eq:fluc.2}) only the current divergence is necessary,
\begin{equation}
   i{\bf q}\cdot{\bf j}_{\bf q} =
   \int{2d{\bf p}\over(2\pi\hbar)^2}
   ie{\bf q}\cdot{\bf v}f_{\bf q}
   = \sigma^\parallel q^2\phi_{\bf q}\;.
\label{eq:appB.4}
\end{equation}
The last equality is the definition of $\sigma^\parallel$. An integration
with respect to angles in Eq.~(\ref{eq:appB.4}) can be performed, but the
integration from Eq.~(\ref{eq:appB.3}) still remains
\begin{eqnarray}
   \sigma^\parallel q^2 &=&
   -{e^2\over\tau}{m\over\pi\hbar^2}
   \int_0^\infty dE {\partial f\over\partial E}
   \int_0^\infty d\theta' e^{-\theta'}
\nonumber\\
   &\times&
   \left[ 1 - J_0\left(
   2{qv\over\Omega_c} \sin{\Omega_c\tau\theta'\over2}
   \right)\right]\;.
\label{eq:appB.5}
\end{eqnarray}
For the case $ql \ll \mbox{max}(\Omega_c\tau, 1)$, this
equation gives the usual classical result,
\begin{equation}
    \sigma^\parallel(B) = {\sigma^\parallel(0) \over 1 + \Omega_c^2\tau^2}
    \;.
\label{eq:appB.6}
\end{equation}

\chapter{ Theory of high-field-domain structures in superlattices. }
\label{chap:domains}
\chaptermark{High-field domains}

A number of experimental works provide evidence for the existence of the 
high-field domains in superlattices when the applied voltage exceeds some 
critical value.  A theoretical description of the structure of such a domain 
is developed. We confine ourselves to the case of narrow-band superlattices, 
where electrons are strongly localized in the wells.  We find that the 
minimum length of the high-field domain can be larger than one superlattice 
period. The maximum current in the oscillating part of the $I$-$V$ 
characteristic can be significantly smaller than the value of the current at 
the voltage where the first instability comes about. The oscillation period 
can be considerably smaller than the value corresponding to the energy 
separation between the first and the second level in a well.  For the case of 
the domain formation at some distance from the anode, we study the field 
distribution in the low-field region downstream of the domain.

\section{ Introduction }

During the past 20 years a number of interesting experimental works have been
performed  in order to investigate transport properties of superlattices in
the growth direction.\cite{esaki-chang,kawamura1,kawamura2,
choi1,choi2,choi3,choi4,vuong,helm,grahn-feb90,sibille7,
merlin-jul94} Under a weak applied bias the superlattice looks like a 
homogeneous medium and exhibits Ohm's law.  Near some critical field $F_{\rm 
th}$ an instability appears and destroys the homogeneous state. As a result 
of the instability the superlattice breaks down into three regions: the 
low-field region with transport in the first mini-band, the high-field 
domain and the low-field region where electrons are injected into the second 
miniband from the high-field domain and then relax down to the first 
miniband; see Fig.~\ref{fig:domains}. An electron can move 1000{}\AA{} in the 
second mini-band before it drops down, \cite{artaki-hess2,herbert,falko} 
because the inter-subband relaxation rate is relatively small.  A further 
increase of the applied bias leads to an expansion of the high-field region 
and the current exhibits an oscillatory behavior. The period of this 
oscillation can be associated with the intersubband space but generally it is 
smaller\cite{choi1,kawamura1}. Under higher biases upper minibands become 
involved in the transport process.

There is no general physical law forbidding domain formation at any place in
the superlattice. However, in the undoped superlattices a domain appears
naturally at the anode\cite{grahn-feb90}, see Figs.~\ref{fig:steps} and
\ref{fig:domainsalt}(a). In the doped superlattices domain can be formed in
the middle of the superlattice, Fig.~\ref{fig:domains}, or near the cathode,
Fig.~\ref{fig:domainsalt}(b). In the last case one has to observe a 
significant increase of the current after the instability point.

\begin{figure}
\begin{center}
\unitlength=1mm
\begin{picture}(87.00,60.00)
\linethickness{0.8pt}
\ifPostScript
\put(-02.00,5.00){
   \put(00.00,27.00){
      $\overbrace{\begin{picture}(30.00,23.00)
         \put(0.00,3.00){\line(1,0){3.00}}
         \put(3.00,3.00){\line(0,1){20.00}}
         \put(3.00,23.00){\line(4,-1){4.00}}
         \put(7.00,22.00){\line(0,-1){20.00}}
         \put(7.00,2.00){\line(1,0){6.00}}
         \put(13.00,2.00){\line(0,1){20.00}}
         \put(13.00,22.00){\line(4,-1){4.00}}
         \put(17.00,21.00){\line(0,-1){20.00}}
         \put(17.00,1.00){\line(1,0){6.00}}
         \put(23.00,1.00){\line(0,1){20.00}}
         \put(23.00,21.00){\line(4,-1){4.00}}
         \put(27.00,20.00){\line(0,-1){20.00}}
         \put(27.00,0.00){\line(1,0){3.00}}
      \end{picture}}^{\mbox{I}}$
   }
   \put(30.00,03.00){
      $\overbrace{\begin{picture}(30.00,44.00)
         \put(0.00,24.00){\line(1,0){3.00}}
         \put(3.00,24.00){\line(0,1){20.00}}
         \put(3.00,44.00){\line(1,-2){4.00}}
         \put(7.00,36.00){\line(0,-1){20.00}}
         \put(7.00,16.00){\line(1,0){6.00}}
         \put(13.00,16.00){\line(0,1){20.00}}
         \put(13.00,36.00){\line(1,-2){4.00}}
         \put(17.00,28.00){\line(0,-1){20.00}}
         \put(17.00,8.00){\line(1,0){6.00}}
         \put(23.00,8.00){\line(0,1){20.00}}
         \put(23.00,28.00){\line(1,-2){4.00}}
         \put(27.00,20.00){\line(0,-1){20.00}}
         \put(27.00,0.00){\line(1,0){3.00}}
         \end{picture}}^{\mbox{II}}$
   }
   \put(60.00,00.00){
      $\overbrace{\begin{picture}(30.00,23.00)
         \put(0.00,3.00){\line(1,0){3.00}}
         \put(3.00,3.00){\line(0,1){20.00}}
         \put(3.00,23.00){\line(4,-1){4.00}}
         \put(7.00,22.00){\line(0,-1){20.00}}
         \put(7.00,2.00){\line(1,0){6.00}}
         \put(13.00,2.00){\line(0,1){20.00}}
         \put(13.00,22.00){\line(4,-1){4.00}}
         \put(17.00,21.00){\line(0,-1){20.00}}
         \put(17.00,1.00){\line(1,0){6.00}}
         \put(23.00,1.00){\line(0,1){20.00}}
         \put(23.00,21.00){\line(4,-1){4.00}}
         \put(27.00,20.00){\line(0,-1){20.00}}
         \put(27.00,0.00){\line(1,0){3.00}}
      \end{picture}}^{\mbox{III}}$
   }
   \linethickness{1.6pt}
   \put(00.00,00.00){
      \put(1.00,02.00){%
         \put(0.00,32.00){\line(1,0){1}}
         \put(2.00,32.00){\line(1,0){1}}
         \put(7.00,31.00){\line(1,0){1.0}}
         \put(9.00,31.00){\line(1,0){1.0}}
         \put(11.00,31.00){\line(1,0){1.0}}
         \put(17.00,30.00){\line(1,0){1.0}}
         \put(19.00,30.00){\line(1,0){1.0}}
         \put(21.00,30.00){\line(1,0){1.0}}
         \put(27.00,29.00){\line(1,0){1.0}}
         \put(29.00,29.00){\line(1,0){1.0}}
         \put(31.00,29.00){\line(1,0){1.0}}
         \put(38.00,31.00){\line(1,0){1.0}}
         \put(40.00,31.00){\line(1,0){1.0}}
         \put(42.00,31.00){\line(1,0){1.0}}
         \put(37.00,21.00){\line(1,0){1.0}}
         \put(39.00,21.00){\line(1,0){1.0}}
         \put(41.00,21.00){\line(1,0){1.00}}
         \put(48.00,23.00){\line(1,0){1.0}}
         \put(50.00,23.00){\line(1,0){1.0}}
         \put(52.00,23.00){\line(1,0){1.0}}
         \put(47.00,13.00){\line(1,0){1.0}}
         \put(49.00,13.00){\line(1,0){1.0}}
         \put(51.00,13.00){\line(1,0){1.0}}
         \put(48.00,23.00){\line(1,0){1.00}}
         \put(50.00,23.00){\line(1,0){1.0}}
         \put(52.00,23.00){\line(1,0){1.00}}
         \put(57.00,5.00){\line(1,0){1.0}}
         \put(59.00,5.00){\line(1,0){1.0}}
         \put(61.00,5.00){\line(1,0){1.0}}
         \put(67.00,4.00){\line(1,0){1.0}}
         \put(69.00,4.00){\line(1,0){1.0}}
         \put(71.00,4.00){\line(1,0){1.0}}
         \put(77.00,3.00){\line(1,0){1.0}}
         \put(79.00,3.00){\line(1,0){1.0}}
         \put(81.00,3.00){\line(1,0){1.0}}
         \put(88.00,2.00){\line(1,0){1.0}}
         \linethickness{0.2pt}
         \put(5.00,32.00){\oval(10.00,10.00)[t]}
         \put(10.00,33.00){\vector(0,-1){2.00}}
         \put(15.00,31.00){\oval(10.00,10.00)[t]}
         \put(20.00,32.00){\vector(0,-1){2.00}}
         \put(25.00,30.00){\oval(10.00,10.00)[t]}
         \put(30.00,31.00){\vector(0,-1){2.00}}
         \put(35.00,32.00){\oval(10.00,10.00)[t]}
         \put(40.00,33.00){\vector(0,-1){2.00}}
         \put(45.00,24.00){\oval(10.00,10.00)[t]}
         \put(50.00,25.00){\vector(0,-1){2.00}}
         \put(55.00,16.00){\oval(10.00,10.00)[t]}
         \put(60.00,17.00){\vector(0,-1){2.00}}
         \put(60.00,14.67){\vector(1,0){20.00}}
         \put(30.00,32.00){\line(0,-1){3.0}}
         \put(40.00,24.00){\line(0,-1){3.0}}
         \put(50.00,16.00){\line(0,-1){3.0}}
         \put(57.00,18.00){\line(1,0){2.0}}
         \put(61.00,18.00){\line(1,0){2.0}}
         \put(65.00,17.00){\line(1,0){2.0}}
         \put(69.00,17.00){\line(1,0){2.0}}
         \put(73.00,17.00){\line(1,0){2.0}}
         \put(77.00,16.00){\line(1,0){2.0}}
         \put(81.00,16.00){\line(1,0){2.0}}
         \put(85.00,15.00){\line(1,0){2.0}}
         \put(89.00,15.00){\line(1,0){1.0}}
         \put(57.00,14.00){\line(1,0){2.0}}
         \put(61.00,14.00){\line(1,0){2.0}}
         \put(65.00,13.00){\line(1,0){2.0}}
         \put(69.00,13.00){\line(1,0){2.0}}
         \put(73.00,13.00){\line(1,0){2.0}}
         \put(77.00,12.00){\line(1,0){2.0}}
         \put(81.00,12.00){\line(1,0){2.0}}
         \put(85.00,11.00){\line(1,0){2.0}}
         \put(89.00,11.00){\line(1,0){1.0}}
      }
      \put(30.00,25.00){\makebox(0,0)[ct]{$\nu=\nu_I$}}
      \put(60.00,01.00){\makebox(0,0)[ct]{$\nu=\nu_{II}$}}
   }
}
\else
\special{em:linewidth 0.8pt}
\put(-02.00,5.00){
   \put(00.00,27.00){
      $\overbrace{\begin{picture}(30.00,23.00)
         \emline{0.00}{3.00}{1}{3.00}{3.00}{2}
         \emline{3.00}{3.00}{3}{3.00}{23.00}{4}
         \emline{3.00}{23.00}{5}{7.00}{22.00}{6}
         \emline{7.00}{22.00}{7}{7.00}{2.00}{8}
         \emline{7.00}{2.00}{9}{13.00}{2.00}{10}
         \emline{13.00}{2.00}{11}{13.00}{22.00}{12}
         \emline{13.00}{22.00}{13}{17.00}{21.00}{14}
         \emline{17.00}{21.00}{15}{17.00}{1.00}{16}
         \emline{17.00}{1.00}{17}{23.00}{1.00}{18}
         \emline{23.00}{1.00}{19}{23.00}{21.00}{20}
         \emline{23.00}{21.00}{21}{27.00}{20.00}{22}
         \emline{27.00}{20.00}{23}{27.00}{0.00}{24}
         \emline{27.00}{0.00}{25}{30.00}{0.00}{26}
      \end{picture}}^{\mbox{I}}$
   }
   \put(30.00,03.00){
      $\overbrace{\begin{picture}(30.00,44.00)
         \emline{0.00}{24.00}{1}{3.00}{24.00}{2}
         \emline{3.00}{24.00}{3}{3.00}{44.00}{4}
         \emline{3.00}{44.00}{5}{7.00}{36.00}{6}
         \emline{7.00}{36.00}{7}{7.00}{16.00}{8}
         \emline{7.00}{16.00}{9}{13.00}{16.00}{10}
         \emline{13.00}{16.00}{11}{13.00}{36.00}{12}
         \emline{13.00}{36.00}{13}{17.00}{28.00}{14}
         \emline{17.00}{28.00}{15}{17.00}{8.00}{16}
         \emline{17.00}{8.00}{17}{23.00}{8.00}{18}
         \emline{23.00}{8.00}{19}{23.00}{28.00}{20}
         \emline{23.00}{28.00}{21}{27.00}{20.00}{22}
         \emline{27.00}{20.00}{23}{27.00}{0.00}{24}
         \emline{27.00}{0.00}{25}{30.00}{0.00}{26}
      \end{picture}}^{\mbox{II}}$
   }
   \put(60.00,00.00){
      $\overbrace{\begin{picture}(30.00,23.00)
         \emline{0.00}{3.00}{1}{3.00}{3.00}{2}
         \emline{3.00}{3.00}{3}{3.00}{23.00}{4}
         \emline{3.00}{23.00}{5}{7.00}{22.00}{6}
         \emline{7.00}{22.00}{7}{7.00}{2.00}{8}
         \emline{7.00}{2.00}{9}{13.00}{2.00}{10}
         \emline{13.00}{2.00}{11}{13.00}{22.00}{12}
         \emline{13.00}{22.00}{13}{17.00}{21.00}{14}
         \emline{17.00}{21.00}{15}{17.00}{1.00}{16}
         \emline{17.00}{1.00}{17}{23.00}{1.00}{18}
         \emline{23.00}{1.00}{19}{23.00}{21.00}{20}
         \emline{23.00}{21.00}{21}{27.00}{20.00}{22}
         \emline{27.00}{20.00}{23}{27.00}{0.00}{24}
         \emline{27.00}{0.00}{25}{30.00}{0.00}{26}
      \end{picture}}^{\mbox{III}}$
   }
   \linethickness{0.4pt}
   \special{em:linewidth 1.6pt}
   \put(00.00,00.00){
      \put(0.00,02.00){%
         \emline{0.00}{32.00}{1}{1.00}{32.00}{2}
         \emline{2.00}{32.00}{3}{3.00}{32.00}{4}
         \emline{7.00}{31.00}{5}{8.00}{31.00}{6}
         \emline{9.00}{31.00}{7}{10.00}{31.00}{8}
         \emline{11.00}{31.00}{9}{12.00}{31.00}{10}
         \emline{17.00}{30.00}{11}{18.00}{30.00}{12}
         \emline{19.00}{30.00}{13}{20.00}{30.00}{14}
         \emline{21.00}{30.00}{15}{22.00}{30.00}{16}
         \emline{27.00}{29.00}{17}{28.00}{29.00}{18}
         \emline{29.00}{29.00}{19}{30.00}{29.00}{20}
         \emline{31.00}{29.00}{21}{32.00}{29.00}{22}
         \emline{37.00}{21.00}{29}{38.00}{21.00}{30}
         \emline{39.00}{21.00}{31}{40.00}{21.00}{32}
         \emline{41.00}{21.00}{33}{42.00}{21.00}{34}
         \emline{47.00}{13.00}{41}{48.00}{13.00}{42}
         \emline{49.00}{13.00}{43}{50.00}{13.00}{44}
         \emline{51.00}{13.00}{45}{52.00}{13.00}{46}
         \emline{57.00}{5.00}{47}{58.00}{5.00}{48}
         \emline{59.00}{5.00}{49}{60.00}{5.00}{50}
         \emline{61.00}{5.00}{51}{62.00}{5.00}{52}
         \emline{67.00}{4.00}{53}{68.00}{4.00}{54}
         \emline{69.00}{4.00}{55}{70.00}{4.00}{56}
         \emline{71.00}{4.00}{57}{72.00}{4.00}{58}
         \emline{77.00}{3.00}{59}{78.00}{3.00}{60}
         \emline{79.00}{3.00}{61}{80.00}{3.00}{62}
         \emline{81.00}{3.00}{63}{82.00}{3.00}{64}
         \emline{88.00}{2.00}{65}{88.99}{2.04}{66}
         \emline{38.00}{31.00}{87}{39.00}{31.00}{88}
         \emline{40.00}{31.00}{89}{41.00}{31.00}{90}
         \emline{42.00}{31.00}{91}{43.00}{31.00}{92}
         \emline{48.00}{23.00}{99}{49.00}{23.00}{100}
         \emline{50.00}{23.00}{101}{51.00}{23.00}{102}
         \emline{52.00}{23.00}{103}{53.00}{23.00}{104}
         \special{em:linewidth 0.2pt}
         \put(5.00,32.00){\oval(10.00,10.00)[t]}
         \put(10.00,33.00){\vector(0,-1){2.00}}
         \put(15.00,31.00){\oval(10.00,10.00)[t]}
         \put(20.00,32.00){\vector(0,-1){2.00}}
         \put(25.00,30.00){\oval(10.00,10.00)[t]}
         \put(30.00,31.00){\vector(0,-1){2.00}}
         \put(35.00,32.00){\oval(10.00,10.00)[t]}
         \put(40.00,33.00){\vector(0,-1){2.00}}
         \emline{30.00}{32.00}{133}{30.00}{29.00}{134}
         \put(45.00,24.00){\oval(10.00,10.00)[t]}
         \put(50.00,25.00){\vector(0,-1){2.00}}
         \emline{40.00}{24.00}{135}{40.00}{21.00}{136}
         \put(55.00,16.00){\oval(10.00,10.00)[t]}
         \put(60.00,17.00){\vector(0,-1){2.00}}
         \emline{50.00}{16.00}{137}{50.00}{13.00}{138}
         \emline{57.00}{18.30}{139}{59.00}{18.11}{140}
         \emline{61.00}{17.91}{141}{63.00}{17.71}{142}
         \emline{65.00}{17.52}{143}{67.00}{17.28}{144}
         \emline{69.00}{17.08}{145}{71.00}{16.88}{146}
         \emline{73.00}{16.69}{147}{75.00}{16.49}{148}
         \emline{77.00}{16.31}{149}{79.00}{16.10}{150}
         \emline{81.00}{15.89}{151}{83.00}{15.72}{152}
         \emline{85.00}{15.51}{153}{87.00}{15.26}{154}
         \emline{89.00}{15.13}{155}{90.00}{14.97}{156}
         \emline{57.00}{14.30}{157}{59.00}{14.11}{158}
         \emline{61.00}{13.91}{159}{63.00}{13.71}{160}
         \emline{65.00}{13.52}{161}{67.00}{13.28}{162}
         \emline{69.00}{13.08}{163}{71.00}{12.88}{164}
         \emline{73.00}{12.69}{165}{75.00}{12.49}{166}
         \emline{77.00}{12.31}{167}{79.00}{12.10}{168}
         \emline{81.00}{11.89}{169}{83.00}{11.72}{170}
         \emline{85.00}{11.51}{171}{87.00}{11.26}{172}
         \emline{89.00}{11.13}{173}{90.00}{10.97}{174}
         \put(60.00,14.67){\vector(1,0){20.00}}
      }
      \put(30.00,25.00){\makebox(0,0)[ct]{$\nu=\nu_I$}}
      \put(60.00,01.00){\makebox(0,0)[ct]{$\nu=\nu_{II}$}}
   }
}
\fi
\end{picture}
\end{center}
\caption{Regions of different conductivity in the superlattice.
I and III are low-field domains, II is a high-field one. Dashed lines show
the level positions. Levels are broadened due to scattering. The
second levels in region III form a miniband and the long dashed lines show
its edges.  Arrows show the hopping of electrons between the levels. In 
region III most of the electrons move in the second mini-band.}
\label{fig:domains}
\end{figure}
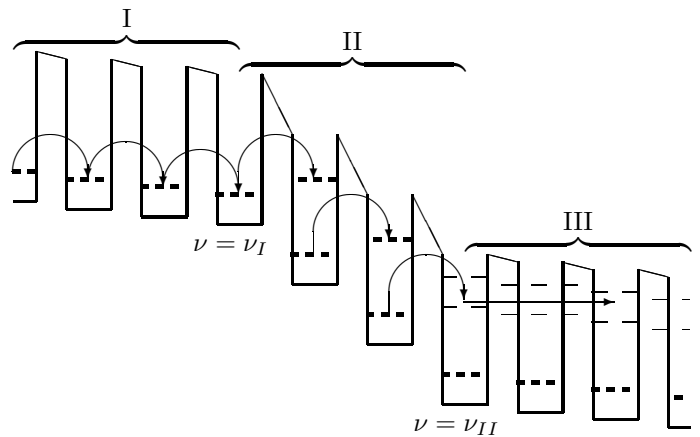

A phenomenological model that described a superlattice by an equivalent
electric circuit was suggested by Laikhtman.\cite{laikhtman1}
In this model each barrier was replaced by a nonlinear resistor parallel to
the capacitor. The model explained current oscillations and the hysteresis
usually observed in the experiment. Prengel, Wacker and
Sch\"oll\cite{scholl-jul94} considered a model for a realistic superlattice
which included electron tunneling between different levels  in the adjacent
wells and the relaxation processes inside one well. They obtained 
multistability of the current-voltage characteristic and various hysteretic 
transitions which arose upon sweeping the applied voltage and which they 
associated with changes in the domain size.

\begin{figure}
\unitlength=1mm
\begin{picture}(88.00,140.00)
   \put(-4.00,60.00){
\unitlength=1mm
\begin{picture}(87.00,80.00)
\ifPostScript
\linethickness{0.8pt}
\put(-02.00,25.0){
   \put(00.00,27.00){
      $\overbrace{\begin{picture}(30.00,23.00)
         \put(0.00,3.00){\line(1,0){3.00}}
         \put(3.00,3.00){\line(0,1){20.00}}
         \put(3.00,23.00){\line(4,-1){4.00}}
         \put(7.00,22.00){\line(0,-1){20.00}}
         \put(7.00,2.00){\line(1,0){6.00}}
         \put(13.00,2.00){\line(0,1){20.00}}
         \put(13.00,22.00){\line(4,-1){4.00}}
         \put(17.00,21.00){\line(0,-1){20.00}}
         \put(17.00,1.00){\line(1,0){6.00}}
         \put(23.00,1.00){\line(0,1){20.00}}
         \put(23.00,21.00){\line(4,-1){4.00}}
         \put(27.00,20.00){\line(0,-1){20.00}}
         \put(27.00,0.00){\line(1,0){3.00}}
      \end{picture}}^{\mbox{I}}$
   }
   \put(30.00,-21.00){
      $\overbrace{\begin{picture}(60.00,68.00)
         \put(0.00,48.00){\line(1,0){3.00}}
         \put(3.00,48.00){\line(0,1){20.00}}
         \put(3.00,68.00){\line(1,-2){4.00}}
         \put(7.00,60.00){\line(0,0){0.00}}
         \put(7.00,60.00){\line(0,-1){20.00}}
         \put(7.00,40.00){\line(1,0){6.00}}
         \put(13.00,40.00){\line(0,1){20.00}}
         \put(13.00,60.00){\line(1,-2){4.00}}
         \put(17.00,52.00){\line(0,-1){20.00}}
         \put(17.00,32.00){\line(1,0){6.00}}
         \put(23.00,32.00){\line(0,1){20.00}}
         \put(23.00,52.00){\line(1,-2){4.00}}
         \put(27.00,44.00){\line(0,-1){20.00}}
         \put(27.00,24.00){\line(1,0){6.00}}
         \put(33.00,24.00){\line(0,1){20.00}}
         \put(33.00,44.00){\line(1,-2){4.00}}
         \put(37.00,36.00){\line(0,-1){20.00}}
         \put(37.00,16.00){\line(1,0){6.00}}
         \put(43.00,16.00){\line(0,1){20.00}}
         \put(43.00,36.00){\line(1,-2){4.00}}
         \put(47.00,28.00){\line(0,-1){20.00}}
         \put(47.00,8.00){\line(1,0){6.00}}
         \put(53.00,8.00){\line(0,1){20.00}}
         \put(53.00,28.00){\line(1,-2){4.00}}
         \put(57.00,20.00){\line(0,-1){20.00}}
         \put(57.00,0.00){\line(1,0){3.00}}
      \end{picture}}^{\mbox{II}}$
   }
   \put(00.00,00.00){
      \put(1.00,02.00){%
\linethickness{1.6pt}
         \put(0.00,32.00){\line(1,0){1}}
         \put(2.00,32.00){\line(1,0){1}}
         \put(7.00,31.00){\line(1,0){1.0}}
         \put(9.00,31.00){\line(1,0){1.0}}
         \put(11.00,31.00){\line(1,0){1.0}}
         \put(17.00,30.00){\line(1,0){1.0}}
         \put(19.00,30.00){\line(1,0){1.0}}
         \put(21.00,30.00){\line(1,0){1.0}}
         \put(27.00,29.00){\line(1,0){1.0}}
         \put(29.00,29.00){\line(1,0){1.0}}
         \put(31.00,29.00){\line(1,0){1.0}}
         \put(38.00,31.00){\line(1,0){1.0}}
         \put(40.00,31.00){\line(1,0){1.0}}
         \put(42.00,31.00){\line(1,0){1.0}}
         \put(37.00,21.00){\line(1,0){1.0}}
         \put(39.00,21.00){\line(1,0){1.0}}
         \put(41.00,21.00){\line(1,0){1.00}}
         \put(48.00,23.00){\line(1,0){1.0}}
         \put(50.00,23.00){\line(1,0){1.0}}
         \put(52.00,23.00){\line(1,0){1.0}}
         \put(47.00,13.00){\line(1,0){1.0}}
         \put(49.00,13.00){\line(1,0){1.0}}
         \put(51.00,13.00){\line(1,0){1.0}}
         \put(48.00,23.00){\line(1,0){1.00}}
         \put(50.00,23.00){\line(1,0){1.0}}
         \put(52.00,23.00){\line(1,0){1.00}}
         \put(57.00,5.00){\line(1,0){1.0}}
         \put(59.00,5.00){\line(1,0){1.0}}
         \put(61.00,5.00){\line(1,0){1.0}}
         \put(67.00,-3.00){\line(1,0){1.0}}
         \put(69.00,-3.00){\line(1,0){1.0}}
         \put(71.00,-3.00){\line(1,0){1.0}}
         \put(77.00,-11.00){\line(1,0){1.0}}
         \put(79.00,-11.00){\line(1,0){1.0}}
         \put(81.00,-11.00){\line(1,0){1.0}}
         \put(87.00,-19.00){\line(1,0){1.0}}
         \put(89.00,-19.00){\line(1,0){1.0}}
         \put(91.00,-19.00){\line(1,0){1.0}}
         \put(58.00,15.00){\line(1,0){1.0}}
         \put(60.00,15.00){\line(1,0){1.0}}
         \put(62.00,15.00){\line(1,0){1.0}}
         \put(68.00,7.00){\line(1,0){1.0}}
         \put(70.00,7.00){\line(1,0){1.0}}
         \put(72.00,7.00){\line(1,0){1.0}}
         \put(78.00,-1.00){\line(1,0){1.0}}
         \put(80.00,-1.00){\line(1,0){1.0}}
         \put(82.00,-1.00){\line(1,0){1.0}}
         \put(88.00,-9.00){\line(1,0){1.0}}
         \put(90.00,-9.00){\line(1,0){1.0}}
         \put(92.00,-9.00){\line(1,0){1.0}}
\linethickness{0.2pt}
         \put(5.00,32.00){\oval(10.00,10.00)[t]}
         \put(10.00,33.00){\vector(0,-1){2.00}}
         \put(15.00,31.00){\oval(10.00,10.00)[t]}
         \put(20.00,32.00){\vector(0,-1){2.00}}
         \put(25.00,30.00){\oval(10.00,10.00)[t]}
         \put(30.00,31.00){\vector(0,-1){2.00}}
         \put(35.00,32.00){\oval(10.00,10.00)[t]}
         \put(40.00,33.00){\vector(0,-1){2.00}}
         \put(30.00,32.00){\line(0,-1){3.0}}
         \put(45.00,24.00){\oval(10.00,10.00)[t]}
         \put(50.00,25.00){\vector(0,-1){2.00}}
         \put(40.00,24.00){\line(0,-1){3.0}}
         \put(55.00,16.00){\oval(10.00,10.00)[t]}
         \put(60.00,17.00){\vector(0,-1){2.00}}
         \put(50.00,16.00){\line(0,-1){3.0}}
         \put(65.00,8.00){\oval(10.00,10.00)[t]}
         \put(70.00,9.00){\vector(0,-1){2.00}}
         \put(60.00,8.00){\line(0,-1){3.0}}
         \put(75.00,0.00){\oval(10.00,10.00)[t]}
         \put(80.00,1.00){\vector(0,-1){2.00}}
         \put(70.00,0.00){\line(0,-1){3.0}}
         \put(85.00,-8.00){\oval(10.00,10.00)[t]}
         \put(90.00,-7.00){\vector(0,-1){2.00}}
         \put(80.00,-8.00){\line(0,-1){3.0}}
      }
      \put(30.00,25.00){\makebox(0,0)[ct]{$\nu=\nu_I$}}
   }
}
\else
\special{em:linewidth 0.8pt}
\linethickness{0.8pt}
\put(-02.00,25.0){
   \put(00.00,27.00){
      $\overbrace{\begin{picture}(30.00,23.00)
         \emline{0.00}{3.00}{1}{3.00}{3.00}{2}
         \emline{3.00}{3.00}{3}{3.00}{23.00}{4}
         \emline{3.00}{23.00}{5}{7.00}{22.00}{6}
         \emline{7.00}{22.00}{7}{7.00}{2.00}{8}
         \emline{7.00}{2.00}{9}{13.00}{2.00}{10}
         \emline{13.00}{2.00}{11}{13.00}{22.00}{12}
         \emline{13.00}{22.00}{13}{17.00}{21.00}{14}
         \emline{17.00}{21.00}{15}{17.00}{1.00}{16}
         \emline{17.00}{1.00}{17}{23.00}{1.00}{18}
         \emline{23.00}{1.00}{19}{23.00}{21.00}{20}
         \emline{23.00}{21.00}{21}{27.00}{20.00}{22}
         \emline{27.00}{20.00}{23}{27.00}{0.00}{24}
         \emline{27.00}{0.00}{25}{30.00}{0.00}{26}
      \end{picture}}^{\mbox{I}}$
   }
   \put(30.00,-21.00){
      $\overbrace{\begin{picture}(60.00,68.00)
         \emline{0.00}{48.00}{1}{3.00}{48.00}{2}
         \emline{3.00}{48.00}{3}{3.00}{68.00}{4}
         \emline{3.00}{68.00}{5}{7.00}{60.00}{6}
         \emline{7.00}{60.00}{7}{7.00}{40.00}{8}
         \emline{7.00}{40.00}{9}{13.00}{40.00}{10}
         \emline{13.00}{40.00}{11}{13.00}{60.00}{12}
         \emline{13.00}{60.00}{13}{17.00}{52.00}{14}
         \emline{17.00}{52.00}{15}{17.00}{32.00}{16}
         \emline{17.00}{32.00}{17}{23.00}{32.00}{18}
         \emline{23.00}{32.00}{19}{23.00}{52.00}{20}
         \emline{23.00}{52.00}{21}{27.00}{44.00}{22}
         \emline{27.00}{44.00}{23}{27.00}{24.00}{24}
         \emline{27.00}{24.00}{25}{30.00}{24.00}{26}
         \emline{30.00}{24.00}{27}{33.00}{24.00}{28}
         \emline{33.00}{24.00}{29}{33.00}{44.00}{30}
         \emline{33.00}{44.00}{31}{37.00}{36.00}{32}
         \emline{37.00}{36.00}{33}{37.00}{16.00}{34}
         \emline{37.00}{16.00}{35}{43.00}{16.00}{36}
         \emline{43.00}{16.00}{37}{43.00}{36.00}{38}
         \emline{43.00}{36.00}{39}{47.00}{28.00}{40}
         \emline{47.00}{28.00}{41}{47.00}{8.00}{42}
         \emline{47.00}{8.00}{43}{53.00}{8.00}{44}
         \emline{53.00}{8.00}{45}{53.00}{28.00}{46}
         \emline{53.00}{28.00}{47}{57.00}{20.00}{48}
         \emline{57.00}{20.00}{49}{57.00}{0.00}{50}
         \emline{57.00}{0.00}{51}{60.00}{0.00}{52}
      \end{picture}}^{\mbox{II}}$
   }
   \linethickness{0.4pt}
   \special{em:linewidth 1.6pt}
   \put(00.00,00.00){
      \put(0.00,02.00){%
         \emline{0.00}{32.00}{1}{1.00}{32.00}{2}
         \emline{2.00}{32.00}{3}{3.00}{32.00}{4}
         \emline{7.00}{31.00}{5}{8.00}{31.00}{6}
         \emline{9.00}{31.00}{7}{10.00}{31.00}{8}
         \emline{11.00}{31.00}{9}{12.00}{31.00}{10}
         \emline{17.00}{30.00}{11}{18.00}{30.00}{12}
         \emline{19.00}{30.00}{13}{20.00}{30.00}{14}
         \emline{21.00}{30.00}{15}{22.00}{30.00}{16}
         \emline{27.00}{29.00}{17}{28.00}{29.00}{18}
         \emline{29.00}{29.00}{19}{30.00}{29.00}{20}
         \emline{31.00}{29.00}{21}{32.00}{29.00}{22}
         \emline{37.00}{21.00}{23}{38.00}{21.00}{24}
         \emline{39.00}{21.00}{25}{40.00}{21.00}{26}
         \emline{41.00}{21.00}{27}{42.00}{21.00}{28}
         \emline{47.00}{13.00}{29}{48.00}{13.00}{30}
         \emline{49.00}{13.00}{31}{50.00}{13.00}{32}
         \emline{51.00}{13.00}{33}{52.00}{13.00}{34}
         \emline{57.00}{5.00}{35}{58.00}{5.00}{36}
         \emline{59.00}{5.00}{37}{60.00}{5.00}{38}
         \emline{61.00}{5.00}{39}{62.00}{5.00}{40}
         \emline{38.00}{31.00}{41}{39.00}{31.00}{42}
         \emline{40.00}{31.00}{43}{41.00}{31.00}{44}
         \emline{42.00}{31.00}{45}{43.00}{31.00}{46}
         \emline{48.00}{23.00}{47}{49.00}{23.00}{48}
         \emline{50.00}{23.00}{49}{51.00}{23.00}{50}
         \emline{52.00}{23.00}{51}{53.00}{23.00}{52}
         \emline{67.00}{-3.00}{53}{68.00}{-3.00}{54}
         \emline{69.00}{-3.00}{55}{70.00}{-3.00}{56}
         \emline{71.00}{-3.00}{57}{72.00}{-3.00}{58}
         \emline{77.00}{-11.00}{59}{78.00}{-11.00}{60}
         \emline{79.00}{-11.00}{61}{80.00}{-11.00}{62}
         \emline{81.00}{-11.00}{63}{82.00}{-11.00}{64}
         \emline{87.00}{-19.00}{65}{88.00}{-19.00}{66}
         \emline{89.00}{-19.00}{67}{90.00}{-19.00}{68}
         \emline{91.00}{-19.00}{69}{92.00}{-19.00}{70}
         \emline{58.00}{15.00}{71}{59.00}{15.00}{72}
         \emline{60.00}{15.00}{73}{61.00}{15.00}{74}
         \emline{62.00}{15.00}{75}{63.00}{15.00}{76}
         \emline{68.00}{7.00}{77}{69.00}{7.00}{78}
         \emline{70.00}{7.00}{79}{71.00}{7.00}{80}
         \emline{72.00}{7.00}{81}{73.00}{7.00}{82}
         \emline{78.00}{-1.00}{83}{79.00}{-1.00}{84}
         \emline{80.00}{-1.00}{85}{81.00}{-1.00}{86}
         \emline{82.00}{-1.00}{87}{83.00}{-1.00}{88}
         \emline{88.00}{-9.00}{89}{89.00}{-9.00}{90}
         \emline{90.00}{-9.00}{91}{91.00}{-9.00}{92}
         \emline{92.00}{-9.00}{93}{93.00}{-9.00}{94}
            \special{em:linewidth 0.2pt}
         \put(5.00,32.00){\oval(10.00,10.00)[t]}
         \put(10.00,33.00){\vector(0,-1){2.00}}
         \put(15.00,31.00){\oval(10.00,10.00)[t]}
         \put(20.00,32.00){\vector(0,-1){2.00}}
         \put(25.00,30.00){\oval(10.00,10.00)[t]}
         \put(30.00,31.00){\vector(0,-1){2.00}}
         \put(35.00,32.00){\oval(10.00,10.00)[t]}
         \put(40.00,33.00){\vector(0,-1){2.00}}
         \emline{30.00}{32.00}{1}{30.00}{29.00}{2}
         \put(45.00,24.00){\oval(10.00,10.00)[t]}
         \put(50.00,25.00){\vector(0,-1){2.00}}
         \emline{40.00}{24.00}{3}{40.00}{21.00}{4}
         \put(55.00,16.00){\oval(10.00,10.00)[t]}
         \put(60.00,17.00){\vector(0,-1){2.00}}
         \emline{50.00}{16.00}{5}{50.00}{13.00}{6}
         \put(65.00,8.00){\oval(10.00,10.00)[t]}
         \put(70.00,9.00){\vector(0,-1){2.00}}
         \emline{60.00}{8.00}{7}{60.00}{5.00}{8}
         \put(75.00,0.00){\oval(10.00,10.00)[t]}
         \put(80.00,1.00){\vector(0,-1){2.00}}
         \emline{70.00}{0.00}{9}{70.00}{-3.00}{10}
         \put(85.00,-8.00){\oval(10.00,10.00)[t]}
         \put(90.00,-7.00){\vector(0,-1){2.00}}
         \emline{80.00}{-8.00}{11}{80.00}{-11.00}{12}
      }
      \put(30.00,25.00){\makebox(0,0)[ct]{$\nu=\nu_I$}}
   }
}
\fi
\end{picture}}
   \put(84.0,105.0){(a)}
   \put(-4.00,00.00){
\unitlength=1mm
\begin{picture}(87.00,85.00)
\ifPostScript
\linethickness{0.8pt}
\put(-02.00,05.00){
   \put(00.00,03.00){
      $\overbrace{\begin{picture}(60.00,68.00)
         \put(0.00,48.00){\line(1,0){3.00}}
         \put(3.00,48.00){\line(0,1){20.00}}
         \put(3.00,68.00){\line(1,-2){4.00}}
         \put(7.00,60.00){\line(0,0){0.00}}
         \put(7.00,60.00){\line(0,-1){20.00}}
         \put(7.00,40.00){\line(1,0){6.00}}
         \put(13.00,40.00){\line(0,1){20.00}}
         \put(13.00,60.00){\line(1,-2){4.00}}
         \put(17.00,52.00){\line(0,-1){20.00}}
         \put(17.00,32.00){\line(1,0){6.00}}
         \put(23.00,32.00){\line(0,1){20.00}}
         \put(23.00,52.00){\line(1,-2){4.00}}
         \put(27.00,44.00){\line(0,-1){20.00}}
         \put(27.00,24.00){\line(1,0){6.00}}
         \put(33.00,24.00){\line(0,1){20.00}}
         \put(33.00,44.00){\line(1,-2){4.00}}
         \put(37.00,36.00){\line(0,-1){20.00}}
         \put(37.00,16.00){\line(1,0){6.00}}
         \put(43.00,16.00){\line(0,1){20.00}}
         \put(43.00,36.00){\line(1,-2){4.00}}
         \put(47.00,28.00){\line(0,-1){20.00}}
         \put(47.00,8.00){\line(1,0){6.00}}
         \put(53.00,8.00){\line(0,1){20.00}}
         \put(53.00,28.00){\line(1,-2){4.00}}
         \put(57.00,20.00){\line(0,-1){20.00}}
         \put(57.00,0.00){\line(1,0){3.00}}
      \end{picture}}^{\mbox{II}}$
   }
   \put(60.00,00.00){
      $\overbrace{\begin{picture}(30.00,23.00)
         \put(0.00,3.00){\line(1,0){3.00}}
         \put(3.00,3.00){\line(0,1){20.00}}
         \put(3.00,23.00){\line(4,-1){4.00}}
         \put(7.00,22.00){\line(0,-1){20.00}}
         \put(7.00,2.00){\line(1,0){6.00}}
         \put(13.00,2.00){\line(0,1){20.00}}
         \put(13.00,22.00){\line(4,-1){4.00}}
         \put(17.00,21.00){\line(0,-1){20.00}}
         \put(17.00,1.00){\line(1,0){6.00}}
         \put(23.00,1.00){\line(0,1){20.00}}
         \put(23.00,21.00){\line(4,-1){4.00}}
         \put(27.00,20.00){\line(0,-1){20.00}}
         \put(27.00,0.00){\line(1,0){3.00}}
      \end{picture}}^{\mbox{III}}$
   }
   \put(00.00,00.00){
      \put(01.00,02.00){
\linethickness{1.6pt}
         \put(27.00,29.00){\line(1,0){1.00}}
         \put(29.00,29.00){\line(1,0){1.00}}
         \put(31.00,29.00){\line(1,0){1.00}}
         \put(37.00,21.00){\line(1,0){1.00}}
         \put(39.00,21.00){\line(1,0){1.00}}
         \put(41.00,21.00){\line(1,0){1.00}}
         \put(47.00,13.00){\line(1,0){1.00}}
         \put(49.00,13.00){\line(1,0){1.00}}
         \put(51.00,13.00){\line(1,0){1.00}}
         \put(57.00,5.00){\line(1,0){1.00}}
         \put(59.00,5.00){\line(1,0){1.00}}
         \put(61.00,5.00){\line(1,0){1.00}}
         \put(67.00,4.00){\line(1,0){1.00}}
         \put(69.00,4.00){\line(1,0){1.00}}
         \put(71.00,4.00){\line(1,0){1.00}}
         \put(77.00,3.00){\line(1,0){1.00}}
         \put(79.00,3.00){\line(1,0){1.00}}
         \put(81.00,3.00){\line(1,0){1.00}}
         \put(88.00,2.00){\line(1,0){1.00}}
         \put(38.00,31.00){\line(1,0){1.00}}
         \put(40.00,31.00){\line(1,0){1.00}}
         \put(42.00,31.00){\line(1,0){1.00}}
         \put(48.00,23.00){\line(1,0){1.00}}
         \put(50.00,23.00){\line(1,0){1.00}}
         \put(52.00,23.00){\line(1,0){1.00}}
         \put(-3.00,53.00){\line(1,0){1.00}}
         \put(-1.00,53.00){\line(1,0){1.00}}
         \put(1.00,53.00){\line(1,0){1.00}}
         \put(7.00,45.00){\line(1,0){1.00}}
         \put(9.00,45.00){\line(1,0){1.00}}
         \put(11.00,45.00){\line(1,0){1.00}}
         \put(17.00,37.00){\line(1,0){1.00}}
         \put(19.00,37.00){\line(1,0){1.00}}
         \put(21.00,37.00){\line(1,0){1.00}}
         \put(18.00,47.00){\line(1,0){1.00}}
         \put(20.00,47.00){\line(1,0){1.00}}
         \put(22.00,47.00){\line(1,0){1.00}}
         \put(28.00,39.00){\line(1,0){1.00}}
         \put(30.00,39.00){\line(1,0){1.00}}
         \put(32.00,39.00){\line(1,0){1.00}}
         \put(8.00,55.00){\line(1,0){1.00}}
         \put(10.00,55.00){\line(1,0){1.00}}
         \put(12.00,55.00){\line(1,0){1.00}}
\linethickness{0.2pt}
         \put(35.00,32.00){\oval(10.00,10.00)[t]}
         \put(40.00,33.00){\vector(0,-1){2.00}}
         \put(45.00,24.00){\oval(10.00,10.00)[t]}
         \put(50.00,25.00){\vector(0,-1){2.00}}
         \put(40.00,24.00){\line(0,-1){3.00}}
         \put(55.00,16.00){\oval(10.00,10.00)[t]}
         \put(60.00,17.00){\vector(0,-1){2.00}}
         \put(50.00,16.00){\line(0,-1){3.00}}
         \put(60.00,14.67){\vector(1,0){20.00}}
         \put(30.00,32.00){\line(0,-1){3.00}}
         \put(5.00,56.00){\oval(10.00,10.00)[t]}
         \put(10.00,57.00){\vector(0,-1){2.00}}
         \put(15.00,48.00){\oval(10.00,10.00)[t]}
         \put(20.00,49.00){\vector(0,-1){2.00}}
         \put(10.00,48.00){\line(0,-1){3.00}}
         \put(25.00,40.00){\oval(10.00,10.00)[t]}
         \put(30.00,41.00){\vector(0,-1){2.00}}
         \put(20.00,40.00){\line(0,-1){3.00}}
         \put(0.00,56.00){\line(0,-1){3.00}}
         \put(57.00,18.00){\line(1,0){2.02}}
         \put(61.00,18.00){\line(1,0){2.03}}
         \put(65.00,17.00){\line(1,0){1.97}}
         \put(69.00,17.00){\line(1,0){1.98}}
         \put(73.00,17.00){\line(1,0){2.00}}
         \put(77.00,16.00){\line(1,0){2.02}}
         \put(81.00,16.00){\line(1,0){2.03}}
         \put(85.00,15.00){\line(1,0){1.97}}
         \put(89.00,15.00){\line(1,0){0.98}}
         \put(57.00,14.00){\line(1,0){2.02}}
         \put(61.00,14.00){\line(1,0){2.03}}
         \put(65.00,13.00){\line(1,0){1.97}}
         \put(69.00,13.00){\line(1,0){1.98}}
         \put(73.00,13.00){\line(1,0){2.00}}
         \put(77.00,12.00){\line(1,0){2.02}}
         \put(81.00,12.00){\line(1,0){2.03}}
         \put(85.00,11.00){\line(1,0){1.97}}
         \put(89.00,11.00){\line(1,0){0.98}}
      }
      \put(60.00,01.00){\makebox(0,0)[ct]{$\nu=\nu_{II}$}}
   }
}
\else
\special{em:linewidth 0.8pt}
\linethickness{0.8pt}
\put(-02.00,05.00){
   \put(00.00,03.00){
      $\overbrace{\begin{picture}(60.00,68.00)
         \emline{0.00}{48.00}{1}{3.00}{48.00}{2}
         \emline{3.00}{48.00}{3}{3.00}{68.00}{4}
         \emline{3.00}{68.00}{5}{7.00}{60.00}{6}
         \emline{7.00}{60.00}{7}{7.00}{40.00}{8}
         \emline{7.00}{40.00}{9}{13.00}{40.00}{10}
         \emline{13.00}{40.00}{11}{13.00}{60.00}{12}
         \emline{13.00}{60.00}{13}{17.00}{52.00}{14}
         \emline{17.00}{52.00}{15}{17.00}{32.00}{16}
         \emline{17.00}{32.00}{17}{23.00}{32.00}{18}
         \emline{23.00}{32.00}{19}{23.00}{52.00}{20}
         \emline{23.00}{52.00}{21}{27.00}{44.00}{22}
         \emline{27.00}{44.00}{23}{27.00}{24.00}{24}
         \emline{27.00}{24.00}{25}{30.00}{24.00}{26}
         \emline{30.00}{24.00}{27}{33.00}{24.00}{28}
         \emline{33.00}{24.00}{29}{33.00}{44.00}{30}
         \emline{33.00}{44.00}{31}{37.00}{36.00}{32}
         \emline{37.00}{36.00}{33}{37.00}{16.00}{34}
         \emline{37.00}{16.00}{35}{43.00}{16.00}{36}
         \emline{43.00}{16.00}{37}{43.00}{36.00}{38}
         \emline{43.00}{36.00}{39}{47.00}{28.00}{40}
         \emline{47.00}{28.00}{41}{47.00}{8.00}{42}
         \emline{47.00}{8.00}{43}{53.00}{8.00}{44}
         \emline{53.00}{8.00}{45}{53.00}{28.00}{46}
         \emline{53.00}{28.00}{47}{57.00}{20.00}{48}
         \emline{57.00}{20.00}{49}{57.00}{0.00}{50}
         \emline{57.00}{0.00}{51}{60.00}{0.00}{52}
      \end{picture}}^{\mbox{II}}$
   }
   \put(60.00,00.00){
      $\overbrace{\begin{picture}(30.00,23.00)
         \emline{0.00}{3.00}{1}{3.00}{3.00}{2}
         \emline{3.00}{3.00}{3}{3.00}{23.00}{4}
         \emline{3.00}{23.00}{5}{7.00}{22.00}{6}
         \emline{7.00}{22.00}{7}{7.00}{2.00}{8}
         \emline{7.00}{2.00}{9}{13.00}{2.00}{10}
         \emline{13.00}{2.00}{11}{13.00}{22.00}{12}
         \emline{13.00}{22.00}{13}{17.00}{21.00}{14}
         \emline{17.00}{21.00}{15}{17.00}{1.00}{16}
         \emline{17.00}{1.00}{17}{23.00}{1.00}{18}
         \emline{23.00}{1.00}{19}{23.00}{21.00}{20}
         \emline{23.00}{21.00}{21}{27.00}{20.00}{22}
         \emline{27.00}{20.00}{23}{27.00}{0.00}{24}
         \emline{27.00}{0.00}{25}{30.00}{0.00}{26}
      \end{picture}}^{\mbox{III}}$
   }
   \linethickness{0.4pt}
   \special{em:linewidth 1.6pt}
   \put(00.00,00.00){
      \put(0.00,02.00){
         \emline{27.00}{29.00}{1}{28.00}{29.00}{2}
         \emline{29.00}{29.00}{3}{30.00}{29.00}{4}
         \emline{31.00}{29.00}{5}{32.00}{29.00}{6}
         \emline{37.00}{21.00}{7}{38.00}{21.00}{8}
         \emline{39.00}{21.00}{9}{40.00}{21.00}{10}
         \emline{41.00}{21.00}{11}{42.00}{21.00}{12}
         \emline{47.00}{13.00}{13}{48.00}{13.00}{14}
         \emline{49.00}{13.00}{15}{50.00}{13.00}{16}
         \emline{51.00}{13.00}{17}{52.00}{13.00}{18}
         \emline{57.00}{5.00}{19}{58.00}{5.00}{20}
         \emline{59.00}{5.00}{21}{60.00}{5.00}{22}
         \emline{61.00}{5.00}{23}{62.00}{5.00}{24}
         \emline{67.00}{4.00}{25}{68.00}{4.00}{26}
         \emline{69.00}{4.00}{27}{70.00}{4.00}{28}
         \emline{71.00}{4.00}{29}{72.00}{4.00}{30}
         \emline{77.00}{3.00}{31}{78.00}{3.00}{32}
         \emline{79.00}{3.00}{33}{80.00}{3.00}{34}
         \emline{81.00}{3.00}{35}{82.00}{3.00}{36}
         \emline{88.00}{2.00}{37}{88.99}{2.04}{38}
         \emline{38.00}{31.00}{39}{39.00}{31.00}{40}
         \emline{40.00}{31.00}{41}{41.00}{31.00}{42}
         \emline{42.00}{31.00}{43}{43.00}{31.00}{44}
         \emline{48.00}{23.00}{45}{49.00}{23.00}{46}
         \emline{50.00}{23.00}{47}{51.00}{23.00}{48}
         \emline{52.00}{23.00}{49}{53.00}{23.00}{50}
         \emline{-3.00}{53.00}{51}{-2.00}{53.00}{52}
         \emline{-1.00}{53.00}{53}{0.00}{53.00}{54}
         \emline{1.00}{53.00}{55}{2.00}{53.00}{56}
         \emline{7.00}{45.00}{57}{8.00}{45.00}{58}
         \emline{9.00}{45.00}{59}{10.00}{45.00}{60}
         \emline{11.00}{45.00}{61}{12.00}{45.00}{62}
         \emline{17.00}{37.00}{63}{18.00}{37.00}{64}
         \emline{19.00}{37.00}{65}{20.00}{37.00}{66}
         \emline{21.00}{37.00}{67}{22.00}{37.00}{68}
         \emline{18.00}{47.00}{69}{19.00}{47.00}{70}
         \emline{20.00}{47.00}{71}{21.00}{47.00}{72}
         \emline{22.00}{47.00}{73}{23.00}{47.00}{74}
         \emline{28.00}{39.00}{75}{29.00}{39.00}{76}
         \emline{30.00}{39.00}{77}{31.00}{39.00}{78}
         \emline{32.00}{39.00}{79}{33.00}{39.00}{80}
         \emline{8.00}{55.00}{81}{9.00}{55.00}{82}
         \emline{10.00}{55.00}{83}{11.00}{55.00}{84}
         \emline{12.00}{55.00}{85}{13.00}{55.00}{86}
                 \special{em:linewidth 0.2pt}
         \put(35.00,32.00){\oval(10.00,10.00)[t]}
         \put(40.00,33.00){\vector(0,-1){2.00}}
         \put(45.00,24.00){\oval(10.00,10.00)[t]}
         \put(50.00,25.00){\vector(0,-1){2.00}}
         \emline{40.00}{24.00}{1}{40.00}{21.00}{2}
         \put(55.00,16.00){\oval(10.00,10.00)[t]}
         \put(60.00,17.00){\vector(0,-1){2.00}}
         \emline{50.00}{16.00}{3}{50.00}{13.00}{4}
         \emline{57.00}{18.30}{5}{59.00}{18.11}{6}
         \emline{61.00}{17.91}{7}{63.00}{17.71}{8}
         \emline{65.00}{17.52}{9}{67.00}{17.28}{10}
         \emline{69.00}{17.08}{11}{71.00}{16.88}{12}
         \emline{73.00}{16.69}{13}{75.00}{16.49}{14}
         \emline{77.00}{16.31}{15}{79.00}{16.10}{16}
         \emline{81.00}{15.89}{17}{83.00}{15.72}{18}
         \emline{85.00}{15.51}{19}{87.00}{15.26}{20}
         \emline{89.00}{15.13}{21}{90.00}{14.97}{22}
         \emline{57.00}{14.30}{23}{59.00}{14.11}{24}
         \emline{61.00}{13.91}{25}{63.00}{13.71}{26}
         \emline{65.00}{13.52}{27}{67.00}{13.28}{28}
         \emline{69.00}{13.08}{29}{71.00}{12.88}{30}
         \emline{73.00}{12.69}{31}{75.00}{12.49}{32}
         \emline{77.00}{12.31}{33}{79.00}{12.10}{34}
         \emline{81.00}{11.89}{35}{83.00}{11.72}{36}
         \emline{85.00}{11.51}{37}{87.00}{11.26}{38}
         \emline{89.00}{11.13}{39}{90.00}{10.97}{40}
         \put(60.00,14.67){\vector(1,0){20.00}}
         \emline{30.00}{32.00}{41}{30.00}{29.00}{42}
         \put(5.00,56.00){\oval(10.00,10.00)[t]}
         \put(10.00,57.00){\vector(0,-1){2.00}}
         \put(15.00,48.00){\oval(10.00,10.00)[t]}
         \put(20.00,49.00){\vector(0,-1){2.00}}
         \emline{10.00}{48.00}{43}{10.00}{45.00}{44}
         \put(25.00,40.00){\oval(10.00,10.00)[t]}
         \put(30.00,41.00){\vector(0,-1){2.00}}
         \emline{20.00}{40.00}{45}{20.00}{37.00}{46}
         \emline{0.00}{56.00}{47}{0.00}{53.00}{48}
      }
      \put(60.00,01.00){\makebox(0,0)[ct]{$\nu=\nu_{II}$}}
   }
}
\fi
\end{picture}}
   \put(84.0,35.0){(b)}
\end{picture}
   \caption{ Particular cases of the potential distribution
   from Fig.~\protect{\ref{fig:domains}}. Domain growth from anode, (a),
   this is the most often observed experimentally case. Domain growth from
   cathode, (b), this situation has never been reported in the literature,
   however, it has very specific fingerprint on the $I$-$V$ curve and it can
   be observed.
   }
\label{fig:domainsalt}
\end{figure}
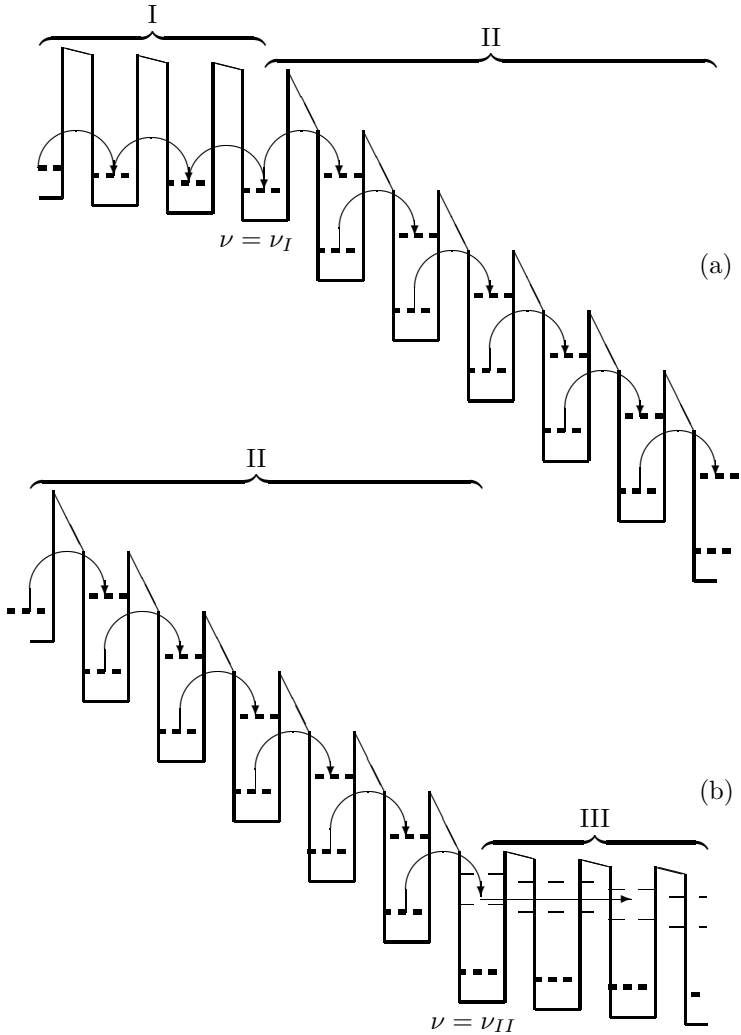

The purpose of the present work is not to simulate the $I$-$V$
characteristics in a specific superlattice but to understand the general
structure of the high-field domain. A diffusion current induced by a charge
accumulation at the domain boundary appears to be very important. We
calculate the field and the carrier distribution in the steady state and get
the main features of the current-voltage characteristic of the superlattice
in some interval of the applied bias.  The size and the position of the
domain is also discussed.

The physical processes characterizing transport in the superlattice
are described briefly in the next section. In Sec.~\ref{sec:coupl} we derive
the equation for the hopping current between two levels in the different
wells. In Secs.~\ref{sec:regI}, \ref{sec:regII}, and
\ref{sec:regIII} we calculate the field distribution in regions I, II, and
III respectively. We discuss the results and make some comparisons with
available experimental data in Sec.~\ref{sec:disc}.

\section{ Physical picture}
\label{sec:pict}

Let us see what happens when the bias applied to the superlattice increases
and goes beyond the instability threshold. If as a result of the instability
development a short high-field domain is spontaneously
generated\cite{esaki-chang} then the field in the low-field regions
is reduced by $F_{\rm H}/N$, where $F_{\rm H}$ is the field in the high-field
region and $N$ is the number of the superlattice periods. The current in
the low-field region is $j\approx 2j_0FF_{\rm th}/(F^2+F_{\rm th}^2)$, where
$F_{\rm th}$ is the threshold field and $j_0$ is the current just before the
instability point\cite{laikhtman2}. So the generation of the short domain
causes the reduction of the current by
$\delta{j}\approx j_0(F_{\rm H}/F_{\rm th}N)^2$. Usually $F_{\rm H}$ is
only 10 or 15 times larger than $F_{\rm th}$, but the number of periods
can range from 50 to 100, and therefore the reduction of the current is
small, $\delta{j}\ll{j}$.

The formation of the high-field domain is
accompanied by the accumulation of electrons in the well just upstream
of the domain and with the depletion of electrons in the well just downstream
of the domain. The accumulation of electrons in one
well gives rise to a diffusion current upstream of this well in the direction
opposite to the total current. Since the total current is the same across all
of the barriers the diffusion current across one barrier has to be
compensated for with the conduction part of the current. 
This compensation may be impossible, because the electric current is too close 
to
its maximum value $j_0$. That is, for a such a value of the total current a
steady state does not exist. A steady state can come about only for a domain
extended enough when the total current isn't too close to its maximum
value~$j_0$.  Therefore there exist a minimal length of the high-field domain
and an upper limit of the total current in a steady state~$j^{\ast}$.

Generally, after the formation of the high-field domain with an increase of
the applied bias, the total current drops below $j^\ast$. A further
increase of the bias leads the growth of the current and when it reaches the
$j^\ast$ the high-field domain expands by one period and the current drops
again; see Fig.~\ref{fig:iv}.

The change of the potential drop across the high-field domain when it
expands by one period is usually associated with the energy spacing $E_g$
between the first and the second levels in a
well.\cite{esaki-chang,kawamura2,choi1,choi2,vuong,helm,grahn-feb90} It is
assumed that in the high-field domain the first level in one well is in 
resonance with the second level in the neighboring well. The number of
electrons in the first $n^{(1)}$ and second $n^{(2)}$ levels can be
found from the simple balance equation,
\begin{equation}
   \frac{n^{(1)}-n^{(2)}}{\tau_{\rm t}}
   = \frac{n^{(2)}-n^{(1)}e^{-{E_g}/T}}{\tau_{21}} \ ,
\label{eq:regII.1}
\end{equation}
where $n^{(1)} + n^{(2)} = \bar n$ is the total concentration. Here
$\tau_{\rm{t}}=\hbar\Gamma/\Lambda_{12}^2$ is the transition time between
adjacent wells\cite{kazarinov72}, $\Gamma$ is the width of the level,
$\Lambda_{12}$ is the overlap between the wave functions of the first level
in one well and the second level in the adjacent well, and $\tau_{21}$ is the
relaxation time from the second level to the first one in the same well. If
$\tau_{\rm{t}}\ll\tau_{21}$, then the current is
$e\bar{n}[1-exp(-{E_g}/T)]/2\tau_{21}$. This current must be smaller
than $j^\ast$, which is not always the case. If $\tau_{\rm t}\gg\tau_{21}$
the current is $e\bar{n}\,\mbox{tanh}({E_g}/T)/\tau_{\rm t}$. However
this quantity is even larger than
$j_0=e\bar{n}\Lambda_{11}^2/\max(E_{\rm{F}},T)$, where $\Lambda_{11}$ is the
overlap between the wave functions of the first levels in adjacent
wells.\cite{laikhtman2}

We see that under resonance conditions the current in the high-field
domain sometimes appears to be larger than the maximum possible current in
region I and such a regime cannot exist. Due to the limitation of the
current in region I, resonance in region II is not reached.
The current in this region is smaller than its resonance value for two
reasons. The tunneling probability is reduced because of a lack of
resonance and not all electrons in the second level in one well have enough
energy to move to the first level in the neighboring well.

If the resonance between the first and the second level in adjacent wells
does not exist, the expansion of the high-field domain by one period requires
a voltage increase smaller than that corresponding to $E_g$.
This is the explanation of a small period of the current oscillations
sometimes observed in experiments\cite{kawamura1,kawamura2,choi1,vuong}.

The well at the boundary between regions II and III is depleted. The
reduction of the electron concentration in this well corresponds to a field
discontinuity between the high-field domain and region III. If the necessary
reduction is larger than the average electron concentration in a well, then
the domain is located near the anode, where a depletion layer is
formed\cite{grahn-feb90}.

In doped superlattices electrons come from the high-field domain to the
second miniband in region III and relax there down to the first miniband.
The relaxation length depends on the relation between the mobilities in the
first and the second minibands and an intersubband relaxation time. The
redistributing of electrons between two minibands can result in a field
inhomogeneity in region III.

\section{ Electric current between adjacent wells.}
\label{sec:coupl}

In this work we consider the case of elastic scattering so strong  that an
electron is scattered in a well before tunneling to the next well, at least
in the first miniband. So for the calculation of the current we need the
transition probabilities between adjacent wells.  Since the widths of levels
are typically much smaller than the energy separations between them, only the
tunneling between those levels that are close to resonance is important.

The general form of the transition probability between such levels is
\begin{equation}
   w \ \propto \frac{\Lambda^2}{\hbar}\ \frac{\Gamma}%
   {\Gamma^2 + \Delta^2} \ ,
\label{eq:pict.1}
\end{equation}
where $\Delta$ is the energy separation between levels. In the low-field
region this equation describes the transition between lowest levels and it is
justified for $\Lambda_{11}\ll\Gamma$; see Ref.~\onlinecite{laikhtman2}.  For
the transition from the first level to the second level in the adjacent well
such an equation was derived by Kazarinov and Suris.\cite{kazarinov72} In
this case Eq.~({\ref{eq:pict.1}})  is justified for an arbitrary relation
between $\Lambda$ and $\Gamma$.

The overlap integral $\Lambda$ and the level width due to an elastic
scattering $\Gamma$ are different for different pairs of levels. The overlap
integral increases with the level number because the penetration length of the 
wave function under the barrier increases with the energy. The
parameter $\Gamma$ in Eq.~({\ref{eq:pict.1}}) is different for the transition
from the first level to the first level and for the transition from the first
level to the second level, because in the latter case the presence of the
first level gives more possibilities for momentum relaxation.

In a part of region III electrons can travel in the second subband,
which can be wide. Therefore in this region instead of Eq.~(\ref{eq:pict.1})
we use Ohm's law.

The transition probability Eq.~(\ref{eq:pict.1}) gives the following
expression for the electric current from the $i$-th level in the $\nu$-th well to
the $i^{\prime}$-th level in the $\nu+1$-th well:
\begin{mathletters}
\begin{eqnarray}
   j_{ii^{\prime}} &=& {e\over\hbar}\int\frac{2d{\bf{p}}}{(2\pi\hbar)^2}
      {2\Gamma\Lambda_{ii^{\prime}}^2\over\Gamma^2+\Delta_{\nu,ii'}^2}
      \bigl(\rho(E_p)-\rho'(E_p+\Delta_{\nu,ii'})\bigr) \;,
\nonumber\\
   \Delta_{\nu,ii'} &> &0\;,
\label{eq:coupl.1a}
\end{eqnarray}
when the level in the $\nu$-th well is higher and
\begin{eqnarray}
   j_{ii^{\prime}} &=& {e\over\hbar}\int\frac{2d{\bf{p}}}{(2\pi\hbar)^2}
      {2\Gamma\Lambda_{ii^{\prime}}^2\over\Gamma^2+\Delta_{\nu,ii'}^2}
      \bigl(\rho(E_p-\Delta_{\nu,ii'})-\rho'(E_p)\bigr)\;,
\nonumber
\\
   \Delta_{\nu,ii'}&<&0\;,
\label{eq:coupl.1b}
\end{eqnarray}
when the level in the $\nu$-th well is lower. In these equations
$\Lambda_{ii^{\prime}}$ is the overlap integral between electron wave
functions of the levels $i$ and $i'$ in the $\nu$-th and $\nu+1$-th wells
correspondingly.  The energy space between these two levels is denoted by
$\Delta_{\nu,ii'}$. The diagonal elements of the electron density matrix
related to these two levels $\rho$ and $\rho'$ can be considered as a
function of the energy $E_p={\bf p}^2/2m$, where $m$ is the effective mass of
the electron, since the in-plane motion of electrons is isotropic and this
density matrix element is independent of the direction of~$\bf p$. Usually
$\Gamma$ is a smooth function of the energy and we assume it to be a
constant.
\label{eq:coupl.1}\end{mathletters}

We see from Eq.~(\ref{eq:coupl.1}) that the value of the current depends on
the shape of the electron distribution function. Equation~(\ref{eq:coupl.1})
is simplified in three cases. The first is the case of weak
electron heating when the electron distribution function is close to the
equilibrium one.

In the second case the electron gas is degenerate and $\Delta$
smaller than the Fermi energy. Then the difference of the
distribution function in the integrands of Eq.~(\ref{eq:coupl.1}) is
proportional to $\Delta$ and the tail of the distribution function above the
Fermi energy does not play any role.

In the third case the electron gas is heated significantly  so that
the electron-electron scattering is very
effective and leads to a fast relaxation of the electron distribution
function to the Fermi function with an effective temperature, $T$, and a
chemical potential $\zeta_\nu^{(i)}$, where $\nu$ is the index of the well
and $i$ is the index of the level. We should note, however, that even a
strong deviation of the electron distribution from the Fermi function does
not change qualitative results of the present work.

\begin{mathletters}
Under these two assumptions the integration in Eq.~(\ref{eq:coupl.1}) results
in
\begin{equation}
   j_{ii^{\prime}} = {2e\Lambda_{ii^{\prime}}^2\over\hbar}
      {\Gamma\over\Gamma^2+\Delta_{\nu,ii'}^2}
      \bigl[ n_\nu^{(i)} - n(\zeta_{\nu+1}^{(i^{\prime})}
      - \Delta_{\nu,ii'})\bigr]  \ ,
\label{eq:coupl.2a}
\end{equation}
when the level in the $\nu$-th well is higher, and
\begin{equation}
   j_{ii^{\prime}} = {2e\Lambda_{ii^{\prime}}^2\over\hbar}
      {\Gamma\over\Gamma^2+\Delta_{\nu,ii'}^2}
      \bigl[ n(\zeta_\nu^{(i)} + \Delta_{\nu,ii'})
      - n_{\nu+1}^{(i^{\prime})}\bigr]  \ ,
\label{eq:coupl.2b}
\end{equation}
when the level in the $\nu$-th well is lower.
\label{eq:coupl.2}
\end{mathletters}
In these two equations
\begin{equation}
   n_\nu^{(i)} = n(\zeta_\nu^{(i)}) \equiv
   g_0 T \log(e^{\zeta_\nu^{(i)}/T}+1) \
\label{eq:coupl.3}
\end{equation}
is the concentration of electrons in the $\nu$-th well at the $i$-th level and
$g_0 = m/\pi\hbar^2$. We will omit the subscripts of $\Delta$ when it does
not lead to a confusion. The barrier for the second level is lower than for
the first level, so one can expect that
$\Lambda_{11}<\Lambda_{12}<\Lambda_{22}$.

The difference in the square brackets in Eq.~(\ref{eq:coupl.2}) can be
simplified,
\begin{eqnarray}
   && n_\nu^{(i)} - n(\zeta_{\nu+1}^{(i^{\prime})} - \Delta_{\nu,ii'})
\nonumber\\
   &=& n_\nu^{(i)} - n_{\nu+1}^{(i^{\prime})} + \Delta_{\nu,ii'}
   \biggl( {\partial n\over\partial\zeta}
   \biggr)_{\zeta_{\nu+1}^{(i^{\prime})}} \ ,
\label{eq:coupl.4}
\end{eqnarray}
in any of two cases, $\zeta - \Delta \gg T$ or
$\zeta \lesssim T, \ \Delta \ll T$.

In the case when the expansion
Eq.~(\ref{eq:coupl.4}) is used one can distinguish between the diffusion
current and the conduction current. The former is proportional to the
concentration difference and the later is proportional to $\Delta$.
Note that
\begin{equation}
   \biggl( {\partial n\over\partial\zeta}
   \biggr)_{\zeta_{\nu+1}^{(i^{\prime})}}
   \equiv  g = g_0\bigl(
     1-e^{-n_{\nu+1}^{(i^{\prime})}/g_0T}\bigr) \ .
\label{eq:coupl.6}
\end{equation}
is not a constant, but depends on the electron concentration, which can be
different for different levels.

\section{ Low-field region upstream of the domain. }
\label{sec:regI}

We assume that in the low-field region upstream of the domain there are
electrons in the first miniband only. Motion of electrons in the narrow
miniband $\Lambda\ll\Gamma$ can be described in terms of hopping between
adjacent wells\cite{laikhtman2}. So, a current via each barrier can be found
from Eq.~(\ref{eq:coupl.2a}), where $i=i^{\prime}=1$, and these indices will
be omitted throughout this section.

The electric field in this low field region is inhomogeneous only near the
boundary with the high-field domain. The field distribution near this
boundary can be calculated from the Poisson equation together with the
condition that the current is the same through all barriers in this region. 
One can see from Eq.~(\ref{eq:coupl.2a}) that the current through a barrier
is a nonlinear function of the electron concentrations near this barrier and
of the field in this barrier, $\Delta/(el)$. For simplicity we consider only
the degenerate electron gas. In this case $\partial{n}/\partial\zeta$ in
Eq.~(\ref{eq:coupl.4}) is a constant; it is equal to $g_0$, see
Eq.~(\ref{eq:coupl.6}).  This restriction is not very strong since additional
electrons come to this region from region III if it exists or from the anode
contact. Therefore the electron gas in region I near the boundary of the
high-field domain is typically degenerate.

In this case the condition that allows us to use the expansion
Eq.~(\ref{eq:coupl.4}), is $\Delta_{\nu}<n_{\nu+1}/g_0$. Since the current in
the superlattice with the high-field domain is smaller than $j_0$, the
potential drop per period far from the domain boundary in  region I is small,
i.e. $\Delta<\Gamma$. On the other hand, the theory is limited by the
condition $\Gamma\;\lesssim\;E_F$, and we see that far from the domain
$\Delta<E_F$.  The field in the barriers increases with the approaching the
high-field domain boundary, however, the electron concentration also
increases and the necessary condition is usually fulfilled. It makes sense to
note that the necessary condition (for $\Delta_\nu$) contains the
concentration downstream of the $\nu$-th barrier where it is larger than that
which is upstream ($n_{\nu}<n_{\nu+1}$).

\begin{mathletters}
We introduce here two quantities which can be measured in practice. The first of
them is the linear conductivity in the low subband $\sigma$. One can find
from Eqs.~(\ref{eq:coupl.2})-(\ref{eq:coupl.6}) that
$\sigma=2e^2gl\Lambda_{11}^2/\hbar\Gamma$, see also
Ref.~\onlinecite{laikhtman2}.  The second one is the critical field
$F_{\rm{th}}$. This field corresponds to the instability of the homogeneous
steady state.  In Ref.~\onlinecite{laikhtman2} it was shown that $F_{\rm
th} \approx\Gamma/el$.  The substitutions of these quantities into
Eq.~(\ref{eq:coupl.2a}) gives \begin{equation}
   j = {\sigma\over l}\ {\phi_{\nu} -  \phi_{\nu+1}
     + (n_{\nu} - n_{\nu+1})/eg
     \over 1 + [( \phi_{\nu} - \phi_{\nu+1}) / F_{\rm th}l]^2 }  \;,
\label{eq:regI.1a}
\end{equation}
where $\nu<\nu_I$, and $\nu_I$ is the number of
the well between the regions I and II, see Fig.~\ref{fig:domains}.

Here $\phi_\nu$ is the diagonal matrix element of the electric potential in
the $\nu$-th well. The term proportional to the concentration difference on
the right hand side of Eq.~(\ref{eq:regI.1a})  is a diffusion current. In
region I the electron concentration grows in the vicinity of the high-field
domain and therefore the direction of the diffusion current is opposite to
the direction of the total current. In terms of these potentials $  
\Delta_{\nu,ii'} = e\phi_\nu-e\phi_{\nu+1} $.

The given definition of the potentials allows us to avoid taking in to 
consideration the well polarization. This effect is taken into account in
Ref. \onlinecite{laikhtman2}, where the integrated Poisson equation was
derived. It provides necessary connection between potentials $\phi_\nu$ and
concentrations $n_\nu$
\begin{equation}
  \Delta_{\nu}[\phi] + { e \over C_{\rm eff} }
  \Delta_{\nu}[n] = - {4 \pi e l \over \bar \epsilon }
  (n_{\nu} - \bar n) \ .
\label{eq:regI.1b}
\end{equation}
Here $C_{\rm eff}$ and $\bar \epsilon$ are constants, which can be calculated
for a given superlattice, $\Delta_{\nu}[f]{\equiv}f(\nu+1)+f(\nu-1)-2f(\nu)$,
and $\bar{n}$ is the average electron concentration. The second term on the
left hand side of Eq.~(\ref{eq:regI.1b}) describes the capacitance of one
well\cite{laikhtman2}. 
\label{eq:regI.1} \end{mathletters}

The system of Eqs.~(\ref{eq:regI.1}) has two boundary conditions. First, when
$\nu$ goes to $-{\infty}$ the difference $\phi_\nu-\phi_{\nu+1}$ goes to
$F_\infty l$. Second, Eq.~(\ref{eq:regI.1b}) for $\nu = \nu_I$ contains the
difference $\phi_{\nu_I}-\phi_{\nu_I+1}$ which is defined by the field in the
high-field region, $F_H$, and therefore it is about $F_Hl$. Far from the
boundary with region II Eq.  (\ref{eq:regI.1a}) becomes
\begin{mathletters}
\begin{equation}
   j = { \sigma F_\infty \over 1+ ( F_\infty / F_{\rm th})^2 } \ .
\label{eq:regI.1c}
\end{equation}
This equation shows that the current in region I is limited from above and reaches
a maximum at $F_\infty=F_{\rm th}$
\begin{equation}
   j{\ }<{\ }j_0 = \frac{1}{2}\sigma F_{\rm th}\ .
\label{eq:regI.1d}
\end{equation}
\label{eq:regI.1e}
\end{mathletters}

It is convenient to introduce a dimensionless field $f_\nu$ and displacement
$h_\nu$ as following,
\begin{mathletters}
\begin{eqnarray}
   f_\nu &=&  \
   \frac{ 2\left[
      (\phi_\nu - \phi_{\nu+1})/l - F_\infty
   \right] F_\infty}%
   {3(F_{\rm th}^2 - F_\infty^2)} \;,
\label{eq:regI.4a} \\
   h_\nu - h_{\nu-1} &=&  \
   \frac{ 8\pi e F_\infty/\bar\epsilon}%
   {3(F_{\rm th}^2 - F_\infty^2)}
   (n_\nu - \bar n )\; ,
\label{eq:regI.4b}
\end{eqnarray}
\label{eq:regI.4}
\end{mathletters}
where $\nu \leq \nu_I $. Equations (\ref{eq:regI.1}) take the form:
\begin{mathletters}
\begin{eqnarray}
   && f_\nu - \lambda_1^{-2}
    \Delta_\nu [h] - (3/2) (f_\nu)^2 = 0 \ ,
\label{eq:regI.5a}   \\
   && h_\nu + \lambda_2^{-2}
    \Delta_\nu [h] - f_\nu = 0 \ ,
\label{eq:regI.5b}
\end{eqnarray}
\label{eq:regI.5}
\end{mathletters}
where
\begin{mathletters}
\begin{eqnarray}
  && \lambda_1^2 = {4 \pi e^2 g l \over \bar \epsilon} \
     {F_{\rm th}^2 - F_\infty^2 \over F_{\rm th}^2 + F_\infty^2} \ ,
\label{eq:regI.6a}   \\
  && \lambda_2^2 =  \frac{4\pi l C_{\rm eff}}{\bar\epsilon} \ .
\label{eq:regI.6b}
\end{eqnarray}
\label{eq:regI.6}
\end{mathletters}

The boundary conditions for Eqs.~(\ref{eq:regI.5}) are
\begin{mathletters}
\begin{eqnarray}
   && \lim_{\nu\rightarrow -\infty}f_\nu = 0 \ ,
\label{eq:regI.7a}   \\
   && f_{\nu_I} =  \
   \frac{ 2(F_{\rm H} - F_\infty)F_\infty}%
   {3(F_{\rm th}^2 - F_\infty^2)}\ .
\label{eq:regI.7b}
\end{eqnarray}
\label{eq:regI.7}
\end{mathletters}

The coefficient $\lambda_2$ depends only on the superlattice parameters.
From the definitions of the $\bar \epsilon$ and $C_{\rm eff}$ in Ref.
\onlinecite{laikhtman2} one can easy get that $\lambda_2 > 2$. The parameter
$\lambda_1$ depends on $F_\infty$ and therefore on the current through the
superlattice $\lambda_1^2\propto\sqrt{1-j^2/j_0^2}$, see Eqs.
(\ref{eq:regI.1e}, \ref{eq:regI.6a}).  Therefore $\lambda_1$ goes to zero
when the current approaches its maximum value.

The analytic solution to nonlinear difference equations Eqs.~(\ref{eq:regI.5})
can be obtained in some limiting cases, but we consider here only one important
example. Later we give a numeric solution in the general case.

For $\lambda_1\ll1,\lambda_2$ the variation of $f_\nu$ and $h_\nu$ from well
to well is small. Hence, the second difference $\Delta_\nu[\ ]$ can be
replaced with the second derivative $d^2/d\nu^2$ and it can be neglected
in Eq.~(\ref{eq:regI.5b}). The resulting differential equation has the
solution
\begin{eqnarray}
   h_\nu =  f_\nu = \frac{1}{\cosh^2(\lambda_1\nu/2+\mbox{const})} \;.
\label{eq:regI.9}
\end{eqnarray}

The constant in Eq.~(\ref{eq:regI.9}) can be found from the boundary
condition Eq.~(\ref{eq:regI.7b}). The important property of this solution
is that it is limited from above. Such a limitation is not connected with a
small value of $\lambda_1$ but is a general property of
Eq.~(\ref{eq:regI.5}). This limitation ultimately results from the limitation
of the current in the first miniband; see e.g., Eq.~(\ref{eq:regI.1c}). In
general, the upper limit of the solution to Eq.~(\ref{eq:regI.5}), which
satisfies boundary condition Eq.~(\ref{eq:regI.7a}), depends on $\lambda_1$
and $\lambda_2$.
\begin{equation}
   f_{\nu_I} < \Upsilon(\lambda_1,\lambda_2) \ ,
\label{eq:regI.10}
\end{equation}

Equations (\ref{eq:regI.5}) can be reduced to the recurrent relation
\begin{equation}
   f_{\nu-1} = {\cal G}(f_{\nu}) \ ,
\end{equation}
where the function ${\cal G}(x)$ depends on the parameters $\lambda_1$ and
$\lambda_2$ and does not depend on $\nu$. This function has to satisfy
the boundary condition Eq.~(\ref{eq:regI.7a}) i.e. ${\cal G}(x)$
vanishes when $x$ goes to zero. For $x\ll1$ function ${\cal G}(x)$ can
be calculated explicitly.

\begin{figure}
\unitlength=1mm
\begin{picture}(87,87)
\put(6,6){
   \ifPostScript%
      \put(0,0){\epsffile{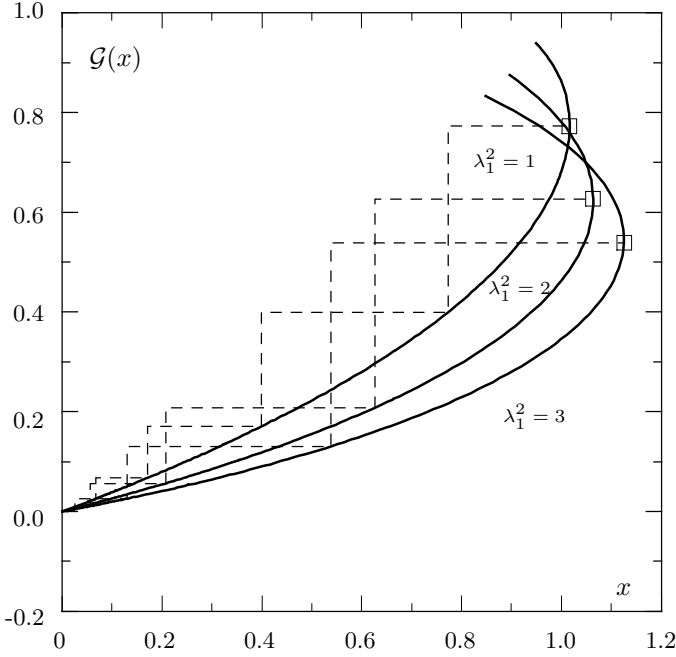}}%
   \else%
      \put(0,80){\special{em:graph graph032.pcx}}%
      \put(40.00,80.00){\special{em:graph squares.pcx}}%
   \fi
   \put(4,76){\makebox(0,0)[lt]{${\cal G}(x)$}}
   \put(76,3){\makebox(0,0)[rb]{$x$}}
   {\scriptsize
      \put(55,60){\makebox(0,0)[lb]{$\lambda_1^2 = 1$}}
      \put(57,43){\makebox(0,0)[lb]{$\lambda_1^2 = 2$}}
      \put(59,26){\makebox(0,0)[lb]{$\lambda_1^2 = 3$}}
   }
   {\small
      \put(00.00,-2.00){\makebox(0,0)[ct]{0}}
      \put(13.33,-2.00){\makebox(0,0)[ct]{0.2}}
      \put(26.66,-2.00){\makebox(0,0)[ct]{0.4}}
      \put(40.00,-2.00){\makebox(0,0)[ct]{0.6}}
      \put(53.33,-2.00){\makebox(0,0)[ct]{0.8}}
      \put(66.66,-2.00){\makebox(0,0)[ct]{1.0}}
      \put(80.00,-2.00){\makebox(0,0)[ct]{1.2}}
      \put(-2.00,00.00){\makebox(0,0)[rc]{-0.2}}
      \put(-2.00,13.13){\makebox(0,0)[rc]{0.0}}
      \put(-2.00,26.26){\makebox(0,0)[rc]{0.2}}
      \put(-2.00,40.40){\makebox(0,0)[rc]{0.4}}
      \put(-2.00,53.53){\makebox(0,0)[rc]{0.6}}
      \put(-2.00,66.66){\makebox(0,0)[rc]{0.8}}
      \put(-2.00,80.80){\makebox(0,0)[rc]{1.0}}
   }
}
\end{picture}
\caption{${\cal G}(x)$ for different values of $\lambda_1^2$ and
   $\lambda_2^2=20$. The quantities $f_\nu$, which are proportional to the
   electric field in the barrier $\nu$, can be obtained by iteration of the
   function ${\cal G}(x)$ and the dashed lines show the example of such an
   iteration. The iteration starts from the value of $f$ that corresponds to
   the field in the high-field domain. Squares show the maximum value of that
   $f$.  }
\label{fig:graph03} \end{figure}

Typical plots for ${\cal G}(x)$ are shown in Fig.~\ref{fig:graph03}. One can
get the sequences of values of $f_\nu$ by iterations of the function ${\cal
G}(x)$. For example, $f_{\nu_I-2}={\cal{G}}({\cal G}(f_{\nu_I}))$,
$f_{\nu_I}$ is given by Eq.~(\ref{eq:regI.7b}). These iterations are shown in
Fig.~\ref{fig:graph03} by dashed lines. One can see that
\begin{equation}
   \Upsilon=\max\left(
      {\cal G}^{-1}(x)
   \right) \ ,
\end{equation}
where ${\cal{G}}^{-1}(x)$ is the function inverse to ${\cal{G}}(x)$,
${\cal{G}}({\cal G}^{-1}(x))=x$.  The position of this maximum is also
marked in Fig.~\ref{fig:graph03} by a square.  In Fig.~\ref{fig:graph18}
this maximum $\Upsilon$ is plotted as a function of $\lambda_1^2$ for
different values of $\lambda_2^2$.

\begin{figure}
\unitlength=1mm
\begin{picture}(87,87)
\put(5,5){
   \insertplot{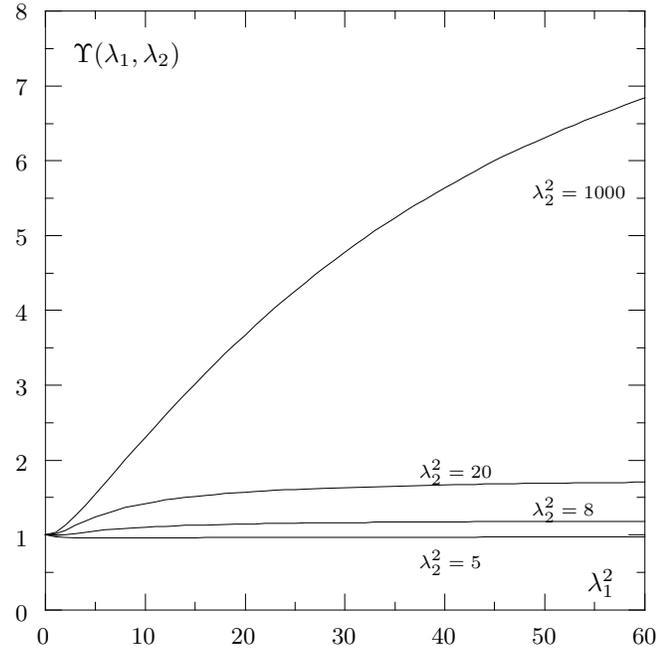}
   \put(4,76){\makebox(0,0)[lt]{$\Upsilon(\lambda_1,\lambda_2)$}}
   \put(76,3){\makebox(0,0)[rb]{$\lambda_1^2$}}
   {\scriptsize
      \put(65,56){\makebox(0,0)[lc]{$\lambda_2^2 = 1000$}}
      \put(50,18){\makebox(0,0)[lb]{$\lambda_2^2 = 20$}}
      \put(65,13){\makebox(0,0)[lb]{$\lambda_2^2 = 8$}}
      \put(50,6){\makebox(0,0)[lb]{$\lambda_2^2 = 5$}}
   }
   {\small
      \put(00.00,-2.00){\makebox(0,0)[ct]{0}}
      \put(13.33,-2.00){\makebox(0,0)[ct]{10}}
      \put(26.66,-2.00){\makebox(0,0)[ct]{20}}
      \put(40.00,-2.00){\makebox(0,0)[ct]{30}}
      \put(53.33,-2.00){\makebox(0,0)[ct]{40}}
      \put(66.66,-2.00){\makebox(0,0)[ct]{50}}
      \put(80.00,-2.00){\makebox(0,0)[ct]{60}}
      \put(-2.00,00.00){\makebox(0,0)[rc]{0}}
      \put(-2.00,10.00){\makebox(0,0)[rc]{1}}
      \put(-2.00,20.00){\makebox(0,0)[rc]{2}}
      \put(-2.00,30.00){\makebox(0,0)[rc]{3}}
      \put(-2.00,40.00){\makebox(0,0)[rc]{4}}
      \put(-2.00,50.00){\makebox(0,0)[rc]{5}}
      \put(-2.00,60.00){\makebox(0,0)[rc]{6}}
      \put(-2.00,70.00){\makebox(0,0)[rc]{7}}
      \put(-2.00,80.00){\makebox(0,0)[rc]{8}}
   }
}
\end{picture}
\caption{$\Upsilon$ as a function of $\lambda_1^2$ is plotted for different
values of $\lambda_2^2$.}
\label{fig:graph18}
\end{figure}

A calculation\cite{laikhtman2} shows that
$\lambda_2^2\ge4(d_{\rm B}\epsilon_{\rm eff}/d_{\rm W}\epsilon+1)$, where
$d_{\rm B}$ and $d_{\rm W}$ are the widths of the barrier and the well
respectively, $\epsilon$ is dielectric constant in the barrier, and
$\epsilon_{\rm eff}$ is the effective dielectric constant in the well. Although
$\epsilon_{\rm eff}$ is proportional to the number of electrons in the well,
usually $\epsilon_{\rm eff}\sim\epsilon$ and $d_{\rm B}\sim d_{\rm W}$. Thus
a typical value of $\lambda_2^2$ is 8--10. It becomes large only in the limit
of extremely narrow wells or highly doped wells. The other parameter,
$\lambda_1^2$ can be estimated by means of Eq.~(\ref{eq:regI.6a}). It gives
$\lambda_1^2 < e^2g_0l = l/24${}\AA. Even if the superlattice period is
500{\AA} $\lambda_1^2<20$. We plot $\Upsilon$ for $\lambda_1^2$ that ranges
from 0 to 60; see Fig.~\ref{fig:graph18}. The estimate for typical
values of $\lambda_1^2$ and $\lambda_2^2$ shows that usually
$\Upsilon\approx1$.

The condition Eq.~(\ref{eq:regI.10}) can be rewritten in terms of the
current $j$ by making use of Eq.~(\ref{eq:regI.7b}) and Eq.~(\ref{eq:regI.1e})
\begin{equation}
   j \ < \ j^\ast  \ ,
\label{eq:regI.11}
\end{equation}
where $j^\ast$ can be found from following equation
\begin{equation}
   \frac{j^\ast}{j_0 + (3\Upsilon-1)\sqrt{j_0^2-(j^\ast)^2}} =
   \frac{F_{\rm th}}{F_{\rm H}} \ .
\label{eq:regI.12}
\end{equation}

The quantity $j^\ast$ is the upper limit of the current in the superlattice
with the high-field domain. The current through the superlattice is also limited
by Eq.~(\ref{eq:regI.1d}), but $j^\ast<j_0$, and therefore
Eq.~(\ref{eq:regI.11}) is a stronger restriction. This restriction comes
from the properties of the boundary between regions~I and~II. Indeed, the
excess of electrons in the well at this boundary caused by a large field
gradient generates  a diffusion current opposite to the
current through the superlattice.  Usually the diffusion backflow of
electrons is compensated by a local increase of the field. This compensation
is possible only if the current through the superlattice is smaller than the
maximum current in region~I.

The condition Eq.~(\ref{eq:regI.11}) also implies that there exists an upper
limit of the field in the low-field region. When this field exceeds that
limit value, the system becomes unstable.  The development of this instability
leads to the expansion of the high-field region, and the field in the
low-field region decreases abruptly.

\section{ Description of the high-field region.  }
\label{sec:regII}

In the high-field region the main contribution to the current is the
electron tunneling from the first level of one well to the second level of
the neighboring well followed by the relaxation from the second level to
the first one. The main mechanism of the energy
relaxation is the emission of optical phonons, if the inter-subband energy
space $E_g$ is larger than the phonon energy $\hbar\Omega_{\rm LO}$. In
this case the relaxation time $\tau_{21}$ ranges from $0.5\times10^{-12}$ to
10$^{-11}$ sec. depending on the inter-subband energy
space\cite{yang-lyon,seilmeier,artaki-hess2,jain-dassarma,stolz88,tatham89,
grahn-may90}. This is larger than the relaxation time in the bulk material,
because the scattering probability is inversely proportional to the transfer
momentum squared and the transfer momentum in the inter-subband relaxation,
$2m\sqrt({E_g} - \hbar\Omega_{\rm LO})$, is larger than that in the bulk
material $2m[\sqrt(E)-\sqrt(E-\hbar\Omega_{\rm{LO}})]
\approx2m\hbar\Omega_{\rm{LO}}/\sqrt(E)$, see Ref.~\onlinecite{herbert}. In
the case of ${E_g} < \hbar\Omega_{\rm LO}$ the main relaxation mechanism
is the electron-electron interaction and $\tau_{21}$ is about 10$^{-10}$ --
10$^{-9}$ sec.\cite{oberli,artaki-hess2,falko}  The transition time
$\tau_{\rm t}$ can vary in a wide range depending on the superlattice parameters.
This time has been measured in 40{\AA}/40{\AA} and in 30{\AA}/30{\AA} GaAs--GaAlAs
superlattices, see Ref.~\onlinecite{deveaud1}, and appeared about
$3.6\times10^{-11}$ sec and $5.3\times10^{-12}$ sec respectively. In a
123{\AA}/21{\AA} superlattice $\tau_{\rm{t}}$ has been found about
$6\times10^{-11}$ sec; see Ref.~\onlinecite{grahn-may90}.

In the case when the first and the second levels in adjacent wells are not
in resonance one can not use the simple balance equation
Eq.~(\ref{eq:regII.1}). The current $j_{12}$ in region II is described by
Eq.~(\ref{eq:coupl.2b}) with $i=1$ and $i^\prime=2$. The energy space between
these levels $\Delta=eF_{\nu}l-{E_g}<0$. The current is equal to the
number of electrons that relax from the second level to the first level
per unit time
\begin{equation}
   j_{12} = e{n^{(2)}\over\tau_{21}}\ .
\label{eq:regII.2}
\end{equation}
For simplicity we neglect in these calculations $\exp(-{E_g}/T)$
compared to unity, because usually ${E_g}\gtrsim{T}$. So, the
generalized balance equations become
\begin{mathletters}
\begin{eqnarray}
   \frac{n(\zeta^{(1)}+\Delta) - n^{(2)} }{\tau_{\rm t}}
   &=& {n^{(2)}\over\tau_{21}}
\label{eq:regII.3a}\\
   n(\zeta^{(1)}) + n^{(2)} &=&\bar{n}
\label{eq:regII.3b}
\end{eqnarray}
In this equation, the inverse transition time
$1/\tau_{\rm{t}}=2\Lambda_{12}^2\Gamma/\hbar(\Delta^2 + \Gamma^2)$.
\label{eq:regII.3}
\end{mathletters}

The elimination of $n^{(2)}$ from Eqs.~(\ref{eq:regII.3}) leads to
\begin{equation}
   j = \frac{e}{\hbar}{\ }
   \frac{2\Lambda_{12}^2\Gamma n(\zeta^{(1)}+\Delta)}%
   {\Delta^2 + \Gamma^2 + %
   2\Lambda_{12}^2\Gamma\tau_{21}/\hbar} \ .
\label{eq:regII.3.add}
\end{equation}
This equation together with Eq.~(\ref{eq:regII.3}) describes the
current-voltage characteristics of the high-field domain, i.e. the dependence
of the current on the potential drop per period, $Fl$. In general, this
quantity is different from the value corresponding to the resonance between
the first level in one well and the second level in the neighboring well,
${E_g}/e$. Usually it is assumed that the deviation from the resonance is
negligibly small\cite{esaki-chang,choi1,vuong}. Actually the levels can be
considered in resonance only if $|\Delta|\lesssim\Gamma$. However, this is not
always the case.

The results of Sec.~\ref{sec:regI} show that the current in the
superlattice with the high-field domain is smaller than~$j^\ast$. If
$\tau_{21}$ is not very large, then
\begin{equation}
   j^\ast\ll{e}\bar{n}/2\tau_{21} \;.
\label{eq:regII.4}
\end{equation}
In the other possible case, $j^\ast\approx{e}\bar{n}/2\tau_{21} $, the levels
have to be in the resonance and therefore $\Delta\approx0$. The
condition, Eq.~(\ref{eq:regII.4}), means that $n^{(2)}\ll\bar{n}$, and the
last inequality is satisfied in two cases,
$n(\zeta^{(1)}-|\Delta|)\ll{\bar{n}}$ or $\tau_{\rm{t}}\gg\tau_{21}$.  In the
first case
\begin{mathletters}
\begin{equation}
   |\Delta|> T,E_{\rm F} \ ,
\label{eq:regII.5a}
\end{equation}
i. e. the deviation from the resonance is rather large. In the second case
$\Delta^2+\Gamma^2\gg2\Lambda_{12}^2\Gamma\tau_{21}/\hbar$.  Usually
$\Gamma\sim$3--5 meV, $\tau_{21}>0.5\times10^{-12}$ sec.,
$\Lambda_{12}\gtrsim$3--5 meV, and therefore
$\Gamma\lesssim\Lambda_{12}^2\tau_{21}/\hbar$. That is, in this case
\begin{equation}
   |\Delta|\gg
   \sqrt{\Lambda_{12}^2\Gamma\tau_{21}/\hbar}
   \gtrsim\Gamma\ .
\label{eq:regII.5b}
\end{equation}
\label{eq:regII.5}
\end{mathletters}
These inequalities for $\Delta$ show that under the condition
Eq.~(\ref{eq:regII.4}), the increase of the applied bias necessary for the
extension of the high-field domain by one period can be considerably smaller
than the resonance value ${E_g}/e$.

The limitation of the current by the value of $j^\ast$ originates from the
boundary between regions I and II, see Sec.~\ref{sec:regI}. When, with the
increase of the bias, the high-field domain extends over the entire
superlattice, this boundary disappears. Then the current jumps up sharply and
reaches the value defined by conditions of the resonant tunneling,
\begin{equation}
   j_{\mbox{res} }
   = {\bar{n}\over\tau_{\rm{t}}+2\tau_{21}}
   =\frac{e}{\hbar}
   {2\Lambda_{12}^2\bar{n}\over\Gamma+4\Lambda_{12}^2\tau_{21}/\hbar}\ .
\label{eq:regII.6}
\end{equation}

\section{ Low-field region downstream of the domain.}
\label{sec:regIII}

In this region, if it exists, electrons are injected into the second miniband
from the high-field domain. They can move 1000{}\AA{} before they
drop down to the first miniband\cite{herbert}. The injection of electrons
into the first miniband can be neglected. Therefore the current in the first
mini-band vanishes near the boundary with the high-field domain and increases
away from it, owning to the relaxation from the second miniband.

In doped superlattices the screening length is about one period. In other
words, the drop of the field at the boundary between regions II and III
causes the depletion of only the one well closest to the high-field domain and
all other wells in the region III can be considered to be electroneutral. The
relaxation of electrons from the second miniband, where their mobility is
high, into the first miniband, where the electron mobility is much smaller,
leads to the fields inhomogeneity on the scale of the relaxation length. We
calculate the field distribution in the most interesting case, when this
length is much larger than the superlattice period.

A relaxation length large compared to the superlattice period and to the
screening length allows us to use the condition of electroneutrality and to
replace difference equations with differential ones. The total current $j$ is
the sum of the currents in the first mini-band, $j_1$, and in the second
mini-band, $j_2$ :
\begin{mathletters}
\begin{eqnarray}
   && j = j_1 + j_2 \ ,
\label{eq:regIII.1a} \\
   && j_1 = {\mu_1 \over l} \left(
      n^{(1)} eF - T\frac{d{n}^{(1)}}{dy}
   \right) \;,
\label{eq:regIII.1b} \\
   && j_2 = {\mu_2 \over l} \left(
      n^{(2)} eF - T\frac{d{n}^{(2)}}{dy}
   \right) \;,
\label{eq:regIII.1c}
\end{eqnarray}
where $\mu_1$ and $\mu_2$ are the mobilities in the first and the second
miniband respectively. Here we assume the Boltzmann distribution in both
mini-bands and neglect the field dependence of the mobilities. The first
assumption is reasonable because one can expect a significant heating of
electrons injected into region III. The second assumption is natural due
to the following circumstance. The field dependence of the mobility is
appreciable when the field is close enough to the instability threshold field
$F_{\rm{th}}$.  Usually the field in the high field domain $F_{\rm{H}}$ is
significantly larger than $F_{\rm{th}}$.  Equation (\ref{eq:regI.12}) shows
that in this case $j^\ast$ is significantly smaller than $j_0$ and therefore
$F_{\infty}$ cannot be close to $F_{\rm th}$. In the part of region III
where the current goes mostly in the second miniband, the field is even
smaller.

The next two equations are the conditions of the electroneutrality and the
continuity equations are
\begin{eqnarray}
   && n = n^{(1)} + n^{(2)} ,
\label{eq:regIII.1d} \\
   &&\frac{d j_2}{dx} = -\frac{en^{(2)}}{\tau_{21}l}
\label{eq:regIII.1e}
\end{eqnarray}
\label{eq:regIII.1}
\end{mathletters}

In the general case the electric field can be found from
Eqs.~(\ref{eq:regIII.1b}) -- (\ref{eq:regIII.1d})
\begin{equation}
   \frac{F}{F_\infty} =
   \frac{j_1}{j} + \frac{\mu_1}{\mu_2}\ \frac{j_2}{j}
   = 1 - \biggl[
      1-\frac{\mu_1}{\mu_2} \biggr]
   \frac{j_2}{j}\ ,
\label{eq:regIII.2}
\end{equation}
where $F_\infty$ is the electric field far from the domain. After
eliminating the current $j_2$ and the concentration $n^{(2)}$ from
Eqs.~(\ref{eq:regIII.1c}), (\ref{eq:regIII.1e}), and (\ref{eq:regIII.2}) one
can get the differential equation for the field in region III
\begin{equation}
   L_{\cal D}^2 \frac{d^2}{dy^2} \frac{F}{F_\infty}
   = L_{\rm Dr} \frac{F}{F_\infty}\frac{d}{dy}\frac{F}{F_\infty}
   + \frac{F}{F_\infty} - 1 \ .
\label{eq:regIII.3}
\end{equation}

The redistributing of electrons between two minibands and the field profile
are characterized by two lengths: a diffusion length $L_{\cal D}$ and a drift
length ${L}_{\rm Dr}$, where
\begin{equation}
   L_{\cal D}^2 = \frac{\mu_2\tau_{21}T}{e} \ ,
\label{eq:regIII.d2}
\end{equation}
and
\begin{equation}
   L_{\rm Dr} = \mu_2\tau_{21}F_\infty \ .
\label{eq:regIII.5}
\end{equation}
The former length is the distance that electrons diffuse in the second
miniband before the relaxation to the first miniband.  The latter length is
the distance that electrons run in the second miniband under the electric
field before the relaxation. Equation~(\ref{eq:regIII.3}) has to be solved
near the boundary for an arbitrary relation between these two lengths. Far
from the boundary we calculate the distribution of electric field
separately in two limiting cases, when one of these two lengths is much
larger than the other.

At $y=0$ there is no current in the first miniband and Eq.~(\ref{eq:regIII.2})
gives the boundary condition for Eq.~(\ref{eq:regIII.3}),
\begin{equation}
   \frac{F}{F_\infty}\biggr|_{x=0} = \frac{\mu_1}{\mu_2} \ .
\label{eq:regIII.6}
\end{equation}
One can see that near the boundary $F/F_\infty$ is small if
$\mu_1\ll\mu_2$. That can be expected because the second miniband is usually
wider than the first miniband. Thus we can neglect $F/F_\infty$ on the
right hand side of Eq.~(\ref{eq:regIII.3}). This simplification allows us
to solve this equation in terms of Airy functions. The result is
\begin{equation}
   \frac{F}{F_\infty} =
   - \left(4 {L_{\cal D}^2 \over L_{\rm Dr}^2} \right)^{1/3}
   \frac{\mbox{Ai}^{\prime}\left(
      (y-y_0)/L_0
   \right)}{\mbox{Ai}\left(
      (y-y_0)/L_0
   \right)} \;,
\label{eq:regIII.7}
\end{equation}
where $\mbox{Ai}^{\prime}(\xi)$ denotes the derivative of the Airy function
with respect to its argument and
$ L_0 = (2L_{\cal D}^4/L_{\rm Dr})^{1/3} $.

The other parameter $y_0$ can be found by substitution of
Eq.~(\ref{eq:regIII.7}) into the boundary condition Eq.~(\ref{eq:regIII.6}).
This gives for $\xi = - y_0/L_0$
\begin{equation}
   q = - \mbox{Ai}^{\prime}(\xi)/\mbox{Ai}(\xi) \ ,
\label{eq:regIII.9}
\end{equation}
where
$ q = ({\mu_1}/{\mu_2}) (L_{\rm Dr}^2/4L_{\cal D}^2)^{1/3} $

The solution Eq.~(\ref{eq:regIII.7}) shows that the electric field increases
away from the boundary. Far from the boundary where $F\sim{F_\infty}$ the
solution Eq.~(\ref{eq:regIII.7}) is not valid any more. In the case of a
short drift length $L_{\rm Dr}{\ll}L_{\cal D}$ one can neglect the
first term on the right hand side of Eq.~(\ref{eq:regIII.3}) because it contains a
small parameter $L_{\rm Dr}/L_{\cal D}$. We can show this by replacing
$y/L_{\cal D}$ by a dimensionless variable. In this case the solution to Eq.
(\ref{eq:regIII.3}) is
\begin{equation}
   \frac{F}{F_\infty} =
   \frac{j_1}{j} + \frac{\mu_1}{\mu_2}\ \frac{j_2}{j}
   = 1 - \biggl[
      1-\frac{\mu_1}{\mu_2} \biggr]
	\exp(-y/L_{\cal D}) \ ,
\label{eq:regIII.d4a}
\end{equation}
where the point $x=0$ corresponds to the boundary between regions II and III.
The characteristic length of the electron relaxation from the second
miniband to the first one is in this case the diffusion length $L_{\cal D}$.

In the opposite case when $L_{\rm Dr}{\gg}L_{\cal D}$, Eq.~(\ref{eq:regIII.3})
can be simplified in the region where
$(L_{\cal{D}}/L_{\rm{Dr}})^{2/3}{\ll}F/F_\infty$, namely, the term on the
left hand side of Eq.~(\ref{eq:regIII.3}) can be neglected. The solution to
the resulting equation is
\begin{equation}
    F/F_\infty - \log(1-F/F_\infty) = (y-y_0)/L_{\rm Dr} \ .
\label{eq:regIII.13}
\end{equation}
Here $y_0$ is the same number as in Eq.~(\ref{eq:regIII.7}). This can be
proven by matching the asymptotes of the both solutions,
Eq.~(\ref{eq:regIII.7}) and Eq.~(\ref{eq:regIII.13}), in the region where
$(L_{\cal{D}}/L_{\rm{Dr}})^{2/3}{\ll}F/F_\infty\ll1$.

\begin{figure}
\begin{picture}(87,87)
\put(6.,6.){
   \insertplot{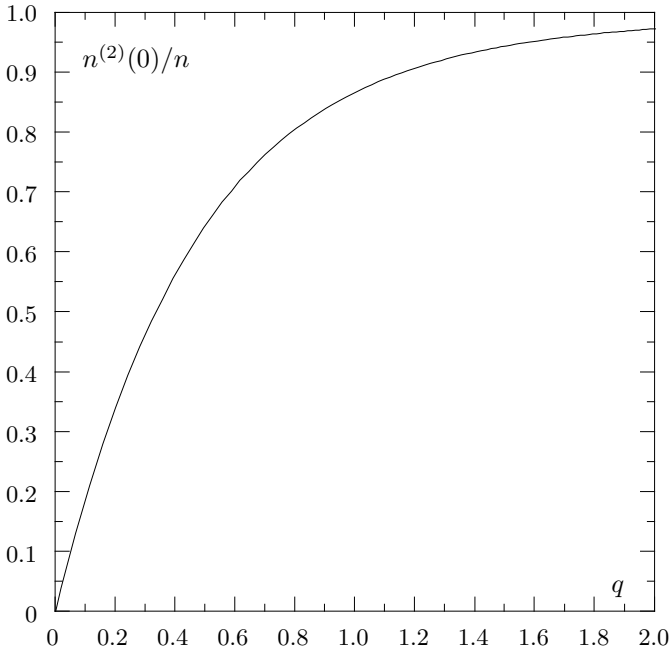}
   \put(4,76){\makebox(0,0)[lt]{$n^{(2)}(0)/n$}}
   \put(76,3){\makebox(0,0)[rb]{$q$}}
   {\small
      \put(00.00,-2.00){\makebox(0,0)[ct]{0}}
      \put(08.00,-2.00){\makebox(0,0)[ct]{0.2}}
      \put(16.00,-2.00){\makebox(0,0)[ct]{0.4}}
      \put(24.00,-2.00){\makebox(0,0)[ct]{0.6}}
      \put(32.00,-2.00){\makebox(0,0)[ct]{0.8}}
      \put(40.00,-2.00){\makebox(0,0)[ct]{1.0}}
      \put(48.00,-2.00){\makebox(0,0)[ct]{1.2}}
      \put(56.00,-2.00){\makebox(0,0)[ct]{1.4}}
      \put(64.00,-2.00){\makebox(0,0)[ct]{1.6}}
      \put(72.00,-2.00){\makebox(0,0)[ct]{1.8}}
      \put(80.00,-2.00){\makebox(0,0)[ct]{2.0}}
      \put(-2.00,00.00){\makebox(0,0)[rc]{0}}
      \put(-2.00,08.00){\makebox(0,0)[rc]{0.1}}
      \put(-2.00,16.00){\makebox(0,0)[rc]{0.2}}
      \put(-2.00,24.00){\makebox(0,0)[rc]{0.3}}
      \put(-2.00,32.00){\makebox(0,0)[rc]{0.4}}
      \put(-2.00,40.00){\makebox(0,0)[rc]{0.5}}
      \put(-2.00,48.00){\makebox(0,0)[rc]{0.6}}
      \put(-2.00,56.00){\makebox(0,0)[rc]{0.7}}
      \put(-2.00,64.00){\makebox(0,0)[rc]{0.8}}
      \put(-2.00,72.00){\makebox(0,0)[rc]{0.9}}
      \put(-2.00,80.00){\makebox(0,0)[rc]{1.0}}
   }
}
\end{picture}
\caption{ The concentration of the electrons in the second mini-band in
   region III of the superlattice close to the boundary with region II is
   plotted versus the parameter
   $q=(e\mu_1^3\tau_{21}F_\infty^2/\mu_2^2T)^{1/3}$.}
\label{fig:popul.2}
\end{figure}

The very important property of Eqs.~(\ref{eq:regIII.1}) is that the number of
electrons in the second miniband explicitly depends on the interminiband
relaxation time. One can get from Eq.~(\ref{eq:regIII.7}) together with
Eqs.~(\ref{eq:regIII.1e}) and (\ref{eq:regIII.2})
\begin{equation}
   n^{(2)}|_{x=0} = 2\left[
      q^3 - q\xi(q)
   \right]n \;.
\label{eq:regIII.15}
\end{equation}
The parameter $q$, which defines the number of electrons in the second
mini-band near the boundary with region II, is proportional to
$\tau_{21}^{1/3}$. When $q$ is small $\xi\approx{-1}$ and $n^{(2)}(0)\approx
2qn$. When $q$ is large the asymptote of the Airy function gives
$q\approx\xi^{-1}/4+\xi^{1/2}$ and $n_2\approx n$.  The concentration
$n^{(2)}$ as a function of $q$ is plotted in Fig.~\ref{fig:popul.2}.

The obtained field distribution is not valid at the distance of about one
screening length from the high-field domain because the electro-neutrality
condition Eq.~(\ref{eq:regIII.1d}) cannot be used there. Thus in the well at
the boundary between regions II and III, see Fig.~\ref{fig:domains}, where
the field changes significantly, the concentration of electrons has to be
found from the Poisson equation. The field change near the next well,
$\nu=\nu_{II}+1$ is much smaller and we can consider it and the rest of 
region III as electro-neutral. Therefore the solution to
Eqs.~(\ref{eq:regIII.1}) is valid up to well number $\nu=\nu_{II}+1$, and
$n_{\nu_{\rm II}+1}^{(2)} = n^{(2)}|_{x=0} $.
The field in the barrier between wells number $\nu_{II}$ and $\nu_{II}+1$
can be easily computed from Eqs.~(\ref{eq:coupl.2a}) and (\ref{eq:regI.1b})
with $\nu=\nu_{II}$. In the case when the relaxation length of electrons
from the second miniband to the first one is much larger than one
superlattice period we can neglect the current between first levels and also
put in the Eq.~(\ref{eq:coupl.2a}) $j_{11}=0$ and $j_{22}=j$.

\section{ Discussion}
\label{sec:disc}

In this section we give a qualitative description of the current-voltage
characteristic of the entire superlattice. Under a small external bias the
superlattice exhibits Ohm's law. At a higher bias the current reaches a
maximum value. The uniform potential distribution in the superlattice is
unstable at this point of the $I$-$V$ characteristic. The theory of
instability was developed by the present authors\cite{laikhtman1,laikhtman2}.
The instability eventually leads to the formation of the high-field domain
and a sharp current drop. In Sec.~\ref{sec:pict} we argued that due to the
charge accumulation at the high-field domain boundary, its length can be more
than one superlattice period. Now we estimate this length.

The threshold value of the bias just before the instability point is
$NF_{\rm{th}}l$, where $N$ is the number of the superlattice periods. Right
after the instability point this bias is a sum of the voltage drop across the
high field region $F_{\rm H}N_{\rm II}l$ and the low field region
$\approx(N-N_{\rm II})F_\infty l$.  Here we neglect a field inhomogeneity
near the domain boundaries.  So we have
\begin{equation}
   NF_{\rm th}l = F_{\rm H}N_{\rm II}l
   + (N-N_{\rm II})F_\infty l \ ,
\label{eq:disc.4a}
\end{equation}
and
\begin{equation}
   N_{\rm II} = N\frac{F_{\rm th}-F_\infty}{F_{\rm H}-F_\infty} \ .
\label{eq:disc.4aa}
\end{equation}
This equation shows that in very long superlattices the length of the
high-field domain is proportional to the length of the superlattice. This
fact is a direct consequence of the upstream diffusion current near the
boundary of the high-field domain. Without the diffusion the minimal length
of the high-field domain is one superlattice period and
$F_{\rm{th}}-F_\infty\propto1/N$. Due to the diffusion current $F_\infty$ is
limited from above by a value independent of the superlattice length. From
Eqs.~(\ref{eq:regI.1c}), (\ref{eq:regI.11}), and (\ref{eq:regI.12}) we have
\begin{equation}
   \sqrt{F_{\rm H}^2+3F_{\rm th}^2} - F_{\rm H} \ {\ge}\ F_\infty\ .
\label{eq:disc.4b}
\end{equation}
For $F_{\rm th}{\ge}F_\infty$ one can easily see that $N_{\rm II}$ decreases
monotonically with increasing $F_\infty$. Then Eqs.~(\ref{eq:disc.4aa}) and
(\ref{eq:disc.4b}) give the following condition for $N_{\rm II}$
\begin{equation}
   N_{\rm II}\ \ge \ N
   \frac{F_{\rm H}+3F_{\rm th}-\sqrt{F_{\rm H}^2+3F_{\rm th}^2}}%
   {3(F_{\rm H}+F_{\rm th})}
\label{eq:disc.4c}
\end{equation}
Actually the field in regions I, II, and III is inhomogeneous. As a result
there is a correction to the right-hand side of Eq.~(\ref{eq:disc.4c}).
Usually this correction can be neglected because it does not contain the
factor $N$.

\begin{figure}
\unitlength=1mm
\linethickness{0.4pt}
\begin{picture}(87.00,87.00)
\put(7.00,7.00){\insertplot{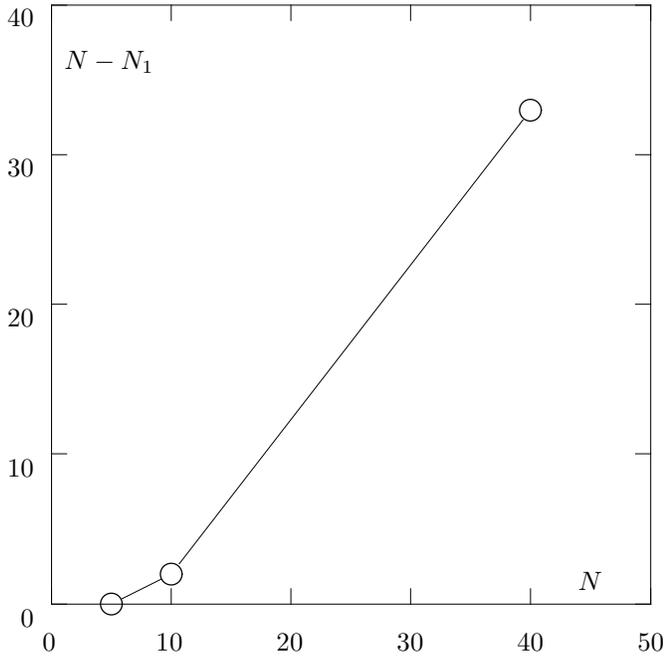}}
\put(7.00,5.00){\makebox(0,0)[ct]{$0$}}
\put(87.00,5.00){\makebox(0,0)[ct]{$50$}}
\put(23.00,5.00){\makebox(0,0)[ct]{$10$}}
\put(39.00,5.00){\makebox(0,0)[ct]{$20$}}
\put(55.00,5.00){\makebox(0,0)[ct]{$30$}}
\put(71.00,5.00){\makebox(0,0)[ct]{$40$}}
\put(79.00,12.00){\makebox(0,0)[cc]{$N$}}
\put(5.00,7.00){\makebox(0,0)[rc]{$0$}}
\put(5.00,27.00){\makebox(0,0)[rc]{$10$}}
\put(5.00,47.00){\makebox(0,0)[rc]{$20$}}
\put(5.00,67.00){\makebox(0,0)[rc]{$30$}}
\put(5.00,87.00){\makebox(0,0)[rc]{$40$}}
\put(15.00,81.00){\makebox(0,0)[cc]{$N-N_1$}}
\end{picture}
   \caption{
   Well number dependence of oscillation repetition number from the
   experimental work of the Y.~Kawamura.
   {\em et all}\protect{\cite{kawamura1}}
   }
\label{fig:oscnum}
\end{figure}

With a further increase of the bias the current nearly periodically increases
and drops down. After Esaki and Chang,\cite{esaki-chang}, each oscillation is
associated with the extension of the high-field domain by one period and the
number of oscillation $N_{\rm{osc}}$ is expected to be equal to $N-1$.
Equation (\ref{eq:disc.4c}) shows that $N_{\rm{osc}}$ can be less than $N-1$
and that really is the case in some experiments. For instance, Kawamura et
al.\cite{kawamura1} observed a current-voltage characteristic with
$N_{\rm{osc}}=35$ in a superlattice that had $N=39$ barriers, see
Fig.~\ref{fig:oscnum}. The difference between $N$ and $N_{\rm{osc}}$ can be
interpreted as the formation of a high-field domain with the minimal length
$N_{\rm II}=4$.  The value of the threshold field, $F_{\rm{th}}l$, in this
experiment can be obtained by dividing the threshold voltage, $\approx$0.5V,
by the number of superlattice barriers. The voltage drop across a barrier
in the high-field domain $F_{\rm{H}}l$ is equal to the period of the
oscillations, 0.14V. The substitution of these values into
Eq.~(\ref{eq:disc.4c}) gives $N_{\rm II}>3.1$ which is in agreement with the
value obtained from the number of the oscillations.

\begin{figure}
\unitlength=1.00mm
\linethickness{0.4pt}
\begin{picture}(87.00,80.00)
\put(0.04,0.00){
   \put(2.00,31.00){\makebox(0,0)[rt]{A}}
   \put(42.00,17.00){\makebox(0,0)[ct]{C}}
   \put(2.00,67.00){\makebox(0,0)[lc]{$eF_\nu d(t)$}}
   \put(42.00,13.00){\vector(2,1){35.93}}
   \put(78.00,29.00){\makebox(0,0)[ct]{$t$}}
   \ifPostScript
   \epsfverbosetrue
   \put(-6.00,13.00){\epsfxsize=226.8445pt
   \epsffile{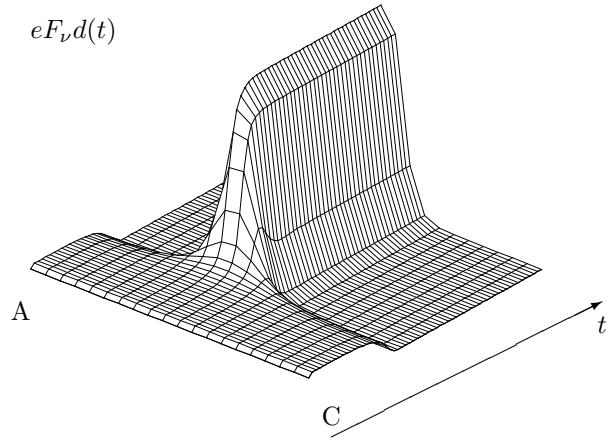}}\else%
   \put(0.00,65.00){\special{em: graph graph110.pcx}}
   \fi
}
\end{picture}

\caption{The time evolution of the voltage drops distribution.
The superlattice has 20 periods, cathode and anode are marked by
C and A respectively. The interlevel spacing in the well is 100meV,
whereas the domain height is only 70meV per period.}
\label{fig:graph11}
\end{figure}

\begin{figure}
\unitlength=1.00mm
\linethickness{0.4pt}
\begin{picture}(87.00,80.00)
\put(0.04,0.00){
   \put(2.00,31.00){\makebox(0,0)[rt]{A}}
   \put(42.00,17.00){\makebox(0,0)[ct]{C}}
   \put(2.00,67.00){\makebox(0,0)[lc]{$n^{(2)}_\nu(t)$}}
   \put(42.00,13.00){\vector(2,1){35.93}}
   \put(78.00,29.00){\makebox(0,0)[ct]{$t$}}
   \ifPostScript
   \epsfverbosetrue
   \put(-6.00,13.00){\epsfxsize=226.8445pt
   \epsffile{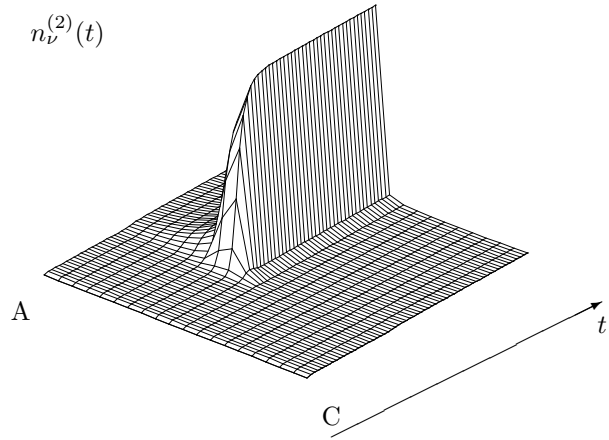}}\else%
   \put(0.00,65.00){\special{em: graph graph120.pcx}}
   \fi
}
\end{picture}

\caption{The time evolution of the concentration at the second level in each
well. The superlattice has 20 periods, cathode and anode are marked by
C and A respectively. One can see the tail of the electron
distribution in region III (behind the domain).}
\label{fig:graph12}
\end{figure}

In order to understand how a large domain comes about, we performed a numeric
calculation of the instability development.  The simulation was made for $20$
period superlattice with the inter-subband energy separation $100$meV, and
the electron concentration corresponding to the Fermi energy 40meV. We assume
some reasonable relations between transition and relaxation times. The
parameters characterizing screening in the superlattice were $\lambda_1^2=4$,
and $\lambda_2^2=8$. The results of the simulation are shown in
Figs.~\ref{fig:graph11} and~\ref{fig:graph12}. One can see that the domain
starts to grow as a large scale instability\cite{laikhtman2} which transforms
into a short domain with a length of three periods.

In the case of low doping there are not enough electrons to form the
depletion layer downstream of the domain. As the result the domain is formed
near the anode\cite{grahn-feb90} and region III does not exist.  The position
of the high-field domain in a highly doped superlattice is determined by
unintentional nonuniformity of the superlattice or by a fluctuation that
initiated the domain.

For fields larger than $F_{\rm{th}}$, the oscillating current is limited from
above by the value of $j^\ast$. It is important to note that this value is
usually smaller than $j_0$, i. e. the current value at $F=F_{\rm{th}}$,
see Sec.~\ref{sec:regI}.  Such a feature of the $I$-$V$ characteristic is
typically observed in experiment\cite{kawamura1,vuong}. We will show now that
the high-field domain expands by joining another period from region I
even in the case when region III does exist.  For this purpose we make
use of an equation similar to Eq.~(\ref{eq:disc.4a}), where we take into account
the nonuniformity of the field at the accumulation layer near the domain
boundary. So instead of Eq.~(\ref{eq:disc.4a}) we have
\begin{eqnarray}
   V&=& el\sum_\nu F_\nu
\nonumber\\
    &=& ((N_{\rm I}-1)F_\infty + F_{\nu_I-1} + N_{\rm II}F_H)l
     + V_{\rm III}\ ,
\label{eq:disc.1}
\end{eqnarray}
where $V$ is the applied bias, $N_{\rm I}$ and $N_{\rm II}$ are the numbers
of periods in regions I and II respectively, $\nu_I$ is the number of the
first barrier in the high-field domain, $V_{\rm III}$ is the potential drop
across region III. Equation (\ref{eq:disc.1}) resembles Eq.
(\ref{eq:disc.4a}), however, it explicitly takes into account the field
inhomogeneity in regions I and III.

As the applied voltage increases, $dV/dj$ goes to infinity which eventually
leads to a current discontinuity and extension of the high field domain by
one period. One can see from Eq.~(\ref{eq:disc.1}) that
$dV/dj\rightarrow\infty$ when in one of the barriers
$dF/dj\rightarrow\infty$. This cannot take place in region II, where the
maximum value of the current is larger than that in region I. The infinite
value of $dF/dj$ or zero value of $dj/dF$ is ultimately connected with the
maximum in the $I$-$V$ characteristic of one barrier. This maximum is reached
only in the regions where current is confined in the first mini-band, i.e. in
region I and in a part of region III. The largest field in these regions is
in the last barrier of region I, $F_{\nu_I-1}$, see the end of the
Sec.~\ref{sec:regI}. The field in this barrier corresponding to
$dF/dj\rightarrow\infty$  is larger than $F_{\rm th}$ for the following
reason. In this barrier $dj/dF=dj_{\rm cond}/dF+dj_{\rm diff}/dF$. Since the
field in  region I grows faster than in  region II, the concentration in the
well between these regions decreases. Therefore the diffusion current also
decreases which means that $dj_{\rm diff}/dF>0$ because the diffusion current
is directed against the conduction current. As a result $dj/dF=0$  not when
$dj_{\rm{cond}}/dF=0$ (i.e.  $F=F_{\rm{th}}$), but when
$dj_{\rm{cond}}/dF<0$, (i.e. $F>F_{\rm{th}}$).

The field in region II is about ${E_g}/el$ and the field in region I
is smaller or about $F_{\rm th}=\Gamma/el$. Usually $\Gamma$ is only a few
meV. So, in experiments with a big energy
separation\cite{kawamura1,kawamura2}, ${E_g}\approx 200$meV, the field
difference between regions I and II is also large and it causes a large
charge accumulation at the boundary between these regions. In these
experiments $eF_{\rm H}l\approx140$meV, $eF_{\rm th}l\approx12.5$meV, so
$F_{\rm th}/F_{\rm H}\approx0.089$ and Eq.~(\ref{eq:regI.12}) gives
$j^{\ast}\approx 0.30j_0$. That value is in good agreement with the ratio
of the maximum of the current in the oscillating region to the peak
value of the current at the instability threshold, see Fig.~2(b) in Ref.
\onlinecite{kawamura1}, reproduced in Fig.~\ref{fig:iv} of the present work.

The oscillations of the current in a superlattice with a high-field
domain caused by the expansion of this domain with an increase of the applied
bias have a period which can be associated with the potential drop per 
superlattice period in the high-field region $F_{\rm H}l$. However,
Eq.~(\ref{eq:regII.5}) shows that the potential drop per period in the
domain is less than the energy separation between levels ${E_g}$ by a
significant quantity, which can reach a few tens of meV.

Such a big difference between the period of the oscillations of the
current-voltage characteristic and the energy separation between the levels
was detected in the experiment of Kawamura et al.~\cite{kawamura2}, see
Fig.~3 in the cited work. One can see that
the difference between the inter-subband energy space and the electric
current oscillation period increases with $E_g$. This has a simple
physical meaning: in samples with larger $E_g$ the upper limit of the
current $j^\ast$ is lower and therefore the resonance in region II is
weaker.  Other examples of such a difference can be found in
Refs.~\onlinecite{choi1,choi3}, and \onlinecite{grahn-apr93}.

The main assumption which was made in our calculation concerns a small
deviation of the electron distribution function from the Fermi function with
effective temperature and chemical potential. Necessary conditions
justifying this assumption (see Sec.III) may not be satisfied in the
high-field domain.  However, the features of the current-voltage
characteristic discussed in the present work do not depend on the detailed shape
of the electron distribution.

In conclusion, we have studied the field distribution and $I$-$V$
characteristic of the superlattice in the voltage region when a high-field
domain exists. The accumulation of electrons at the domain boundary causes a
strong limitation of the current. Due to this limitation the minimal length
of the high-field domain can be not one but a few superlattice periods. The
current limitation also results in the reduction of the period of the current
oscillations compared to that corresponding to the energy separation between
the first and second levels in a well. These results are in good
agreement with available experimental data. The field distribution in the
region downstream of the high-field domain (if it exists) is nonuniform due
to electron relaxation from the second miniband to the first miniband.


\end{document}